# Aspects of the history, anatomy, taxonomy and palaeobiology of sauropod dinosaurs

**Michael P. Taylor**

This thesis is submitted in partial fulfilment of the requirements for the award of the degree of Doctor of Philosophy of the University of Portsmouth

Palaeobiology Research Group
School of Earth and Environmental Sciences
Burnaby Building, Burnaby Road
University of Portsmouth
Portsmouth, PO1 3QL
Hampshire
ENGLAND

9 February 2009

## Declaration

Whilst registered as a candidate for the above degree, I have not been registered for any other research award. The results and conclusions embodied in this thesis are the work of the named candidate and have not been submitted for any other academic award.

## Acknowledgements

First of all, from the bottom of my heart, I must thank my family for allowing me to spend so much time and money on what is ultimately something that I do only for fun. My wife Fiona has been unfailingly forbearing as I have repeatedly taken time away from her and our three sons, Daniel, Matthew and Jonno, to visit museums and attend conferences. Looking further back, I thank my mum, Marion, for encouraging me to be interested in everything around me, and my dad, Keith, for introducing me to the wonder of dinosaurs at an early age. I greatly regret that he is no longer with us to see the fruits of my labours; I know he would have been proud.

I am indebted to my supervisor Dave Martill, not just for his advice and support during the last four years, but more important still for taking a punt on me in the first place. He had the insight and instinct to take me on back when I had nothing to recommend me as a Ph.D student – no publications, no training, no qualifications beyond a fifteen-year-old degree in an irrelevant subject that I'd forgotten 90% of. I will always be grateful to him for giving me this opportunity, and I hope that my work in this dissertation and elsewhere has repaid his faith in me.

That I was able to approach Dave with even the little knowledge I did have is largely due to the education-by-stealth that I received from members of the Internet's Dinosaur Mailing List (http://dinosaurmailinglist.org) – most notably Tom Holtz, who has a rare gift for explaining complex concepts simply and briefly. I thank Mickey Rowe and Mary Kirkaldy for running the list so efficiently and providing a forum that has been so valuable to me.

A palaeontologist is nothing without specimens, and I gratefully thank all those who have allowed me access to sauropod material, including but not limited to Sandra Chapman and Lorna Steel (Natural History Museum, London, UK), David Unwin, Wolf-Dieter Heinrich and Daniela Schwarz-Wings (Humboldt Museum für Naturkunde, Berlin, Germany), Nick Czaplewski and Jeff Person (Oklahoma Museum of Natural History, Norman, Oklahoma, USA) and Bill Simpson (Field Museum of Natural History, Chicago, Illinois, USA).

I also thank the various journal editors who have handled my papers, and the many reviewers whose suggestions have so improved my work. Among reviewers, Jerry Harris deserves special mention: his unique brand of super-detailed yet always good-natured pedantry ensures that every manuscript that passes through his hands is much the better for it, and it has been my good fortune to spend many a long evening working through the literally hundreds of tiny corrections and comments that he includes in his reviews.

My experiences of working in palaeontology have been overwhelmingly positive, with nearly everyone keen to encourage newcomers, share information, allow access to specimens, give credit where it is due and generally avoid treading on others' toes. Sadly this is not true of absolutely everyone in our field, and in the rare cases when ethics are violated, too many palaeontologists maintain a code of silence. During a recent such episode, Kevin Padian did not hesitate to speak up clearly and repeatedly – in contrast to too many other senior vertebrate palaeontologists. I thank him for his courage and clarity: it is people like him that keep this field so good to work in.

Finally, I come to my colleagues. More people than I can list here have discussed interesting problems with me, supplied me with photocopies and PDFs of important papers, and generally provided an exhilarating sense of being part of a wider community. I will mention only two by name: my co-authors on the Sauropod Vertebra Picture of the Week blog (http://svpow.wordpress.com/), Darren Naish and Matt Wedel.

I first encountered Darren on the Dinosaur Mailing List, where from his posts he seemed terrifyingly well-informed and ferociously opinionated. When I subsequently met him in person, I found that my initial impressions were spot on – he was and is all that and more. At the same time, he is excellent company (at least, to anyone interested in tetrapods), and it was on his recommendation that Dave Martill first considered accepting me as a student. Over the last few years, I have worked with Darren on several projects, and he is a truly outstanding collaborator – quick, conscientious, and with an appallingly comprehensive mastery of the literature, by which I mean basically everything published on any tetrapod since 1700. Only Darren's total inability to handle numbers keeps him down here with us mortals; were he able to perform simple addition, he would no doubt be a world-dominating evil genius by now.

Last and absolutely not least, it my joy to thank Matt Wedel for his enormous contribution to getting me where I am today. Since that fateful day (9th January 2001) when I emailed him to ask for a copy of the *Sauroposeidon* paper, he has not only become one of my very best friends (second only to Fiona), but also an unofficial (and unpaid) additional supervisor, inducting me into the ways of sauropod science through a stream of informative and amusing emails while also maintaining a steady barrage on the subjects of Star Wars, ducks, Catholicism, biodiversity, the publishing business, killer rabbits, and why Sauron was the innocent victim of unprovoked Gondorian aggression. Matt's nerdy exterior conceals a mind that is not only relentlessly curious on the subject of palaeontology but also erudite, thoughtful and well informed on a startling range of other topics. My own enthusiasm for sauropod vertebrae is matched only by his, which makes him absolutely the best person to visit a collection with. He and I can happily sit for hours discussing the laminae on the single vertebra. He also, unlike Darren, appreciates the wonder of sushi.

The bottom line is that without Matt's provocation, encouragement and occasional abuse, I would never have become any kind of scientist; and it is to him that I gratefully dedicate this dissertation.

# Contents

## Assistance received

The introduction, chapters 1, 2, 4 and the appendices of this dissertation are entirely my own work, except that Mathew J. Wedel of Western University of Health Sciences contributed part B of figure 4 of chapter 4. Chapters 3 and 5 were co-authored, with myself as lead author in both cases. The work for these chapters breaks down as follows.

Chapter 3 (description of *Xenoposeidon*): I was responsible for the anatomical part of the introduction, the systematic palaeontology section, description, comparisons and interpretation, phylogenetic analysis, length and mass calculations, diversity discussion, references, figures with their captions except figure 2, and both tables. Darren Naish of the University of Portsmouth was responsible for the geological and historical part of the introduction, the historical taxonomy section, and figure 2.

Chapter 5 (evolution of long necks): this chapter was written by me as a consequence of a series of discussions with Mathew J. Wedel. Dr. Wedel also contributed figure 5.

## Abstract

Although the sauropod dinosaurs have been recognised for more than a hundred and sixty years, much remains to be discovered and understood about their functional anatomy and palaeobiology. The taxonomy of older genera requires revision, and new taxa await description. The characteristic long necks of sauropods are mechanically perplexing and their evolution is obscure. All these issues are addressed herein.

The well-known genus *Brachiosaurus* is represented by the American type species *B. altithorax* and the better known African species *B. brancai*. However, the referral of the latter to the genus *Brachiosaurus* was based mostly on symplesiomorphies, and an element-by-element examination of the overlapping material shows 26 differences between the species. *B. brancai* must be removed from *Brachiosaurus* and referred to the genus *Giraffatitan*, which has been previously proposed for it.

*Xenoposeidon proneneukos* is a new neosauropod from the Lower Cretaceous Hastings Bed Group of the Wealden Supergroup, known from a single partial dorsal vertebra. Although such scant remains would usually be non-diagnostic, the excellent preservation of the *Xenoposeidon* holotype reveals six unique characters. The distinctive morphology suggests that *Xenoposeidon* may represent a new sauropod family, extending sauropod disparity as well as bringing to four the number of sauropod families known from the Wealden.

A second new Early Cretaceous genus is also described, a titanosauriform from the Ruby Ranch Member of the Cedar Mountain Formation, in Utah. This taxon is known from at least two individuals, and adult and a juvenile, which together provide vertebrae, ribs, a scapula, sternal plates and an ilium. The ilium is particularly unusual, exhibiting five unique features.

The longest sauropod necks were five times as long as those of the next longest terrestrial animals, and four separate lineages with very different cervical vertebrae evolved necks exceeding 10 m. Elongation was enabled by sheer size, vertebral pneumaticity and the relative smallness of sauropod heads, despite aspects of cervical osteology that are difficult to understand on mechanical first principles.

## Introduction

Sauropods were a taxonomically diverse and ecologically important group of saurischian dinosaurs. They were very large quadrupedal herbivores characterised by small heads carried on extremely long necks, relatively compact torsos and long tails, although the relative proportions of the body parts varied greatly between taxa – morphological disparity has traditionally been underestimated. The clade Sauropoda, together with a paraphyletic series of more basal relatives known as "prosauropods", formed the clade Sauropodomorpha, which is the sister group of Theropoda, the clade of bipedal carnivorous dinosaurs, including birds. Sauropods first evolved in the late Triassic (Norian or perhaps Carnian), and persisted until the extinction of all non-avian dinosaurs at the end of the Cretaceous – a range of 155 million years. A good general overview of sauropod anatomy and systematics is given by Upchurch et al. (2004).

The first sauropod to be named, *Cetiosaurus*, was described by Richard Owen in 1841, on the basis of isolated and eroded vertebrae, and was initially interpreted by him as a marine reptile. The distinctive sauropod body plan did not begin to be understood until the description of *Cetiosaurus oxoniensis* Phillips, 1871 thirty years later. This species was named on the basis of material from three individuals in a single quarry, and included dorsal vertebrae and many appendicular elements. Phillips was able to recognise his species as a dinosaur, and tentatively interpreted it as terrestrial. The understanding of sauropods quickly improved with the discovery of the dinosaur-bearing Morrison Formation of the western U.S.A in the 1870s, and the consequent description of genera such as *Camarasaurus* Cope, 1877 and *Apatosaurus* Marsh, 1877. In particular, the characteristic long neck of sauropods was recognised for the first time, as the various *Cetiosaurus* specimens did not include any useful cervical material. A more detailed survey of the history of sauropod palaeontology is given in chapter 1.

Despite more than 160 years of study of sauropods, much remains to be discovered, and much of what we think we know stands in need of revision, clarification or verification.

## Background to this dissertation

My interest in sauropod dinosaurs has focussed on four broad areas: their phylogeny, including its application in phylogenetic nomenclature; diversity, i.e. the number of valid genera in various sauropod clades; morphological disparity; and palaeobiology. I am particularly interested in these areas as they apply to brachiosaurids and to the sauropods of the Wealden Supergroup of England. My earlier published papers, and in-press and in-review manuscripts, reflect these interests:

- Taylor and Naish (2005) reviewed the phylogenetic nomenclature of the large and important sauropod clade Diplodocoidea, and recommended a coherent, stable set of names and definitions for twelve related clades.
- Taylor (2006) analysed the diversity of dinosaurs, both in terms of the breakdown by clade and the changing level of diversity throughout the Mesozoic, using hand-built analysis programs.
- Taylor (2007) discussed how best to apply the tenets of phylogenetic nomenclature in the absence of a formal code governing its use, in the interrim period before the Phylocode (Cantino and de Queiroz, 2006) is implemented.
- Taylor and Naish (2007) described and named the new sauropod *Xenoposeidon* from the Wealden, based on a single highly diagnostic dorsal vertebra. This paper is included as chapter 3 of this thesis.

Work in press includes:

- Taylor (in press a), a history of sauropod studies, was submitted as a chapter in a forthcoming Geological Society book and is included as chapter 1 of this thesis.
- Taylor (in press b), a taxonomic review of the genus *Brachiosaurus*, for the *Journal of Vertebrate Paleontology*, is included as chapter 2 of this thesis.
- Taylor et al. (in press a, b) are the chapters on the clades Sauropoda and Sauropodomorpha for the forthcoming Phylocode companion volume (as lead author on a team including Paul Upchurch, Adam Yates, Mathew J. Wedel and Darren Naish).
- Upchurch et al. (in press) is a formal petition to the ICZN to establish

*Cetiosaurus oxoniensis* as the type species of the sauropod genera *Cetiosaurus* (as third author with Paul Upchurch and John Martin).

- Taylor (in press c) will be *Encyclopaedia Britannica*'s article on the sauropod *Europasaurus*.

My in-review work includes two palaeobiological studies of sauropod necks (one co-authored with Mathew J. Wedel, submitted to *Paleobiology* and included as chapter 5 of this thesis; the other co-authored with Mathew J. Wedel and Darren Naish, and submitted to *Proceedings of the Royal Society B*).

In addition to published, in-press and in-review papers, I have given talks on sauropod-related topics to several conferences and symposia, and an invited talk:

- As SVPCA 2004, I presented an early version of my dinosaur diversity work.
- At Progressive Palaeontology 2005, I presented a study on upper limits on the mass of land animals estimated through the articular area of limb-bone cartilage.
- At SVPCA 2005, I presented preliminary results of my work on the Natural History Museum's Tendaguru brachiosaurid specimen, which I do not believe belongs to either *Brachiosaurus altithorax* or "*Brachiosaurus*" *brancai*. This work remains ongoing: no manuscript has yet been submitted, as work was delayed during the re-evaluation of *Brachiosaurus* mentioned above.
- At Progressive Palaeontology 2006, I presented the Wealden specimen that would subsequently be described as *Xenoposeidon*, demonstrating its uniqueness and showing that its position within Neosauropoda cannot be resolved.
- At the Ninth International Symposium on Mesozoic Terrestrial Ecosystems and Biota, I presented a graphical analysis of diversity patterns across clades and through time. (An earlier version of this work was published in the conference proceedings as Taylor (2006)).
- At Progressive Palaeontology 2007, I presented in-progress work on resolving the affinities of fragmentary sauropod specimens from the Wealden. This project, too, continues but is not yet ready for publication.
- At the symposium *Dinosaurs: a historical perspective* in 2008, I presented a

summary of the history of dinosaur research. This work was condensed from the account which became Taylor (in press a).

- At the symposium *Biology of the sauropod dinosaurs: the evolution of gigantism* in 2008, I gave two presentations: one discussed the likely habitual posture of sauropod necks in light of the behaviour of extant animals (based on an in-review manuscript); the other was a repeat of the 2005 talk on articular cartilage in sauropod limbs.
- In 2008 I gave an invited talk at the Humboldt Museum für Naturkunde, Berlin, on the generic separation of the two "*Brachiosaurus*" species and the need to rename their central exhibit *Giraffatitan*. This was condensed from Taylor (in press b).

## Contents of this dissertation

The five chapters herein address several of the issues highlighted above, as follows.

Chapter 1 provides a historical context for the novel work by reviewing the progress of sauropod palaeontology from 1841 to 2008.

The foundation of all palaeontology is careful description: material that is not properly understood cannot be the basis of correct palaeobiological inferences. Many older sauropod genera need revision in light of recent discoveries, and chapter 2 provides such a revision for the important genus *Brachiosaurus*. This genus was initially described and subsequently monographed by Riggs (1903, 1904) on the basis of the type species *B. altithorax*. Janensch (1914) named a second species, *B. brancai*, and subsequently described numerous specimens in great detail (Janensch, 1922, 1929, 1935-1936, 1947, 1950, 1961). However, his referral of the new species to *Brachiosaurus* was based mostly on symplesiomorphies, and an element-by-element comparison of the two species shows that they must be considered generically separate. This has important implications for sauropod palaeobiogeography (the Morrison and Tendaguru formations in which the two animals are found had been considered faunally similar) and, indirectly, for sauropod biomechanics, a field in which "*Brachiosaurus*" is widely used as a test-case taxon. This paper is also an important foundation for my continuing work on the Natural History Museum's Tendaguru brachiosaur (Taylor,

2005).

An equally important application of description is in the recognition of new taxa, which in turn affect our understanding of diversity and disparity. Accordingly, chapters 3 and 4 describe two new sauropods: *Xenoposeidon*, from the Hastings Beds Group of the Wealden Supergroup of England; and the Hotel Mesa sauropod from the Cedar Mountain Formation of Colorado, USA, whose name is not included in this dissertation and awaits formal publication. The informal designation "Hotel Mesa sauropod" is based on the name of the type locality. Although both are known from incomplete remains, each is characterised by a suite of distinct and diagnosable novelties which not only justify the new taxa but also extend our understanding of the morphological range of sauropods. Further, *Xenoposeidon* increases the diversity of the Hastings Beds Group of Wealden strata, from which four distinct sauropod families are now known. The Hotel Mesa sauropod contributes to our understanding of Early Cretaceous North American sauropod diversity, which until recently had been thought greatly reduced since the Late Jurassic, but is now recognised to have equalled or surpassed that of the Late Jurassic Morrison Formation.

Finally, chapter 5 addresses some palaeobiological, biomechanical and evolutionary issues related to the long necks of sauropods. The morphology of sauropod cervical vertebrae is reviewed and compared with that of birds and crocodilians, the ventral bracing hypothesis of Martin et al. (1998) is refuted, some perplexing aspects of vertebral morphology are discussed, and the evolution of long necks in sauropods and other terrestrial animals is discussed.

Chapters 2, 3 and 4 each contain a phylogenetic analysis based on that of Harris (2006a) – that of chapter 2 breaking the compound "*Brachiosaurus*" OTU of Harris into separate *Brachiosaurus altithorax* and *Giraffatitan brancai* OTUs, and those of chapters 3 and 4 adding the new taxa individually to Harris's matrix. These three analyses are kept separate rather than integrated into a single analysis because the sparse material of both new taxa causes their trees to lose resolution, so that the result of an integrated analysis is an uninformative neosauropod polytomy.

Each chapter is formatted according to the guidelines of the book or journal where it was submitted.

**Bibliography**

Blows, W. T. 1995. The Early Cretaceous brachiosaurid dinosaurs *Ornithopsis* and *Eucamerotus* from the Isle of Wight, England. Palaeontology 38:187–197.

Cantino, P. D., and K. de Queiroz. 2006. PhyloCode: A Phylogenetic Code of Biological Nomenclature (Version 4b, September 12, 2007). http://www.ohiou.edu/phylocode/PhyloCode4b.pdf

Cope, E. D. 1877. On a gigantic saurian from the Dakota Epoch of Colorado. Paleontology Bulletin 25:5–10.

Harris, J. D. 2006a. The significance of *Suuwassea emiliae* (Dinosauria: Sauropoda) for flagellicaudatan intrarelationships and evolution. Journal of Systematic Palaeontology 4:185–198.

Janensch, W. 1914. Übersicht über der Wirbeltierfauna der Tendaguru-Schichten nebst einer kurzen Charakterisierung der neu aufgefuhrten Arten von Sauropoden. Archiv für Biontologie 3:81–110.

Janensch, W. 1922. Das Handskelett von *Gigantosaurus robustus* u. *Brachiosaurus Brancai* aus den Tendaguru-Schichten Deutsch-Ostafrikas. Centralblatt für Mineralogie, Geologie und Paläontologie 15:464–480.

Janensch, W. 1929. Material und Formengehalt der Sauropoden in der Ausbeute der Tendaguru-Expedition. Palaeontographica (Suppl. 7) 2:1–34.

Janensch, W. 1935-1936. Die Schadel der Sauropoden *Brachiosaurus*, *Barosaurus* und *Dicraeosaurus* aus den Tendaguru-Schichten Deutsch-Ostafrikas. Palaeontographica (Suppl. 7) 2:147–298.

Janensch, W. 1947. Pneumatizitat bei Wirbeln von Sauropoden und anderen Saurischien. Palaeontographica (Suppl. 7) 3:1–25.

Janensch, W. 1950. Die Wirbelsaule von *Brachiosaurus brancai*. Palaeontographica (Suppl. 7) 3:27–93.

Janensch, W. 1961. Die Gliedmaszen und Gliedmaszengürtel der Sauropoden der Tendaguru-Schichten. Palaeontographica (Suppl. 7) 3:177–235.

Marsh, O. C. 1877. Notice of new dinosaurian reptiles from the Jurassic Formation.

American Journal of Science, Series 3, 14:514–516.

Martin, J., V. Martin-Rolland, and E. Frey. 1998. Not cranes or masts, but beams: the biomechanics of sauropod necks. Oryctos 1:113–120.

Owen, R. 1841. A description of a portion of the skeleton of the *Cetiosaurus*, a gigantic extinct Saurian Reptile occurring in the Oolitic formations of different portions of England. Proceedings of the Geological Society of London 3:457–462.

Phillips, J. 1871. Geology of Oxford and the Valley of the Thames. Clarendon Press, Oxford, 529 pp.

Riggs, E. S. 1903. *Brachiosaurus altithorax*, the largest known dinosaur. American Journal of Science 15:299–306.

Riggs, E. S. 1904. Structure and relationships of opisthocoelian dinosaurs. Part II, the Brachiosauridae. Field Columbian Museum, Geological Series 2:229–247.

Taylor, M. P. 2005. Sweet seventy-five and never been kissed: the Natural History Museum's Tendaguru brachiosaur; pp. 25 in P. M. Barrett (ed.), Abstracts volume for 53rd Symposium of Vertebrae Palaeontology and Comparative Anatomy. The Natural History Museum, London.

Taylor, M. P. 2006. Dinosaur diversity analysed by clade, age, place and year of description; pp. 134–138 in P. M. Barrett (ed.), Ninth international symposium on Mesozoic terrestrial ecosystems and biota, Manchester, UK. Cambridge Publications, Cambridge, UK.

Taylor, M. P. 2007. Phylogenetic definitions in the pre-PhyloCode era; implications for naming clades under the PhyloCode. PaleoBios( 27(1):1–6.

Taylor, M. P. In press a. Sauropod dinosaur research: a historical review; in R. Moody, E. Buffetaut, D. M. Martill, and D. Naish (eds.), Dinosaurs (and other extinct saurians): a historical perspective. The Geological Society, London.

Taylor, M. P. In press b. A re-evaluation of *Brachiosaurus altithorax* Riggs 1903 (Dinosauria, Sauropoda) and its generic separation from *Giraffatitan* brancai (Janensch 1914). Journal of Vertebrate Paleontology 29.

Taylor, M. P. In press c. *Europasaurus*; in Encyclopaedia Britannica. Encyclopaedia

Britannica, Inc., Chicago, IL.

Taylor, M. P., and D. Naish. 2005. The phylogenetic taxonomy of Diplodocoidea (Dinosauria: Sauropoda). PaleoBios 25(3):1–7.

Taylor, M. P., and D. Naish. 2007. An unusual new neosauropod dinosaur from the Lower Cretaceous Hastings Beds Group of East Sussex, England. Palaeontology 50:1547–1564.

Taylor, M. P., P. Upchurch, A. M. Yates, M. J. Wedel, and D. Naish. In press a. Sauropoda; in K. de Queiroz, P. D. Cantino, and J. A. Gauthier (eds.), Phylonyms: a Companion to the PhyloCode. University of California Press, Berkeley, CA.

Taylor, M. P., P. Upchurch, A. M. Yates, M. J. Wedel, and D. Naish. In press b. Sauropodomorpha; in K. de Queiroz, P. D. Cantino, and J. A. Gauthier (eds.), Phylonyms: a Companion to the PhyloCode. University of California Press, Berkeley, CA.

Upchurch, P., and J. Martin. 2003. The anatomy and taxonomy of *Cetiosaurus* (Saurischia, Sauropoda) from the Middle Jurassic of England. Journal of Vertebrate Paleontology 23:208–231.

Upchurch, P., P. M. Barrett, and P. Dodson. 2004. Sauropoda; pp. 259–322 in D. B. Weishampel, P. Dodson, and H. Osmólska (eds.), The Dinosauria, 2nd edition. University of California Press, Berkeley and Los Angeles.

Upchurch, P., J. Martin, and M. P. Taylor. In press. Case 3472: *Cetiosaurus* Owen, 1841 (Dinosauria, Sauropoda): proposed conservation of usage by designation of *Cetiosaurus oxoniensis* Phillips, 1871 as the type species. Bulletin of Zoological Nomenclature 66.

Chapter 1 follows. This paper has been accepted and is in press as a chapter in the forthcoming book, *Dinosaurs (and other extinct saurians): a historical perspective*, published by The Geological Society.

# Sauropod dinosaur research: a historical review

MICHAEL P. TAYLOR

*Palaeobiology Research Group, School of Earth and Environmental Sciences, University of Portsmouth, Burnaby Road, Portsmouth PO1 3QL, UK (e-mail: dino@miketaylor.org.uk)*

**Abstract:** In the 167 years since Owen named a tooth as *Cardiodon*, the study of sauropod dinosaurs has gone through several distinct periods. In the early years, a sequence of descriptions of isolated skeletal elements gave rise to a gradually emerging understanding of the animals that would later be known as sauropods. The second phase began in 1871 with Phillips's description of *Cetiosaurus oxoniensis*, the first reasonably complete sauropod, and continued with the Marsh-Cope Bone Wars and the description of the nearly-complete sauropods *Camarasaurus* and "*Brontosaurus*" (= *Apatosaurus*). As these and other genera became better known, a third phase began, exploring not just the remains but the lives of these giants, with arguments about posture and habitat to the fore, and with the public becoming increasingly aware of sauropods due to skeletal mounts. A "dark age" followed during and after the Second World War, with sauropods considered uninteresting evolutionary dead ends and largely ignored. This was brought to an end by the "dinosaur renaissance" that began in the late 60s, since when work has recommenced with new vigour, and the public has been introduced to a more vigorous and terrestrial image of sauropods through film and television. Both diversity and disparity of sauropods continue to increase through new descriptive work, and the group is now seen as more fascinating and worthy of study than ever before.

Sauropod dinosaurs are the terrestrial superlative: they were not just the largest animals ever to have walked on land, but an order of magnitude heavier than their nearest rivals, the hadrosaurid dinosaurs and proboscidean and indricotherian mammals. Although the first genera now recognised as sauropods were named in 1841, the nature of the animals was not understood for some time, and many aspects of their palaeobiology remained controversial for considerably longer – some, including habitual neck posture, remain unresolved to this day. Throughout the 167 years of research into sauropods, an increasingly clear picture has gradually emerged. This paper traces the process of discovery through five distinct eras: an initial period of studies restricted to isolated elements, the period in which near-complete specimens first became available, the age of interpretation and controversy, the "dark ages" and the modern renaissance.

Institutional abbreviations: **AMNH**, American Museum of Natural History, New York, New York; **BMNH**, Natural History Museum, London, UK; **CM**, Carnegie Museum of Natural History, Pittsburgh, Pennsylvania; **HMN**, Humboldt Museum für Naturkunde, Berlin, Germany; **OUMNH**, Oxford University Museum of Natural History, Oxford, UK; **USNM**, National Museum of Natural History, Washington, DC; **YPM**, Yale Peabody Museum, New Haven, Connecticut.

## Stage 1: early studies, isolated elements (1841–1870)

It was only seventeen years after the naming of the first dinosaur recognised by science, *Megalosaurus* Buckland 1824, and a year before the coinage of the name Dinosauria Owen 1842, that the first sauropods were named: *Cardiodon* Owen 1841a and *Cetiosaurus* Owen 1841b. The former was named on the basis of a single tooth crown from the Middle Jurassic Forest Marble Formation of Bradford-on-Avon, Wiltshire. It was later figured by Owen (1875a, plate IX, figs. 2-5), and has since been lost (Fig. 1a). A second tooth crown, BMNH R1527, was referred to this genus by Lydekker (1890, p. 236), and was later figured by Barrett (2006, fig. 2a-b). These two teeth are the only elements to have been assigned to *Cardiodon*, and this genus – the first sauropod – is now all but forgotten. Various workers have suggested that *Cardiodon* might be a senior synonym of *Cetiosaurus*, but this putative synonymy was refuted by Upchurch & Martin (2003, p. 214-215).

It is with the genus *Cetiosaurus*, named later that same year, that the story of sauropods really begins. Owen (1841b) used a wide variety of specimens from six different localities as the basis for the new genus *Cetiosaurus*, for which no specific name was initially given. Despite the large amount of material, most of it was rather poor, consisting largely of partial caudal vertebrae and appendicular fragments. Owen noted that in their size, and in the size and proportions of their neural spines and chevron articulations, the vertebrae resembled those of whales; but that the concavity of their articular surfaces and high position of the transverse processes suggested a reptilian affinity. Accordingly he named the new genus *Cetiosaurus*, or “whale lizard” (Fig. 1b).

It is often said that Owen (1841b) described *Cetiosaurus* as a gigantic crocodilian, but in fact this assignment came later. In his initial description, Owen (1841b, p. 462) explicitly separated his new animal from crocodiles, concluding that “the surpassing bulk and strength of the *Cetiosaurus* were probably assigned to it with carnivorous habits, that it might keep in check the Crocodilians and Plesiosauri”. What is certain is that when, a year later, Owen (1842, p. 103) created the name Dinosauria, he omitted *Cetiosaurus* from it, limiting its initial content to “the gigantic Crocodile-lizards of the dry land”, *Megalosaurus*, *Iguanodon* Mantell 1825 and *Hylaeosaurus* Mantell 1833. *Cetiosaurus*, then thought aquatic, was explicitly excluded.

In subsequent years, a total of thirteen species of *Cetiosaurus* were named by Owen and others on the basis of British material, although nearly all of these are now considered nomina nuda or nomina dubia (Upchurch & Martin 2003, p. 209-215). It was not until 1871 that truly informative *Cetiosaurus* remains would be described. Before this, though, several more historically important sauropods would be named on the basis of isolated elements.

The first of these, and the first sauropod to be named on the basis of appendicular material, was *Pelorosaurus* Mantell 1850 (Fig. 1c), based on a humerus from the Early Cretaceous Wealden Supergroup that at the time seemed “stupendous” (p. 379) at a length of four and a half feet – although this is little more than 60% the length of the humeri of the subsequently described brachiosaurids *Brachiosaurus altithorax* Riggs 1903a and *Brachiosaurus brancai* Janensch 1914, animals which if they were isometrically similar to *Pelorosaurus* would have weighed four times as much as it did. The significance of *Pelorosaurus* is that it was the first-named sauropod that was recognised by its describer as being terrestrial – ironically, due to the possession of a medullary cavity, a feature that seems to be unique among sauropods. Although Owen (1859a, p. 40) tried to portray Mantell as having mistaken the “anterior for the posterior of the bone”, it is clear from Mantell's description, and particularly his correct identification of the deltoid process (deltopectoral crest), that he oriented the humerus correctly and that the error was only in the caption of Mantell's plate XXI. Mantell subsequently described a second species, *Pelorosaurus becklesii* Mantell 1852, which in fact is not closely related to the type species (Upchurch 1995, p. 380). The type specimen of “*Pelorosaurus*” *becklesii*, BMNH R1868, is important because as well as a humerus, radius and ulna, it includes a skin impression – the first known from any sauropod, and still one of only very few sauropod skin impressions. Because Mantell referred to *Pelorosaurus* the same caudal vertebrae that Owen (1842) used as the type specimen for *Cetiosaurus brevis* Owen 1842, the taxonomy of *Cetiosaurus* and *Pelorosaurus* is complex and intertwined. This situation is being addressed by a petition to the ICZN (Upchurch et al. 2009). *Pelorosaurus*, including the misassigned species “*Pelorosaurus*” *becklesii*, is being re-studied to better determine its affinities (Taylor and Upchurch, in prep.), but the type material appears to represent a basal titanosauriform, possibly a brachiosaurid (Upchurch & Martin 2003, p. 210).

As with dinosaurs in general, England was very much the home of sauropods during the early days of their study. The first sauropod named from outside England was *Aepisaurus* Gervais 1852, based on a subsequently lost humerus of which the proximal part has since been found; it is now considered a nomen dubium. The first sauropod from outside Europe was *Astrodon* Johnston 1859, which, like *Cardiodon*, was named on the basis of a single tooth crown and not initially given a specific name. Six years later, the tooth was referred to the new species *Astrodon johnstoni* Leidy 1865, although this is often misspelled as A. *johnsoni* (e.g. Carpenter & Tidwell 2005). (*Pleurocoelus* Marsh 1888, based on mostly juvenile vertebral centra, has sometimes been considered separate from *Astrodon*, but is now generally considered a junior synonym of that genus despite the inadequate *Astrodon* type material – see overview in Carpenter & Tidwell 2005.)

Another significant find was *Ornithopsis* Seeley 1870, named on the basis of two partial presacral vertebrae from different localities which are now known to belong to sauropods (probably two different sauropod taxa) but thought by Seeley to be “of the Pterodactyle kind” (Fig. 1d). Seeley's mistake was based on his recognition of pneumatic features in the bones – internal air-spaces giving rise to a honeycombed internal structure, and lateral foramina through which air entered these spaces from the sides of the bones. At the time of Seeley's writing, almost all animals known to have pneumatised bones in their postcranial skeletons were birds and pterosaurs, the only exception being the theropod *Becklespinax altispinax* Paul 1988b, then thought to belong to *Megalosaurus* (Naish, this volume). Since both birds and pterosaurs are flying vertebrates, Seeley's assumption that an animal with postcranial skeletal pneumaticity (PSP) was closely related to, or even intermediate between, the flying vertebrate groups was perfectly sensible. We now know that PSP also occurs in sauropods, non-avian theropods and in some basal sauropodomorphs (Wedel 2006), and possibly in some crocodile-line archosaurs (Gower 2001; Nesbitt & Norell 2006, p. 3). Sauropod pneumaticity has been subsequently studied by Longman (1933) and Janensch (1947), but thereafter remained largely overlooked until the more recent work of Britt (1993) and Wedel (2003a, b, 2005). A picture has now emerged of a complex range of pneumatic features, encompassing everything from gentle lateral depressions in basal sauropods such as *Barapasaurus* Jain *et al.* 1975, via large internal spaces in basal neosauropods such as *Camarasaurus* Cope 1877a, to the dense, irregularly honeycombed internal structure of derived titanosaurs such as *Saltasaurus* Bonaparte & Powell 1980.

**Stage 2: the emerging picture (1871–1896)**

Understanding of sauropods took a giant leap forward with the description of *Cetiosaurus oxoniensis* Phillips 1871 (Fig. 2), a Middle Jurassic sauropod from England, described and illustrated in detail by Phillips in fifty pages of his book on the geology of Oxford and the Thames Valley. Phillips described remains from several localities, all near Oxford, and there is no compelling reason not to accept his assessment that they all belong to the same species. Most important are the associated remains of several individuals from Kirtlington Station, north of Oxford, of which the largest is also the best represented and was accordingly nominated by Upchurch & Martin (2003, p. 216) as the lectotype. Material described and figured by Phillips included a tooth; dorsal, sacral and caudal vertebrae; dorsal ribs; sternal plate, coracoids and scapulae; humeri and ulnae; ilium, pubis and ischium; femora, tibiae and fibula; metatarsals and pedal phalanges. The only parts of the skeleton not represented were the skull, cervical vertebrae, radius and manus – although recent work by Galton & Knoll (2006) has tentatively agreed with Woodward's (1910) and Huene's (1926) assignment of the isolated saurischian braincase OUMNH J13596 to *Cetiosaurus oxoniensis*. Given the lack of prior information about sauropods, Phillips's identification of the various bones was impressively accurate. He made only two errors: he interpreted the sole recovered sternal plate as a median element rather then as one of a pair, and he interpreted the ischiadic and pubic articular surfaces of the pubis and ischium respectively as articulating with the ilium. Phillips did not attempt a skeletal reconstruction – unfortunately, as it would have been of great historical importance.

Armed with all this material, Phillips was able to envisage the sauropod body-plan for the first time (although he could not have known about the long neck and small head), recognising it as capable of terrestrial locomotion and possessing erect posture: "all the articulations [of the limb bones] are definite, and made so as to correspond to determinate movements in particular directions, and these are such as to be suited for walking. In particular, the femur, by its head projecting freely from the acetabulum, seems to claim a movement of free stepping more parallel to the line of the body, and more approaching to the vertical than the sprawling gait of the crocodile." (pp. 293-294). However, Phillips hedged his bets with regard to lifestyle, concluding that "we have, therefore, a marsh-loving or river-side animal" (p. 294). Phillips was also first to suggest the dinosaurian affinities of *Cetiosaurus*, albeit tentatively: "The [femur] is nearly straight, in this respect differing much from the crocodilian, and approaching towards the deinosaurian type" (p. 280); "a lizard of such vast proportions would seem to claim easy admission to the deinosaurians, and to take its place naturally with megalosaurus or iguanodon ... but its fore-limbs are more crocodilian, its pelvic girdle more lacertilian, while its vertebral system is of a peculiar type".

Phillips's work on *Cetiosaurus* marked a significant step forward, giving the first meaningful window on the morphology and ecology of a sauropod dinosaur. However, his work was to be largely superseded just six years later, by a sequence of important announcements in 1877: the first recognised Gondwanan sauropod, *Titanosaurus* Lydekker 1877; the onset of the Bone Wars, with the descriptions of the sauropods *Camarasaurus*, *Apatosaurus* Marsh 1877b, *Atlantosaurus* Marsh 1877b, *Amphicoelias* Cope 1877b and *Dystrophaeus* Cope 1877c; and the first skeletal reconstruction of a sauropod.

*Titanosaurus* was named by Lydekker (1877) on the basis of a partial femur and two incomplete caudal vertebrae, and was diagnosed by only a single character – procoelous caudal vertebrae (i.e. having centra that are concave anteriorly and pronouncedly convex posteriorly). Although the original *Titanosaurus* material was from India, similar procoelous caudal vertebrae from other countries were subsequently referred to the genus, eventually resulting in a total of fourteen species! It has since been shown by Wilson & Upchurch (2003, p. 152) that the type species of *Titanosaurus*, *T. indicus* Lydekker 1877 is invalid as it can no longer be diagnosed: the single diagnostic character identified by Lydekker, procoelous caudal vertebrae, is now recognised as synapomorphic of the much larger clade Titanosauria, which at the last count encompasses more than fifty valid genera. Lydekker's initial naming of *Titanosaurus* on the basis of this morphology remains historically significant, however, as not only the first recognition of the important group now known as Titanosauria but also as the first sauropod recognised from the Gondwanan supercontinent (Table 1).

The year 1877 also marked the beginning of the Bone Wars – a period of intense, aggressive competition between Othniel Charles Marsh and his great rival Edward Drinker Cope to find and name dinosaurs from the newly discovered Morrison Formation of the western United States (Colbert 1997). Besides such well-known non-sauropod dinosaurs as *Allosaurus* Marsh 1877b and *Stegosaurus* Marsh 1877c, this year saw the establishment of two classic sauropods in *Apatosaurus* and *Camarasaurus*, as well as the less well known sauropod genera, *Amphicoelias*, *Atlantosaurus* (probably synonymous with *Apatosaurus ajax* Marsh 1877b, Berman & McIntosh 1978, p. 11) and *Dystrophaeus* (probably a nomen dubium). Unfortunately, in their haste to beat each other to press, both Marsh and Cope published rushed and inadequate descriptions, often without illustrations, most of which would not be considered taxonomically valid if published today. Synonymies also abounded: for example, Marsh's genus

*Atlantosaurus* was first published under the name *Titanosaurus montanus* Marsh 1877a, until Marsh became aware of Lydekker's slightly earlier use of this generic name, and so renamed it *Atlantosaurus*; and this is now thought to be probably synonymous with *Apatosaurus*, as is the slightly later *Brontosaurus* Marsh 1879. While the Marsh-Cope rivalry undoubtedly benefited palaeontology by catalysing work that would not otherwise have been done so quickly, the net results of this race were negative, yielding a set of specimens with very poor locality documentation, and a trail of shoddy scientific work that had to be re-done subsequently (Barbour 1890): so while, for example, Marsh is credited with the names *Apatosaurus* and *Brontosaurus*, most of his publications on these animals are now of purely historical interest, while the subsequent monographs on this genus by Riggs (1903b) and Gilmore (1936) are still widely used.

The year after the initial Morrison “Dinosaur Rush”, *Camarasaurus* became the first sauropod to be adequately figured (Cope 1878), but prior to this it had already been made the subject of the first attempt to reconstruct the skeleton of a sauropod: that of Dr. John Ryder, executed in 1877 under the direction of Cope (Fig. 3a). Astonishingly, the reconstruction was life-sized, “over fifty feet in length” (Osborn & Mook 1921, p. 252), and was based on material from several individuals. Although it was exhibited at a meeting of the American Philosophical Society on December 21, 1877, and subsequently exhibited at the AMNH, it was not published until 37 years later (Mook 1914), and is now best known from the excellent reproduction in the monograph of Osborn & Mook (1921, plate LXXXII). In the light of subsequent work, Ryder's reconstruction can be seen to be replete with mistakes: the head is a complete fiction, the neck is too short, the vertebrae in the region of the pectoral girdle are coalesced like the sacrum, there are far too many dorsal vertebrae, the tail is clearly modelled on those of aquatic animals, being dorsoventrally tall for much of its length but not in the proximal region, and the manus does not at all resemble the correct arrangement in sauropods, with the distinctive vertical arcade of near-parallel metacarpals. Nevertheless, Ryder's work remains admirable in some respects: the animal depicted is immediately recognisable as a sauropod, having the distinctive long neck and erect posture, and the dorsal vertebrae are recognisable as those of *Camarasaurus*.

It was not until a year after Ryder's reconstruction that the group Sauropoda got its name – at the fourth attempt. Owen (1859b, p. 164-165) had previously proposed the name Opisthocoelia for the group consisting of *Cetiosaurus* and *Streptospondylus* Meyer 1832, and as the first suprageneric taxon containing a genus now recognised as a sauropod, this name has some claim to priority. A second candidate name for this group, Ceteosauria [sic], was raised by Seeley (1874, p. 690) in a paper describing the partial dorsal neural arch of a stegosaur which he misinterpreted as part of the braincase of a sauropod, but this name has been mostly overlooked. Marsh (1877b, p. 514) ignored both of these prior names and instead referred his genera *Atlantosaurus* and *Apatosaurus* to the new family Atlantosauridae, diagnosed by pneumatic vertebra and the absence of the third trochanter on the femur. Finally, the very next year, Marsh (1878b, p. 412) subsumed this family within yet another new taxon, Sauropoda: “A well marked group of gigantic Dinosaurs ... has been characterized by the writer as a distinct family, Atlantosauridae, but they differ so widely from typical Dinosauria, that they belong rather in a suborder, which may be called Sauropoda, from the general character of the feet.” The name is a strange one, as the feet of sauropods do not resemble those of lizards, but it was quickly adopted. Marsh's diagnosis consisted of ten characters, and while most of these are now known to be plesiomorphies characterising a larger clade, two or three remain diagnostic. Marsh's name did not immediately win unanimous acceptance: Osborn (1898, p. 227) used the name Cetiosauria, listing twelve included genera that encompass diplodocoids, camarasaurs and titanosaurs; Riggs (1903b, p. 166-169) discussed the names Opisthocoelia, Cetiosauria and Sauropoda in detail, concluding that “the three terms are essentially co-ordinate and co-extensive. 'Opisthocoelia' has priority, and is entitled to preference”; and Matthew (1915) also preferred the name Opisthocoelia. However, Hatcher (1903b, p. 47-48), considered the name Cetiosauria “of subordinal rank only” (i.e. less inclusive than Sauropoda), and also rejected Owen's Opisthocoelia on the grounds that “it was initially proposed as a suborder of the Crocodilia” and that Owen “did not adequately define his proposed suborder and did not recognize its real relationships as being with the Dinosauria rather than the Crocodilia”. Instead, Hatcher concluded that “Sauropoda, proposed and defined by Marsh ... should be accepted as the first adequately defined name for this group of dinosaurs”, and this usage has since been followed almost unanimously.

*Diplodocus* Marsh 1878a was described in the same year as the name Sauropoda was first used, and *Brontosaurus* a year later. Both would become the subjects of important developments, *Brontosaurus* as the first sauropod to be satisfactorily reconstructed and *Diplodocus* as the first sauropod for which a complete skull was described (Marsh 1884). Both would also become among the most iconic of sauropods, due to the discovery of complete or near-complete skeletons, and the erection of famous mounts in museums around the world. Marsh (1883) reconstructed *Brontosaurus* far more accurately than Ryder had been able to do with *Camarasaurus* six years earlier, correctly depicting the anterior dorsals as not coalesced, reducing the trunk to ten dorsal vertebrae, greatly increasing the height of the sacral neural spines, showing the tail as decreasing evenly in height along its length and wrapping the coracoids around the anterior part of the trunk (Fig. 4a). Marsh also gave a reasonably accurate estimate of the mass of *Brontosaurus* as “more than twenty tons” (Marsh 1883, p. 82). Some important mistakes were made, though: most importantly, the wrong skull was used, based on that of a camarasaur (YPM 1911) rather than that of a diplodocid; only eleven cervical vertebrae were included, rather than fifteen; the forelimbs were posed in a strongly flexed posture, with the humeri at 25° and 55° from the vertical, and the manus was reconstructed as plantigrade, like the pes, rather than with a vertical arcade of metacarpals. Marsh's errors in the forelimb and manus resulted in the shoulder girdle, and hence the cervicodorsal transition, being much too low, and therefore in the neck leaving the shoulders anteroventrally so that even pronounced extension of the neck resulted only in the head being at the same height as the scapula. Eight years later, Marsh (1891) provided a revised reconstruction of *Brontosaurus* (Fig. 4b), but while this correctly increased the number of cervicals, it also incorrectly increased the dorsal count from 10 to 14, and failed to correct the skull even though the new reconstruction's skull was based on a different specimen, YPM 1986 (now USNM 5730), now thought to belong to *Brachiosaurus* Riggs 1903a (Carpenter & Tidwell 1998). Osborn (1899, p. 213) criticised Marsh's reconstructions for making the mid-dorsal vertebrae the highest point of the axial column rather than the sacrum, thereby relegating the tail to being “an appendage of the body instead of an important locomotor organ of the body”, and provided his own reconstruction of the posterior dorsals, sacrum and tail of *Diplodocus* (Osborn 1899, fig. 1), the only parts of that animal then available to him. (The articulation of the sauropod manus would not be properly understood until 21 years later, when Osborn (1904, p. 181) began a paper with the refreshingly honest statement,

"my previous figures and descriptions of the manus are all incorrect", and figured a correctly articulated manus.)

Having already named the first Gondwanan sauropod, the globe-trotting Englishman Richard Lydekker (1893) also named the first sauropods from South America, which has subsequently become a very important region for sauropods: two new species of his genus *Titanosaurus*, *T. australis* and *T. nanus*, and two new genera, *Argyrosaurus* and *Microcoelus*. Of these taxa, only *Argyrosaurus* remains valid, with *T. australis* having been referred to the new titanosaurian genus *Neuquensaurus* Powell 1992, and *Microcoelus* and *T. nanus* being nomina dubia (Powell 2003, p. 44; Wilson & Upchurch 2003, p. 140). Huene (1929a, fig. 10) would go on to provide the first reconstruction of a titanosaur; and in the same year, Huene (1929b, p. 497) was also to provide what was probably the first life restoration of a titanosaur. This figure is remarkable not so much for the rather poorly proportioned main individual as for the sketch of two more individuals fighting in the background, one of them rearing on its hind legs.

**Stage 3: interpretation and controversy (1897–1944)**

By the end of the 19th century, sauropod osteology was sufficiently well understood that it had become possible to make palaeobiological inferences. Three controversies have dominated discussions of sauropod palaeobiology ever since: habitat, athleticism and neck posture. Although early illustrations of sauropods used a variety of neck postures, the subject was not explicitly discussed until relatively recently, beginning with the work of Martin (1987). By contrast, arguments about habitat and athleticism date right back to Phillips's comments in his 1871 book.

Ballou (1897) included, as one of his six figures, the first published life restoration of a sauropod, executed by Knight under the direction of Cope (Fig. 5a). This illustration, subsequently republished by Osborn & Mook (1921, fig. 127), depicted four *Amphicoelias* individuals in a lake, two of them entirely submerged and two with only their heads above the water. The skins were shown with a bold mottled pattern like that of some lizards, which would not be seen again in a sauropod restoration for the best part of a century.

Later the same year came what may still be the most immediately recognisable of all sauropod depictions: Charles Knight's 1897 painting of *Brontosaurus* (Fig. 6a), executed under the direction of Osborn and reproduced by Matthew (1905, fig. 4). The centrepiece of Knight's painting was an amphibious *Brontosaurus* in right anterolateral aspect, its legs, tail and most of its torso submerged, with its back projecting above the surface of the water and its neck nearly vertical. In the background, a *Diplodocus* grazed on the lake shore, shown in lateral view. Both animals were a uniform dull grey. Knight was unwittingly setting the template for how sauropods would be depicted for the next three quarters of a century, not least in the Jurassic part of Zallinger's mural (see below). In Knight's world, sauropods were clumsy, lumbering behemoths, barely able to support their weight out of water: even the terrestrial *Diplodocus*, lighter than its swamp-bound cousin, looks ponderous and inert. A dramatically different opinion, at least as regards *Diplodocus*, was offered by Osborn (1899, p. 213-214), who considered sauropods much more athletic and not restricted to an aquatic lifestyle – though still at least partially aquatic by habit: “The animal was capable not only of powerful but of very rapid movements. In contrast with *Brontosaurus* it was essentially long and light-limbed and agile. Its tail was a means of defence upon land and a means of rapid escape by water from its numerous carnivorous foes”. Osborn also asserted that *Diplodocus* was capable of rearing to feed: “the tail ... functioned as a lever to balance the weight of the dorsals, anterior limbs, neck, and head, and to raise the entire forward portion of the body upwards. This power was certainly exerted while the animal was in the water, and possibly also while upon land. Thus the quadrupedal Dinosaurs occasionally assumed the position characteristic of the bipedal Dinosaurs – namely, a tripodal position, the body supported upon the hind feet and the tail” (p. 213). Ironically, it was the same artist, Knight, who was to depict this more nimble *Diplodocus*, in his painting of 1907 (Fig. 6b), created as a cover image for Scientific American to celebrate the AMNH's donation of one of its *Diplodocus* skeletons to the Senckenberg Museum in Frankfurt, Germany. In this painting, the animal is depicted with its torso raised about 60 degrees from the horizontal, its forefeet raised to knee height and its neck high in the air – well above the foliage that it seems to be trying to eat, in fact. Even this athletic *Diplodocus*, however, is accompanied by the traditional aquatic counterpart, whose head and neck are visible peering into the frame from the body of water on the right of the picture.

One of the most important sauropod workers of the early 20th century was Elmer S.

Riggs of the Field Columbian Museum (now the Field Museum of Natural History, Chicago). Riggs (1903a) named and briefly described *Brachiosaurus*, which had been found by the expedition that he led to Grand Junction, Colorado in 1900. It was at that time the largest known dinosaur. In the same year as the description of *Brachiosaurus*, Riggs published an important monograph on *Apatosaurus* which argued that Marsh's genus *Brontosaurus* was synonymous with his own earlier *Apatosaurus*, and that the difference in the number of sacral vertebrae between the two genera was an ontogenetic character, the latter having been described from a juvenile specimen in which not all the sacral vertebrae had fused by the time of death (Riggs 1903b). Although Riggs's argument has since proven conclusive for most palaeontologists, so that the older name *Apatosaurus* takes priority over its junior synonym, the more euphonious and resonant name *Brontosaurus* continued to be used in scientific publication for some time after Riggs's work, and remains popular with the public even today (e.g. Chapman & Cleese 1989). The next year, Riggs (1904) published a full monographic description of *Brachiosaurus*, erecting the family Brachiosauridae to contain this genus and *Haplocanthosaurus* Hatcher 1903a. This work was also important for its forceful argument in favour of a terrestrial lifestyle for sauropods: “There is no evidence among [sauropods] of that shortening or angulation of limb, or the broadening of foot, which is common to amphibious animals. Nor is there anything in the structure of the opisthocoelians [i.e. sauropods] which is not found in some terrestrial forms. The straight hind leg occurs in quadrupeds only among those forms which inhabit the uplands ... The short, stout metapodials and blunted phalanges ... would be as ill adapted for propulsion in water or upon marsh lands as are those of the elephant ... In short, if the foot structure of these animals indicates anything, it indicates specialization for terrestrial locomotion” (p. 244-245). Riggs also argued that, while *Apatosaurus* and *Diplodocus* were capable of rearing on their hind limbs, *Brachiosaurus* would have found this much more difficult – a finding consonant with current thinking.

February 1905 saw the unveiling of the mounted skeleton of *Brontosaurus* at the American Museum of Natural History, its posture based on the results of dissections of alligators and other reptiles to elucidate the functioning of the joints (Matthew 1905). This mount, the first of a sauropod, consisted primarily of the remains of a single individual, AMNH 460, with some elements from AMNH 222, AMNH 339 and AMNH 592, and the remainder cast or modelled in plaster. Most important among these constructed elements was the *Camarasaurus*-like skull, modelled after the reconstructions of Marsh (1883, 1891) discussed above. Osborn's thoughts on *Brontosaurus* have not aged well: he estimated the mass of the mounted specimen as “not less than ninety tons” (p. 64), and its age as “some eight millions of years” (p. 66), and followed Owen and Cope in considering sauropods as “spending their lives entirely in shallow water, partly immersed, wading about on the bottom or, perhaps, occasionally swimming, but unable to emerge entirely upon dry land” (p. 67), “Hence we can best regard the *Brontosaurus* as a great, slow-moving animal-automaton” (p. 69). Based on the mounted skeleton, Knight modelled a 1:16 scale life restoration of *Brontosaurus*, illustrated by Matthew (1905, fig. 3), and at Osborn's request, Gregory (1905) used this model to calculate the mass of *Brontosaurus* more rigorously, using the volume of water displaced by the model. Gregory's estimate of 38 tons was the first scientifically calculated mass estimate for a sauropod. While much better than Osborn's, the estimate is still rather high: this is partly because it was based on the assumption that *Brontosaurus* was 10% more dense than water – an assumption now known to be incorrect because of the increased understanding of the pneumatic cavities in the skeleton and soft tissue. Gregory's volume estimate was 31.13 $m^3$, which, using a density of 0.8 (Wedel 2005, p. 220) would yield a mass of 24900 kg, corresponding well to more recent estimates such 26000 kg (Anderson *et al.* 1985) and 23000 kg (Paul 1988a) for comparable specimens.

The AMNH *Brontosaurus* mount was followed only three months later by the second mounted sauropod, that of *Diplodocus carnegii* Hatcher 1901. The type and cotype specimen of this species (CM 84 and CM 94 respectively) had been discovered at Sheep Creek, Albany County, Wyoming, and collected by J. L. Wortman and O. A. Peterson in expeditions funded by Andrew Carnegie. Hatcher's (1901) description was based on both of these specimens, and included a skeletal reconstruction (Hatcher 1901, plate XIII) based primarily on these two individuals, but with the missing forelimbs provided by an AMNH specimen which subsequently proved to be from *Camarasaurus*. A cast of the combined skeleton was prepared under the direction of first Hatcher and then, after his death, Holland. At the request of the king of England, this was sent to the BMNH in January 1905, assembled there in April and unveiled on May 12 (Holland 1905, p.

443-446). Further casts of the same material were subsequently sent to museums in Berlin, Paris, Vienna, Madrid, St. Petersburg, Bologna, La Plata, Mexico City and Munich, and the original material mounted at the Carnegie Museum in 1907 (McIntosh 1981, p. 20), making this perhaps the single most viewed skeleton of any animal in the world.

The availability of the skeleton of *Diplodocus carnegii* provoked much speculation about its lifestyle. Hay (1908) proposed that it sprawled like a crocodile: “The mammal-like pose attributed to the Sauropoda is one that is not required by their anatomy and one that is improbable” (p. 677); “The weight of *Diplodocus* and *Brontosaurus* furnishes a strong argument against their having had a mammal-like carriage” (p. 679-680); “*Diplodocus* ... could creep about on land, with perhaps laborious effort” (p. 681). Tornier (1909), also rejected Hatcher's mammal-like erect-legged posture for *Diplodocus*, despite its pedigree going all the way back to Phillips, in favour of an interpretation in which *Diplodocus* sprawled like a lizard. Tornier (1909, plate II) provided a bizarre skeletal reconstruction of *Diplodocus* (Fig. 7) in which the scapulae were vertical and articulated with the last cervical rather than the first few dorsals, the glenoid faced directly to the posterior with no ventral component, the radius and ulna formed an acute angle with the humerus, the tibia and fibula formed an acute angle with the femur, and the neck was so flexible that the fifth most proximal cervical was vertical, C6-C10 were inclined backwards, and the skull was held directly dorsal to the shoulder. Hay (1910) reaffirmed and amplified his position, concluding his paper with a drawing by Mary Mason, executed under his instruction, which depicted four *Diplodocus* individuals. In the foreground, two individuals sprawl on dry land, one of them trailing its right leg painfully behind it. Further back, a nearly submerged individual swims towards them; further back still, a fourth lies absolutely flat on a distant shore, its neck, torso and tail all lying on the ground.

The unconventional posture suggested independently by Hay and Tornier was rebutted by Holland (1910), whose paper combined solid anatomical analysis with devastating sarcasm and rhetoric to convincingly demonstrate that the sprawling posture was impossible for *Diplodocus*, and other sauropods, to adopt: “It was a bold step for [Tornier] immediately to transfer the creature from the order Dinosauria, and evidently with the skeleton of a *Varanus* and a *Chameleon* before him, to proceed with the help of a pencil, the powerful tool of the closet-naturalist, to reconstruct the skeleton upon the study of which two generations of American paleontologists have expended considerable time and labor, and squeeze the animal into the form which his brilliantly illuminated imagination suggested" (p. 262). Holland demonstrated that Tornier's posture requires the greater trochanter of the femur to articulate with the ischiadic peduncle of the ilium, "thus locking the femur into a position utterly precluding all motion whatsoever" and that it disarticulates the knee, leaving the distal articular surface of the femur unused, and the tibia and fibula articulating with the posterior edges of the condyles. He commented on Tornier's skeletal reconstruction that "As a contribution to the literature of caricature the success achieved is remarkable" (p. 264). Holland (1910, fig. 9) showed that, were the Tornierian posture actually achieved, the chest and belly of *Diplodocus* would be much lower than its feet, so this it would have required deep grooves in the ground to walk along. Although Hay (1911) attempted to counter Holland's arguments, the debate was effectively over. Whatever doubt may have remained was dispelled by the description of a complete and articulated juvenile *Camarasaurus* by Gilmore (1925), which clearly showed that the posture advocated by Holland was correct, and by the fossilised sauropod trackways later described by Bird (1939, 1941, 1944).

The years 1909 to 1912 saw what was perhaps the most ambitious palaeontological undertaking in history: the German expeditions to collect fossils from the Tendaguru region of German East Africa (now Tanzania), under the leadership of Werner Janensch and subsequently Hans Reck (Maier 2003). The scale of the undertaking was immense: the Germans recruited 170 native labourers for the 1909 season, rising to 400 and then 500 in subsequent years. In total, 235 tonnes of fossils were shipped back to Germany, having been carried from Tendaguru to the port of Lindi in 5,400 four-day marches. Much of this material remains unprepared nearly a century later, but the prepared specimens include some of the most spectacular sauropod material in the world, including the *Brachiosaurus brancai* type specimen HMN SII (officially MB.R.2181), which is the largest known reasonably complete skeleton of any terrestrial animal. Other new sauropods recognised from the Tendaguru fossils include *Dicraeosaurus* Janensch 1914, *Tornieria* Sternfeld 1911, *Janenschia* Wild 1991, *Tendaguria* Bonaparte *et al.* 2000 and *Australodocus* Remes 2007 – all but the first of which were previously subsumed under the name *Gigantosaurus* Fraas 1908, which was abandoned when found to be a synonym of the nomen dubium *Gigantosaurus* Seeley 1869. The Tendaguru sauropods have a complex nomenclatural history which is only now being resolved (e.g. Remes 2006; Taylor, in press). These sauropods represent several groups: Brachiosauridae (*B. brancai*), Dicraeosauridae (*Dicraeosaurus*), Diplodocinae (*Tornieria* and *Australodocus*) and probably Titanosauria (*Janenschia* and *Tendaguria*, although the former may instead represent a camarasaurid or an apatosaurine, and the latter is enigmatic, known only from a few presacral vertebrae that do not closely resemble those of any other known sauropod). Together with the theropods, ornithopods and stegosaurs of Tendaguru, these taxa constitute one of the richest known dinosaur faunas – all the more amazing in light of the difficult working conditions in which the fossils were excavated and the scarcity of materials such as plaster for jacketing. Janensch devoted much of his career to an exhaustive series of detailed monographs on the sauropods of Tendaguru (Janensch 1922, 1929a, 1935-1936, 1947, 1950a, 1961) so that his work on these sauropods spanned more than half a century. Between 1919 and 1930, the British Museum (Natural History) mounted a series of under-resourced expeditions to Tendaguru, but the results were disappointing, with only one good specimen recovered and even that not properly described. A very brief preliminary report was provided by the expedition leader, Migeod (1931), but a full description and analysis of this specimen is only now under way (Taylor 2005), with preliminary results suggesting that Migeod's specimen may represent yet another new taxon.

Matthew (1915) wrote the first book about dinosaurs for non-specialists, which included (fig. 24) the first attempt to reconstruct the skeleton of *Brachiosaurus*, based on both the American *B. altithorax* and the German *B. brancai* material. Given that it was executed only one year after Janensch's (1914) initial, brief report of the German brachiosaur material, this reconstruction is impressively accurate: it is instantly recognisable as *Brachiosaurus*, and has all the proportions essentially correct. Unfortunately, sauropods otherwise receive short shrift in Matthew's book, the relevant chapter of which consists primarily of a reprint of his own (1905) account of the mounting of the AMNH *Brontosaurus*, and includes a reproduction of Knight's 1897 *Brontosaurus* painting. The book undoubtedly helped to establish swamp-bound sauropods as conventional wisdom, despite the earlier opposite conclusions of Phillips (1871), Osborn (1899), Riggs (1904) and others. This perception, once established, would prove difficult to shake off.

The 1920s opened with the publication of the sauropod monograph that stands alone: the detailed redescription of *Camarasaurus* by Osborn & Mook (1921). In 141 pages, 127 stunningly detailed figures and 25 large plates, and working from excellent and abundant material, Osborn and Mook did in detail the work that Cope had rushed through so inadequately 40 years earlier (Fig. 3b). So exhaustive was their work that, nearly 90 years on, it remains the most comprehensive guide not only to *Camarasaurus* but to sauropod anatomy in general. The monograph also redescribed *Amphicoelias*, resolved some synonymies and other nomenclatural issues, and reproduced important earlier figures, including the pioneering 1877 *Camarasaurus* reconstruction of Ryder. While palaeobiological hypotheses have come and gone, and as papers that were once highly regarded are now seen as hopelessly wrong, Osborn and Mook's careful and comprehensive descriptive work remains as relevant as ever. Four years later, Gilmore (1925) described the marvellously preserved juvenile *Camarasaurus* CM 11338 in great detail, and was able to correct the vertebral formula and other minor errors of Osborn and Mook. Gilmore presented a skeletal reconstruction in his plate XVII, which was the first reconstruction of a sauropod based on the remains of a single individual. Also significant in the 1920s was the description of *Helopus* Wiman (1929), the first of many Chinese sauropods. Like Gilmore, Wiman was fortunate enough to work from material so complete that it would have been the envy of earlier workers such as Owen and Seeley: the skull, axial and appendicular elements are all figured in multiple views. Like *Amphicoelias* before it, *Helopus* was conceived as a snorkeler (Fig. 5b). (The name *Helopus* was preoccupied, and so this genus is now known as *Euhelopus* Romer 1956.)

Around 1930, during an economic slump in Germany precipitated in part by the Wall Street Crash, plans were made to mount the skeleton of the *Brachiosaurus brancai* type specimen HMN SII at the Humboldt Museum in Berlin (Maier 2003, p. 260-268). Original plans to mount cast and replica bones were superseded by the yet more ambitious goal of using original bones (from SII and referred specimens) for all but the skull, the fragile presacral vertebrae and a few other minor bones. The herculean effort took seven years to complete, and the mounted skeleton was unveiled, to a backdrop of swastika banners, in August 1937 – the year after the Berlin Olympics and just two years before the start of the Second World War. The war would interrupt further work on the Tendaguru material so that it would be a further thirteen years before a paper describing the skeletal mount could be published (Janensch 1950b).

Bird (1939, 1941, 1944) was the first to describe sauropod tracks, from several sites including Glen Rose and Davenport Ranch, both in Texas. Bird (1944, p. 65) noted that, at the Davenport Ranch site, all 23 individual trackways were headed in the same direction, and concluded “this suggests that they passed in a single herd, an important conclusion, borne out by the consistency of the preserved tracks”. Equally significantly, despite assuming that the tracks were made on a stream bed, Bird (1944, p. 65) noted that “if the the smallest animals in the herd were wading, as the depth of their tracks indicates, then, by comparison, the larger creatures were progressing well out of water. The question 'Could *Brontosaurus* walk on land?' can be answered in all probability in the affirmative.” This evidence of a terrestrial lifestyle continued to be widely overlooked, however, as in Zdeněk Burian's widely reproduced 1941 painting of three snorkelling *Brachiosaurus* individuals – a painting that seems directly descended from Knight's 1897 *Amphicoelias* drawing. In the foreground and background, two of the animals are standing on the bottom of a lake, with only their heads and the anterior part of their necks protruding above water; between them, the third has lowered its neck to eat vegetation growing on the lake bed, and is entirely submerged. This kind of lifestyle was later proved impossible by Kermack (1951), who pointed out that snorkelling cannot be achieved by means of a long neck as water pressure would make it impossible to ventilate lungs below a certain depth.

**Stage 4: the dark ages (1945–1967)**

Understandably, little effort was put into palaeontology during the Second World War (1939-1945); more surprisingly, the study of dinosaurs, including sauropods, did not resume after the war, as dinosaurs were perceived as an evolutionary dead end and mammal palaeontology was perceived as more interesting and important (Bakker 1975, p. 58). Despite the huge popular appeal of Rudolf F. Zallinger's gigantic *Age of Reptiles* mural at the Yale Peabody Museum, completed in 1947 and reproduced in Life Magazine's 1952 series *The World We Live In*, it can only have helped reinforce the popular perception of dinosaurs in general, and sauropods in particular, as sluggish and unathletic. The Jurassic part of the mural, which contains its sauropods, owes a massive debt to Knight's 1897 *Brontosaurus* painting, both compositionally and in terms of the palaeobiology that it represents. Like Knight's image, Zallinger's has as its principal subject an amphibious *Brontosaurus*, in right anterolateral aspect, submerged to the shoulders in a lake and with its neck raised to a near-vertical posture. Also like Knight's painting, the mural depicts a *Diplodocus* in the background, on land, in lateral view and with a horizontal neck. As with Knight, both sauropods are an undistinguished grey colour. Half a century of palaeobiological work had resulted in absolutely no visible progress in how sauropods were perceived. That Zallinger had a tendency to repeat himself as well as recycle others' compositions was demonstrated by his 1966 painting of *Brachiosaurus*, published in Watson (1966, p. 20-21). Once more, the principal subject was depicted in right anterolateral view, up to its shoulders in water, with a steeply inclined neck, in dull grey, and with a second sauropod (this time, another *Brachiosaurus* individual) shown in the background, standing on the shore of the lake. In both the Zallinger paintings, a small, red ramphorynchoid pterosaur flies with the tip of its left wing in front of the principal subject's neck. Outdated ideas were further propagated by a stream of children's books, such as *The How and Why Wonder Book of Dinosaurs* (Geis 1960) with its grotesquely fat sauropods in poses recycled from the work of Knight.

Apart from work mentioned above (e.g. Janensch's monographs on the Tendaguru sauropods and Bird's work on tracks), little significant research was published on sauropods during this period. One exception was the recognition of the first rebbachisaurid, *Rebbachisaurus* Lavocat 1954, from Morocco, although this specimen has never been properly described; another was the description of *Mamenchisaurus* Young 1954, from China, although the extreme neck elongation in this genus would not be recognised until the subsequent description of the referred species *Mamenchisaurus hochuanensis* Young & Zhao 1972.

Of more general interest was the work of Colbert (1962) on dinosaur masses, the first systematic attempt to estimate and compare the masses of different dinosaurs. Colbert used a variation on the method of Gregory (1905), measuring the volumes of scale models by the amount of sand displaced, and multiplying up by the scale to determine the volume of the modelled animal, and by an estimated density of 0.9 kg/l to determine its mass. Colbert (1962, p. 10) obtained values of 27.87 and 32.42 tonnes for *Brontosaurus* (using two different models, of which he favoured the heavier), 10.56 tonnes for *Diplodocus* and 78.26 tonnes for *Brachiosaurus* – the latter figure being widely quoted in popular books. Since Colbert's efforts, several further surveys have been made of the masses of various dinosaurs, among which those of Alexander (1985, 1989) and Anderson *et al*. (1985) are of particular interest – the former based on the volumes of models and the latter based on regression equations that relate limb-bone measurements to mass in extant animals, and extrapolate them to yield the masses of sauropods whose limb bones are known. Mass estimation has progressed significantly in recent years, especially with the growing understanding of how important pneumaticity was for weight reduction. Table 2 presents a summary of the history of mass estimates for *Brachiosaurus brancai*, a much studied taxon due to its large size and the existence of an excellent near-complete skeleton. Several trends are evident: first, the improvement in methods, from simple gestalt estimates via volume measurements of physical models to computer models; second, a tendency to assume lower densities in recent years; third, generally decreasing estimates of volume, due to the use of more scientifically rigorous models than the grossly obese models available to the earlier studies. The net result of the last two of these is that modern estimates tend to be much lower than older ones, especially if the aberrant result of Gunga *et al*. (1995) is ignored due to its use of circular rather than elliptical conic sections in its model. This trend towards lower mass estimates also applies to other sauropods, though it is more difficult to quantify in the case of, for example, *Apatosaurus*, due to different authors' use of different specimens.

**Stage 5: the modern renaissance (1968–present)**

Having fallen into dormancy, dinosaur palaeontology reawakened dramatically as the 1960s closed. The beginnings of the “dinosaur renaissance” (Bakker 1975) are usually attributed to the description of the bird-like theropod *Deinonychus* Ostrom 1969a and its full osteology (Ostrom 1969b), which pointed out many aspects of its anatomy indicative of an active lifestyle. However, the first shoots of revival had appeared a year earlier, in Bakker's article *The Superiority of Dinosaurs*, in the magazine of the Yale Peabody Museum (Bakker 1968). Bakker (1968, p. 14-20) discussed sauropods specifically and at length, advocating a vigorous, endothermic, terrestrial lifestyle on the basis of limb articulations, torso shape, neck length and palaeoenvironmental evidence, and included a revolutionary life restoration (Bakker 1968, fig. 4) showing two individuals of *Barosaurus* Marsh 1890, heads held high and alert, striding briskly across dry land. It is difficult, forty years on, to appreciate how radical this image seemed at the time: the visual impact of *Jurassic Park*, *Walking With Dinosaurs* and the new generation of palaeoartists has brought such images so firmly into the mainstream that Bakker's drawing no longer surprises. But against the then ubiquitous backdrop of swamp-bound, sluggish sauropods exemplified by the art of Knight, Zallinger and Burian, it was a remarkable departure. As indicated by the title of a subsequent paper (Bakker 1980) and a popular book (Bakker 1986), Bakker was preaching Dinosaur Heresies, and old views were not quick to change – for example, Weaver (1983) argued that *Brachiosaurus* would be physically unable to gather food quickly enough to support the metabolic demands of endothermy, although this study was flawed by its assumption that the head of *Brachiosaurus* was only the size of that of a giraffe; and Dodson (1990) continued to advocate ectothermy for sauropods, with correspondingly long life-spans of multiple centuries.

The first shots had been fired in the battle to bring sauropods out of the swamps, and Coombs (1975) provided many compelling arguments for sauropod terrestriality. In a careful study which found that some anatomical evidence was equivocal, Coombs found that the tall and relatively narrow sauropod torso both resembles that of terrestrial rather than amphibious extant species, and is mechanically optimised for load-bearing. Using this and several other lines of evidence (e.g. lack of secondary palate, weight-reduction through pneumaticity, straight-limbed posture, compact feet, and the terrestrial sediments in which sauropod remains occur) he concluded that sauropods were primarily terrestrial, though they likely spent some time in water as do elephants.

McIntosh & Berman (1975) reconsidered the problem of the skull of *Apatosaurus*, which had long been thought, following the reconstructions of Marsh (1883, 1891), to resemble the robust skull of *Camarasaurus*. On reviewing the historical evidence concerning the large *Diplodocus*-like skull CM 11162, they concurred with the earlier suggestion of Holland (1915) that it belonged to *Apatosaurus*. This conclusion has now been widely accepted, though in Holland's time it had been rejected due to the disagreement of Osborn. It is widely believed that the use of the name *Apatosaurus* for the animal previously known as *Brontosaurus* is related to the recognition of the correct skull, but in fact no such connection exists.

Jensen (1985) formally described and named three new giant sauropods, although he had been referring to them informally in print since the late 1970s: *Supersaurus* Jensen 1985, *Dystylosaurus* Jensen 1985 and *Ultrasaurus* Jensen 1985. These attracted much media attention because of the enormous sizes attributed to them: in particular, *Ultrasaurus*, considered a brachiosaurid on the basis of a referred scapulocoracoid, was estimated to weigh as much as 180 tonnes (McGowan 1991, p. 118) – a ludicrously inflated estimate that was based on Colbert's (1962) 78-tonne estimate for *Brachiosaurus*, scaled for an animal 32% larger in linear dimension. Unfortunately, spectacular though they are, Jensen's finds have not proven to be all that he claimed. First, it became apparent that *Ultrasaurus* Jensen 1985 was a junior homonym of *Ultrasaurus* Kim 1983, and so it was given the rather inelegant replacement name *Ultrasauros* Olshevsky 1991. Next, Curtice *et al.* (1996) showed that the dorsal vertebra that was the holotype of *Ultrasauros* belonged to the same individual as the *Supersaurus* holotype, so that *Ultrasauros* was synonymised with *Supersaurus*. This meant that the brachiosaurid scapulocoracoid that had been considered to belong to *Ultrasauros* could not belong to the same animal as the diplodocid *Ultrasauros* = *Supersaurus*. Curtice *et al.* (1996) also showed that this scapulocoracoid was not larger than the largest Tendaguru brachiosaur specimens. Finally, Curtice & Stadtman (2001) showed that the *Dystylosaurus* holotype and only specimen, a dorsal vertebra, also belonged to the same individual as the *Supersaurus* holotype, so that this name became another junior synonym. In short, all Jensen's three giant sauropods proved to be a single sauropod, with only the referred scapulocoracoid belonging to a different taxon. Nevertheless, *Supersaurus* remains a gigantic animal: its neck is longer than any other for which there is osteological evidence, probably about 15 m in length.

With the debate about sauropod terrestriality having been effectively settled by the mid-1980s, neck posture and flexibility became the next point of contention. From the early days of sauropod palaeontology, it had been assumed that the long necks of sauropods were flexible: for example, “The slender skull ... was supported by a very long and flexible neck which permitted of an almost unlimited variety of movements throughout a considerable arc” (Hatcher 1901, p. 57). Skeletal reconstructions had shown necks held in a variety of postures. Horizontal and near-horizontal postures had been illustrated by, among others, Ryder for his 1877 *Camarasaurus*, Marsh (1883, 1891) for *Brontosaurus* (= *Apatosaurus*), Hatcher (1901, plate XIII) for *Diplodocus* and Gilmore (1936, plate XXXIV) for *Apatosaurus*. Upward-inclined and near-vertical necks had been depicted by Osborn & Mook (1921, plate LXXXIV) for *Camarasaurus*, Wiman (1929, fig. 3) for *Helopus* (= *Euhelopus*), Janensch (1950b, plate VIII) for *Brachiosaurus brancai* and Bakker (1968, fig. 4) for *Barosaurus*. However, since it was generally assumed that sauropod necks were very flexible, it is not clear how much importance these authors attached to the illustrated postures: they probably considered each illustrated posture to be just one of many that were habitually adopted. In contradiction to this, Martin (1987), having investigated the range of motion between adjacent cervical vertebrae during the mounting of the Rutland specimen of *Cetiosaurus* at the Leicester City Museum, concluded that the neck would have been much less flexible than previously assumed – only just able to lower the head to the ground, and able to lift the head only about a meter above shoulder height. Martin also found horizontal flexibility to be limited to only a 4.5 m arc. These findings were later corroborated by the work of Stevens & Parrish (1999) on DinoMorph. a computer program for modelling such articulations digitally. Stevens & Parrish (1999, p. 799) found that both *Apatosaurus louisae* CM 3018 and *Diplodocus carnegii* CM 84 were limited in their ability to raise their heads, but that their osteology did not prevent them from lowering their heads well below ground-level – an adaptation that they interpreted as facilitating browsing on aquatic plants from the shore. This interpretation has been opposed by, among others, Paul (1998), who disputed the morphological evidence; Upchurch (2000), who pointed out that the *Apatosaurus* reconstruction was based on badly damaged vertebrae; and Christian & Heinrich (1998) and Christian & Dzemski (2007), who argued from the pattern of stresses in the intervertebral joints that *Brachiosaurus brancai* held its neck erect. The issue is not yet settled.

The release of the film *Jurassic Park* in 1993 marked a turning point in public perception of dinosaurs, and particularly sauropods. Until then, the dinosaur renaissance of Bakker, Ostrom and others, while challenging the traditional views of paleontologists, had had little impact on non-specialists. The terrestrial and athletic *Brachiosaurus* that is the first dinosaur clearly seen in the film brought this revolution to a far wider audience.

Similarly, the depiction of sauropods in the BBC's 1999 documentary series *Walking with Dinosaurs* helped to publicise new ideas, including both the relatively inflexible and horizontal necks advocated by Stevens and Parrish, and rearing in order to feed and to mate. Subsequent films, including the *Jurassic Park* sequels, and TV programmes including *When Dinosaurs Roamed America*, have continued to present a view of sauropods that is largely in keeping with current thought.

The evolutionary relationships of sauropods were very poorly understood up until the mid-1990s, and their classification had not progressed beyond the establishment of a handful of families – Diplodocidae, Brachiosauridae, Titanosauridae, Cetiosauridae – whose content was unstable and whose interrelationships were obscure, and indeed largely unexplored. For example, the evolutionary diagram of Bonaparte (1986) consisted only of a Prosauropoda block leading to a central block representing Cetiosauridae, and with branches leading from it to further undifferentiated and unrelated blocks for Brachiosauridae, Camarasauridae, Diplodocidae and Dicraeosauridae. Against this backdrop, Russell & Zheng (1993) performed the first phylogenetic analysis on sauropods, as part of their paper describing the new species *Mamenchisaurus sinocanadorum*. Their analysis consisted of only 21 characters applied to nine taxa, and produced a tree that, in light of more recent work, appears wrong in placing the basal eusauropods *Mamenchisaurus*, *Omeisaurus* Young 1939 and *Shunosaurus* Dong *et al.* 1983 as closely related to the diplodocoids *Dicraeosaurus* and *Apatosaurus*. However, their analysis was quickly followed by others using more characters and taxa, notably those of Upchurch (1995), using 174 characters and 27 taxa; Wilson and Sereno (1998) using 109 characters and 10 taxa; Upchurch (1998), using 205 characters and 26 taxa; Wilson (2002), using 234 characters and 29 taxa; and Upchurch *et al.* (2004), using 309 characters and 47 taxa. The results of Wilson's and Upchurch's independent series of analyses are largely in agreement, with only the position of *Euhelopus* and the nemegtosaurids differing greatly between them. A subsequent collaboration between the authors of these studies (Wilson & Upchurch, in press) has established a consensus phylogeny, in which a sequence of basal sauropods leads to the great clade Neosauropoda, which comprises Diplodocoidea (Diplodocidae, Dicraeosauridae and Rebbachisauridae) and Macronaria (Camarasauridae, Brachiosauridae and Titanosauria). Although some work remains to be done, this basic structure now seems quite well established.

The advent of rigorous phylogenetic methods has dramatically affected the field of sauropod palaeontology by placing classification on a sound theoretical basis and making it possible to trace the evolution of particular features. Before the pioneering studies of the early and mid 1990s, much sauropod work was undertaken by non-specialists, and ideas about the group's classification were arbitrary and often contradictory. Since then, the establishment of a consensus on sauropod phylogeny has made it possible for the first time to do meaningful work on palaeobiogeography, diversity and palaeoecology, and these opportunities have attracted a crop of specialist workers who continue to expand the boundaries of sauropod science.

Until relatively recently, discussions of the feeding strategy of sauropods have been speculative, and dominated by then-prevailing ideas about sauropod habitats – hence the claim of Hatcher (1901:60) and many others that sauropods subsisted on “tender, succulent aquatic or semi-aquatic plants”. This began to change in 1994, with the publication of two papers in the same volume (Calvo, 1994; Barrett & Upchurch, 1994) on feeding mechanisms. These papers established the modern approach by forsaking analogies with extant megaherbivores, instead relying on the direct evidence of functional anatomy, tooth wear and stomach contents when available. These and subsequent studies have yielded a consensus view that sauropods used minimal oral processing, though various groups seem to have differed in details of feeding strategy.

Chiappe *et al.* (1998) reported the first known sauropod embryos, those of titanosaurs, from the Auca Mahuevo site of Patagonia. The site covers more than a square kilometre, and has furnished many hundreds of specimens – for example, 200 whole eggs in a single 25 $m^2$ area (Chiappe *et al.* 2000). The preservation of the embryos is also excellent, including skin as well as bone, and articulated near-complete skulls (Chiappe *et al.* 2001), the first known from any titanosaur.

Curry (1999) applied the techniques of bone histology to sauropod remains for the first time, yielding insights into the growth history of *Apatosaurus*. By sampling bones from juvenile, sub-adult and adult specimens, she determined that growth was rapid and not seasonal, and that near-adult size was attained in about ten years. Sander (2000) analysed the microstructure of a wide selection of bones from four different Tendaguru sauropods, and was able to demonstrate that the bones of different taxa can be differentiated on histological features alone. He also found two distinct types of histology in the bones of "*Barosaurus*" *africanus* Fraas 1908 (probably *Tornieria* sensu Remes 2006), which he tentatively interpreted as representing sexual dimorphism.

The recognition and description of new sauropod taxa has continued and accelerated in recent years, with significant new genera including *Rapetosaurus* Curry Rogers & Forster 2001, from Madagascar, a titanosaur much more complete than any known up till that time. The association of its skull with an unquestionably titanosaurian postcranial skeleton finally established the nature of titanosaur skulls, and resolved the phylogenetic position of nemegtosaurids as titanosaurs closely related to *Rapetosaurus*.

**Today and tomorrow**

As with other dinosaurs (Taylor 2006), the rate at which new sauropods are being recognised, described and named is far greater now than at any previous time. Of the 137 valid sauropod genera known at the end of 2006, more than half had been named in the previous 13 years, and all six of the most fruitful years have fallen since 1999. Fig. 8 shows the rate of accumulation of valid sauropod genera, broken down by clade and in total. The general trend is towards exponential growth – not a trend that can be maintained indefinitely, but one that shows no signs of slowing yet. While brachiosaurid and diplodocid genera began to accumulate early in the history of sauropod palaeontology, it is only relatively recently that recognised titanosaur diversity has begun to climb, primarily due to the growth of work in South America. Titanosauria now represents one third of valid sauropod genera, whereas of the 20 valid sauropod genera that had been named by 1921, only a single titanosaur genus had been named that is still considered valid today, *Argyrosaurus*. (*Titanosaurus* and *Microcoelus* had also been named, but are no longer considered valid.)

Not only is sauropod diversity rising steeply, so is sauropod disparity – that is, the degree of morphological variation between different sauropods. The sauropod body-plan has traditionally been described as conservative, but this prejudice is breaking down in light of the many bizarre forms that have been described in recent years. These include the following:

- *Amargasaurus* Salgado & Bonaparte 1991 is an Argentinian dicraeosaurid with enormously elongated forked neural spines on the cervical and dorsal vertebrae. These spines may have appeared in life as individual spikes, or may have supported long, tall, parallel sails.
- *Nigersaurus* Sereno *et al.* 1999 is an African rebbachisaurid whose well-preserved skull has a distinctive dentary with a completely straight, transversely oriented tooth row, extending further laterally than the posterior part of the skull does. The skull is also extraordinarily lightly built, even by sauropod standards (Sereno *et al.* 2007, fig. 1E).
- *Agustinia* Bonaparte 1999 is an armoured sauropod from Argentina, with spiked dorsal osteoderms which would have made the animal somewhat resemble *Stegosaurus*. Bonaparte found *Agustinia* so distinctive that he raised the new monogeneric family Agustiniidae to contain it, though it is probably a titanosaur.
- *Tendaguria*, from the Tendaguru Formation of Tanzania, is represented only by two dorsal vertebrae, one of which was figured by Janensch (1929b, fig. 11) as "*Gigantosaurus*" *robustus* Fraas 1908. They are unique in having neural spines so low as to be all but absent, so that they are much broader than they are tall. Bonaparte *et al.* (2000, p. 47) considered these vertebrae sufficiently distinct to merit another monogeneric family, Tendaguriidae, perhaps related to Camarasauridae.
- *Brachytrachelopan* Rauhut *et al.* 2005 is an Argentinian dicraeosaurid unique among known sauropods in having a proportionally short neck, so that in profile it more closely resembles an ornithopod than a classic sauropod.
- Conversely, *Erketu* Ksepka & Norell 2006 seems likely to have had the proportionally longest neck of any known sauropod, since the anterior cervical vertebrae from which it is principally known are more elongate even than the mid-cervicals of *Sauroposeidon* Wedel *et al.* 2000a.
- *Europasaurus* Mateus, Laven and Knötschke in Sander *et al.* 2006 is a German titanosauriform somewhat resembling *Brachiosaurus* except in its diminutive size: it is the smallest of all known sauropods, with adults measuring up to 6.2 m, and weighing perhaps 500 kg, about the mass of a cow.

- At the other end of the size scale, *Futalognkosaurus* Calvo *et al.* 2007 joins its fellow Argentinian titanosaurs *Argentinosaurus* Bonaparte & Coria 1993 and *Puertasaurus* Novas *et al.* 2005 as one of the largest known sauropods. All three of these animals would have massed in the region of 50-100 tonnes.
- *Xenoposeidon* Taylor & Naish 2007, a British neosauropod, is known from a single partial dorsal vertebra, but has several features unique among all sauropods (e.g. neural arch is taller than centrum, covers dorsal surface of centrum, slopes forward by 35° and has featureless areas of unlaminated flat bone on its lateral surfaces). *Xenoposeidon* may represent a major new group of sauropods, of which further specimens are greatly to be desired.

The study of sauropods has come a long way since Owen named the tooth of *Cardiodon* 167 years ago, and the future looks very bright: with new sauropods being named at an ever-increasing rate, new techniques being applied to their study, and old specimens being re-evaluated in the light of new knowledge, our understanding of sauropod morphology, ecology and phylogeny seems set to grow in richness and scope for the foreseeable future. At the same time, a great deal of work remains to be done. New specimens are being found and excavated more quickly than they can be described, and many sauropods named in recent years still await the monograph to follow up an often inadequate preliminary description. Also, many historical genera are long overdue for revision: for example, no modern analysis exists of the various species of *Diplodocus* or *Camarasaurus*. Much is being done, and much must be done in the future. Although they have been dead for 65 million years, history continues to roll relentlessly on for sauropods.

This article would never have been written without the opportunity offered by the editors of this volume, R. Moody, E. Buffetaut, D. Martill and D. Naish, all of whom I thank for their enlightened interest in the history of our discipline. My work would have been shapeless without F. Taylor's invaluable advice on fitting all the information into a coherent structure. In an undertaking of this kind, old literature is indispensable, and I thank M. Wedel, D. Naish, R. Irmis, S. Werning and D. Fowler for their aid in obtaining many crucial papers. M. Wedel also provided a helpful review of an earlier draft. Reviews of the submitted manuscript by P. Upchurch and D. Schwarz-Wings were detailed and constructive.

**References**

Alexander, R.M. 1985. Mechanics of posture and gait of some large dinosaurs. *Zoological Journal of the Linnean Society*, **83**, 1–25.

Alexander, R.M. 1989. *Dynamics of Dinosaurs and Other Extinct Giants.* Columbia University Press, New York.

Anderson, J.F., Hall-Martin, A. & Russell, D.A. 1985. Long-bone circumference and weight in mammals, birds and dinosaurs. *Journal of Zoology*, **207**, 53–61.

Bakker, R.T. 1968. The superiority of dinosaurs. *Discovery: magazine of the Peabody Museum of Natural History*, **3**, 11–22.

Bakker, R.T. 1975. Dinosaur renaissance. *Scientific American*, **232**, 58–78.

Bakker, R.T. 1980. Dinosaur heresy, dinosaur renaissance: why we need endothermic archosaurs for a comprehensive theory of bioenergetic evolution. *In:* Thomas, R.D.K. & Olsen, E.C. (eds) *A cold look at the warm-blooded dinosaurs: AAAS selected symposia series, no. 28*. American Association for the Advancement of Science, Washington, D.C, 351–462.

Bakker, R.T. 1986. *The Dinosaur Heresies.* Morrow, New York.

Ballou, W.H. 1897. Strange creatures of the past: gigantic saurians of the reptilian age. *Century Magazine*, **November 1897**, 15–23.

Barbour, E.H. 1890. Scientific news 5: notes on the paleontological laboratory of the United States Geological Survey under Professor Marsh. *The American Naturalist*, **24**, 388–400.

Barrett, P.M. 2006. A sauropod dinosaur tooth from the Middle Jurassic of Skye, Scotland. *Transactions of the Royal Society of Edinburgh: Earth Sciences*, **97**, 25–29.

Berman, D.S. & McIntosh, J.S. 1978. Skull and Relationships of the Upper Jurassic sauropod *Apatosaurus* (Reptilia, Saurischia). *Bulletin of the Carnegie Museum*, **8**, 1–35.

Bird, R.T. 1939. Thunder in his footsteps. *Natural History*, **43**, 254–261.

Bird, R.T. 1941. A dinosaur walks into the museum. *Natural History*, **47**, 74–81.

Bird, R.T. 1944. Did *Brontosaurus* ever walk on land? *Natural History*, **53**, 60–67.

Blows, W.T. 1995. The Early Cretaceous brachiosaurid dinosaurs *Ornithopsis* and *Eucamerotus* from the Isle of Wight, England. *Palaeontology*, **38**, 187–197.

Bonaparte, J.F. 1986. The early radiation and phylogenetic relationships of the Jurassic sauropod dinosaurs, based on vertebral anatomy. *In:* Padian, K. (ed) *The Beginning of the Age of Dinosaurs*. Cambridge University Press, Cambridge, UK, 247–258.

Bonaparte, J.F. 1999. An armoured sauropod from the Aptian of northern Patagonia, Argentina. *In:* Tomida, Y., Rich, T.H. & Vickers-Rich, P. (eds) *Proceedings of the Second Gondwanan Dinosaur Symposium*. National Science Museum, Tokyo, 1–12.

Bonaparte, J.F. & Powell, J.E. 1980. A continental assemblage of tetrapods from the Upper Cretaceous beds of El Brete, northwestern Argentina (Sauropoda-Coelurosauria-Carnosauria-Aves). *Mémoires de la Société Géologique de France, Nouvelle Série*, **139**, 19–28.

Bonaparte, J.F. & Coria, R.A. 1993. Un nuevo y gigantesco sauropodo titanosaurio de la Formacion Río Limay (Albiano-Cenomaniano) de la Provincia de Neuquén, Argentina. *Ameghiniana*, **30**, 271–282.

Bonaparte, J.F., Heinrich, W.-D. & Wild, R. 2000. Review of *Janenschia* Wild, with the description of a new sauropod from the Tendaguru beds of Tanzania and a discussion on the systematic value of procoelous caudal vertebrae in the Sauropoda. *Palaeontographica A*, **256**, 25–76.

Britt, B.B. 1993. *Pneumatic postcranial bones in dinosaurs and other archosaurs (Ph.D. dissertation)*. University of Calgary, Calgary.

Broom, R. 1904. On the occurrence of an opisthocoelian dinosaur (*Algoasaurus bauri*) in the Cretaceous beds of South Africa. Geological Magazine, **1 (Ser. 5)**, 445–447.

Buckland, W. 1824. Notice on the *Megalosaurus* or great fossil lizard of Stonesfield. *Transactions of the Geological Society of London*, **21**, 390–397 and plates 40-44.

CALVO, J.O., PORFIRI, J.D., GONZÁLEZ-RIGA, B.J. & KELLNER, A.W.A. 2007. A new Cretaceous terrestrial ecosystem from Gondwana with the description of a new sauropod dinosaur. *Anais da Academia Brasileira de Ciências*, **79**, 529–541.

CARPENTER, K. & TIDWELL, V. 1998. Preliminary description of a *Brachiosaurus* skull from Felch Quarry 1, Garden Park, Colorado. *Modern Geology*, **23**, 69–84.

CARPENTER, K. & TIDWELL, V. 2005. Reassessment of the Early Cretaceous Sauropod *Astrodon johnsoni* Leidy 1865 (Titanosauriformes). *In:* TIDWELL, V. & CARPENTER, K. (eds) *Thunder Lizards: the Sauropodomorph Dinosaurs*. Indiana University Press, Bloomington, Indiana, 78–114.

CHAPMAN, G. & CLEESE, J. 1989. Anne Elk's Theory on Brontosauruses. *In:* CHAPMAN, G., CLEESE, J., GILLIAM, T., IDLE, E., JONES, T. & PALIN, M. (eds) *Just the Words, volume 2*. Methuen, London, 118–120.

CHIAPPE, L.M., SALGADO, L. & CORIA, R.A. 2001. Embryonic skulls of titanosaur sauropod dinosaurs. *Science*, **293**, 2444–2446.

CHIAPPE, L.M., CORIA, R.A., DINGUS, L., JACKSON, F., CHINSAMY, A. & FOX, M. 1998. Sauropod dinosaur embryos from the Late Cretaceous of Patagonia. *Nature*, **396**, 258–261.

CHIAPPE, L.M., DINGUS, L., JACKSON, F., GRELLET-TINNER, G., ASPINALL, R., CLARKE, J., CORIA, R.A., GARRIDO, A. & LOOPE, D. 2000. Sauropod eggs and embryos from the Late Cretaceous of Patagonia. *In:* BRAVO, A.M. & REYES, T. (eds) *First International Symposium on Dinosaur Eggs and Babies – Extended Abstracts*. Isona I Conca Dellà Catalonia, Spain, 23–29.

CHRISTIAN, A. & HEINRICH, W.-D. 1998. The neck posture of *Brachiosaurus brancai*. *Mitteilungen aus dem Museum für Naturkunde, Berlin, Geowissenschaften*, **1**, 73–80.

CHRISTIAN, A. & DZEMSKI, G. 2007. Reconstruction of the cervical skeleton posture of *Brachiosaurus brancai* Janensch, 1914 by an analysis of the intervertebral stress along the neck and a comparison with the results of different approaches. *Fossil Record*, **10**, 38–49.

Christiansen, P. 1997. Locomotion in sauropod dinosaurs. *Gaia*, **14**, 45–75.

Colbert, E.H. 1962. The weights of dinosaurs. *American Museum Novitates*, **2076**, 1–16.

Colbert, E.H. 1997. North American dinosaur hunters. *In:* Farlow, J.O. & Brett-Surman, M.K. (eds) *The Complete Dinosaur*. Indiana University Press, Bloomington, Indiana, 24–33.

Coombs, W.P. 1975. Sauropod habits and habitats. *Palaeogeography, Palaeoclimatology, Palaeoecology*, **17**, 1–33.

Cope, E.D. 1877a. On a gigantic saurian from the Dakota Epoch of Colorado. *Paleontology Bulletin*, **25**, 5–10.

Cope, E.D. 1877b. On *Amphicoelias*, a genus of saurians from the Dakota Epoch of Colorado. *Paleontology Bulletin*, **27**, 1–5.

Cope, E.D. 1877c. On a dinosaurian from the Trias of Utah. *Proceedings of the American Philosophical Society*, **16**, 579–584.

Cope, E.D. 1878. On the vertebrata of the Dakota Epoch of Colorado. *Proceedings of the American Philosophical Society*, **17**, 233–247.

Curry, K.A. 1999. Ontogenetic histology of *Apatosaurus* (Dinosauria: Sauropoda): new insights on growth rates and logevity. *Journal of Vertebrate Paleontology*, **19**, 654–665.

Curry Rogers, K. & Forster, C.A. 2001. Last of the dinosaur titans: a new sauropod from Madagascar. *Nature*, **412**, 530–534.

Curtice, B.D. & Stadtman, K.L. 2001. The demise of *Dystylosaurus edwini* and a revision of *Supersaurus vivianae*. *Western Association of Vertebrate Paleontologists and Mesa Southwest Paleontological Symposium, Mesa Southwest Museum Bulletin*, **8**, 33–40.

Curtice, B.D., Stadtman, K.L. & Curtice, L.J. 1996. A reassessment of *Ultrasauros macintoshi* (Jensen, 1985). *Museum of Northern Arizona Bulletin*, **60**, 87–95.

Dodson, P. 1990. Sauropod Paleoecology. *In:* Weishampel, D.B., Dodson, P. & Osmólska,

H. (eds) *The Dinosauria*. University of California Press, Berkeley and Los Angeles, 402–407.

DONG, Z., ZHOU, S. & ZHANG, Y. 1983. The dinosaurian remains from Sichuan Basin, China. *Palaeontologica Sinica (Series C)*, **23**, 1–145.

FRAAS, E. 1908. Ostafrikanische Dinosaurier. *Palaeontographica*, **55**, 105–144.

GEIS, D. 1960. *The How and Why Wonder Book of Dinosaurs.* Price Stern Sloan, Los Angeles, California.

GERVAIS, P. 1852. *Zoologie et paléontologie française (animaux vertébrés).* A. Bertrand, Paris.

GALTON, P.M. & KNOLL, F. 2006. A saurischian dinosaur braincase from the Middle Jurassic (Bathonian) near Oxford, England: from the theropod *Megalosaurus* or the sauropod *Cetiosaurus*? *Geological Magazine*, **143**, 905-921.

GILMORE, C.W. 1925. A nearly complete articulated skeleton of *Camarasaurus*, a saurischian dinosaur from the Dinosaur National Monument, Utah. *Memoirs of the Carnegie Museum*, **10**, 347–384 and plates 13-17.

GILMORE, C.W. 1936. Osteology of *Apatosaurus*, with special reference to specimens in the Carnegie Museum. *Memoirs of the Carnegie Museum*, **11**, 175–298 and plates XXI-XXXIV.

GOWER, D.J. 2001. Possible postcranial pneumaticity in the last common ancestor of birds and crocodilians: evidence from *Erythrosuchus* and other Mesozoic archosaurs. *Naturwissenschaften*, **88**, 119–122.

GREGORY, W.K. 1905. The weight of *Brontosaurus*. *Science*, **22**, 572–572.

GUNGA, H.-C., KIRSCH, K.A., BAARTZ, F., ROCKER, L., HEINRICH, W.-D., LISOWSKI, W., WIEDEMANN, A. & ALBERTZ, J. 1995. New data on the dimensions of *Brachiosaurus brancai* and their physiological implications. *Naturwissenschaften*, **82**, 190–192.

GUNGA, H.-C., SUTHAU, T., BELLMANN, A., STOINSKI, S., FRIEDRICH, A., TRIPPEL, T., KIRSCH, K. & HELLWICH, O. 2008. A new body mass estimation of *Brachiosaurus brancai* Janensch, 1914 mounted and exhibited at the Museum of Natural History (Berlin,

Germany). *Fossil Record*, **11**, 28–33.

Hatcher, J.B. 1901. *Diplodocus* (Marsh): its osteology, taxonomy and probable habits, with a restoration of the skeleton. *Memoirs of the Carnegie Museum*, **1**, 1–63.

Hatcher, J.B. 1903a. A new name for the dinosaur *Haplocanthus* Hatcher. *Proceedings of the Biological Society of Washington*, **16**, 100.

Hatcher, J.B. 1903b. Osteology of *Haplocanthosaurus* with description of a new species, and remarks on the probable habits of the Sauropoda and the age and origin of the Atlantosaurus beds. *Memoirs of the Carnegie Museum*, **2**, 1–72 and plates I-V.

Hay, O.P. 1908. On the habits and the pose of the sauropodous dinosaurs, especially of *Diplodocus*. *The American Naturalist*, **42**, 672–681.

Hay, O.P. 1910. On the manner of locomotion of the dinosaurs especially *Diplodocus*, with remarks on the origin of the birds. *Proceedings of the Washington Academy of Science*, **12**, 1–25.

Hay, O.P. 1911. Further observations on the pose of the sauropodous dinosaurs. *The American Naturalist*, **45**, 396–412.

Henderson, D.M. 2004. Tipsy punters: sauropod dinosaur pneumaticity, bouyancy and aquatic habits. *Proceedings of the Royal Society of London, Series B (Biology Letters)*, **271**, S180–S183.

Henderson, D.M. 2006. Burly Gaits: centers of mass, stability, and the trackways of sauropod dinosaurs. *Journal of Vertebrate Paleontology*, **26**, 908–921.

Holland, W.J. 1905. The presentation of a reproduction of *Diplodocus carnegiei* to the trustees of the British Museum. *Annals of the Carnegie Museum*, **3**, 443–452 and plates 17-18.

Holland, W.J. 1910. A review of some recent criticisms of the restorations of sauropod dinosaurs existing in the museums of the United States, with special reference to that of *Diplodocus carnegiei* in the Carnegie museum. *American Naturalist*, **44**, 259–283.

Holland, W.J. 1915. Heads and tails: a few notes relating to the structure of the sauropod dinosaurs. *Annals of the Carnegie Museum*, **9**, 273–278.

Huene, F.v. 1926. The carnivorous Saurischia in the Jura and Cretaceous formations principally in Europe. *Revista del Museo de La Plata*, **29**, 35–114.

Huene, F.v. 1929a. Los Saurisquios y Ornitisquios del Cretaceo Argentina. *Annales Museo de La Plata*, **3**, Serie 2a. Museo de La Plata, Argentina.

Huene, F.v. 1929b. Die Besonderheit der Titanosaurier. *Centralblatt für Mineralogie, Geologie und Paläontologie*, **1929B**, 493–499.

Jain, S.L., Kutty, T.S. & Roy-Chowdhury, T.K. 1975. The sauropod dinosaur from the Lower Jurassic Kota Formation of India. *Proceedings of the Royal Society of London A*, **188**, 221–228.

Janensch, W. 1914. Übersicht über der Wirbeltierfauna der Tendaguru-Schichten nebst einer kurzen Charakterisierung der neu aufgefuhrten Arten von Sauropoden. *Archiv für Biontologie*, **3**, 81–110.

Janensch, W. 1922. Das Handskelett von *Gigantosaurus robustus* u. *Brachiosaurus Brancai* aus den Tendaguru-Schichten Deutsch-Ostafrikas. *Centralblatt für Mineralogie, Geologie und Paläontologie*, **15**, 464–480.

Janensch, W. 1929a. Die Wirbelsaule der Gattung *Dicraeosaurus*. *Palaeontographica (Suppl. 7)*, **2**, 35–133.

Janensch, W. 1929b. Material und Formengehalt der Sauropoden in der Ausbeute der Tendaguru-Expedition. *Palaeontographica (Suppl. 7)*, **2**, 1–34.

Janensch, W. 1935-1936. Die Schadel der Sauropoden *Brachiosaurus*, *Barosaurus* und *Dicraeosaurus* aus den Tendaguru-Schichten Deutsch-Ostafrikas. *Palaeontographica (Suppl. 7)*, **2**, 147–298.

Janensch, W. 1938. Gestalt und Größe von *Brachiosaurus* und anderen riesigwüchsigen Sauropoden. *Der Biologe*, **7**, 130–134.

Janensch, W. 1947. Pneumatizitat bei Wirbeln von Sauropoden und anderen Saurischien. *Palaeontographica (Suppl. 7)*, **3**, 1–25.

Janensch, W. 1950a. Die Wirbelsaule von *Brachiosaurus brancai*. *Palaeontographica (Suppl. 7)*, **3**, 27–93.

Janensch, W. 1950b. Die Skelettrekonstruktion von *Brachiosaurus brancai*. *Palaeontographica (Suppl. 7)*, **3**, 97–103 and plates VI-VIII.

Janensch, W. 1961. Die Gliedmaszen und Gliedmaszengürtel der Sauropoden der Tendaguru-Schichten. *Palaeontographica (Suppl. 7)*, **3**, 177–235 and plates XV-XXIII.

Jensen, J.A. 1985. Three new sauropod dinosaurs from the Upper Jurassic of Colorado. *Great Basin Naturalist*, **45**, 697–709.

Johnston, C. 1859. Note on odontography. *American Journal of Dental Science*, **9**, 337–343.

Kermack, K.A. 1951. A note on the habits of sauropods. *Annals and Magazine of Natural History, Series 12*, **4**, 830–832.

Kim, H.M. 1983. Cretaceous dinosaurs from Korea. *Journal of the Geology Society of Korea*, **19**, 115–126.

Ksepka, D.T. & Norell, M.A. 2006. *Erketu ellisoni*, a long-necked sauropod from Bor Guve (Dornogov Aimag, Mongolia). *American Museum Novitates*, **3508**, 1–16.

Lavocat, R. 1954. Sur les Dinosauriens du continental intercalaire des Kem-Kem de la Daoura. *Comptes Rendus 19th Intenational Geological Congress 1952*, **1**, 65–68.

Leidy, J. 1865. Cretaceous reptiles of the United States. *Smithsonian Contribution to Knowledge*, **192**, 1–135.

Longman, H.A. 1926. A giant dinosaur from Durham Downs, Queensland. *Memoirs of the Queensland Museum*, **8**, 183–194.

Longman, H.A. 1933. A new dinosaur from the Queensland Cretaceous. *Memoirs of the Queensland Museum*, **10**, 131–144.

Lydekker, R. 1877. Notices of new and other Vertebrata from Indian Tertiary and Secondary rocks. *Records of the Geological Survey of India*, **10**, 30–43.

Lydekker, R. 1890. *Catalogue of the fossil Reptilia and Amphibia in the British Museum, part IV, containing the orders Anomodontia, Ecaudata, Caudata and Labyrinthodontia.* British Museum of Natural History, London.

Lydekker, R. 1893. The dinosaurs of Patagonia. *Anales del Museo de La Plata*, **2**, 1–14.

Maier, G. 2003. *African Dinosaurs Unearthed: The Tendaguru Expeditions.* Indiana University Press, Bloomington and Indianapolis.

Mantell, G.A. 1825. Notice on the *Iguanodon*, a newly discovered fossil reptile, from the sandstone of Tilgate Forest, in Sussex. *Philosophical Transactions of the Royal Society*, **115**, 179–186.

Mantell, G.A. 1833. *The Geology of the South-East of England.* Longman Ltd, London.

Mantell, G.A. 1850. On the *Pelorosaurus*: an undescribed gigantic terrestrial reptile, whose remains are associated with those of the *Iguanodon* and other saurians in the strata of Tilgate Forest, in Sussex. *Philosophical Transactions of the Royal Society of London*, **140**, 379–390.

Mantell, G.A. 1852. On the structure of the *Iguanodon* and on the fauna and flora of the Wealden Formation. *Notice: Proceedings of the Royal Institute of Great Britain*, **1**, 141–146.

Marsh, O.C. 1877a. Notice of a new and gigantic dinosaur. *American Journal of Science and Arts*, **14**, 87–88.

Marsh, O.C. 1877b. Notice of new dinosaurian reptiles from the Jurassic Formation. *American Journal of Science and Arts*, **14**, 514–516.

Marsh, O.C. 1877c. A new order of extinct Reptilia (Stegosauria) from the Jurassic of the Rocky Mountains. *American Journal of Science and Arts*, **14**, 513–514.

Marsh, O.C. 1878a. Principal characters of American Jurassic dinosaurs. Part I. *American Journal of Science, Series 3*, **16**, 411–416.

Marsh, O.C. 1878b. Notice of new dinosaurian reptiles. *American Journal of Science, Series 3*, **15**, 241–244.

Marsh, O.C. 1879. Notice of new Jurassic reptiles. *American Journal of Science, Series*

*3*, **18**, 501–505.

Marsh, O.C. 1883. Principal characters of American Jurassic dinosaurs. Pt. VI. Restoration of *Brontosaurus*. *American Journal of Science, Series 3*, **26**, 81–85 and plate 1.

Marsh, O.C. 1884. Principal characters of American Jurassic dinosaurs. Pt. VII. On the Diplodocidae, a new family of the Sauropoda. *American Journal of Science, Series 3*, **27**, 161–167 and plates 3-4.

Marsh, O.C. 1888. Notice of a new genus of Sauropoda and other new dinosaurs from the Potomac Formation. *American Journal of Science, Series 3*, **35**, 89–94.

Marsh, O.C. 1890. Description of new dinosaurian reptiles. *American Journal of Science, Series 3*, **39**, 81–86 and plate I.

Marsh, O.C. 1891. Restoration of *Brontosaurus*. *American Journal of Science, Series 3*, **41**, 341–342.

Martin, J. 1987. Mobility and feeding of *Cetiosaurus* (Saurischia: Sauropoda) – why the long neck? *Occasional Papers of the Tyrrell Museum of Palaeontology (Fourth Symposium on Mezozoic Terrestrial Ecosystems)*, **3**, 154–159.

Matthew, W.D. 1905. The mounted skeleton of *Brontosaurus*. *The American Museum Journal*, **5**, 62–70.

Matthew, W.D. 1915. *Dinosaurs, with special reference to the American Museum collections.* American Museum of Natural History, New York.

McGowan, C. 1991. *Dinosaurs, spitfires and sea dragons.* Harvard University Press, Cambridge, Massachusetts.

McIntosh, J.S. 1981. Annotated catalogue of the dinosaurs (Reptilia, Archosauria) in the collections of Carnegie Museum of Natural History. *Bulletin of the Carnegie Museum*, **18**, 1–67.

McIntosh, J.S. & Berman, D.S. 1975. Description of the palate and lower jaw of the sauropod dinosaur *Diplodocus* (Reptilia: Saurischia) with remarks on the nature of the skull of *Apatosaurus*. *Journal of Paleontology*, **49**, 187–199.

Meyer, H.v. 1832. *Palaeologica zur Geschichte der Erde und ihrer Geschöpfe.* Schmerber, Frankfurt am Main.

Migeod, F.W.H. 1931. British Museum East Africa Expedition: Account of the work done in 1930. *Natural History Magazine*, **3**, 87–103.

Mook, C.C. 1914. Notes on *Camarasaurus* Cope. *Annals of the New York Academy of Sciences*, **24**, 19–22.

Nesbitt, S.J. & Norell, M.A. 2006. Extreme convergence in the body plans of an early suchian (Archosauria) and ornithomimid dinosaurs (Theropoda). *Proceedings of the Royal Society of London B*, **273**, 1045–1048.

Novas, F.E., Salgado, L., Calvo, J. & Agnolin, F. 2005. Giant titanosaur (Dinosauria, Sauropoda) from the Late Cretaceous of Patagonia. *Revista del Museo Argentino dei Ciencias Naturales, Nuevo Serie*, **7**, 37–41.

Olshevsky, G. 1991. A revision of the parainfraclass Archosauria Cope, 1869, excluding the advanced Crocodylia. *Mesozoic Meanderings*, **2**, 1–196.

Osborn, H.F. 1898. Additional characters of the great herbivorous dinosaur *Camarasaurus*. *Bulletin of the American Museum of Natural History*, **10**, 219–233.

Osborn, H.F. 1899. A skeleton of *Diplodocus*. *Memoirs of the American Museum of Natural History*, **1**, 189–214 and plates 24-28.

Osborn, H.F. 1904. Manus, sacrum and caudals of Sauropoda. *Bulletin of the American Museum of Natural History*, **20**, 181–190.

Osborn, H.F. & Mook, C.C. 1921. *Camarasaurus*, *Amphicoelias* and other sauropods of Cope. *Memoirs of the American Museum of Natural History, new series*, **3**, 247–387 and plates LX-LXXXV.

Ostrom, J.H. 1969a. A new theropod dinosaur from the Lower Cretaceous of Montana. *Postilla*, **128**, 1–17.

Ostrom, J.H. 1969b. Osteology of *Deinonychus antirrhopus*, an unusual theropod from the Lower Cretaceous of Montana. *Bulletin of the Peabody Museum of Natural History*, **30**, 1–165.

Owen, R. 1841a. *Odontography, Part II.* Hippolyte Bailliere, London.

Owen, R. 1841b. A description of a portion of the skeleton of the *Cetiosaurus*, a gigantic extinct Saurian Reptile occurring in the Oolitic formations of different portions of England. *Proceedings of the Geological Society of London*, **3**, 457–462.

Owen, R. 1842. Report on British fossil reptiles, Part II. *Reports of the British Association for the Advancement of Science*, **11**, 60–204.

Owen, R. 1859a. Monograph on the fossil Reptilia of the Wealden and Purbeck Formations. Supplement no. II. Crocodilia (*Streptospondylus*, &c.). *Palaeontographical Society Monograph*, **11**, 20–44.

Owen, R. 1859b. On the orders of fossil and recent Reptilia, and their distribution in time. *Report on the British Association for the Advancement of Science, 29th Meeting*, **1859**, 153–166.

Owen, R. 1875a. Monograph of the Mesozoic Reptilia, part 2: Monograph on the genus *Bothriospondylus*. *Palaeontolographical Society Monograph*, **29**, 15–26.

Owen, R. 1875b. Monograph of the Mesozoic Reptilia, part 2: Monograph on the genus *Cetiosaurus*. *Palaeontolographical Society Monograph*, **29**, 27–43.

Paul, G.S. 1988a. The brachiosaur giants of the Morrison and Tendaguru with a description of a new subgenus, *Giraffatitan*, and a comparison of the world's largest dinosaurs. *Hunteria*, **2**, 1–14.

Paul, G.S. 1988b. *Predatory Dinosaurs of the World.* Simon & Schuster, New York, New York.

Paul, G.S. 1998. Terramegathermy and Cope's Rule in the land of titans. *Modern Geology*, **23**, 179–217.

Phillips, J. 1871. *Geology of Oxford and the Valley of the Thames.* Clarendon Press, Oxford.

Powell, J.E. 1992. Osteología de *Saltasaurus loricatus* (Sauropoda-Titanosauridae) del Cretácico Superior del Noroeste Argentino. *In:* Sanz, J.L. & Buscalioni, A.D. (eds) *Los Dinosaurios y su Entorno Biotico. Actas del Segundo Curso de*

*Paleontologia en Cuenca*. Instituto Juan de Valdés, Ayuntamiento de Cuenca, 165–230.

Powell, J.E. 2003. Revision of South American Titanosaurid dinosaurs: palaeobiological, palaeobiogeographical and phylogenetic aspects. *Records of the Queen Victoria Museum*, **111**, 1–94.

Rauhut, O.W., Remes, K., Fechner, R., Cladera, G. & Puerta, P. 2005. Discovery of a short-necked sauropod dinosaur from the Late Jurassic period of Patagonia. *Nature*, **435**, 670–672.

Remes, K. 2006. Revision of the Tendaguru sauropod dinosaur *Tornieria africana* (Fraas) and its relevance for sauropod paleobiogeography. *Journal of Vertebrate Paleontology*, **26**, 651–669.

Remes, K. 2007. A second Gondwanan diplodocid dinosaur from the Upper Jurassic Tendaguru beds of Tanzania, East Africa. *Palaeontology*, **50**, 653–667.

Riggs, E.S. 1903a. *Brachiosaurus altithorax*, the largest known dinosaur. *American Journal of Science*, **15**, 299–306.

Riggs, E.S. 1903b. Structure and relationships of opisthocoelian dinosaurs. Part I, *Apatosaurus* Marsh. *Field Columbian Museum, Geological Series*, **2**, 165–196.

Riggs, E.S. 1904. Structure and relationships of opisthocoelian dinosaurs. Part II, the Brachiosauridae. *Field Columbian Museum, Geological Series*, **2**, 229–247.

Romer, A.S. 1956. *Osteology of the Reptiles.* University of Chicago Press, Chicago.

Russell, D.A. & Zheng, Z. 1993. A large mamenchisaurid from the Junggar Basin, Xinjiang, People's Republic of China. *Canadian Journal of Earth Sciences*, **30**, 2082–2095.

Russell, D.A., Beland, P. & McIntosh, J.S. 1980. Paleoecology of the dinosaurs of Tendaguru (Tanzania). *Memoires de la Société Geologique de France*, **139**, 169–175.

Salgado, L. & Bonaparte, J.F. 1991. Un nuevo sauropodo Dicraeosauridae, *Amargasaurus cazaui* gen. et sp. nov., de la Formacion La Amarga, Neocomiano de la Provincia del Neuquen, Argentina. *Ameghiniana*, **28**, 333–346.

Sander, P.M. 2000. Longbone histology of the Tendaguru sauropods: implications for growth and biology. *Paleobiology*, **26**, 466–488.

Sander, P.M., Mateus, O., Laven, T. & Knötschke, N. 2006. Bone histology indicates insular dwarfism in a new Late Jurassic sauropod dinosaur. *Nature*, **441**, 739–741.

Seeley, H.G. 1869. *Index to the fossil remains of Aves, Ornithosauria, and Reptilia, from the Secondary System of Strata, arranged in the Woodwardian Museum of the University of Cambridge.* Deighton, Bell, and Co., Cambridge.

Seeley, H.G. 1870. On *Ornithopsis*, a gigantic animal of the Pterodactyle kind from the Wealden. *Annals of the Magazine of Natural History, series 4*, **5**, 279–283.

Seeley, H.G. 1874. On the base of a large lacertian cranium from the Potton Sands, presumably dinosaurian. *Quarterly Journal of the Geological Society, London*, **30**, 690–692.

Sereno, P.C., Wilson, J.A., Witmer, L.M., Whitlock, J.A., Maga, A., Ide, O. & Rowe, T.A. 2007. Structural Extremes in a Cretaceous Dinosaur. *PLoS ONE*, **2**, e1230 (9 pages).

Sereno, P.C., Beck, A.L., Dutheil, D.B., Larsson, H.C.E., Lyon, G.H., Moussa, B., Sadleir, R.W., Sidor, C.A., Varricchio, D.J., Wilson, G.P. & Wilson, J.A. 1999. Cretaceous sauropods from the Sahara and the uneven rate of skeletal evolution among dinosaurs. *Science*, **282**, 1342–1347.

Sternfeld, R. 1911. Zur Nomenklatur der Gattung *Gigantosaurus* Fraas. *Sitzungsberichte der Gesellschaft Naturforschender Freunde zu Berlin*, **1911**, 398.

Stevens, K.A. & Parrish, J.M. 1999. Neck Posture and Feeding Habits of Two Jurassic Sauropod Dinosaurs. *Science*, **284**, 798–800.

Taylor, M.P. 2005. Sweet seventy-five and never been kissed: the Natural History Museum's Tendaguru brachiosaur. *In:* Barrett, P.M. (ed) *Abstracts volume for 53rd Symposium of Vertebrae Palaeontology and Comparative Anatomy*. The Natural History Museum, London, 25.

Taylor, M.P. 2006. Dinosaur diversity analysed by clade, age, place and year of description. *In:* Barrett, P.M. (ed) *Ninth international symposium on Mesozoic terrestrial ecosystems and biota, Manchester, UK.* Cambridge Publications, Cambridge, UK, 134–138.

Taylor, M.P. In press. A re-evaluation of *Brachiosaurus altithorax* Riggs 1903 (Dinosauria, Sauropoda) and its generic separation from *Giraffatitan brancai* (Janensch 1914). *Journal of Vertebrate Paleontology.*

Taylor, M.P. & Naish, D. 2007. An unusual new neosauropod dinosaur from the Lower Cretaceous Hastings Beds Group of East Sussex, England. *Palaeontology*, **50**, 1547–1564.

Tornier, G. 1909. Wie war der *Diplodocus carnegii* wirklich gebaut? *Sitzungsbericht der Gesellschaft naturforschender Freunde zu Berlin*, **4**, 193–209.

Upchurch, P. 1995. The evolutionary history of sauropod dinosaurs. *Philosophical Transactions of the Royal Society of London B*, **349**, 365–390.

Upchurch, P. 1998. Phylogenetic relationships of sauropod dinosaurs. *Zoological Journal of the Linnean Society*, **124**, 43–103.

Upchurch, P. 2000. Neck posture of sauropod dinosaurs. *Science*, **287**, 547b.

Upchurch, P. & Martin, J. 2003. The anatomy and taxonomy of *Cetiosaurus* (Saurischia, Sauropoda) from the Middle Jurassic of England. *Journal of Vertebrate Paleontology*, **23**, 208–231.

Upchurch, P., Martin, J. & Taylor, M. P. 2009. Case 3472: *Cetiosaurus* Owen, 1841 (Dinosauria, Sauropoda): proposed conservation of usage by designation of *Cetiosaurus oxoniensis* Phillips, 1871 as the type species. *Bulletin of Zoological Nomenclature*, **66**, 51–55.

Upchurch, P. & Wilson, J.A. 2007. *Euhelopus zdanksyi* and its bearing on the evolution of East Asian sauropod dinosaurs. *In:* Liston, J. (ed) *Abstracts volume for 55th Symposium of Vertebrae Palaeontology and Comparative Anatomy*. University of Glasgow, Glasgow, 30–30.

Upchurch, P., Barrett, P.M. & Dodson, P. 2004. Sauropoda. *In:* Weishampel, D.B.,

DODSON, P. & OSMÓLSKA, H. (eds) *The Dinosauria, 2nd edition*. University of California Press, Berkeley and Los Angeles, 259–322.

WATSON, J.W. 1966. *Dinosaurs and other prehistoric reptiles.* Hamlyn, London, New York, Sydney and Toronto.

WEAVER, J.C. 1983. The improbable endotherm: the energetics of the sauropod dinosaur *Brachiosaurus*. *Paleobiology*, **9**, 173–182.

WEDEL, M.J. 2003a. Vertebral pneumaticity, air sacs, and the physiology of sauropod dinosaurs. *Paleobiology*, **29**, 243–255.

WEDEL, M.J. 2003b. The evolution of vertebral pneumaticity in sauropod dinosaurs. *Journal of Vertebrate Paleontology*, **23**, 344–357.

WEDEL, M.J. 2005. Postcranial skeletal pneumaticity in sauropods and its implications for mass estimates. *In:* WILSON, J.A. & CURRY-ROGERS, K. (eds) *The Sauropods: Evolution and Paleobiology*. University of California Press, Berkeley, 201–228.

WEDEL, M.J. 2006. Origin of postcranial skeletal pneumaticity in dinosaurs. *Integrative Zoology*, **2**, 80–85.

WEDEL, M.J., CIFELLI, R.L. & SANDERS, R.K. 2000a. *Sauroposeidon proteles*, a new sauropod from the Early Cretaceous of Oklahoma. *Journal of Vertebrate Paleontology*, **20**, 109–114.

WILD, R. 1991. *Janenschia* n. g. *robusta* (E. Fraas 1908) pro *Tornieria robusta* (E. Fraas 1908) (Reptilia, Saurischia, Sauropodomorpha). *Stuttgarter Beiträge zur Naturkunde, Serie B (Geologie und Paläontologie)*, **173**, 1–4.

WILSON, J.A. 2002. Sauropod dinosaur phylogeny: critique and cladistic analysis. *Zoological Journal of the Linnean Society*, **136**, 217–276.

WILSON, J.A. & SERENO, P.C. 1998. Early evolution and higher-level phylogeny of sauropod dinosaurs. *Journal of Vertebrate Paleontology*, Memoir **5**, 1–68.

WILSON, J.A. & UPCHURCH, P. 2003. A revision of *Titanosaurus* Lydekker (Dinosauria – Sauropoda), the first dinosaur genus with a 'Gondwanan' distribution. *Journal of Systematic Palaeontology*, **1**, 125–160.

Wilson, J.A. & Upchurch, P. In press. Redescription and reassessment of the phylogenetic affinities of *Euhelopus zdanskyi* (Dinosauria: Sauropoda) from the Early Cretaceous of China. *Journal of Systematic Palaeontology*.

Wiman, C. 1929. Die Kreide-Dinosaurier aus Shantung. *Palaeontologia Sinica (Series C)*, **6**, 1–67 and plates 1–9.

Woodward, A.S. 1910. On a skull of *Megalosaurus* from the Great Oolite of Minchinhampton (Gloucestershire). *Quarterly Journal of the Geological Society of London*, **66**, 111–115.

Young, C.-C. 1937. A new dinosaurian from Sinkiang. *Palaeontologia Sinica*, Series C, **1937.2**, 1–25,

Young, C.-C. 1939. On a new Sauropoda, with notes on other fragmentary reptiles from Szechuan. *Bulletin of the Geological Society of China*, **19**, 279–315.

Young, C.-C. 1954. On a new sauropod from Yiping, Szechuan, China. *Acta Scientia Sinica*, **3**, 491–504.

Young, C.-C. & Zhao, X. 1972. [*Mamenchisaurus*. In Chinese: description of the type material of *Mamenchisaurus hochuanensis*]. *Institute of Vertebrate Paleontology and Paleoanthropology Monograph Series I*, **8**, 1–30.

**Table 1**. First sauropods named from each continent.

| **Continent** | **First named genus**<br>**Earliest still valid** | **Author and date** | **Clade** |
|---|---|---|---|
| Europe | *Cardiodon*[1] | Owen (1841a) | ?Cetiosauridae |
| | *Cetiosaurus* | Owen (1841b) | Cetiosauridae |
| North America | *Astrodon* | Johnston (1859) | Titanosauriformes |
| Asia | *Titanosaurus*[2] | Lydekker (1877) | Titanosauria |
| | *Tienshanosaurus*[3] | Young (1937) | Eusauropoda |
| South America | *Argyrosaurus* | Lydekker (1893) | Titanosauria |
| Africa | *Algoasaurus*[4] | Broom (1904) | Sauropoda |
| | *Tornieria* | Sternfeld (1911) | Diplodocinae |
| Australasia | *Rhoetosaurus* | Longman (1926) | Sauropoda |
| Antarctica | (None named) | | |

[1]The type specimen of *Cardiodon* is lost, and the referred specimen is not diagnosable.

[2]*Titanosaurus* was diagnosed by a character that now characterises the large clade Titanosauria (see text).

[3]The Chinese genus *Helopus* Wiman 1929 predates *Tienshanosaurus*, but because the name *Helopus* was preoccupied by a bird, the genus was renamed *Euhelopus* Romer 1956.

[4]*Algosaurus* is not diagnosable.

**Table 2**. Changing mass estimates for *Brachiosaurus brancai*.

| **Author and date** | **Method** | **Volume (l)** | **Density (kg/l)** | **Mass (kg)** |
|---|---|---|---|---|
| Janensch 1938 | Not specified | — | — | “40 t” |
| Colbert 1962 | displacement of sand | 86953 | 0.9 | 78258 |
| Russell *et al*. 1980 | limb-bone allometry | — | — | 13618[1] |
| Anderson *et al*. 1985 | limb-bone allometry | — | — | 29000 |
| Paul 1988a | displacement of water | 36585 | 0.861 [2] | 31500 |
| Alexander 1989[3] | weighing in air and water | 46600 | 1.0 | 46600 |
| Gunga *et al*. 1995 | computer model | 74420 | 1.0 | 74420 |
| Christiansen 1997 | weighing in air and water | 41556 | 0.9 | 37400 |
| Henderson 2004 | computer model | 32398 | 0.796 | 25789 |
| Henderson 2006 | computer model | — | — | 25922 |
| Gunga *et al*. 2008 | computer model | 47600 | 0.8 | 38000 |
| Taylor in press | graphic double integration | 29171 | 0.8 | 23337 |

[1]Russell et al. give the mass as “14.9t”, which has usually been interpreted as representing metric tonnes, e.g. 14900 kg. However, they cite “the generally accepted figure of 85 tons” (p. 170), which can only be a reference to Colbert (1962). Colbert stated a mass of 85.63 U.S. tons as well as the metric version, so we must assume that Russell et al. were using U.S. tons throughout.

[2]Paul used a density of 0.9 kg/l for most of the model, and 0.6 kg/l for the neck, which was measured separately and found to constitute 13% of the total volume, yielding an aggregate density of (0.9 × 87%) + (0.6 × 13%) = 0.861 kg/l.

[3]Alexander did not state which Brachiosaurus species his estimate was for, only that it was based on the BMNH model. This model is simply stamped “*Brachiosaurus*”.

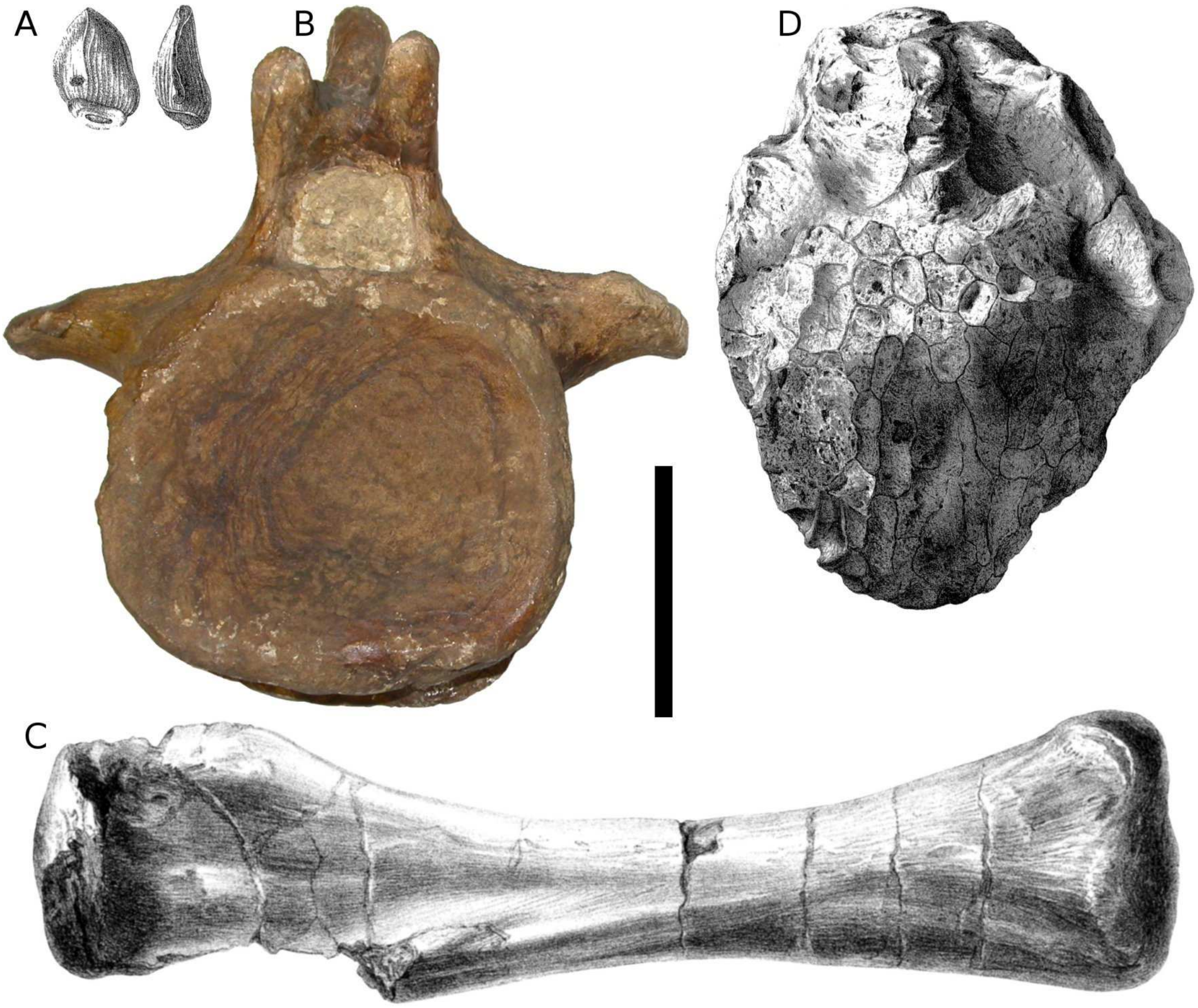

**Fig. 1.** Historically significant isolated sauropod elements. (**a**) the holotype tooth of *Cardiodon* in labial and distal views, modified from Owen (1875a, plate IX, fig. 2–3); (**b**), anterior caudal vertebra of *Cetiosaurus brevis* in anterior view, part of the holotype, photograph by author; (**c**), holotype right humerus of *Pelorosaurus* in anterior view, modified from Mantell (1850, plate XXI, fig. 1b); (**d**), lectotype dorsal vertebra of *Ornithopsis* (see Blows 1995, p. 188) in anterior view, exposing pneumatic cavities due to erosion of the anterior articular surface, modified from Owen (1875a, plate IX, fig. 1). Scale bar is 5 cm for (**a**), 10 cm for (**b**) and (**d**), and 30 cm for (**c**).

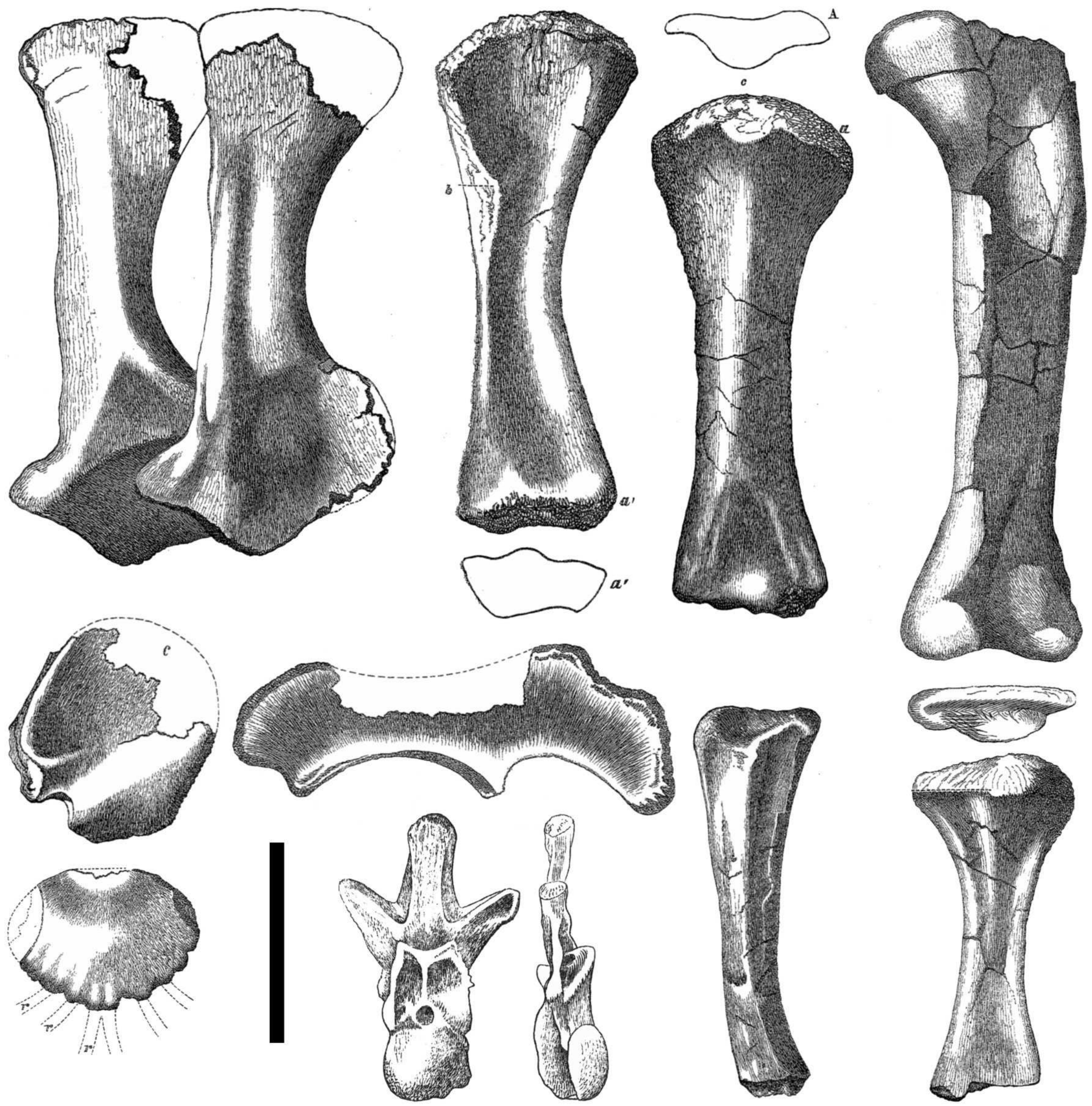

**Fig. 2.** Elements of *Cetiosaurus oxoniensis*. Top row, left to right: right scapula in lateral view and left scapula in medial view; right humerus in anterior and distal views, and left humerus in proximal and posterior views; left femur in anterior view. Bottom row, left to right: left coracoid in medial view and ?left sternal plate in ?dorsal view; right ilium in lateral view and ?fourth dorsal vertebra in anterior and right lateral views; ?right ulna in ?posterolateral view; right tibia in proximal and posterolateral views. Dorsal vertebra modified from Phillips (1871, fig. 86), other elements modified from Owen (1875b, fig. 1-9), which were reproduced from Phillips (1871). Scale bar is 50 cm.

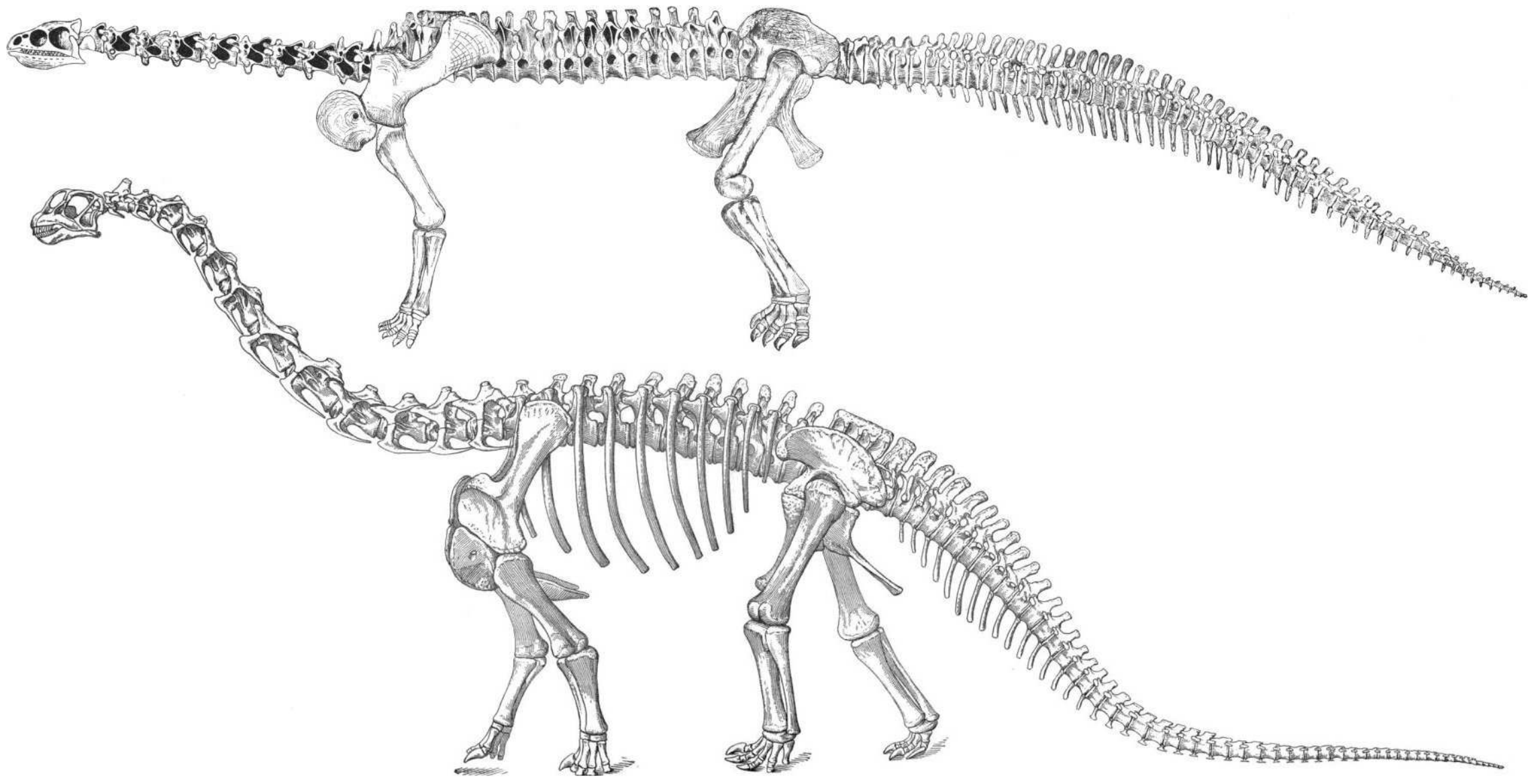

**Fig. 3.** Early reconstructions of *Camarasaurus*. Top, Ryder's 1877 reconstruction, the first ever made of any sauropod, modified from Osborn & Mook (1921, plate LXXXII); bottom, Osborn and Mook's own reconstruction. modified from Osborn & Mook (1921, plate LXXXIV).

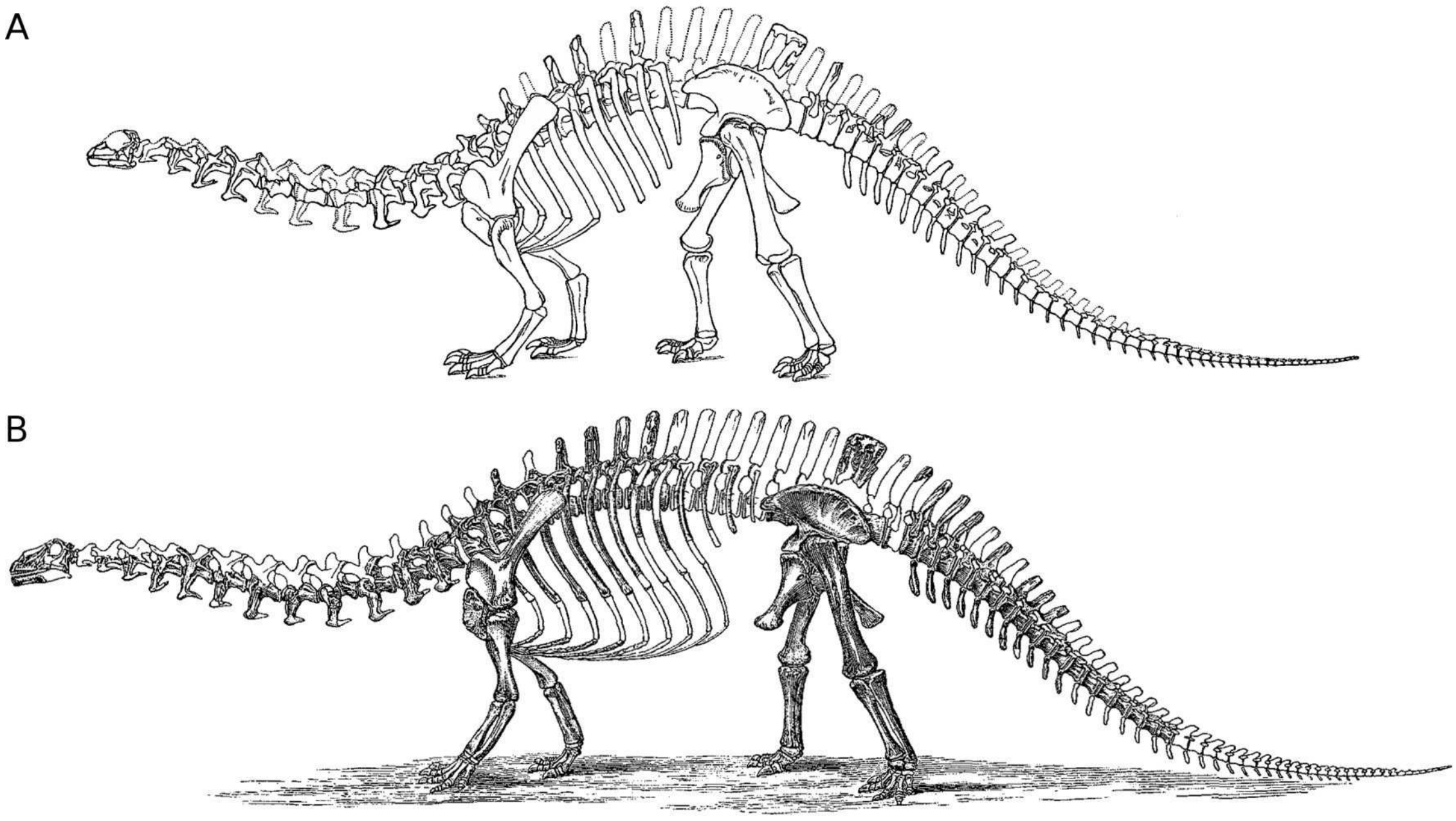

**Fig. 4.** Marsh's reconstructions of "*Brontosaurus*" (now *Apatosaurus*). Top, first reconstruction, modified from Marsh (1883, plate I); bottom, second reconstruction, modified from Marsh (1891, plate XVI).

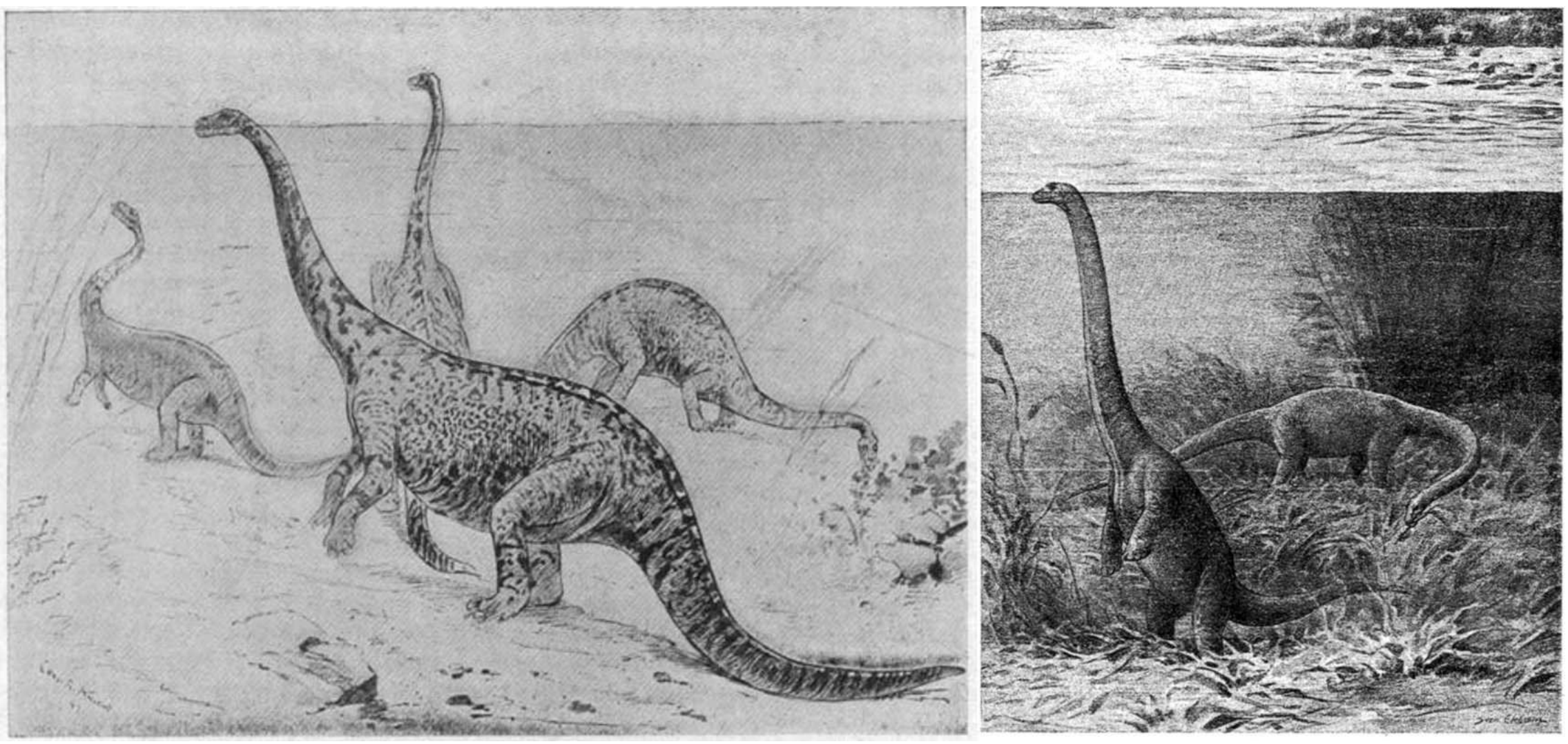

**Fig. 5.** Snorkelling sauropods. Left, the first ever life restoration of a sauropod, Knight's drawing of *Amphicoelias*, published by Ballou (1897), modified from Osborn & Mook (1921, fig. 127); right, a similar scene with "*Helopus*" (now *Euhelopus*), modified from Wiman (1929, fig. 5).

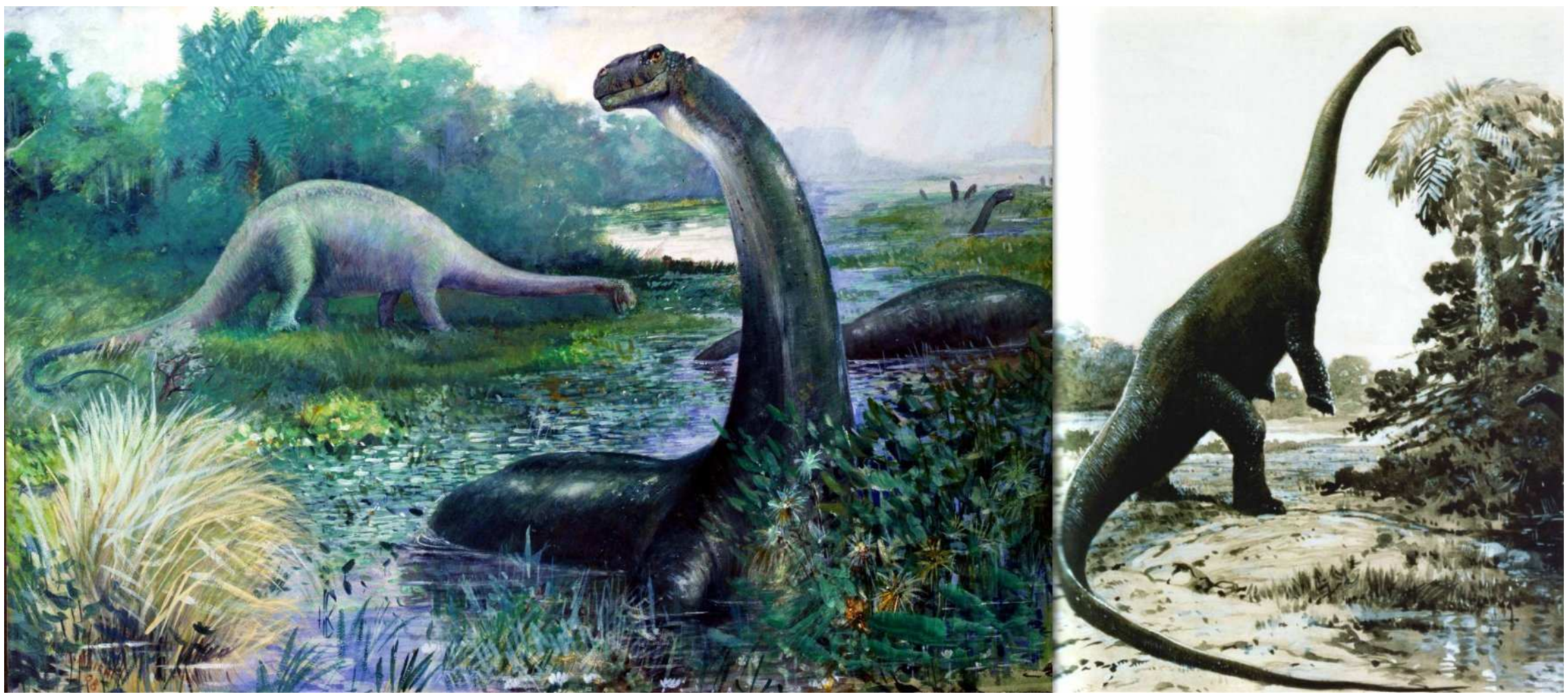

**Fig. 6.** Two classic sauropod paintings by Knight. Left, swamp-bound "*Brontosaurus*" (now *Apatosaurus*), painted in 1897, with static terrestrial *Diplodocus* in background. Right, athletic *Diplodocus*, painted in 1907.

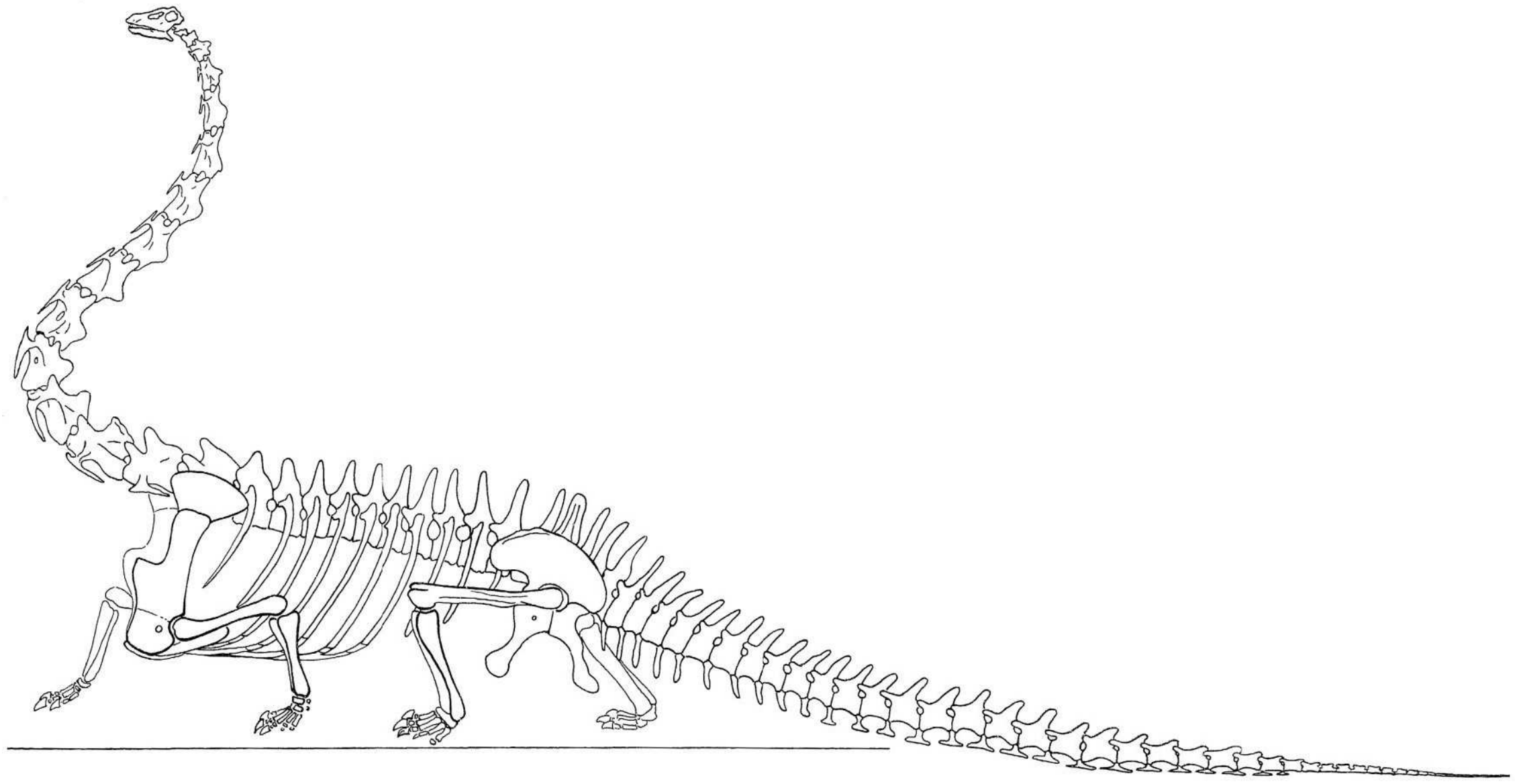

**Fig. 7.** Tornier's sprawling, disarticulated reconstruction of *Diplodocus*, modified from Tornier (1909, plate II).

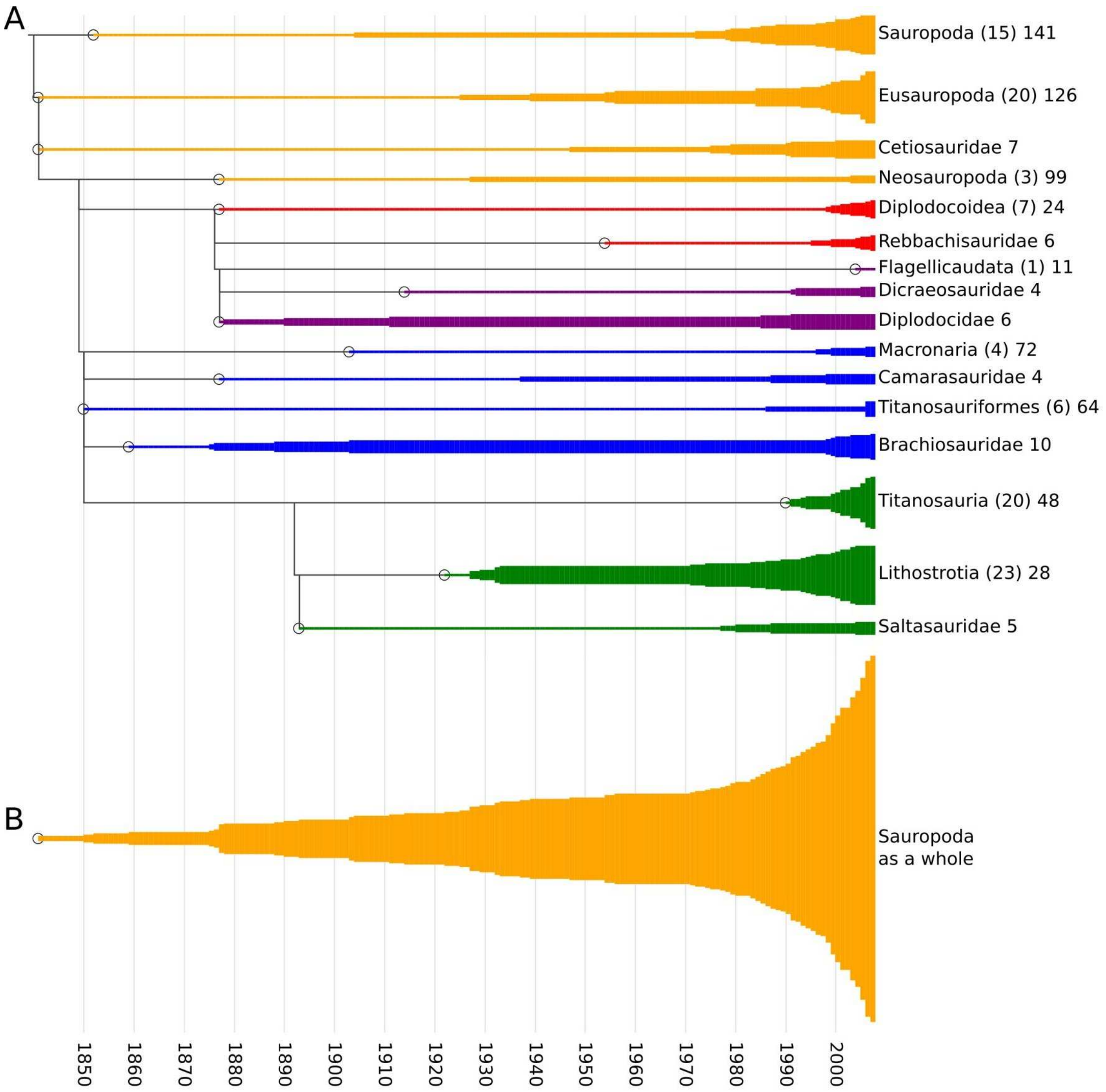

**Fig. 8.** Growing recognition of sauropod diversity through history. Only genera now considered valid are included. A, broken down by clade. Vertical thickness of lines is proportional to number of genera; earliest valid genus in each clade is marked by circle. Terminal clades have simple counts; for non-terminal clades, parentheses enclose the number of basal genera, i.e. not members of depicted subclades, and are followed by total counts which include those of all subclades. B, total recognised diversity.

Chapter 2 follows. This paper has been accepted and is in press at the Journal of Vertebrate Paleontology, published by The Society of Vertebrate Paleontology.

# A RE-EVALUATION OF *BRACHIOSAURUS ALTITHORAX* RIGGS 1903 (DINOSAURIA, SAUROPODA) AND ITS GENERIC SEPARATION FROM *GIRAFFATITAN BRANCAI* (JANENSCH 1914)

MICHAEL P. TAYLOR

Palaeobiology Research Group, School of Earth and Environmental Sciences,
University of Portsmouth, Burnaby Road, Portsmouth PO1 3QL, United Kingdom,
dino@miketaylor.org.uk

ABSTRACT—Although the macronarian sauropod *Brachiosaurus* is one of the most iconic dinosaurs, its popular image is based almost entirely on the referred African species *Brachiosaurus brancai* rather than the North American type species *Brachiosaurus altithorax*. Reconsideration of Janensch's referral of the African species to the American genus shows that it was based on only four synapomorphies and would not be considered a convincing argument today. Detailed study of the bones of both species show that they are distinguished by at least 26 characters of the dorsal and caudal vertebrae, coracoids, humeri, ilia, and femora, with the dorsal vertebrae being particularly different between the two species. These animals must be therefore be considered generically separate, and the genus name *Giraffatitan* Paul 1988 must be used for "*Brachiosaurus*" *brancai*, in the combination *Giraffatitan brancai*. A phylogenetic analysis treating the two species as separate OTUs nevertheless recovers them as sister taxa in all most parsimonious trees, reaffirming a monophyletic Brachiosauridae, although only one additional step is required for *Giraffatitan* to clade among somphospondylians to the exclusion of *Brachiosaurus*. The American *Brachiosaurus* is shown to be somewhat different from *Giraffatitan* in overall bodily proportions: it had a longer and deeper trunk and probably a longer and taller tail, carried a greater proportion of its mass on the forelimbs, and may have had somewhat sprawled forelimbs. Even though it was overall a larger animal than the *Giraffatitan* lectotype, the *Brachiosaurus* holotype was probably immature, as its coracoids were not fused to its scapulae.

## INTRODUCTION

The sauropod dinosaur *Brachiosaurus* Riggs, 1903 is one of the most iconic of all prehistoric animals, immediately recognizable by its great size, tall shoulders, long neck and helmet-like skull. However, much of the distinctive morphology attributed to *Brachiosaurus* is known only from the referred species, *B. brancai* Janensch, 1914, and not from the type species *B. altithorax* Riggs, 1903. That *B. brancai* belongs to *Brachiosaurus* was asserted but not convincingly demonstrated by Janensch (1914), and contradicted but not disproved by Paul (1988). This study reviews the history of the two species, assesses the similarities and differences between them, assesses their relationships within a broader phylogenetic context, and discusses the implications for the phylogenetic nomenclature of sauropods.

**Anatomical Nomenclature—**The term Gracility Index (GI) is introduced to quantify the gracility of the humeri and other long bones discussed in this study, and is defined as the ratio between the proximodistal length of the bone and its minimum transverse width.

Many different sets of directions have been used to describe sauropod coracoids, with the edge furthest from the scapular articulations having been variously described as median (e.g., Seeley, 1882), inferior (Riggs, 1904), anteromedial (Powell, 1992), distal (Curry Rogers, 2001) and anterior (Upchurch et al., 2004), and the designation of the other directions varying similarly. I follow Upchurch et al. (2004) in describing the coracoid as though the scapulocoracoid were oriented horizontally: the scapular articular surface is designated posterior, so that the glenoid surface of the coracoid is considered to face posteroventrally.

Nomenclature for vertebral laminae follows that of Wilson (1999).

**Anatomical Abbreviations—ACDL**, anterior centrodiapophyseal lamina; **PCPL**, posterior centroparapophyseal lamina; **PODL**, postzygadiapophyseal lamina; **PPDL**, paradiapophyseal lamina; **PRPL**, prezygaparapophyseal lamina; **SPPL**, spinoparapophyseal lamina.

**Institutional Abbreviations—BMNH**, Natural History Museum, London, United

Kingdom; **BYU**, Brigham Young University, Provo, Utah; **FMNH**, Field Museum of Natural History, Chicago, Illinois; **HMN**, Humboldt Museum für Naturkunde, Berlin, Germany; **OMNH**, Oklahoma Museum of Natural History, Norman, Oklahoma; **USNM**, National Museum of Natural History, Smithsonian Institution, Washington D.C.

## HISTORICAL BACKGROUND

### Initial finds

The type species of the genus *Brachiosaurus* is *Brachiosaurus altithorax*, founded on a partial skeleton collected from the Grand River valley of western Colorado by the Field Columbian Museum paleontological expedition of 1900 under the leadership of Elmer S. Riggs (now accessioned as specimen FMNH P 25107). It comprises the last seven dorsal vertebrae, sacrum, the first two caudal vertebrae (one in very poor condition), left coracoid, right humerus, ilium and femur, fragmentary left ilium, and dorsal ribs. The type specimen does not contain any material from the skull, neck, anterior dorsal region, median or posterior parts of the tail, distal parts of the limbs or feet; nor has such material been confidently referred to the species (although see below).

*Brachiosaurus altithorax* was first reported, unnamed, by Riggs (1901), and subsequently named and briefly described on the basis of some but not all elements, the dorsal vertebrae not yet having been prepared (Riggs, 1903). After preparation was complete, the material was more fully described and figured in a monograph on the new family Brachiosauridae (Riggs, 1904). Riggs (1903:299, 1904:230) assigned the coracoid to the right side, but it is from the left: the orientation of the scapular margin indicates that the coracoid as figured by Riggs (1903:fig. 3, 1904:pl. LXXV, fig. 4) and as displayed in the FMNH collection is either a right coracoid in medial view or a left coracoid in lateral view; and the posteromedial-anterolateral orientation of the coracoid foramen and lateral flaring of the bone to support the glenoid articular surface (observed in photos provided by Phil Mannion) show that it is the latter.

*Brachiosaurus* is also known from a second species, *Brachiosaurus brancai*, excavated from Tendaguru in Tanzania by the German expeditions of 1909-1912 (Maier, 2003). Janensch (1914) initially also named a second Tendaguru species, *Brachiosaurus* *fraasi* Janensch, 1914, but subsequently synonymized this species with *B. brancai* (Janensch, 1929:5, 1935-1936:153, 1950a:31). Unlike the type species, *B. brancai* is known from many specimens of varying degrees of completeness, in total including almost all skeletal elements. The original type specimen, "Skelett S" (Janensch, 1914:86) was subsequently found (e.g., Janensch, 1929:8) to consist of two individuals, which were designated SI (the smaller) and SII (the larger and more complete). Janensch never explicitly designated these two specimens as a syntype series or nominated either specimen as a lectotype; I therefore propose HMN SII as the lectotype specimen of *Brachiosaurus brancai*. The skull, with its distinctive nasal arch, and the very long neck, are known only from *B. brancai*, and it is primarily from this species that nearly all previous skeletal reconstructions and life restorations have been executed, beginning with that of Matthew (1915: fig. 24). The sole exception is the partial reconstruction that Paul (1998: fig. 1B) included in a montage of skeletal reconstructions.

Janensch provided comprehensive descriptions of the Tendaguru elements in a series of monographs on the manus (Janensch, 1922), skull (Janensch, 1935-1936), axial skeleton (Janensch, 1950a) and appendicular skeleton (Janensch, 1961), as well as a discussion of pneumatic structures in the vertebrae (Janensch, 1947) and an account of the reconstruction of the mounted skeleton (Janensch, 1950b). In consequence, *B. brancai* is the most comprehensively described of all sauropods, although the papers are not widely read as they were written in High German.

## Additional Material

**Migeod's Tendaguru Brachiosaur—**After the German Tendaguru expeditions were ended by the First World War, Tanzania became a British territory, and a series of expeditions were sent to Tendaguru by the British Museum (Natural History) (now the Natural History Museum) from 1919 to 1931 (Maier, 2003). Although the British were in Tendaguru for much longer than the Germans had been, their expeditions were under-funded and lacked the excellent scientific leadership of the earlier efforts. As a consequence, most of the material recovered by the British was unimpressive, consisting only of disarticulated elements. The sole exception was a nearly complete brachiosaurid sauropod skeleton, BMNH R5937, collected by F. W. H. Migeod in the 1930 field season. Migeod (1931) briefly reported on this specimen, which was said to include a complete and mostly articulated set of vertebrae from the fifth cervical through to the ninth caudal, together with cervical and dorsal ribs. Other material considered to be part of this specimen included three teeth, a scapulocoracoid, two humeri, an ilium, a partial pubis, a broken ischium, an incomplete femur, parts of a second femur and a calcaneum. Unfortunately, the association of some of this material was uncertain, much of it appears to have been lost, and more remains unprepared, although further preparation work is now under way. The preservation of the prepared material varies considerably: a pair of posterior dorsal vertebrae are in excellent condition, while most cervical vertebrae are lacking nearly all processes and lamina. An initial assessment of the material indicates that it probably represents a second distinct Tendaguru brachiosaur (Taylor, 2005).

Apart from the specimens recovered by the German expeditions, Migeod's specimen is the only Tendaguru brachiosaur material to have been been reported. However, several later finds of sauropod material in the U.S.A. have been referred, with varying degrees of certainty, to *Brachiosaurus*.

**Potter Creek Humerus—**As recounted by Jensen (1985, 1987), Eddie and Vivian Jones collected a large left humerus from the Uncompahgre Upwarp of Colorado and donated it to the Smithsonian Institution where it is accessioned as USNM 21903. It was designated *Brachiosaurus* (Anonymous, 1959) although no reason for this assignment was published; it was subsequently described very briefly and inadequately by Jensen (1987:606-607). Although its great length of 213 cm (pers. obs.) is compatible with a brachiosaurid identity, it is in some other respects different from the humeri of both *B. altithorax* and *B. brancai*, although some of these differences may be due to errors in the significant restoration that this element has undergone. The bone may well represent *Brachiosaurus altithorax*, but cannot be confidently referred to this species, in part because its true proportions are concealed by restoration (Wedel and Taylor, in prep.). It can therefore be discounted in terms of contributing to an understanding of the relationship between *B. altithorax* and *B. brancai*.

**Other Potter Creek Material**—Further brachiosaurid material was recovered from the Potter Creek quarry in 1971 and 1975 (Jensen, 1987:592-593), including a mid-dorsal vertebra, incomplete left ilium, left radius and right metacarpal. This material is accessioned as BYU 4744 (BYU 9754 of Jensen's usage). The material that overlaps with that of the *B. altithorax* type specimen appears very similar to it, and can be confidently assigned to that species. Preservation is supposedly very good (Jensen, 1987:599), but because the material was restored before figuring, its quality is difficult to assess. Further study is needed.

**Dry Mesa Material**—Jensen (1985) described further brachiosaurid material from the Dry Mesa quarry, erecting the new genus and species *Ultrasaurus macintoshi* Jensen, 1985 to receive it. It subsequently became apparent that Kim (1983), seemingly unaware of Jensen's informal prior use of the name "*Ultrasaurus*," had used this name for an indeterminate Korean sauropod which therefore has priority. Olshevsky (1991) therefore proposed the replacement genus name *Ultrasauros*, and it is this spelling that will be used herein. The type specimen of *U. macintoshi* is the dorsal vertebra BYU 9044 (BYU 5000 of Jensen's usage); referred specimens included a mid-cervical vertebra BYU 9024 (BYU 5003 of Jensen's usage), an anterior caudal vertebra BYU 9045 (BYU 5002 of Jensen's usage) and a scapulocoracoid BYU 9462 (BYU 5001 of Jensen's usage). Jensen (1987:603) subsequently asserted that the scapulocoracoid was the *U. macintoshi* holotype, but the original designation must stand. Jensen (1987:602) also recovered a large rib, which he considered to belong to *Brachiosaurus* (Jensen, 1987: caption to fig. 1). Unfortunately, little of Jensen's *Ultrasauros* material is actually brachiosaurid. Jensen (1987:600-602) recognized that the cervical vertebra, having a bifid neural spine, could not be brachiosaurid and instead tentatively referred it to Diplodocidae. Curtice (1995) subsequently referred the caudal vertebra to *Supersaurus* Jensen, 1985, leaving only the type dorsal and the scapulocoracoid. Curtice et al. (1996:88) asserted incorrectly that Jensen (1987) had referred the cervical specifically to *Supersaurus* rather than more generally to Diplodocidae, and this identification has been followed subsequently (e.g., Curtice and Stadtman, 2001; Wedel, 2006). Most importantly, Curtice et al. (1996) demonstrated that type type specimen of *U. macintoshi*, the dorsal vertebra, was not an anterior dorsal from a brachiosaurid as Jensen had believed, but a posterior dorsal from a diplodocid. Curtice et al.(1996) referred this specimen, too, to *Supersaurus*, making *Ultrasauros* a junior subjective synonym of that name. The result of this is that only the scapulocoracoid BYU 9462 is recognized as brachiosaurid. Curtice et al. (1996:95) referred this element to *Brachiosaurus* sp., citing the narrow scapular neck, distal blade expansion and irregular shape of the coracoid as brachiosaurid characters (Curtice et al., 1996:93), and Paul, (1988:6-7) referred it specifically to *B. altithorax*. Its coracoid, however, does not closely resemble that of the *B. altithorax* holotype, lacking the the latter's distinctively strong lateral deflection of the glenoid. Neither is the scapula very similar to that of *B. brancai*, having a less pronounced acromion process – compare Curtice et al. (1996:fig. 1a) with Janensch (1961:pl. XV figs. 1 and 3a). As shown by Curtice et al. (1996:table 1), the coracoid of the "*Ultrasauros*" scapulocoracoid is smaller in both length and breadth than that of the *Brachiosaurus altithorax* holotype FMNH P 25107 (Riggs 1904:241); so the Dry Mesa brachiosaur, often cited as unusually large, was most likely rather smaller than the holotype. In conclusion, none of the Dry Mesa material described by Jensen can be confidently referred to *Brachiosaurus altithorax*.

Curtice and Stadtman (2001) briefly described BYU 13023, a pair of articulated dorsal vertebrae from Dry Mesa. They referred them to *Brachiosaurus ?altithorax*, and figured one of the pair in anterior, right lateral and dorsal views (Curtice and Stadtman, 2001:fig. 1B, 2C, 5B). The figured vertebra resembles those of the *B. altithorax* type specimen in general construction and lamina topology, but is proportionally very short anteroposteriorly: total height is about 4.2 times centrum length (including condyle), compared with values of no more than 2.2 in the *B. altithorax* holotype. This discrepancy might be accounted for by anteroposterior crushing, but since the diapophysis appears unaffected and there is no shearing, this seems unlikely. Therefore, a species-level referral cannot be confidently supported.

**Jensen/Jensen Material**—Jensen (1987:594-595) very briefly reported "several brachiosaur elements including a rib 2.75m (9 ft) long ..., a distal cervical vertebra, the proximal half of a scapula, and a coracoid" from a locality near Jensen, Utah, but did not describe any of this material and figured only a cast of the rib. The cervical vertebra, if correctly identified, would be particularly significant due to the paucity of North American brachiosaur cervical material.

**Felch Quarry Skull**—In 1883, a large sauropod skull (81 cm in length) was found in Felch Quarry 1, Garden Park, Colorado. It was shipped to O. C. Marsh in Yale that year and an illustration of the skull was used in the restoration of *Brontosaurus* Marsh, 1879 (= *Apatosaurus* Marsh, 1877) (Marsh, 1891: pl. 16). The skull was subsequently transferred to the National Museum of Natural History, where it was accessioned as USNM 5730. McIntosh and Berman (1975:195-198) recognized that the skull did not pertain to *Apatosaurus*, but described it as being "of the general *Camarasaurus* [Cope, 1877] type" (p. 196). McIntosh subsequently identified the skull tentatively as *Brachiosaurus* (Carpenter and Tidwell, 1998:70) and it was later described by Carpenter and Tidwell (1998), who considered it intermediate between the skulls of *Camarasaurus* and *Brachiosaurus brancai*, and referred it to *Brachiosaurus* sp. The skull may be that of *B. altithorax*, but this is currently impossible to test due to the lack of comparable parts (Carpenter and Tidwell, 1998:82).

Near this skull was a 99 cm cervical vertebra, probably of *Brachiosaurus*, but this was destroyed during attempts to collect it (McIntosh and Berman, 1975:196).

**OMNH Metacarpal**—Bonnan and Wedel (2004) described an isolated metacarpal, OMNH 01138, from Kenton Pit 1, Cimarron County, Oklahoma. This element, previously believed to belong to *Camarasaurus*, was referred to *Brachiosaurus* sp. on the basis of its elongation and slenderness.

**BYU Cervicals**—Cervical vertebrae in the BYU collection have been identified as *Brachiosaurus*, and found indistinguishable from those those of *B. brancai* (Wedel et al., in prep.). Two of these vertebrae, BYU 12866 and 12867, were figured by Wedel et al. (2000:fig. 10D, E, 12A-D) and Wedel (2005:fig. 7.2A). These may be the cervicals of *B. altithorax*, or may represent an as-yet unrecognized form more closely related to *B. brancai*.

Besides the material discussed here, Foster (2003:23) briefly reported a *Brachiosaurus* caudal vertebra from the Freezeout Hills of Wyoming, and Turner and Peterson (1999) mentioned, without discussion, *Brachiosaurus* material from Lower Split Rock Site 1, Mesa County, Colorado (p. 109), Callison's Quarries and Holt's Quarry, both Mesa County, Colorado (p. 110) and Bone Cabin Quarry E, Albany County, Wyoming (p. 144). Further North American specimens of *Brachiosaurus* remain for the moment unavailable, being unprepared, unpublished, or privately held. Discounting these unavailable specimens, very little of the available North American brachiosaur material can be confidently identified as *B. altithorax*, due to the absence of articulated or even associated elements. The Potter Creek radius and metacarpal may perhaps be considered to belong to this species, but their association with elements that overlap with the type material is not made clear in the publications that describe them (Jensen, 1985, 1987). In conclusion, comparisons of *B. altithorax* with the African brachiosaur material can only be safely made on the basis of the type specimen FMNH P 25107 described by Riggs (1904).

**Janensch's Referral of *B. brancai* to *Brachiosaurus***

Although Janensch corresponded extensively with palaeontologists around the world, including America, there is no record that he ever visited America (G. Maier, pers. comm., 2007), so he would never have seen the *Brachiosaurus altithorax* type material. Therefore his referral of the Tendaguru brachiosaur material to this genus was based exclusively on the published literature – and perhaps private correspondence, although I have not been able to locate any.

The basis of Janensch's initial referral of his two new species to *Brachiosaurus* was not explicit: “Both species are so close to the genus *Brachiosaurus*, so far as the present state of preparation allows a judgement, that there was no recognizable reason to hold them separate from *Brachiosaurus*” (Janensch, 1914:83). [Here and elsewhere, quotes from Janensch are in English-language translations provided by Gerhard Maier.] This was elaborated as follows: “The referral here of both species to the American genus *Brachiosaurus* Riggs will be based on the description of *B. Fraasii* [sic] below” (p. 94). “All the relationships of the humerus of *Brachiosaurus altithorax* ... are very similar to our species” (p. 97), although “the width at the proximal end of [the humerus of] our species is indeed relatively still somewhat larger than in the American sauropods. Above all, the contour of the proximal end is different in so far as it ascends sharply medially from the lateral side” (p. 97). “A left ilium was found with Skeleton J from the Upper Saurian Marl, which resembles to quite an extraordinary degree that of *Brachiosaurus altithorax* ... A caudal vertebra of the same skeleton exhibits exactly the same form as that of the second caudal vertebra of the American species ... This similarity of the ilium and the caudal vertebra further render it quite likely that the species under consideration cannot be generically separated from *Brachiosaurus*” (p. 97-98). Finally, “A comparison of the East African forms with that of *Brachiosaurus altithorax* Riggs allows very major similarities to be recognized, as cited above, particularly in the description of the individual skeletal elements. This is valid above all for the dorsal vertebrae of the American sauropods and those of *Br. Brancai* [sic, here and elsewhere]. ... The similarity in relation to the humerus is particularly great between *Br. altithorax* and *Br. Fraasi* [sic, here and elsewhere]; the ilium in these two species has a nearly entirely identical form. Furthermore, the caudal vertebrae of all three species are very similar. Finally, the agreement in the enormous dimensions, which exists especially between *Br. altithorax* and *Br. Brancai*, can also be cited. For all these reasons, it did not appear to me to be justified to hold the two described East African species under consideration generically separate from the cited North American genus" (p. 98).

Because this assessment did not describe specific derived characters shared between the Tendaguru forms and *Brachiosaurus altithorax*, it would not be considered a valid justification for the referral if published today. Lull (1919:42) commented that "Unless the German author, Janensch, actually made a comparison of the dorsals of the Tendaguru genus with those of the American *Brachiosaurus* and found sufficient agreement, I see no reason for including the African form in this genus merely on the ground of the elongated fore limbs, as we have no reason to know that *Brachiosaurus* had huge cervical vertebrae," although he noted that "Further evidence from Berlin, if such were available, might serve ... to clarify the relationships."

Janensch (1929:20) made a more specific comparison: "The contrasting condition of ... particularly low neurapophyses of the anterior caudal vertebrae is found in the genus *Brachiosaurus*, and indeed is in complete agreement with the American species *Br. altithorax* Riggs (1904 Pl. 75 Fig. 1, 2) and the East African *Br. Brancai* JAN (including *Br. Fraasi* JAN), as the illustration (Fig. 15) shows. The harmony stressed above also exists, in *Brachiosaurus*, in the low height of the neurapophyses of the anterior caudal vertebrae of this genus, and in those of the sacrum."

The subsequent monograph on the axial skeleton of *B. brancai* (Janensch, 1950a) provided a more rigorous justification for the referral: "The dorsal vertebrae of the African *Brachiosaurus brancai* correspond extensively to those of *Brachiosaurus altithorax* ... The vertebrae in the two species exhibit extensive pleurocentral excavations and undivided, dorsally widened neurapophyses, which are relatively low in the posterior dorsal vertebrae, but which become taller from the sacrum up to just before the mid-trunk; in addition there are horizontally or almost horizontally oriented diapophyses that are of considerable size prior to the mid-trunk. The considerable increase in the height of the neurapophysis from the sacrum to just before mid-trunk is a characteristic that is found in no other sauropod genus in the same manner; it is also particularly characteristic for *Brachiosaurus*" (p. 72), although "differences between both species can be confirmed, that concern the overall morphology. Thus the centra of the dorsal vertebrae of *B. altithorax* are noticeably longer. In *B. brancai* the neurapophysis and the entire vertebra of what is probably the eighth-last presacral vertebra is taller and the diapophyses longer than in the seventh-last presacral vertebra of *B. altithorax*" (p. 72). Features of the sacrum also contributed to the referral: "The extensive, triangular first sacral rib is completely similar in both species. The long extension by which the sacral rib of the second sacral vertebrae attaches to the first and second centrum is also to be found in the American forms and indeed apparently somewhat more so. The characteristically great length of the transverse processes, that confers the sacrum its significant width in comparison to other genera, is again conformable" (p. 76). The caudal vertebrae were also mentioned: "In its construction the second caudal vertebra of *B. altothorax* [sic] that Riggs (1904) illustrated resembles the corresponding vertebra of *Br. brancai* extraordinarily," although in *B. altithorax*, "a lateral depression is not indicated ... the neurapophysis is particularly thickened block-like dorsally, and ... the wedge at the ventral end of the postzygapophyses has a stronger zygosphenal character" (p. 76). Finally, Janensch drew attention to pneumaticity in the ribs of both species: "Cavernous construction can be confirmed in the head of the most robust dorsal ribs ... I interpret these depressions as manifestations that developed through the formative pressure of air sacs. In *Brachiosaurus altithorax* (Riggs 1904) a large foramen even sits in the upper section of the shaft, which leads to an internal cavity and is to be interpreted as pneumatic" (p. 87), although "The circumstance that the anterior ribs of *Brachiosaurus altithorax* are even wider than those of *B. brancai* is to be considered. That may be related to the fact that the dorsal vertebrae of the American species are noticeably larger than those of the African" (p. 90).

Janensch's final publication on *Brachiosaurus brancai* was a monograph on the limbs and limb girdles (Janensch, 1961). Here, Janensch provided further arguments for the assignment of his species to *Brachiosaurus* as follows: “The humerus of the type species of the genus *Brachiosaurus altithorax* Riggs (1904) from the Morrison Formation, is in broad terms so similar in outline to *Br. brancai* that a detailed comparison is unnecessary; with a length of 204 cm the proximal width amounts to 65 cm, that is 32% of the length. The distal end is not preserved in its width; the smallest shank width of 28 cm (= 14% of the overall length) is insignificantly larger than in the East African species” (p. 187). “The ilium of *Brachiosaurus altithorax* Riggs corresponds well with the ilium of the African *Brachiosaurus* in the characteristic features of the strong development of the anterior wing of the blade and the compressed shape of the pubic peduncle, so that thereby the assumption of generic association is strongly supported. The differences are not significant. In the American species the posterior wing of the blade, which does not extend over the ischiadic peduncle, is less tapered; the forward wing still somewhat more highly developed” (p. 200). “The outline of the 2.03 cm [sic] long femur of the type species of the genus is very similar [to that of *B. brancai*]. The more exact shape of the distal articular end and its condyles is not presented, therefore cannot be compared” (p. 207). (Janensch was mistaken regarding the preservation of the *B. altithorax* femur: while the distal end of its humerus is eroded, that of the femur is intact.) “*Br. brancai* is very similar to the North American *Br. altithorax* in the form of the humerus and the ilium, it is also similar in the humerus and astragalus to the Portuguese *Br. botalaiensis* [sic]” (p. 231). This final reference is to “*Brachiosaurus*” *atalaiensis* Lapparent and Zbyszewski, 1957, a probable brachiosaurid which was considered by Upchurch et al. (2004:308) to be distinct from *Brachiosaurus* as its ischium has a less steeply inclined distal shaft, and which has been subsequently referred to its own genus, *Lusotitan* Antunes and Mateus, 2003. Similarities between this species and either *B. altithorax* or *B. brancai* cannot be taken to indicate similarities between the latter two species.

Disregarding statements of general similarity, then, Janensch advanced a total of 13 putative shared characters in support of the referral of the Tendaguru species to *Brachiosaurus*, none of which pertain to the coracoid, humerus or femur (Table 1). Of these, one is invalid (does not apply to *B. brancai*), six diagnose more inclusive clades than Brachiosauridae, two are difficult to evaluate, and four appear to be valid synapomorphies: anterior dorsal vertebrae with long diapophyses; neural spines low in posterior dorsals and taller anteriorly; ilium with strongly developed anterior wing; and ilium with compressed pubic peduncle. While four synapomorphies constitute good evidence of a relationship between the two species, they are not in themselves compelling evidence for congenericity.

**Paul's Separation of *Giraffatitan* from *Brachiosaurus***

Although Lull (1919:42) had suggested seventy years earlier that *B. brancai* may not belong to *Brachiosaurus*, this idea was ignored in both scientific and popular literature until Paul (1988) executed a new skeletal reconstruction of *B. brancai* and thereby recognized proportional differences between the two species. Although believing that "the caudals, scapula, coracoid, humerus, ilium, and femur of *B. altithorax* and *B. brancai* are very similar" (p. 7), Paul argued that "it is in the dorsal column and trunk that the significant differences occur ... the dorsal column of *B. altithorax* is about 25-30% longer relative to the humerus or femur than that of *B. brancai* ... the longest dorsal rib [in *B. altithorax*] is some 10% longer relative to the humerus than in *B. brancai* ... All the dorsal centra of *B. altithorax* have pleurocoels that are about 50% larger than those of *B. brancai* ... The neural arches are taller and longer in *B. altithorax* [sic; probably a typo for *B. brancai* since the opposite is in fact the case], but are much narrower. The transverse processes form a shallow V in *B. brancai*; in *B. altithorax* they appear to be flatter ... Excepting the centrum, dorsal 4 [of *B. brancai*] differs greatly from the posterior dorsals in being much taller and wider. In the upper portions, the anterior dorsals of *B. altithorax* differ relatively little from from the more posterior vertebrae ... In HMN SII ... the anterior dorsals are about the same length as the posterior dorsals. In FMNH P 25107 the mid dorsal centra are abut 50% longer than those of the posterior dorsals." Although the axial variation in centrum lengths is indeed greater in *B. altithorax* than in *B. brancai*, the difference is nowhere near as great as Paul suggests (Table 3).

As recognized by Wilson and Sereno (1998:21), Paul's comparisons were in part based on the wrongly referred specimen BYU 9044 (BYU 5000 of his usage). Paul followed Jensen (1985) in considering this element to be brachiosaurid, but went further in referring it to *B. altithorax* (Paul, 1988:6) and using it "to bolster our knowledge of the shoulder of *B. altithorax*"; but as described above, Curtice et al. (1996) demonstrated convincingly that this element is diplodocid, and referred it to *Supersaurus*, so it can tell us nothing about *Brachiosaurus*. Paul (1988:6-7) also referred the "*Ultrasauros*" scapulocoracoid BYU 9462 to *B. altithorax*, and used its supposed similarity to *B. brancai* scapular material as evidence of the close relationship between the two *Brachiosaurus* species; but this referral is not justified because the fused coracoid that is part of this "*Ultrasauros*" scapulocoracoid is different from that of the *B. altithorax* holotype, lacking that specimen's characteristically strong lateral deflection of the glenoid facet, and having the glenoid facet more nearly continuous with the scapular suture rather than at about 60° to it as in FMNH P 25107.

Despite these errors, however, the differences between the dorsal columns of the two *Brachiosaurus* species highlighted by Paul (1988) are mostly correct, and do not depend heavily on the wrongly referred "*Ultrasauros*" dorsal (contra Wilson and Sereno, 1998:21). In particular, the different ways in which the dorsal vertebrae vary along the column in the two species are striking: in *B. altithorax* the more anterior dorsals are more anteroposteriorly elongate but not significantly taller than the more posterior dorsals, whereas in *B. brancai* they are taller but not not significantly more anteroposteriorly elongate. More generally, the seven preserved dorsal vertebrae of the *B. altithorax* holotype form a clear sequence with only small and smooth changes in proportions and morphology between adjacent vertebrae, while the preserved dorsal vertebrae of HMN SII vary much more dramatically, even when corrected for distortion.

Having demonstrated differences between the two *Brachiosaurus* species, however, Paul (1988:8) was circumspect about separating them: "The incompleteness of the remains of *B. altithorax* makes it difficult to prove full generic separation, as does the small sample size of Morrison and Tendaguru dorsal columns. Therefore only a separation at the subgeneric level is proposed." Paul therefore introduced the subgenus *Giraffatitan* Paul, 1988 to contain the species *brancai*, yielding the new combinations *Brachiosaurus* (*Brachiosaurus*) *altithorax* and *Brachiosaurus* (*Giraffatitan*) *brancai*. Subgenera are almost unknown in dinosaur taxonomy, and these combinations have not been used in any subsequent publication. The only subsequent mention of the subgenus *Giraffatitan* was that of McIntosh (1990b:66), who mentioned it only to indicate that he considered subgeneric separation unwarranted. However, Olshevsky (1991:238) raised the subgenus *Giraffatitan* to generic rank, commenting only that “the above genus, initially described as a subgenus of *Brachiosaurus*, is separable thereform on the basis of the vertebral column figured by Paul (1988).” (Olshevsky's listing gives the publication authority of the name *Giraffatitan* as Paul 1987 [nomen nudum], but he cannot remember what the publication was (G. Olshevsky, pers. comm., 2007); Paul, however, does not recall any published use of the name *Giraffatitan* prior to its 1988 formal erection (G. Paul, pers. comm., 2007)).

Although popular on the Internet, the name *Giraffatitan* has been very little used in the scientific literature: even Paul himself has reverted to using the name *Brachiosaurus* for the Tendaguru brachiosaur (e.g., Paul, 1994:246, 2000:93). Unfortunately, the only subsequent uses of this genus in the literature, in the taxonomic lists of McIntosh (1990a:347) and Upchurch et al. (2004:267), wrongly listed it as containing the species *altithorax* rather than *brancai* – a situation that would be impossible under ICZN rules since *altithorax* is the type species of *Brachiosaurus*.

## COMPARISONS

To determine whether the two *Brachiosaurus* species belong in the same genus, an element-by-element comparison is presented here. Even discounting questionable referred *Brachiosaurus altithorax* material such as the “*Ultrasauros*” scapulocoracoid, sufficient elements of the *B. altithorax* holotype are preserved to allow comparison, including both axial and appendicular elements. Except where noted, the following comparisons are based on personal observation of the type material of both species.

### Dorsal vertebrae (Figure 1)

As pointed out by Paul (1988:2), Janensch did not give his reasons for assuming that the dorsal column of *Brachiosaurus brancai* consisted of eleven vertebrae, and since Migeod's brachiosaur BMNH R5937 has twelve dorsal vertebrae, this should be considered the most likely number in other brachiosaurids. Although this specimen probably does not belong to *B. brancai* as assumed by Paul (Taylor, 2005), it is important as it includes the only complete dorsal column of any described brachiosaurid. Accordingly, I follow Paul in considering both *Brachiosaurus* species to have had twelve dorsal vertebrae: this means that the seven posterior vertebrae designated by Riggs (1903, 1904) as “presacrals 1-7” (counted forward from the sacrum) are here interpreted as dorsals 6-12, and the posterior dorsal vertebrae of HMN SII designated by Janensch (1950a) as presacrals 20, 22, 23 and 24 (i.e., dorsals 7, 9, 10 and 11) are here reinterpreted as dorsals 8, 10, 11 and 12. Janensch's presacral 17 (dorsal 4) is provisionally retained in this designation, although its true position cannot presently be determined.

As discussed above, the manner in which dorsal vertebrae vary along the column differs between *B. altithorax* and *B. brancai*: in the former, the more anterior dorsal vertebrae are only a little taller than the posterior dorsals but much longer anteroposteriorly, whereas in the latter, the more anterior dorsal vertebrae are much taller than the posterior dorsals but only a little longer anteroposteriorly. The dorsal vertebrae of *B. brancai* also differ from those of the type species in the following characters:

- Centra are broader transversely than dorsoventrally, rather than subcircular in cross-section.
- As noted by Paul (1988:7), the centra are proportionally less elongate.
- Lateral foramina of centra are proportionally smaller, especially in anterior to middle dorsals.
- Lateral processes are dorsally inclined rather than horizontal.
- Lateral processes are terminated by distinct triangular articular surfaces.
- Neural spines are inclined posterodorsally about 25° in the more anterior vertebrae, rather than vertical.
- Each neural spine is nearly constant in anteroposterior width through much of its height, rather than pronouncedly triangular in lateral view with the base about twice as wide as the narrowest point.
- Each anterior and middle dorsal neural spine is roughly constant in transverse width for much of its height, flaring suddenly rather than gradually at the top.
- The rugosities on the anterior and posterior faces of neural spines are limited to

shallow semicircles at the dorsal extremities, rather than the deep inverted triangular rugosities on both faces of the *B. altithorax* dorsals.

- Neural spines bear postspinal laminae.
- Spinodiapophyseal and spinopostzygapophyseal laminae do not contact each other: spinodiapophyseal laminae continue up the neural spine to the lateral flaring near the top rather than merging into the spinopostzygapophyseal laminae half way up the spine.

In addition to these features that apply to the HMN SII dorsals in general, D8 of that specimen (presacral 17 of Janensch's usage) has several other features not observed in any other sauropod, all of them preserved on both sides of the vertebra and therefore probably not pathological. Some of these features may also have existed in the other dorsals of the sequence, but the preservation of the relevant parts of the vertebrae is insufficient to determine this.

- D8 has spinoparapophyseal laminae. This novel lamina, distinct from the "accessory spino-diapophyseal lamina" (ASDL) of Salgado et al. (1997:22-23), is here assigned the standard abbreviation SPPL in accordance with the system of nomenclature for vertebral laminae proposed by Wilson (1999).
- The anterior centroparapophyseal laminae of D8 are unusually broad and flat, and perforated just below the horizontal lamina complex of PRPL, PPDL and PODL. The perforations appear not to be breakage, as the bone is finished around them.
- The horizontal lamina complex is supported ventrally by a dorsally forked lamina which cannot be designated as an ACDL, PCPL or even a generic infradiapophyseal lamina because it does not reach the diapophysis. Instead, the forked dorsal extremities of this lamina meet the horizontal lamina complex either side of the diapophysis, the anterior branch supporting the PPDL and the posterior branch supporting the PODL.

In summary, while the dorsal vertebrae of the two *Brachiosaurus* species are superficially similar, they vary in so many characters that they cannot be considered to support congenericity.

**Dorsal Ribs**

The dorsal rib heads of *B. altithorax* (figured by Riggs, 1904:pl. LXXV, fig. 5) and *B. brancai* (figured by Janensch, 1950a:fig. 107-108) are similarly proportioned, although the greater curvature of the latter suggests that it may be from a more posterior position. The locations of the pneumatic foramina are notably different, however: in the rib of *B. altithorax*, the rib is invaded on the anterior side by a small foramen in the proximal part of the rib shaft; in that of *B. brancai*, foramina are present on both the anterior and posterior aspects of the tuberculum, very close to the articular surface. The significance of this difference is difficult to assess, however, because the ribs of sauropods vary serially and the serial positions of the figured elements are not known; and also because pneumatic features are generally variable between individuals, between adjacent elements and even between the two sides of a single element – e.g., in *Xenoposeidon* Taylor and Naish, 2007: see Taylor and Naish (2007:1552-1553). Personal observation of the *B. altithorax* type material suggests that at least one rib head of that individual has a large pneumatic opening in its tuberculum similar to that figured by Janensch for *B. brancai*.

Paul (1988:7) stated that the longest dorsal rib of *B. altithorax* is 10% longer than that of *B. brancai*. However, Janensch (1950a:88) gave the length of HMN SII left rib 3 as 2.63 m, so the measurement of 2.75 m for an anterior rib of *Brachiosaurus altithorax* (Riggs, 1904:239) is only 4% longer, a difference that is probably not very significant.

**Sacrum (Figure 2)**

The sacra of *B. altithorax* and *B. brancai* are difficult to compare because no good material exists of the latter: the sacrum is the only part of the skeleton in which the type species is better represented than the referred species. While the sacrum of FMNH P 25107 has been subjected to some dorsoventral crushing, it is essentially complete, while the two sacra known for *B. brancai* are unsatisfactory: HMN Aa is distorted and missing the centra of all its sacral vertebrae, and HMN T is juvenile, incomplete and only partly ossified. The sacra of both species are transversely broad, and they share unusually short neural spines, especially when compared to the sacra of diplodocids (Hatcher, 1903b:pl. IV, figs. 1-2), although the spines are not very much shorter proportionally than those of *Camarasaurus* (Osborn and Mook, 1921:fig. 87) or *Haplocanthosaurus* Hatcher, 1903a (Hatcher, 1903b:pl. IV, fig. 3). Apart from these proportional similarities, poor preservation prevents the identification of further similarities between the sacra of the two *Brachiosaurus* species beyond the retention of plesiomorphies such as the sacricostal yoke.

**Caudal vertebrae (Figure 3)**

The *Brachiosaurus altithorax* type specimen includes the first two caudal vertebrae, of which the first consists only of a heavily crushed centrum and is uninformative but the second is well preserved. In contrast, several nearly complete caudal sequences of *B. brancai* are known, of which that of HMN Aa is best preserved (Janensch, 1950a:60) and includes the second caudal. The corresponding caudals of the two species resemble each other in their gently amphicoelous centra, absence of lateral foramina, short and simple lateral processes, and neural spines that are short and simple, rectangular in lateral view and somewhat swept back.

The caudal vertebrae of the two species also differ in several respects, however: the caudal of *B. altithorax*, although very nearly the same length anteroposteriorly as that of *B. brancai*, is about 30% taller, due to a relatively taller neural arch and spine. The articular face is also broader in *B. altithorax*, so that the total articular area is about 55% greater than in *B. brancai*. While the neural spines of all *B. brancai* caudals are laterally compressed, that of *B. altithorax* expands dorsally to about three times its minimum transverse width. The neural spine of the *B. altithorax* caudal is posteriorly inclined more steeply than those of *B. brancai* – about 30° as opposed to about 20°. The caudal ribs of *B. brancai* are swept back by about 30° while those of *B. altithorax* project directly laterally. The *B. altithorax* caudal has a distinct, block-like hyposphene whereas those of *B. brancai* have at most a slender hyposphenal ridge. While the postzygapophyseal facets of both species face ventrolaterally, those of *B. altithorax* are closer to a ventral orientation but those of *B. brancai* are more nearly laterally oriented. Finally, while the caudal vertebra of *B. altithorax* has no lateral depressions at all, the anterior caudals of *B. brancai* have pronounced lateral fossae, distinctly visible in the tail HMN Fund no that is incorporated in the mounted skeleton at the Humboldt Museum.

**Coracoid (Figure 4)**

The coracoids of the two *Brachiosaurus* species resemble each other in being somewhat taller dorsoventrally than they are broad anteroposteriorly, in their roughly semicircular shape, and in the possession of an indentation in the anterodorsal margin. However, the coracoids of *B. brancai* differ from that of *B. altithorax* in having a less straight scapular suture, a more pronounced anteroventral expansion in front of the glenoid, and a more slender, almost pointed, dorsal extremity. Most significantly, the glenoid surface of the coracoid of *B. altithorax* is strongly deflected laterally rather than facing directly posteroventrally – a feature not found in *B. brancai* or indeed in any other sauropod. The glenoid surface is also mediolaterally broader in *B. altithorax*, extending laterally on a thick buttress which is lacking in the coracoid of *B. brancai.*

**Humerus (Figure 5)**

As noted above, Janensch did not identify any synapomorphies between the humeri of the two *Brachiosaurus* species, observing only that “the similarity in relation to the humerus is particularly great between *Br. altithorax* and *Br. Fraasi* [= *B. brancai*]” (Janensch, 1914:98). The superficial similarity is indeed striking, the humeri of both species being more gracile than those of any other sauropods. Discarding a single outlier, the ratio of proximodistal length to minimum transverse width (Gracility Index or GI) in humeri of *B. brancai* varies between 7.86 for the right humerus HMN F2 and 9.19 for the left humerus HMN J12, with the type specimen's right humerus scoring 8.69, slightly more gracile than the middle of the range. (It is notable that the juvenile left humerus HMN XX19 has a GI of 8.63, and so is as gracile as the humeri of adult specimens, corroborating in *B. brancai* the findings of Carpenter and McIntosh (1994:277) for *Apatosaurus*, Ikejiri et al. (2005:176) for *Camarasaurus,* and Tidwell and Wilhite (2005) for *Venenosaurus* Tidwell, Carpenter and Meyer, 2001 that sauropod limb bones, unlike their vertebrae, scale isometrically during ontogeny.) For the *B. altithorax* type specimen, the GI is 8.50, based on the length of 204 cm and the minimum transverse width of 24 cm reported by Riggs (1904:241). However, the *B. altithorax* humerus looks rather less gracile to the naked eye than that of *B. brancai*, and careful measurement from Riggs's plate LXXIV yields a GI of 7.12, indicating that the true value of the minimum transverse width is closer to 28.5 cm. As noted by Riggs (1903:300-301), the surface of the distal end of this humerus has flaked away in the process of weathering. Careful comparison of the humeral proportions with those of other sauropods (Taylor and Wedel, in prep.) indicates that the missing portion of this bone would have extended approximately a further 12 cm, extending the total length to 216 cm and so increasing the GI to 7.53 – still less gracile than any *B. brancai* humerus except the outlier, but more gracile than any other sauropod species except *Lusotitan atalaiensis* (8.91), and much more gracile than the humerus of any non-brachiosaurid sauropod (e.g., *Diplodocus* Marsh, 1878 sp., 6.76; *Malawisaurus dixeyi* Jacobs, Winkler, Downs and Gomani, 1993, 6.20; *Mamenchisaurus constructus* Young, 1958, 5.54; *Camarasaurus supremus* Cope, 1877, 5.12; *Opisthocoelicaudia skarzynskii* Borsuk-Bialynicka, 1977, 5.00 – see Taylor and Wedel, in prep.) The humeri of the two *Brachiosaurus* species are also alike in their deltopectoral crests: although that of the *B. altithorax* humerus is broken near its tip, enough remains to indicate that, like that of *B. brancai*, it was sharply pronounced, located about one third of the way down the shaft, and oriented directly in a distal direction rather than sloping distomedially across the anterior face of the shaft as in some other sauropods.

However, the profiles of the lateral edges of the humeri are rather different, progressing smoothly upwards in *B. brancai* to the rounded proximolateral corner, whereas in *B. altithorax* there is a low but distinct lateral bulge one fifth of the way down the humerus, proximal to which the lateral margin is directed somewhat proximomedially rather than continuing in its gently proximolateral trajectory (Fig. 5B, H). Inspection of the bone shows that this is a genuine osteological feature, not caused by erosion, breakage or distortion. The maximum width of the *B. altithorax* humerus proximally is about 10% greater than that of *B. brancai*, and the reconstructed distal end (Taylor and Wedel, in prep.) is similarly broader than in *B. brancai*. Taken together with the 16% broader minimum width, these measurements show the *B. altithorax* humerus to be altogether more robust than that of *B. brancai*. The anteroposterior width of the *B. altithorax* humerus is presently impossible to measure accurately because the bone is half enclosed in a plaster jacket, but inspection of a cast of this element incorporated in the FMNH's mounted *Brachiosaurus* skeleton indicates that it is at least as anteroposteriorly broad as in *B. brancai*, perhaps a little more so.

**Ilium (Figure 4)**

As noted by Janensch (1961:200), the ilia of the two *Brachiosaurus* species resemble one another in the great development of the anterior wing and the “compressed” pubic peduncle, which I understand to mean elongate and gracile. The peduncle appears more recurved in *B. altithorax*, but this may be an error in reconstruction, since the right ilium figured by Riggs (1904:pl. LXXV) did not have its pubic peduncle preserved, and it was restored in the figure after the public peduncle of the otherwise uninformative left ilium (Riggs, 1904:238). A photograph of the preserved ilium shows how poor a condition it is in compared with the illustration produced by Riggs (Fig. 4A, C). Other proportional differences may therefore also be less significant than they appear: for example, the different trajectories of the dorsal borders of the ilia of the two species may be due to the distortion mentioned by Riggs (1904:238). With this caveat, however, there remain several potentially important differences between the ilia. The ischiadic peduncle of the *B. altithorax* ilium, though not pronounced, extends further ventrally than that of *B. brancai*, so that a line projected through the most ventral portions of both peduncles passes some distance ventrally of the posterior extremity of the *B. altithorax* ilium but is coincident with this extremity in *B. brancai*. In *B. brancai*, there is a distinct and acute notch between the ischiadic peduncle and the posterior extremity whereas *B. altithorax* has a much less pronounced indentation. Finally, the dorsal surface of the postacetabular region of the *B. altithorax* ilium bears a distinct tubercle which *B. brancai* lacks, and which also seems not to be present in any other sauropod.

**Femur (Figure 5)**

The femora of *B. altithorax* and *B. brancai* are similar in most respects, sharing a prominent medially directed head, a flat proximal end, a sharply defined proximolateral corner, a fourth trochanter projecting somewhat medially and therefore visible in anterior view, and extreme eccentricity with the mediolateral width being more than twice the anteroposterior diameter for most of the length of the shaft.

As with the humerus, the femur is somewhat more gracile in *B. brancai* than in *B. altithorax* (GI = 6.21 compared with 5.49). The fourth trochanter of *B. altithorax* is more prominent in anterior view than that of *B. brancai*, and is located more distally, at the half-way point rather than about 40% of the way down the shaft. The distal condyles do not project as far posteriorly in *B. altithorax* as in *B. brancai* (Figs. 5F, N), and while the tibial and fibular condyles are equally wide in the former, the fibular condyle is rather wider than the tibial condyle in the latter. Finally, the femur of *B. brancai* has a prominent lateral bulge one quarter of the way down its lateral margin, which *B. altithorax* lacks. This bulge was proposed as a titanosauriform synapomorphy by Salgado et al. (1997:16) but its distribution appears to be more complex.

**Summary**

Although McIntosh (1990b:65) felt that "the coracoid, femur, and sacrum of the two species are in complete accord," differences exist in both coracoid and femur, as well as the humerus, ilium, caudals, and most significantly the dorsal vertebrae. Since poor preservation prevents detailed comparison of the sacra, and lack of information about ribs makes it impossible to evaluate the significance of observed differences, these elements are therefore uninformative for comparative purposes. All elements sufficiently well preserved in both species, then, exhibit distinct differences, and generic separation is warranted since the two species are more different from each other than, for example, *Diplodocus* and *Barosaurus* Marsh, 1890. Accordingly, the name *Giraffatitan* will be used in the remainder of this paper.

## SYSTEMATIC PALEONTOLOGY

DINOSAURIA Owen, 1842

SAURISCHIA Seeley, 1888

SAUROPODA Marsh, 1878

NEOSAUROPODA Bonaparte, 1986

MACRONARIA Wilson and Sereno, 1998

TITANOSAURIFORMES Salgado, Coria and Calvo, 1997

BRACHIOSAURIDAE Riggs, 1904

**Revised Diagnosis**—Ratio of humerus:femur length ≥ 0.90 (character 206); centroprezygapophyseal lamina on middle and posterior dorsal vertebrae undivided at upper end (character 138); anterior dorsal vertebrae with long diapophyses (Janensch, 1950a:72); neural spines low in posterior dorsals, taller anteriorly (Janensch, 1950a:72); ilium with strongly developed anterior wing (Janensch, 1961:200); ilium with compressed pubic peduncle (Janensch, 1961:200); ratio of mediolateral:anteroposterior diameter of femur at midshaft ≥ 1.85 (character 284).

*BRACHIOSAURUS* Riggs, 1903

*BRACHIOSAURUS ALTITHORAX* Riggs, 1903

(Figs. 1-5 in part, 7)

*Brachiosaurus* (*Brachiosaurus*) *altithorax* (Riggs): Paul, 1988:8, figs. 2A, 3F, 4B.

**Holotype—**FMNH P 25107, partial skeleton comprising last seven dorsal vertebrae, sacrum, first two caudal vertebrae, left coracoid, right humerus, ilium and femur, fragmentary left ilium, and dorsal ribs.

**Referred Specimens—**USNM 21903, left humerus; BYU 4744, dorsal vertebra, left ilium and radius, right metacarpal III; BYU 9462, right scapulocoracoid, dorsal rib; BYU 13023, two dorsal vertebrae; USNM 5730, nearly complete skull; OMNH 01138, left metacarpal II; BYU 12866 and 12867, mid-cervical vertebrae; undescribed specimens. Not all referrals are certain.

**Occurrence and Distribution**—Morrison Formation, North America (Colorado, Utah, Oklahoma).

**Age—**Latest Jurassic (Kimmeridgian-Tithonian), 155.6-145.5 Mya.

**Revised Diagnosis**—Postspinal lamina absent from dorsal vertebrae (character 130); distal ends of transverse processes of dorsal vertebrae transition smoothly onto dorsal surfaces of transverse processes (character 142); spinodiapophyseal and spinopostzygapophyseal laminae on middle and posterior dorsal vertebrae contact each other (character 146); posterior dorsal centra subcircular in cross-section (character 151); posterior dorsal neural spines progressively expand mediolaterally through most of their length ("petal" or "paddle" shaped) (character 155); mid-dorsals about one third longer than posterior dorsals (see Paul, 1988:7); mid-dorsals only about 20% taller than posterior dorsals (see Paul, 1988:8); dorsals centra long (Janensch, 1950a:72) so that dorsal column is over twice humerus length (Paul, 1988:8); transverse processes of dorsal vertebrae oriented horizontally (Paul, 1988:8); dorsal neural spines oriented close to vertical in lateral view; dorsal neural spines triangular in lateral view, diminishing smoothly in anteroposterior width from wide base upwards; deep inverted triangular ligament rugosities on anterior and posterior faces of neural spines; hyposphenes present on anterior caudal vertebrae (character 178); anterior caudal vertebrae have transversely widened neurapophyses (Janensch, 1950a:76); anterior caudal vertebrae lack lateral fossae (Janensch, 1950a:76); glenoid articular surface of coracoid oriented somewhat laterally; glenoid articular surface of coracoid mediolaterally broad, extending laterally onto thick buttress; humerus relatively robust (GI = 7.5); lateral margin of humerus with low bulge one fifth of the way down; projected line connecting articular surfaces of ischiadic and pubic processes of ilium passes ventral to ventral margin of postacetabular lobe (character 264); subtle posterior notch between ischiadic peduncle and postacetabular lobe of ilium; tubercle present on dorsal margin of postacetabular lobe of ilium; distal tip of fourth trochanter lies at femoral midshaft height (character 282); lateral margin of femoral shaft in anterior or posterior view straight (character 285); femur relatively robust (GI = 5.5); tibial and fibular condyles of femur equal in width.

*GIRAFFATITAN* Paul, 1988

*GIRAFFATITAN BRANCAI* (Janensch, 1914)

(Figs. 1-5 in part)

*Brachiosaurus brancai* Janensch, 1914:86, figs. 1-4 (original description)

*Brachiosaurus fraasi* Janensch, 1914:94, figs. 5-6 (original description)

*Brachiosaurus* (*Giraffatitan*) *brancai* (Janensch): Paul, 1988:9, figs. 1, 2B, 3G, 4B.

**Lectotype**—HMN SII, partial skeleton comprising skull fragments including dentaries, eleven cervical vertebrae, cervical ribs, seven dorsal vertebrae, nearly complete set of dorsal ribs, distal caudal vertebrae, chevrons, left scapula, both coracoids and sternal plates, right forelimb and manus, left humerus, ulna and radius, both pubes, partial left femur, right tibia and fibula (Janensch, 1950b). Contra Janensch, the right femur of the mounted skeleton is complete while the left femur has a reconstructed shaft, and is presumably the partial femur of SII while the right femur is that of the referred specimen from locality Ni.

**Paralectotype**—HMN SI, skull and cervical vertebrae 2-7.

**Referred Specimens**—As listed by Janensch (1929:7-9) for "Brachiosaurus Brancai und Br. Fraasi Janensch."

**Occurrence and Distribution**—Middle and Upper Saurian Members, Tendaguru Formation, Tanzania, east Africa.

**Age**—Latest Jurassic (Kimmeridgian-Tithonian), 155.6-145.5 Mya.

**Revised Diagnosis**—Postspinal laminae present on dorsal vertebrae (character 130); distal ends of transverse processes of dorsal vertebrae possess distinctive, elevated areas with dorsally-facing surface that is connected to the dorsal surface of the remaining process only by a sloping region (character 142); spinodiapophyseal and spinopostzygapophyseal laminae on middle and posterior dorsal vertebrae do not contact each other (character 146); posterior dorsal centra dorsoventrally compressed in cross-section (character 151); posterior dorsal neural spines rectangular for most of their length with little or no lateral expansion except at distal end (character 155); mid-dorsals only about one quarter longer than posterior dorsals (see Paul, 1988:7); mid-dorsals about 40% taller than posterior dorsals (see Paul, 1988:9); dorsal centra short (Janensch, 1950a:72) so that dorsal column is less than twice humerus length (Paul, 1988:9); transverse processes of dorsal vertebrae oriented dorsolaterally (Paul, 1988:9); dorsal neural spines oriented posterodorsally in lateral view; dorsal neural spines are nearly constant in anteroposterior width through much of their height; ligament rugosities on anterior and posterior faces of neural spines limited to shallow semicircles at the dorsal extremity; spinoparapophyseal laminae on some dorsal vertebrae; broad, perforated anterior centroparapophyseal laminae on some dorsal vertebrae; horizontal lamina complex (PRPL, PPDL and PODL) supported from below by forked lamina that does not contact the diapophysis in some dorsal vertebrae; hyposphenal ridge weakly developed or absent from anterior caudal vertebrae (character 178); anterior caudal vertebrae have transversely narrow neural spines (Janensch, 1950a:76); anterior caudal vertebrae with lateral fossae (Janensch, 1950a:76); glenoid articular surface of coracoid oriented directly posteroventrally; glenoid articular surface of coracoid mediolaterally narrow; humerus very gracile (GI = 8.7); lateral margin of humerus straight; projected line connecting articular surfaces of ischiadic and pubic processes of ilium passes through ventral edge of postacetabular lobe (character 264); acute posterior notch between ischiadic peduncle and postacetabular lobe of ilium; tubercle absent from dorsal margin of postacetabular lobe of ilium; distal tip of fourth trochanter lies above midshaft height (character 282); lateral margin of femoral shaft in anterior view with distinct bulge 1/3 down (character 285); femur relatively gracile (GI = 6.2); fibular condyle of femur wider than tibial condyle.

## PHYLOGENETIC ANALYSIS

Salgado and Calvo (1997:43) suggested that the two “*Brachiosaurus*” species may not form a clade, as they were unable to identify any unequivocal synapomorphies linking the species. However, they did not test their own hypothesis by codifying the two species as separate OTUs in the phylogenetic analysis of the companion paper, Salgado et al. (1997). Neither have subsequent studies done so: in the phylogenetic analysis of Harris (2006), and those of Wilson (2002) and Upchurch et al. (2004) on which that of Harris is largely based, and in Wilson and Sereno (1998), a single *Brachiosaurus* OTU is used, the scoring for which represents a combination of states observed in the two species (J. Harris, J. Wilson, P. Upchurch; pers. comms., 2007). As no published phylogenetic analysis treats the two species separately, there is no numerical evidence either for or against the paraphyly proposed by Salgado and Calvo (1997).

To address this deficiency, I adapted the matrix of Harris (2006) by splitting the composite *Brachiosaurus* OTU into two separate OTUs representing *Brachiosaurus altithorax* and *Giraffatitan brancai*, yielding a matrix of 31 taxa (29 ingroups and two outgroups) and 331 characters. While rescoring the two brachiosaurid species, it became apparent that the composite *Brachiosaurus* OTU of Harris (2006) was incorrectly scored for several characters, having been assigned states that do not occur in either species (Table 2). These were corrected for both new OTUs.

Following Harris (2006), PAUP* 4.0b10 (Swofford, 2002) was used to perform a heuristic search using random stepwise addition with 50 replicates and with maximum trees = 500000. The analysis yielded 72 equally parsimonious trees with length = 791, consistency index (CI) = 0.5196, retention index (RI) = 0.6846 and rescaled consistency index (RC) = 0.3557. The statistics indicate a slightly less consistent tree than that of Harris (CI = 0.526, RI = 0.687) because additional homoplasy is introduced by splitting the *Brachiosaurus* OTU.

The strict consensus tree (Figure 6) is identical to that of Harris (2006) in the relationships of the 29 taxa that they share. (It does not appear identical to the strict consensus tree figured by Harris (2006:fig. 5a) due to a drawing error that resulted in the positions of *Haplocanthosaurus* and *Losillasaurus* Casanovas, Santafé and Sanz, 2001 being exchanged in that figure, as is confirmed by both the text and the majority rule tree in fig. 5b.) Within the current study's strict consensus tree, Neosauropoda is fully resolved except for a trichotomy of the three rebbachisaurids, *Nigersaurus* Sereno, Beck, Dutheil, Larsson, Lyon, Moussa, Sadleir, Sidor, Varricchio, Wilson and Wilson, 1999, *Rebbachisaurus* Lavocat, 1954 and *Limaysaurus* Salgado, Garrido, Cocca and Cocca, 2004. The two "*Brachiosaurus*" species form a clade to the exclusion of all other sauropods, and together occupy the same position as did the composite *Brachiosaurus* OTU in Harris (2006), as basal titanosauriforms forming the outgroup to the (*Euhelopus* [Romer, 1956]+ Titanosauria) clade. This result does not rule out the possibility that other brachiosaurid species, if included in the analysis, might break up the *Brachiosaurus*-*Giraffatitan* clade, but does argue against the possibility that *Giraffatitan* is more closely related to titanosaurs than to *Brachiosaurus* as was suggested by Naish et al. (2004:793). However, only a single further step is required for *Giraffatitan* to fall closer to titanosaurs than to *Brachiosaurus*; strict consensus of all most parsimonious trees under this constraint maintains the topology (*Camarasaurus* (*Brachiosaurus* (*Giraffatitan* (*Euhelopus*, titanosaurs)))). Four further steps are required for *Brachiosaurus* to fall closer to titanosaurs than to *Giraffatitan*, and the strict consensus of trees satisfying this constraint also keeps *Brachiosaurus* outside the (*Euhelopus* + titanosaurs) clade. Further phylogenetic work including more brachiosaurid OTUs is needed.

In the analysis of Taylor and Naish (2007), which used the composite *Brachiosaurus* OTU of Harris (2006), *Xenoposeidon* emerged as the sister taxon to "*Brachiosaurus*" in 72 of 1089 trees (6.6%). When it is added to the new analysis, it is never sister taxon to *Brachiosaurus altithorax*, but is sister to *Giraffatitan* in 72 of 1014 trees (7.1%). Splitting the brachiosaurs, then, reduces the number of most parsimonious trees by 75 (6.9%); in the initial analysis *Xenoposeidon* was attracted to the *Giraffatitan* component of the composite OTU rather than the *Brachiosaurus altithorax* component.

## DISCUSSION

### Association of the *Giraffatitan* lectotype material

As noted by Paul (1988:7), the anterior dorsal vertebra considered by Janensch to be D4 of the *Giraffatitan* lectotype HMN SII differs markedly from the more posterior dorsal vertebrae of the same specimen, being much taller, having a more slender neural spine, and bearing notably broad diapophyses. However, another possibility should be considered: that the aberrant anterior dorsal vertebra does not in fact belong to HMN SII. As noted by Janensch (1950a:33), the excavation at Site S yielded presacral material from two individuals, designated SI and SII. The material assigned to SI consists of a partial skull and an articulated sequence of cervicals 2-7, with all remaining Site S material assigned to SII. However, the dorsal vertebrae posterior to the third were disarticulated, isolated from one another and jumbled together with other skeletal elements. Although Janensch (1929:8) had previously considered it possible that some of the Site S dorsal vertebrae belonged to specimen SI, he subsequently asserted that "These individually embedded vertebrae are far too large to have belonged to the smaller *Brachiosaurus* SI ... In size they completely match the articulated vertebral series and can thus be associated with Skeleton SII without hesitation" (Janensch, 1950a:33). However, while the overall size of D4 is commensurate with that of D8, his statement is misleading because its centrum is significantly smaller and its processes much longer. The association of D4 with SII, then, cannot be considered certain. Janensch's preserved field sketches, reproduced by Heinrich (1999:figs. 16, 18) do not indicate the relative positions of the vertebrae, and his field notes subsequent to his first week at Tendaguru are lost (G. Maier, pers. comm., 2007), so further information will probably not be forthcoming.

A dorsal neural spine that is part of Migeod's specimen BMNH R5937 closely resembles that of the vertebra D4 assigned to HMN SII (Taylor, in prep.) Since this specimen, though brachiosaurid, does not belong to *Giraffatitan* (Taylor 2005), it must be possible that the anterior dorsal vertebra assigned to SII actually belongs to SI, and that SI belongs to the same taxon as BMNH R5937. If this is correct, then the cervical vertebrae of this taxon very closely resemble those of *Giraffatitan*. Since the North American cervical vertebrae BYU 12866 and 12867, which may belong to *Brachiosaurus altithorax*, are also indistinguishable from those of *Giraffatitan*, it is possible that cervical morphology is highly conserved in brachiosaurids while more variation is found in the dorsal column. If so, this would be the converse of the situation among diplodocids, among which *Diplodocus*, *Apatosaurus* and *Barosaurus* have rather similar dorsal vertebrae but very different cervicals.

**Differences in body proportions**

Having made a careful element-by-element comparison between the two brachiosaurid species, it is now possible to consider how the osteological differences between the species might have been reflected in differences in gross bodily proportions.

First, as stated by Paul (1988:7), the trunk is proportionally longer in *Brachiosaurus* than in *Giraffatitan* due to the greater length of its dorsal centra. Paul states that the difference is “25-30%” on the basis of his figure 2. Independent calculation of the lengths of the sequences of dorsals 6-12 in both species corroborates this, finding that the posterior dorsal centra of *Brachiosaurus* are about 23% longer then those of *Giraffatitan* (Table 3). This is a significant proportional difference, apparent to the naked eye.

Paul (1988:8) argued that *Brachiosaurus* lacked the “withers” (tall neural spines over the shoulders) of *Giraffatitan*. This cannot be substantiated, however, since the anterior dorsal vertebrae of *Brachiosaurus* are not known, the putative fourth dorsal vertebra of *Giraffatitan* being from a location two places forward of the most anterior known *Brachiosaurus* dorsal. Bearing in mind that the association of the supposed fourth dorsal of *Giraffatitan* may not be secure, it is apparent that nothing can be confidently said about differences between the genera in the anterior dorsal region.

More significant are the differences between the single known caudal vertebra of *Brachiosaurus* and those of *Giraffatitan*. Many caudals of the latter are known, and are remarkably consistent in morphology, while the single known caudal of the former is unambiguously associated with the remainder of the specimen and differs from those of *Giraffatitan* in two mechanically significant ways: first, although it is from a similarly sized animal as the *G. brancai* type specimen, and is comparable in anteroposterior length, it is taller in both the centrum and the neural arch (Fig. 3B, D); and second, the transverse broadening of the neural spine towards its extremity allows a much greater area for ligament attachment – about 2.25 times as great. The former character certainly indicates that the tail was taller in the American taxon, and the latter suggests that it was longer, perhaps by about 20-25%.

Since *Brachiosaurus* had both a longer trunk and tail than *Giraffatitan*, it is tempting to wonder whether its neck was also longer, contra the suggestion of Paul (1988:8) that it was shorter. However, the example of *Diplodocus* and *Barosaurus* demonstrates that even closely related sauropods may vary unpredictably in proportions: the longer tail of *Diplodocus*, taken alone, might be thought to imply that it also had a longer neck than its cousin, but the opposite is the case. Therefore, conclusions about the neck of *Brachiosaurus* cannot be drawn from elongation in other parts of the body; and indeed the North American brachiosaur cervicals BYU 12866 and 12867, if correctly referred to *Brachiosaurus*, indicate that its neck proportions were identical to those of *Giraffatitan*.

One of the most distinctive osteological features of *Brachiosaurus* is the strong lateral deflection of the glenoid surface of its coracoid, which in other sauropods including *Giraffatitan* faces directly posteroventrally. This may indicate that the humeri were also directed somewhat laterally, again in contrast to the parasagittally oriented forelimbs of other sauropods. Janensch restored the skeleton of *Giraffatitan* with somewhat sprawling upper arms, reasoning that “In the forelimb the humerus [...] displays characters that are similar to the conditions of the humerus of lacertilians, crocodylians and *Sphenodon*, even if pronounced to a lesser degree, which, however, show that, in the type of motion of the upper arm, a component of lateral splaying was included" (Janensch, 1950b:99). Ironically, while it is now established that sauropods in general held their limbs vertically, it seems possible that *Giraffatitan*'s sister taxon *Brachiosaurus* may have been the sole exception to this rule. If correct, this would be surprising: the bending stress on a sprawled humerus would greatly exceed the compressive stress on one held vertically (Alexander, 1985:18), and the proportionally slender humeri of *Brachiosaurus* would seem particularly unsuited to such a posture.

Finally, while slender, the humeri of *Brachiosaurus* are less so than those of *Giraffatitan*, having a GI of 7.12 compared with 8.69. The femora of the two species, however are proportionally very similar. Since the humerus of *Brachiosaurus*, then, is more robust in comparison with its femur than in *Giraffatitan*, it is possible that the American species carried a greater proportion of its weight on its forelimbs than the African species.

In conclusion, the osteological evidence suggests that *Brachiosaurus* differed from the popular *Giraffatitan*-based conception of the genus in that its trunk was 23% longer, its tail 20-25% longer and thicker, its forelimbs were possibly somewhat sprawled, and a greater concentration of its mass was probably above the forelimbs. Taking these differences into account, I prepared a skeletal reconstruction of *Brachiosaurus altithorax* (Figure 7), including the holotype and all referred skeletal elements, with the remaining elements modified from Paul's (1988) reconstruction of *Giraffatitan*. Comparison with Paul's *Giraffatitan* clearly illustrates the differences.

Martin et al. (1998:120) argued that "*Brachiosaurus*" (i.e. *Giraffatitan*) "had front and hind limbs of roughly equal length," that "the now 'traditional' disparity of the fore- and hind-limb proportions (about 1.2:1) has been based on the 1937 mounted skeleton in the Humboldt-Museum, Berlin, which is a composite reconstruction," and that "other taxa referred to *Brachiosaurus* (including *B. altithorax* Riggs, 1903 and *B. atalaiensis* Lapparent and Zbyszewski, 1957) appear, as far as the evidence permits us to say, to have had front and hind limbs of roughly equal length." While it is true that the Humboldt mount is a composite, Janensch (1950b:99) explained that it includes a forelimb, complete except for one carpal, and the tibia, fibula and partial femur, all from the same individual (HMN SII), and that the femur's reconstructed length was "calculated from other finds." Therefore the limb proportions of this skeleton are reliable. It is true that the humeri and femora are nearly identical in length in both *Brachiosaurus* and *Giraffatitan*, but this does not mean that the torso was held horizontal for three reasons: first, the vertically oriented metacarpal arcade of sauropods causes the wrist to be held higher than the ankle, especially in the case of brachiosaurids, which have particularly elongate metacarpals; second, as shown by Janensch (1950b:pl. 8), the lower forelimb (ulna and radius) is longer than the lower hind-limb (tibia and fibula), causing the shoulder to be higher than the hip; third, the shoulder joint is mounted much lower on the rib-cage than the hip joint is on the sacrum, so that shoulder vertebrae must have been higher than hip vertebrae. In conclusion, the forelimbs of brachiosaurids were indeed longer than their hind-limbs, and their backs were strongly inclined anterodorsally, as reconstructed by Janensch (1950b), Paul (1988), Wedel (2000) and others.

### Masses of *Brachiosaurus* and *Giraffatitan*

In order to determine the effects that these proportional differences would have had on the mass of *Brachiosaurus* as compared with its better known cousin, I estimated the volumes of the type specimens of both species using Graphic Double Integration (Jerison, 1973; Hurlburt, 1999; Murray and Vickers-Rich, 2004). For the lateral silhouette of *Brachiosaurus*, I used the reconstruction of Fig. 7; for the corresponding dorsal, anterior and posterior silhouettes, I modified Paul's (1998:fig. 1) reconstruction of *Giraffatitan* as follows: head, neck, forelimbs and hind-limbs I left unmodified. I stretched the dorsal view of the torso by 12.8% to match the length of the lateral view, and conservatively increased the transverse width by half this proportion, 6.4%. Similarly, I stretched the dorsal view of the tail by 20% to match the lateral view, and increased transverse width by 10%. The selected increases in transverse width are unavoidably arbitrary, because the ribs of *Brachiosaurus* are not sufficiently well known to inform a rigorous dorsal-view reconstruction. As pointed out by Murray and Vickers-Rich (2004:211), such guesswork is unavoidable in mass estimation: the best we can do is to be explicit about what the assumptions are, to facilitate repeatability and the subsequent construction of better models.

The results are summarized, by body-part, in Table 4. Most significantly, the volume, and hence mass, of *Brachiosaurus* is calculated to be 23% greater than that of *Giraffatitan*, whereas Paul (1988:3) found *Brachiosaurus* to be only 11% heavier than *Giraffatitan* (35000 kg vs. 31500 kg). This is partly explained by the larger tail of *Brachiosaurus* in this reconstruction, where Paul assigned it a similarly sized tail to that of *Giraffatitan*, but Paul must also have modelled the torso of *Brachiosaurus* as much narrower that I have. The absolute masses calculated here are significantly lower than those of Paul (1988:3) – 28688 kg for *Brachiosaurus* is 82% of Paul's 35000 kg, and 23337 kg for *Giraffatitan* is only 74% of Paul's 31500 kg. One reason for this is that I have assumed a density of 0.8 kg/l based on the average density of *Diplodocus* calculated by Wedel (2005:220) whereas Paul (1988:10) used 0.6 for the neck and 0.9 for the remainder of the animal, yielding an average density of 0.861. However, even using Paul's higher value for density, my mass estimates would be only 88% and 80% of Paul's. This may be because the models used by Paul (1988:10) for his mass estimates were sculpted separately from his execution of the skeletal reconstructions that I used as the basis of my calculations, and may have been bulkier. The mass of 23337 kg for *Giraffatitan*, while surprisingly light for so large an animal, compares well with the 25789 kg of Henderson (2004:S181).

**Size of the largest brachiosaurid sauropods**

The largest brachiosaurid sauropods known from reasonably complete remains are still the type specimens of *Brachiosaurus altithorax* and *Giraffatitan brancai*, which are of very similar sizes: their humeri differ in length by only 3 cm (1.4%) and their femora by 8 cm (4%). As noted by Janensch (1950b:102) and Paul (1988:10), the fibula HMN XV2 is about 13% longer than that of the type specimen, indicating that *Giraffatitan* grew significantly larger than the type specimen. Curtice et al. (1996:93) noted that the "*Ultrasauros*" scapulocoracoid BYU 9462 belonged to an animal no larger than the largest Tendaguru specimens. It has not been noted, however, that while the scapula and coracoid that constitute BYU 9462 are fully fused, with the suture obliterated, the coracoid of the *B. altithorax* type specimen is unfused, indicating that it belonged to a subadult individual. It is possible that this individual would have grown significantly larger had it survived.

**Phylogenetic nomenclature**

The genus *Brachiosaurus* is important in three widely used phylogenetic definitions: those of the clades Brachiosauridae, Somphospondyli, and Titanosauriformes. In all formulations of these three clades together, they form a node-stem triplet with the first two as sisters to each other within the last; therefore the same two specifiers should be used in their definitions.

Although the name Brachiosauridae has been in use as a “family” since Riggs (1904), its earliest phylogenetic definition is that of Wilson and Sereno (1998:20) as “titanosauriforms more closely related to *Brachiosaurus* than to *Saltasaurus*.” This same definition was also proposed by Sereno (1998:63); no other definition has been published.

Wilson and Sereno (1998:53) erected the taxon Somphospondyli as the sister group to Brachiosauridae, defining it as “Titanosauriformes more closely related to *Saltasaurus* than to *Brachiosaurus*”; this definition was affirmed by Sereno (1998:63) and no alternative has been published. This clade was proposed in the context of a scheme in which Titanosauria was defined as “Titanosauriforms more closely related to *Saltasaurus* than to either *Brachiosaurus* or *Euhelopus*” (Wilson and Sereno, 1998:22). Upchurch et al. (2004:308), however, noting that this definition is confusing in its use of three reference taxa and that the distinction between the Somphospondyli and Titanosauria of Wilson and Sereno (1998) depended on the controversial position of *Euhelopus* as a basal somphospondylian, instead dispensed with the name Somphospondyli altogether and defined Titanosauria as “Titanosauriformes more closely related to *Saltasaurus* than to *Brachiosaurus*.” At the time of that writing, *Euhelopus* was indeed controversial, having been recovered as a mamenchisaurid (euhelopodid of his usage) by Upchurch (1995, 1998), as a basal somphospondylian by Wilson and Sereno (1998) and Wilson (2002), and as a near outgroup of Neosauropoda by Upchurch et al. (2004). However, a subsequent joint study between these two schools of sauropod phylogeny (Upchurch and Wilson, 2007) has more firmly established *Euhelopus* as more closely related to titanosaurs sensu stricto than to brachiosaurids, so the name Somphospondyli retains some utility. For this reason, and because the precise definition of Titanosauria remains controversial, I recommend the retention of the name

Somphospondyli as part of the Titanosauriformes-Brachiosauridae node-stem triplet, and leave the matter of the definition of Titanosauria to others.

Titanosauriformes was initially defined by Salgado et al. (1997:12) as “the clade including the most recent common ancestor of *Brachiosaurus brancai*, *Chubutisaurus insignis* and Titanosauria and all its descendants.” In accordance with recommendation 2 of Taylor (2007:2), this definition should be interpreted according to the apparent intentions of the author, and it seems obvious that Salgado et al. intended to indicate the least inclusive clade containing the specified taxa rather than any of the more inclusive clades that also do. Subsequent redefinitions have all been similar to this one. Wilson and Sereno (1998:51) redefined this taxon as “*Brachiosaurus*, *Saltasaurus*, their common ancestor, and all of its descendants,” which improves on the original definition by omitting the unstable and poorly represented specifier *Chubutisaurus insignis*, but which uses genera rather than species.

All of these definitions are in need of revision in order to comply with the requirements of the draft PhyloCode (Cantino and de Queiroz, 2006) which requires that species rather than genera must be used as specifiers, and that species used as the internal specifier in a phylogenetic definition of a clade whose name is based on a genus must be the type of that genus (article 11.7). So, for example, Sereno's (2005) refinement of Titanosauriformes as “The least inclusive clade containing *Brachiosaurus brancai* Janensch 1914 and *Saltasaurus loricatus* Bonaparte and Powell 1980” (published only on the Internet) falls into the same trap as the original definition in anchoring on the non-type species *B. brancai*, with the result that, under some topologies, the type species *Brachiosaurus altithorax* is excluded from Titanosauriformes.

In order to avoid this eventuality, then, I offer the following triplet of definitions:

**Titanosauriformes** = the most recent common ancestor of *Brachiosaurus altithorax* Riggs 1903 and *Saltasaurus loricatus* Bonaparte and Powell 1980 and all its descendants.

**Brachiosauridae** = all taxa more closely related to *Brachiosaurus altithorax* Riggs 1903 than to *Saltasaurus loricatus* Bonaparte and Powell 1980.

**Somphospondyli** = all taxa more closely related to *Saltasaurus loricatus* Bonaparte and Powell 1980 than to *Brachiosaurus altithorax* Riggs 1903.

## CONCLUSIONS

The popular image of *Brachiosaurus* is based on *Giraffatitan*, a generically distinct animal that is separated from *Brachiosaurus* by at least 26 osteological characters. The two genera remain closely related within Brachiosauridae, but would have appeared distinctly different in life.

## ACKNOWLEDGMENTS

I thank W. F. Simpson (FMNH) for access to the *Brachiosaurus altithorax* type material, D. M. Unwin and W.-D. Heinrich (then of the HMN) for access to *Giraffatitan brancai* material, and S. D. Chapman (BMNH) for access to Migeod's specimen. W. F. Simpson also provided a high-quality scan of Riggs (1904). P. Mannion (University College London) made available his photographs of those elements of the *Brachiosaurus altithorax* material that I had neglected to photograph. I also thank those who allowed me to cite personal communications.

Indispensable translations of Janensch (1914, 1950a, b) were provided by G. Maier, who also provided invaluable assistance in other points relating to Janensch's German. In all cases where I have quoted translated portions of Janensch's papers, I used Maier's translations. Those translations are now freely available at the Polyglot Paleontology web-site, http://ravenel.si.edu/paleo/paleoglot/, from which translations of other papers were also obtained.

M. Wedel and J. Harris provided excellent and very detailed reviews, and H.-D. Sues handled the manuscript with admirable tact and efficiency. E. Schweizerbart'sche Verlagsbuchhandlung (http://www.schweizerbart.de/) kindly gave permission for Janensch's Palaeontographica figures to be reproduced for the comparative figures.

Finally, I beg forgiveness from all brachiosaur lovers, that so beautiful an animal as "*Brachiosaurus*" *brancai* now has to be known by so inelegant a name as *Giraffatitan*.

## LITERATURE CITED

Alexander, R. M. 1985. Mechanics of posture and gait of some large dinosaurs. Zoological Journal of the Linnean Society 83:1–25.

Anonymous. 1959. *Brachiosaurus* exhibit at the Smithsonian Institution. Nature 183:649–650.

Antunes, M. T., and O. Mateus. 2003. Dinosaurs of Portugal. Comptes Rendus Palevol 2:77–95.

Bonaparte, J. F. 1986. Les dinosaures (Carnosaures, Allosauridés, Sauropodes, Cétiosauridés) du Jurassique moyen de Cerro Cóndor (Chubut, Argentina). Annales de Paléontologie 72:325–386.

Bonnan, M. F., and M. J. Wedel. 2004. First occurrence of *Brachiosaurus* (Dinosauria: Sauropoda) from the Upper Jurassic Morrison Formation of Oklahoma. PaleoBios 24:13–21.

Borsuk-Bialynicka, M. 1977. A new camarasaurid sauropod *Opisthocoelicaudia skarzynskii*, gen. n., sp. n., from the Upper Cretaceous of Mongolia. Palaeontologica Polonica 37:5–64.

Cantino, P. D., and K. de Queiroz. 2006. PhyloCode: A Phylogenetic Code of Biological Nomenclature (Version 4b, September 12, 2007). http://www.ohiou.edu/phylocode/PhyloCode4b.pdf

Carpenter, K., and J. S. McIntosh. 1994. Upper Jurassic sauropod babies from the Morrison Formation; pp. 265–278 in K. Carpenter, K. F. Hirsch, and J. R. Horner (eds.), Dinosaur Eggs and Babies. Cambridge University Press, Cambridge.

Carpenter, K., and V. Tidwell. 1998. Preliminary description of a *Brachiosaurus* skull from Felch Quarry 1, Garden Park, Colorado. Modern Geology 23:69–84.

Casanovas, M. L., J. V. Santafé, and J. L. Sanz. 2001. *Losillasaurus giganteus*, un neuvo saurópodo del tránsito Jurásico-Cretácico de la cuenca de "Los Serranos" (Valencia, España). Paleontologia i Evolució 32-33:99–122.

Cope, E. D. 1877. On a gigantic saurian from the Dakota epoch of Colorado.

Paleontology Bulletin 25:5–10.

Curry Rogers, K. 2001. The evolutionary history of the Titanosauria (Ph.D. dissertation). State University of New York, Stony Brook, 573 pp.

Curtice, B. D. 1995. A description of the anterior caudal vertebrae of *Supersaurus vivianae*. Journal of Vertebrate Paleontology 15:3–25A.

Curtice, B. D., and K. L. Stadtman. 2001. The demise of *Dystylosaurus edwini* and a revision of *Supersaurus vivianae*. Western Association of Vertebrate Paleontologists and Mesa Southwest Paleontological Symposium, Mesa Southwest Museum Bulletin 8:33–40.

Curtice, B. D., K. L. Stadtman, and L. J. Curtice. 1996. A reassessment of *Ultrasauros macintoshi* (Jensen, 1985). Museum of Northern Arizona Bulletin 60:87–95.

Foster, J. R. 2003. Paleoecological analysis of the vertebrate fauna of the Morrison Formation (Upper Jurassic), Rocky Mountain Region, U.S.A. (NMMNHS bulletin 23). New Mexico Museum of Natural History and Science, Albuquerque, New Mexico, 95 pp.

Harris, J. D. 2006. The significance of *Suuwassea emiliae* (Dinosauria: Sauropoda) for flagellicaudatan intrarelationships and evolution. Journal of Systematic Palaeontology 4:185–198.

Hatcher, J. B. 1903a. A new name for the dinosaur *Haplocanthus* Hatcher. Proceedings of the Biological Society of Washington 16:100.

Hatcher, J. B. 1903b. Osteology of *Haplocanthosaurus* with description of a new species, and remarks on the probable habits of the Sauropoda and the age and origin of the Atlantosaurus beds. Memoirs of the Carnegie Museum 2:1–72.

Heinrich, W.-D. 1999. The taphonomy of dinosaurs from the Upper Jurassic of Tendaguru (Tanzania) based on field sketches of the German Tendaguru Expedition (1909-1913). Mitteilungen aus dem Museum für Naturkunde, Berlin, Geowissenschaften, Reihe 2:25–61.

Henderson, D. M. 2004. Tipsy punters: sauropod dinosaur pneumaticity, buoyancy and aquatic habits. Proceedings of the Royal Society of London B, 271(Suppl. 4): S180-

S183.

Hurlburt, G. R. 1999. Comparison of body mass estimation techniques, using Recent reptiles and the pelycosaur *Edaphosaurus boanerges*. Journal of Vertebrate Paleontology 19:338–350.

Ikejiri, T., V. Tidwell, and D. L. Trexler. 2005. new adult specimens of *Camarasaurus lentus* highlight ontogenetic variation within the species; pp. 154–179 in V. Tidwell, and K. Carpenter (eds.), Thunder Lizards: the Sauropodomorph Dinosaurs. Indiana University Press, Bloomington, Indiana.

Jacobs, L. L., D. A. Winkler, W. R. Downs, and E. M. Gomani. 1993. New material of an Early Cretaceous titanosaurid sauropod dinosaur from Malawi. Palaeontology 36:523–534.

Janensch, W. 1914. Übersicht über der Wirbeltierfauna der Tendaguru-Schichten nebst einer kurzen Charakterisierung der neu aufgefuhrten Arten von Sauropoden. Archiv fur Biontologie 3:81–110.

Janensch, W. 1922. Das Handskelett von *Gigantosaurus robustus* u. *Brachiosaurus Brancai* aus den Tendaguru-Schichten Deutsch-Ostafrikas. Centralblatt für Mineralogie, Geologie und Paläontologie 15:464–480.

Janensch, W. 1929. Material und Formengehalt der Sauropoden in der Ausbeute der Tendaguru-Expedition. Palaeontographica (Suppl. 7) 2:1–34.

Janensch, W. 1935-1936. Die Schadel der Sauropoden *Brachiosaurus*, *Barosaurus* und *Dicraeosaurus* aus den Tendaguru-Schichten Deutsch-Ostafrikas. Palaeontographica (Suppl. 7) 2:147–298.

Janensch, W. 1947. Pneumatizitat bei Wirbeln von Sauropoden und anderen Saurischien. Palaeontographica (Suppl. 7) 3:1–25.

Janensch, W. 1950a. Die Wirbelsaule von *Brachiosaurus brancai*. Palaeontographica (Suppl. 7) 3:27–93.

Janensch, W. 1950b. Die Skelettrekonstruktion von *Brachiosaurus brancai*. Palaeontographica (Suppl. 7) 3:97–103.

Janensch, W. 1961. Die Gliedmaszen und Gliedmaszengürtel der Sauropoden der Tendaguru-Schichten. Palaeontographica (Suppl. 7) 3:177–235.

Jensen, J. A. 1985. Three new sauropod dinosaurs from the Upper Jurassic of Colorado. Great Basin Naturalist 45:697–709.

Jensen, J. A. 1987. New brachiosaur material from the Late Jurassic of Utah and Colorado. Great Basin Naturalist 47:592–608.

Jerison, H. J. 1973. Evolution of the brain and intelligence. Academic Press, New York, 482 pp.

Kim, H. M. 1983. Cretaceous dinosaurs from Korea. Journal of the Geology Society of Korea 19:115–126.

Lapparent, A. F. d., and G. Zbyszewski. 1957. Mémoire no. 2 (nouvelle série): les dinosauriens du Portugal. Services Géologiques du Portugal, Lisbon, Portugal, 63 pp.

Lavocat, R. 1954. Sur les Dinosauriens du continental intercalaire des Kem-Kem de la Daoura. Comptes Rendus 19th Intenational Geological Congress 1952, 1:65–68.

Lull, R. S. 1919. The sauropod dinosaur *Barosaurus* Marsh. Memoirs of the Connecticut Academy of Arts and Sciences 6:1–42.

Maier, G. 2003. African Dinosaurs Unearthed: The Tendaguru Expeditions. Indiana University Press, Bloomington and Indianapolis, 380 pp.

Marsh, O. C. 1877. Notice of new dinosaurian reptiles from the Jurassic formation. American Journal of Science and Arts 14:514–516.

Marsh, O. C. 1878. Principal characters of American Jurassic dinosaurs. Part I. American Journal of Science, Series 3, 16:411–416.

Marsh, O. C. 1879. Notice of new Jurassic reptiles. American Journal of Science, Series 3, 18:501–505.

Marsh, O. C. 1890. Description of new dinosaurian reptiles. American Journal of Science, Series 3, 39:81–86 and plate I.

Marsh, O. C. 1891. Restoration of *Triceratops*. American Journal of Science, Series 3, 41:339–342.

Martin, J., V. Martin-Rolland, and E. Frey. 1998. Not cranes or masts, but beams: the biomechanics of sauropod necks. Oryctos 1:113–120.

Matthew, W. D. 1915. Dinosaurs, with special reference to the American Museum collections. American Museum of Natural History, New York, 164 pp.

McIntosh, J. S. 1990a. Sauropoda; pp. 345–401 in D. B. Weishampel, P. Dodson, and H. Osmólska (eds.), The Dinosauria. University of California Press, Berkeley and Los Angeles.

McIntosh, J. S. 1990b. Species determination in sauropod dinosaurs with tentative suggestions for the their classification; pp. 53–69 in K. Carpenter, and P. J. Currie (eds.), Dinosaur Systematics: Approaches and Perspectives. Cambridge University Press, Cambridge.

McIntosh, J. S., and D. S. Berman. 1975. Description of the palate and lower jaw of the sauropod dinosaur *Diplodocus* (Reptilia: Saurischia) with remarks on the nature of the skull of *Apatosaurus*. Journal of Paleontology 49:187–199.

Migeod, F. W. H. 1931. British Museum East Africa Expedition: Account of the work done in 1930. Natural History Magazine 3:87–103.

Murray, P. F. and Vickers-Rich, P. 2004. Magnificent mihirungs. Indiana University Press, Bloomington, Indiana, 410 pp.

Naish, D., D. M. Martill, D. Cooper, and K. A. Stevens. 2004. Europe's largest dinosaur? A giant brachiosaurid cervical vertebra from the Wessex Formation (Early Cretaceous) of southern England. Cretaceous Research 25:787–795.

Olshevsky, G. 1991. A revision of the parainfraclass Archosauria Cope, 1869, excluding the advanced Crocodylia. Mesozoic Meanderings 2:1–196.

Osborn, H. F., and C. C. Mook. 1921. *Camarasaurus*, *Amphicoelias* and other sauropods of Cope. Memoirs of the American Museum of Natural History, n.s. 3:247–387.

Owen, R. 1842. Report on British fossil reptiles, Part II. Reports of the British Association for the Advancement of Sciences 11:60–204.

Paul, G. S. 1988. The brachiosaur giants of the Morrison and Tendaguru with a description of a new subgenus, *Giraffatitan*, and a comparison of the world's largest dinosaurs. Hunteria 2:1–14.

Paul, G. S. 1994. Dinosaur reproduction in the fast lane: implications for size, success and extinction; pp. 244–255 in K. Carpenter, K. F. Hirsch, and J. R. Horner (eds.), Dinosaur Eggs and Babies. Cambridge University Press, Cambridge.

Paul, G. S. 1998. Terramegathermy and Cope's rule in the land of titans. Modern Geology 23:179–217.

Paul, G. S. 2000. Restoring the life appearances of dinosaurs; pp. 78–106 in G. S. Paul (ed.), The Scientific American book of dinosaurs. St. Martin's Press, New York.

Powell, J. E. 1992. Osteología de *Saltasaurus loricatus* (Sauropoda-Titanosauridae) del Cretácico Superior del Noroeste Argentino; pp. 165–230 in J. L. Sanz, and A. D. Buscalioni (eds.), Los Dinosaurios y su Entorno Biotico. Actas del Segundo Curso de Paleontologia en Cuenca. Instituto Juan de Valdés, Ayuntamiento de Cuenca.

Riggs, E. S. 1901. The largest known dinosaur. Science 13:549–550.

Riggs, E. S. 1903. *Brachiosaurus altithorax*, the largest known dinosaur. American Journal of Science 15:299–306.

Riggs, E. S. 1904. Structure and relationships of opisthocoelian dinosaurs. Part II, the Brachiosauridae. Field Columbian Museum, Geological Series 2, 6:229–247.

Romer, A. S. 1956. Osteology of the Reptiles. University of Chicago Press, Chicago, 772 pp.

Salgado, L., and J. O. Calvo. 1997. Evolution of titanosaurid sauropods. II: the cranial evidence. Ameghiniana 34:33–48.

Salgado, L., R. A. Coria, and J. O. Calvo. 1997. Evolution of titanosaurid sauropods. I: Phylogenetic analysis based on the postcranial evidence. Ameghiniana 34:3–32.

Salgado, L., A. Garrido, S. E. Cocca, and J. R. Cocca. 2004. Lower Cretaceous rebbachisaurid sauropods from Cerro Aguada Del Leon (Lohan Cura Formation), Neuquen Province, Northwestern Patagonia, Argentina. Journal of Vertebrate Paleontology 24:903–912.

Seeley, H. G. 1882. On a remarkable dinosaurian coracoid from the Wealden of Brook in the Isle of Wight, preserved in the Woodwardian Museum of Cambridge, probably referable to *Ornithopsis*. Quarterly Journal of the Geological Society, London 38:367–371.

Seeley, H. G. 1888. On the classification of the fossil animals commonly named Dinosauria. Proceedings of the Royal Society of London 43:165–171.

Sereno, P. C. 1998. A rationale for phylogenetic definitions, with application to the higher-level taxonomy of Dinosauria. Neues Jahrbuch für Geologie und Paläontologie, Abhandlungen 210:41–83.

Sereno, P. C., A. L. Beck, D. B. Dutheil, H. C. E. Larsson, G. H. Lyon, B. Moussa, R. W. Sadleir, C. A. Sidor, D. J. Varricchio, G. P. Wilson, and J. A. Wilson. 1999. Cretaceous sauropods from the Sahara and the uneven rate of skeletal evolution among dinosaurs. Science 282:1342–1347.

Swofford, D. L. 2002. PAUP*: Phylogenetic Analysis Using Parsimony (* and Other Methods). Sinauer Associates, Sunderland, Massachusetts.

Taylor, M. P. 2005. Sweet seventy-five and never been kissed: the Natural History Museum's Tendaguru brachiosaur; pp. 25–25 in P. M. Barrett (ed.), Abstracts volume for 53rd Symposium of Vertebrae Palaeontology and Comparative Anatomy. The Natural History Museum, London.

Taylor, M. P. 2007. Phylogenetic definitions in the pre-PhyloCode era; implications for naming clades under the PhyloCode. PaleoBios 27:1–6.

Taylor, M. P., and D. Naish. 2007. An unusual new neosauropod dinosaur from the Lower Cretaceous Hastings Beds Group of East Sussex, England. Palaeontology 50:1547–1564.

Tidwell, V., and D. R. Wilhite. 2005. Ontogenetic variation and isometric growth in the forelimb of the Early Cretaceous sauropod *Venenosaurus*; pp. 187–198 in V. Tidwell, and K. Carpenter (eds.), Thunder Lizards: the Sauropodomorph Dinosaurs. Indiana University Press, Bloomington, Indiana.

Tidwell, V., K. Carpenter, and S. Meyer. 2001. New Titanosauriform (Sauropoda) from the Poison Strip Member of the Cedar Mountain Formation (Lower Cretaceous), Utah; pp. 139–165 in D. H. Tanke, and K. Carpenter (eds.), Mesozoic Vertebrate Life: New Research inspired by the Paleontology of Philip J. Currie. Indiana

University Press, Bloomington and Indianapolis, Indiana.

Turner, C. E., and F. Peterson. 1999. Biostratigraphy of dinosaurs in the Upper Jurassic Morrison Formation of the Western Interior, U.S.A; pp. 77–114 in D. D. Gillette (ed.), Vertebrate Paleontology in Utah (Utah Geological Survey Miscellaneous Publication 99-1). Utah Geological Survey, Salt Lake City, Utah.

Upchurch, P. 1995. The evolutionary history of sauropod dinosaurs. Philosophical Transactions of the Royal Society of London, Series B 349:365–390.

Upchurch, P. 1998. The phylogenetic relationships of sauropod dinosaurs. Zoological Journal of the Linnean Society 124:43–103.

Upchurch, P., and J. A. Wilson. 2007. *Euhelopus zdanksyi* and its bearing on the evolution of East Asian sauropod dinosaurs; pp. 30–30 in J. Liston (ed.), Abstracts volume for 55th Symposium of Vertebrae Palaeontology and Comparative Anatomy. University of Glasgow, Glasgow.

Upchurch, P., P. M. Barrett, and P. Dodson. 2004. Sauropoda; pp. 259–322 in D. B. Weishampel, P. Dodson, and H. Osmólska (eds.), The Dinosauria, 2nd edition. University of California Press, Berkeley and Los Angeles.

Wedel, M. J. 2000. Reconstructing *Brachiosaurus*. Prehistoric Times 42:47.

Wedel, M. J. 2005. Postcranial skeletal pneumaticity in sauropods and its implications for mass estimates; pp. 201–228 in J. A. Wilson, and K. Curry-Rogers (eds.), The Sauropods: Evolution and Paleobiology. University of California Press, Berkeley.

Wedel, M. J. 2006. Pneumaticity, neck length, and body size in sauropods. Journal of Vertebrate Paleontology 26:3–137A.

Wedel, M. J., R. L. Cifelli, and R. K. Sanders. 2000. Osteology, paleobiology, and relationships of the sauropod dinosaur *Sauroposeidon*. Acta Palaeontologica Polonica 45:343–388.

Wilson, J. A. 1999. A nomenclature for vertebral laminae in sauropods and other saurischian dinosaurs. Journal of Vertebrate Paleontology 19:639–653.

Wilson, J. A. 2002. Sauropod dinosaur phylogeny: critique and cladistic analysis. Zoological Journal of the Linnean Society 136:217–276.

Wilson, J. A., and P. C. Sereno. 1998. Early evolution and higher-level phylogeny of sauropod dinosaurs. Society of Vertebrate Paleontology Memoir 5:1–68.

Young, C.-C. 1958. New sauropods from China. Vertebrata Palasiatica 2:1–28.

TABLE 1. Characters used by Janensch (1929, 1950a, 1961) in support of the referral of the species *Brachiosaurus brancai* to the genus *Brachiosaurus*, with their corresponding character numbers in the analysis of Harris (2006) and their distribution as presently understood.

| Character | Reference | Corresponding character in Harris (2006) | Distribution |
|---|---|---|---|
| Dorsal vertebrae with extensive lateral foramina | Janensch (1950a: 72) | 123 (state 2) | Neosauropoda + *Haplocanthosaurus* + *Jobaria* |
| Dorsal vertebrae with undivided neural spines | Janensch (1950a: 72) | 120 (state 0) | Saurischia |
| Dorsal vertebrae with neural spines that broaden dorsally | Janensch (1950a: 72) | 148 (state 2) | Macronaria (but reverts to state 1 in Somphospondyli) |
| Dorsal vertebrae with horizontal diapophyses | Janensch (1950a: 72) | -- | (Not present in *B. brancai*, so irrelevant.) |
| Anterior dorsal vertebrae with long diapophyses | Janensch (1950a: 72) | -- | Brachiosauridae (i.e., *B. altithorax* and *B. brancai*) |
| Neural spines low in posterior dorsals, taller anteriorly | Janensch (1950a: 72) | -- | Brachiosauridae |
| Sacrum with extensive triangular first sacral rib | Janensch (1950a: 72) | -- | (Cannot be assessed) |
| Second sacral rib with extensive attachment to first and second sacral centra | Janensch (1950a: 72) | -- | (Cannot be assessed) |
| Sacrum with long transverse processes | Janensch (1950a: 72) | 258 | Neosauropoda + *Haplocanthosaurus* + *Jobaria* |
| Sacrum and proximal caudal vertebrae with low neural spines | Janensch (1929: 20) | 164 (state 0) | Saurischia |
| Dorsal ribs with pneumatic foramina | Janensch (1950a: 87) | 197 | Titanosauriformes |
| Ilium with strongly developed anterior wing | Janensch (1961: 200) | -- | Brachiosauridae |
| Ilium with compressed pubic peduncle | Janensch (1961: 200) | -- | Brachiosauridae |

TABLE 2. Corrected character codings for the compound "*Brachiosaurus*" OTU in the analysis of Harris (2006). Owenian anatomical nomenclature is used in place of the avian nomenclature of Harris.

| Character in Harris (2006) | Coding in Harris (2006) | Corrected coding | Comments |
|---|---|---|---|
| 128. Dorsal vertebrae with spinodiapophyseal lamina | On posterior dorsals only (1) | On middle and posterior dorsals (2) | This lamina is clearly visible in middle dorsals of both *Brachiosaurus* (Riggs, 1904: plate LXXII) and *Giraffatitan* (Janensch, 1950a: figs. 53 and 54). |
| 133. Morphology of posterior margins of lateral fossae on anterior dorsal vertebrae | Acute (1) | Unknown (?) in *Brachiosaurus*; rounded (0) in *Giraffatitan*. | Figured for *Giraffatitan* by Janensch (1950a: fig. 53) |
| 134. Morphology of ventral surfaces of anterior dorsal centra | With sagittal crest (creating two ventrolaterally facing surfaces) (2) | Unknown (?) in *Brachiosaurus*; variable (?) in *Giraffatitan*. | In *Giraffatitan*, the ventral morphology of the centra changes along the vertebral column as described by Janensch (1950a: 44-46)[1] |
| 138. Morphology of centroprezygapophyseal lamina on middle and posterior dorsal vertebral arches | Bifurcate toward upper end (= infraprezyga-pophyseal fossa present) (1) | Single (0) | No infraprezygapophyseal fossa is present in any dorsal vertebra of either species. |

1: "With the transition [from the neck] to the trunk in the 13th presacral vertebra, and increasingly in the two subsequent first dorsal vertebrae, the flatness of the ventral surface behind the parapophyses gives way to vaulting; in the 15th presacral vertebra the middle section is completely cylindrical ventrally ... The centrum of the 15th to 17th presacral vertebrae exhibits a somewhat depressed ventral field bordered by two rounded margins; in the remaining dorsal vertebrae in contrast, as in the three SII 121-123 (essentially preserved only as centra), they exhibit a median ventral ridge" (translated from Janensch, 1950a:44-46).

| | | | |
|---|---|---|---|
| 141. Posterior centroparapophyseal lamina on middle and posterior dorsal vertebral arches | Present (1) | Uncertain (?) in *Brachiosaurus*; variable (?) in *Giraffatitan*. | Preservation is not good enough in *Brachiosaurus* to be sure about this character. There are no PCPLs in D8 of the *Giraffatitan* type specimen HMN SII, but they are present in the last two dorsals. |
| 143. Lamination on anterior face of (non-bifid) neural spine of middle and posterior dorsal vertebrae | Both prespinal and spinoprezyga-pophyseal laminae present and connected to each other either directly (merging) or via accessory laminae (3) | Prespinal lamina absent, spinoprezyga-pophyseal laminae present (2) | No prespinal lamina is present in any dorsal vertebra of either species. |
| 154. Ratio of mediolateral width to anteroposterior length of posterior (non-bifid) dorsal neural spines | <=1.0 (longer than wide) (0) | >1.0 (wider than long) (1) | Only the neural spine of the last dorsal of *Brachiosaurus* is narrower transversely than anteroposteriorly. |
| 160. Ratio of height of sacral neural spines to anteroposterior length of centrum | 2.0-3.49 (1) | <2.0 (0) | The partial sacrum of *Giraffatitan* figured by Janensch (1950a: fig. 74) shows the height of the sacral spines to be less than 1.5 times the average sacral centrum length. |
| 169. Morphology of articular surfaces in anterior caudal centra | Dorsoventrally compressed (1) | Subcircular (0) | Circular anterior caudal articular surfaces are figured for both *Brachiosaurus* (Riggs, 1904: plate LVVI, fig. 1) and *Giraffatitan* (Janensch, 1950a: plates II and III). |
| 182. Morphology of anterior centrodiapophyseal lamina on anterior caudal transverse processes | Single (0) | Inapplicable (?) | These laminae do not exist in the caudal vertebrae of either species. |
| 218. Morphology of anterodorsal margin of coracoid | Rectangular (meet at abrupt angle) (1) | Rounded (anterior and dorsal margins grade into one another) (0) | See Riggs (1904: plate LXXV, fig. 4). |
| 284. Ratio of mediolateral to anteroposterior diameter of femur at midshaft | 1.25-1.50 (1) | 1.85 (2) | This ratio is 2.1 in *Brachiosaurus* (pers. obs. of cast) and 2.3 in *Giraffatitan* (Janensch, 1961: Beilage J, figs. 1a, c). |
| 287. Relative mediolateral breadth of distal femoral condyles | Tibial condyle much broader than fibular condyle (1) | Subequal (0) | The condyles are subequal in width in *Brachiosaurus* (Riggs, 1903: fig. 4, 1904: plate LXXIV, fig. 2); in *Giraffatitan*, the fibular condyle is about 1.5 times as wide as the tibial condyle (Janensch, 1961: Beilage J, fig. 1b). |

TABLE 3. Functional lengths of dorsal centra of *Brachiosaurus* and *Giraffatitan*, omitting condyles. Measurements for *Brachiosaurus* taken from Riggs (1904:pl. LXXII) and scaled according to total vertebra heights as given by Riggs (1904:234). Measurements for *Giraffatitan* taken from Janensch (1950a:44) for D4 and D8, scaled from Janensch (1950a:fig. 62) for D11 and D12, and linearly interpolated for D5-D7, D9 and D10.

| | *Brachiosaurus altithorax* | *Giraffatitan brancai* |
|---|---|---|
| Dorsal | Length (cm) | Length (cm) |
| D4 | — | 28.5 |
| D5 | — | 28.7 |
| D6 | 37 | 29 |
| D7 | 38 | 29.2 |
| D8 | 34 | 29.4 |
| D9 | 32 | 27.3 |
| D10 | 35 | 25.1 |
| D11 | 28 | 23 |
| D12 | 22 | 20 |
| Total D6–D12 | 226 | 183 |

TABLE 4. Volumes and masses of *Brachiosaurus* and *Giraffatitan*, estimated by Graphic Double Integration and broken down by body part. Volume ratio indicates *Brachiosaurus* volumes as a proportion of corresponding *Giraffatitan* volumes. Masses assume a density of 0.8 kg/l (Wedel 2005:220).

| | *Brachiosaurus* | | *Giraffatitan* | | |
|---|---|---|---|---|---|
| Body part | Volume (l) | %Total | Volume (l) | %Total | Volume ratio |
| Head | 140 | 0.39 | 140 | 0.48 | |
| Neck | 4117 | 11.48 | 4117 | 14.11 | |
| Forelimbs (pair) | 1344 | 3.75 | 1344 | 4.61 | |
| Hindlimbs (pair) | 1462 | 4.08 | 1462 | 5.01 | |
| Torso | 26469 | 73.81 | 20588 | 70.58 | 1.29 |
| Tail | 2328 | 6.49 | 1520 | 5.21 | 1.53 |
| Total volume | 35860 | | 29171 | | 1.23 |
| Total mass (kg) | 28688 | | 23337 | | |

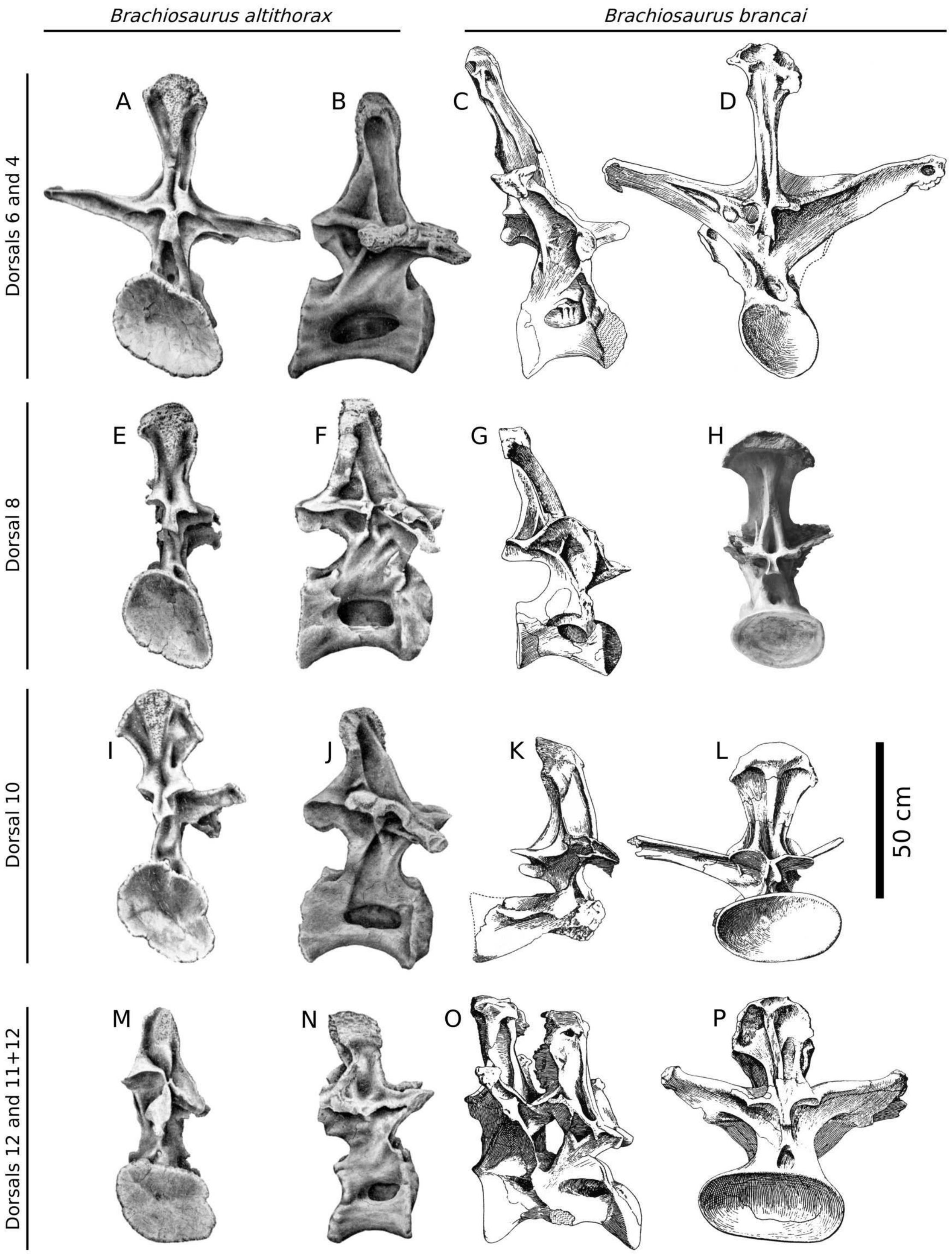

FIGURE 1. Dorsal vertebrae of *Brachiosaurus altithorax* and *Brachiosaurus brancai* in posterior and lateral views, equally scaled. **A**, **B**, **E**, **F**, **I**, **J**, **M**, **N**, *B. altithorax* holotype FMNH P 25107, modified from Riggs (1904:pl. LXXII); **C**, **D**, **G**, **H**, **K**, **L**, **O**, **P**, *B. brancai* lectotype HMN SII, modified from Janensch (1950a:figs. 53, 54, 56, 60-62, 64) except **H**, photograph by author. Neural arch and spine of **K** sheared to correct for distortion. **A**, **D**, **E**, **H**, **I**, **L**, **M**, **P**, posterior; **B**, **F**, **G**, **J**, **N**, right lateral; **C**, **K**, **O**, left lateral reflected. **A**, **B**, dorsal 6; **C**, **D**, dorsal 4; **E-H**, dorsal 8; **I-L**, dorsal 10; **M**, **N**, **P**, dorsal 12; **O**, dorsals 11 and 12. Corresponding vertebrae from each specimen are shown together except that dorsal 4 is not known from *B. altithorax* so dorsal 6, the most anterior known vertebra, is instead shown next to dorsal 4 of *B. brancai*. Scale bar equals 50 cm.

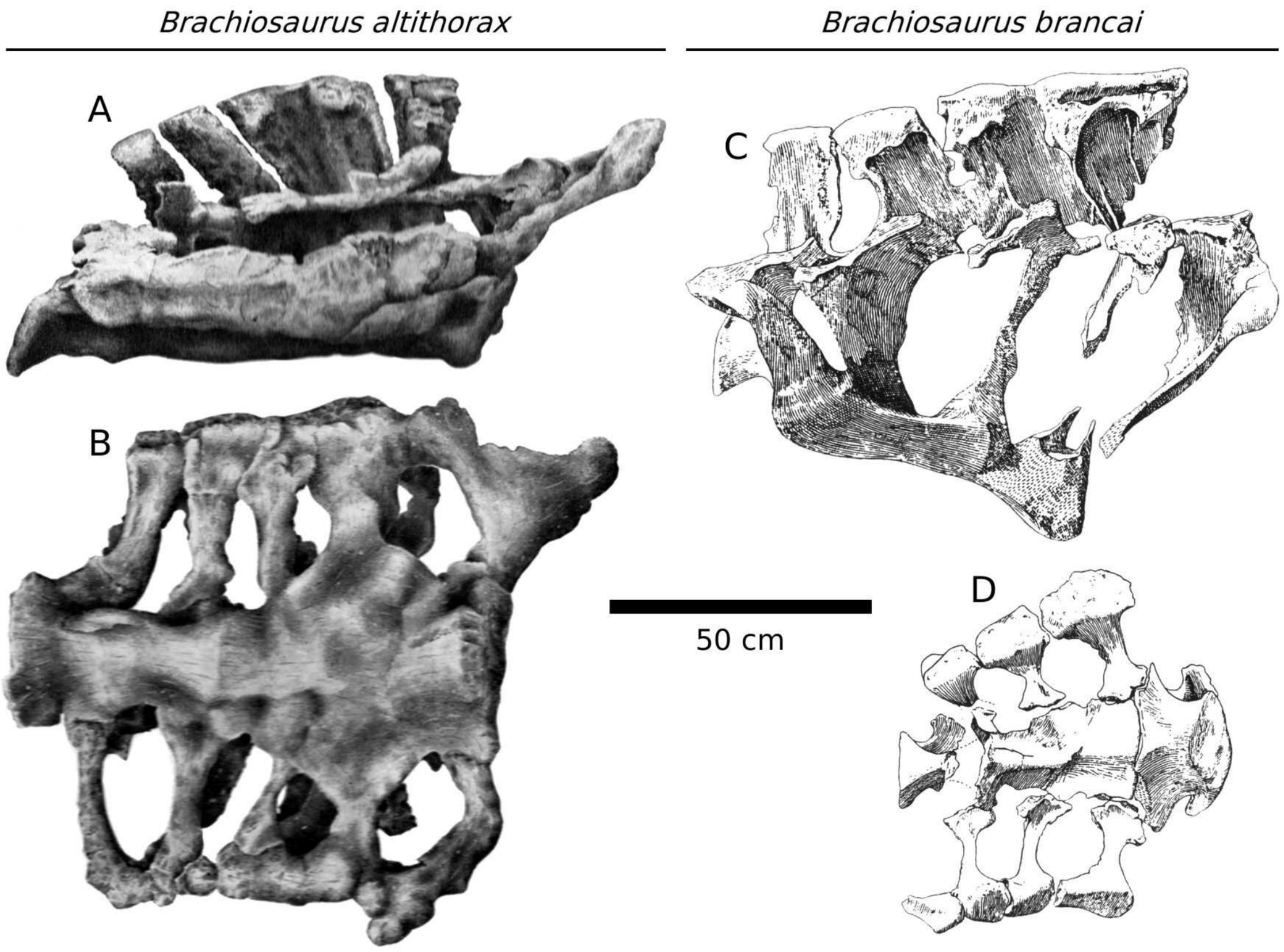

FIGURE 2. Sacra of *Brachiosaurus altithorax* and *Brachiosaurus brancai*, equally scaled. **A**, **B**, *B*. *altithorax* holotype FMNH P 25107; **C**, *B*. *brancai* referred specimen HMN Aa; **D**, juvenile *B*. *brancai* referred specimen HMN T. **A**, **C**, right lateral; **B**, **D**, ventral. **A**, **B** modified from Riggs (1904:pl. LXXIII); **C**, **D** modified from Janensch (1950a: figs. 74 and 76). Scale bar equals 50 cm.

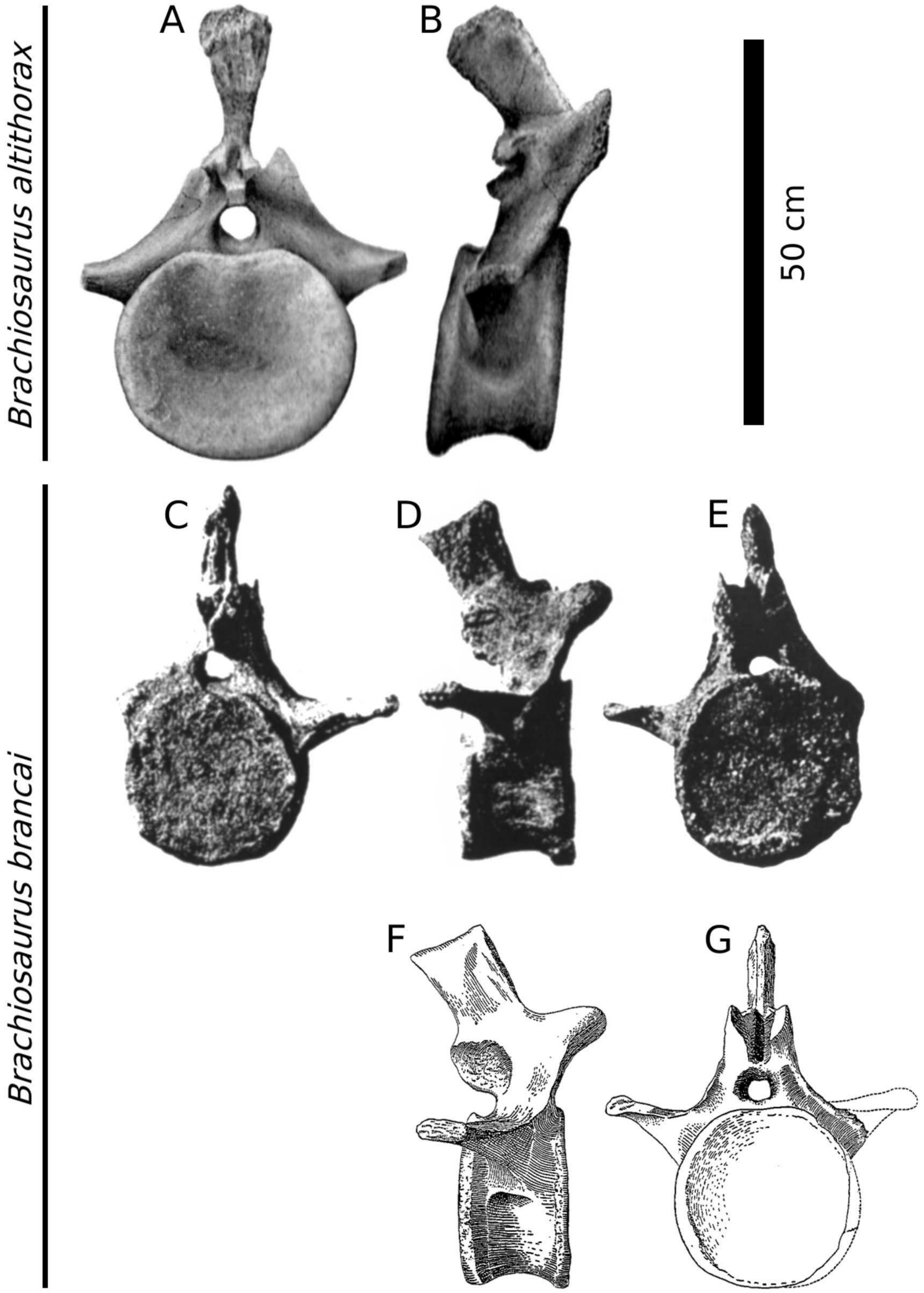

FIGURE 3. Second caudal vertebrae of *Brachiosaurus altithorax* and *Brachiosaurus brancai*, equally scaled. **A**, **B**, *B*. *altithorax* holotype FMNH P 25107; **C**-**G**, *B*. *brancai* referred specimen HMN Aa. **A**, **C**, posterior; **B**, **D**, **F**, right lateral; **E**, **G**, anterior. **A**-**B** modified from Riggs (1904:pl. LXXV); **C**-**E** modified from Janensch (1950a:pl. 2), **F**-**G**

modified from Janensch (1929:fig. 15). Scale bar equals 50 cm.

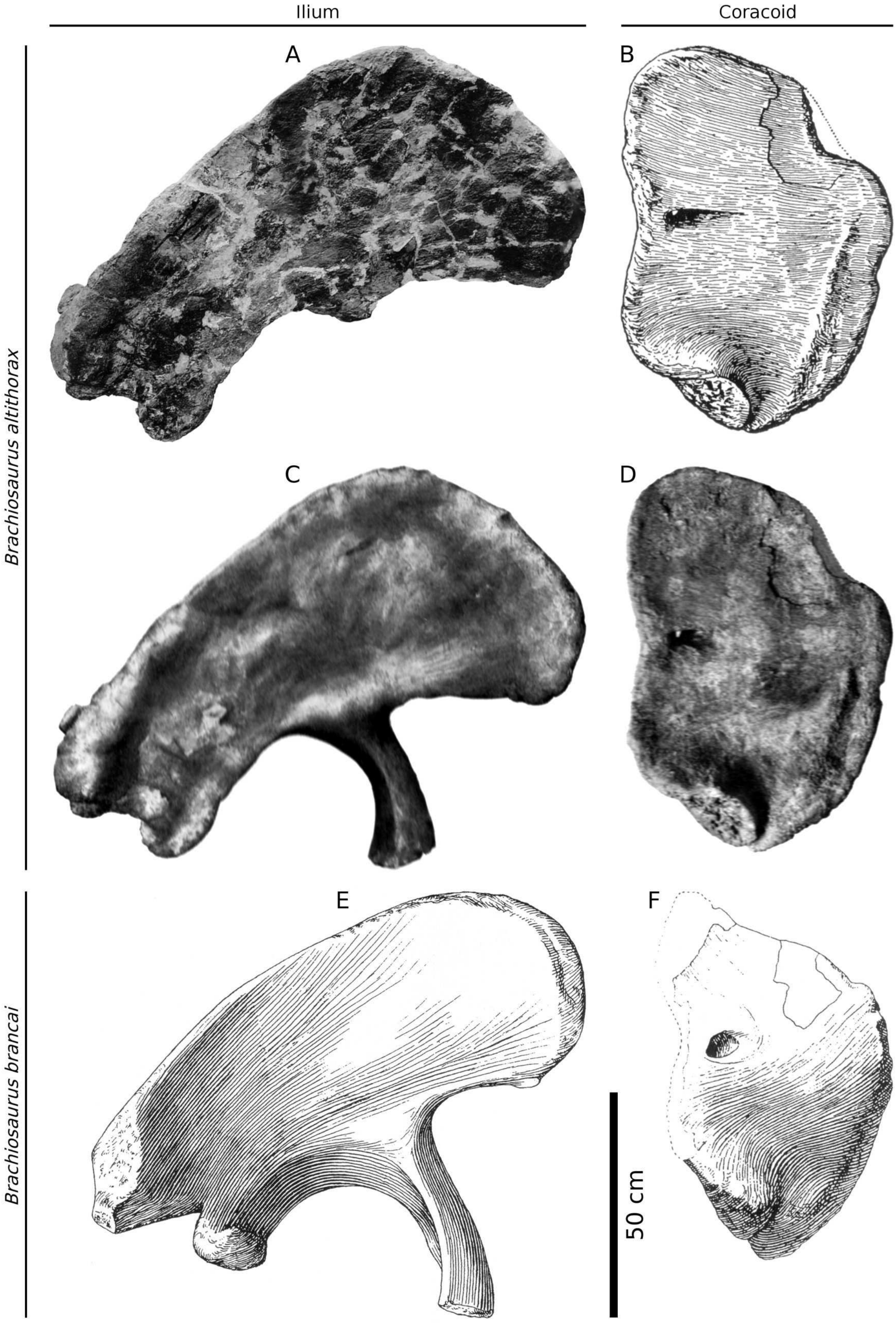

FIGURE 4. Limb girdle bones of *Brachiosaurus altithorax* and *Brachiosaurus brancai*, equally scaled. **A**, **C**, right ilium of *B*. *altithorax* holotype FMNH P 25107; **B**, **D**, left

coracoid of same, reflected; **E**, right ilium of *B. brancai* referred specimen Aa 13, scaled to size of restored ilium of *B. brancai* lectotype HMN SII as estimated by Janensch (1950b:99); **F**, right coracoid of *B. brancai* lectotype HMN SII. **A** modified from FMNH neg. #GEO-16152, showing poor preservation and absence of public peduncle; **B** modified from Riggs (1903:fig. 3); **C**, **D** modified from Riggs (1904:pl. LXXV); **E** modified from Janensch (1961:Beilage E, fig. 1a); **F** modified from Janensch (1961:fig. 1a). Scale bar equals 50 cm.

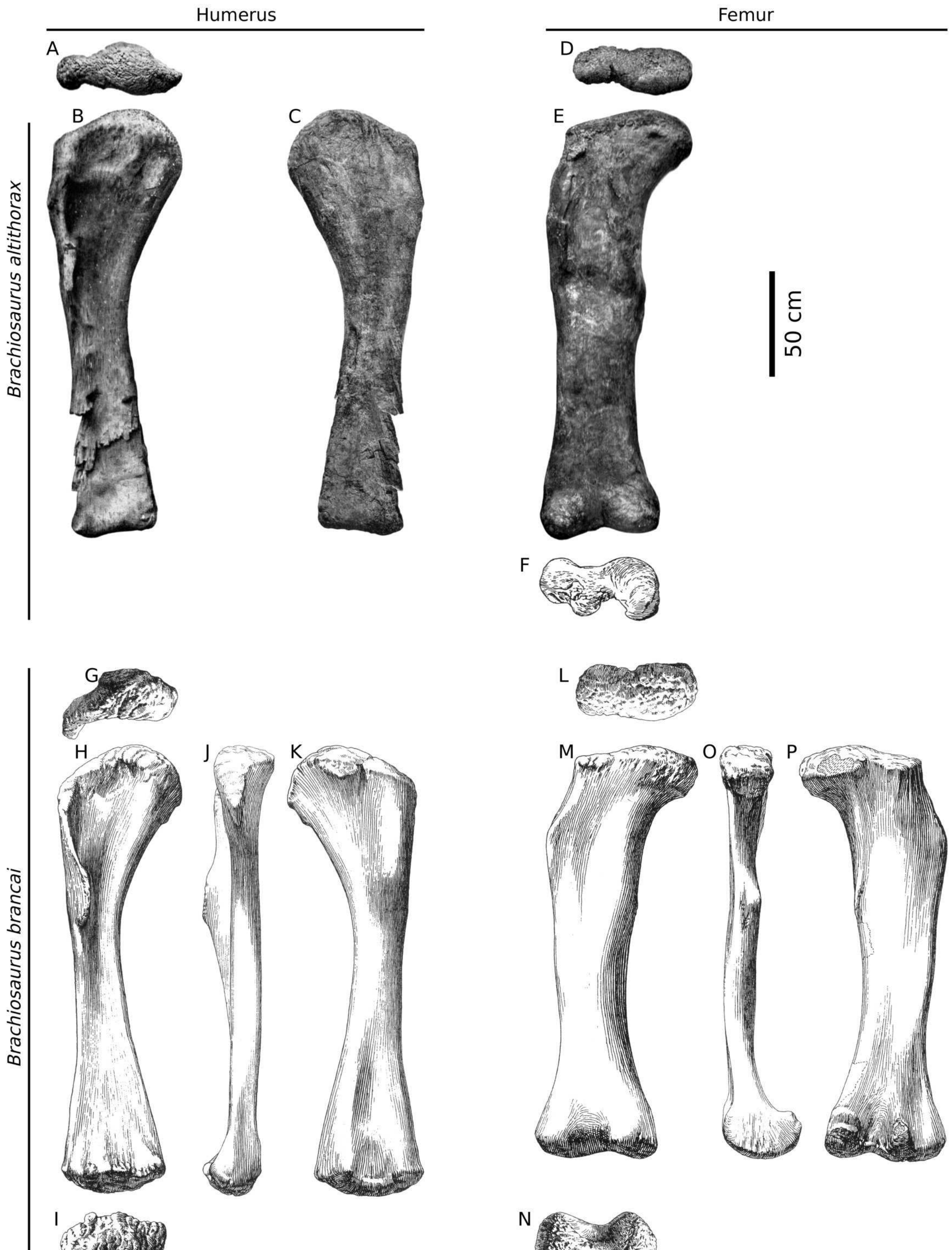

FIGURE 5. Right limb bones of *Brachiosaurus altithorax* and *Brachiosaurus brancai*, equally scaled. **A**-**C**, humerus of *B*. *altithorax* holotype FMNH P 25107; **D**-**F**, femur of same; **G**-**K**, humerus of *B*. *brancai* lectotype HMN SII; **L**-**P**, femur of *B*. *brancai* referred specimen HMN St 291, scaled to size of restored femur of HMN SII as estimated by Janensch (1950b:99). **A**, **D**, **G**, **L**, proximal; **B**, **E**, **H**, **M**, anterior; **C**, **K**, **P**,

posterior; **J**, **O**, medial; **F**, **I**, **N**, distal. **A**, **B**, **D**, **E** modified from Riggs (1904:pl. LXXIV); **C** modified from Riggs (1904:fig. 1); **F** modified from Riggs (1903:fig. 7); **G-K** modified from Janensch (1961:Beilage A); **L-P** modified from Janensch (1961:Beilage J). Scale bar equals 50 cm.

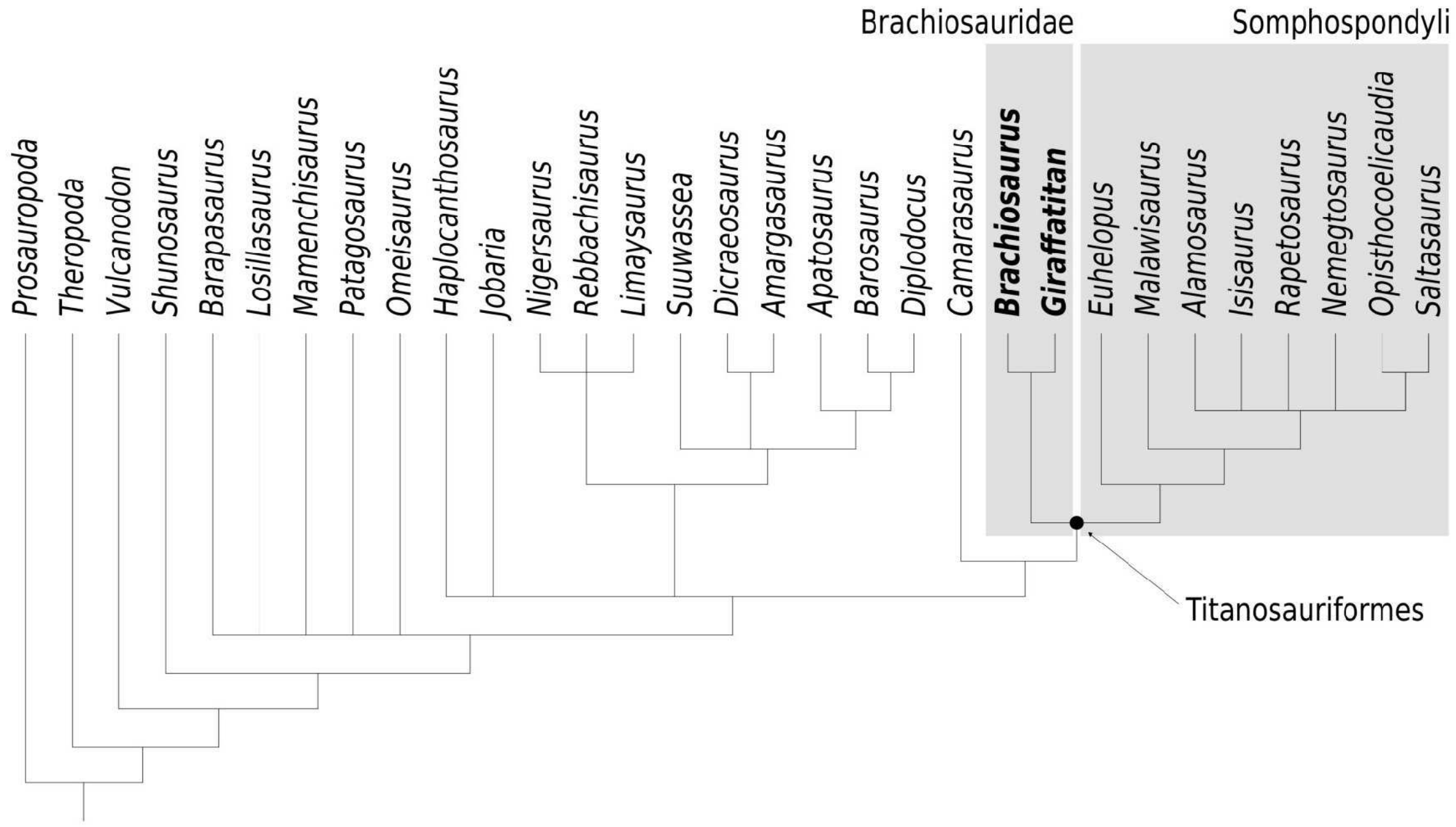

FIGURE 6. Phylogenetic relationships of *Brachiosaurus* and *Giraffatitan*, produced using PAUP* 4.0b10 on the matrix of Harris (2006) modified by splitting the composite "*Brachiosaurus"* OTU into two separate OTUs for the two species, having 31 taxa and 331 characters. Strict consensus of 72 most parsimonious trees (length = 791; CI = 0.5196; RI = 0.6846, RC = 0.3557). Three clades forming a node-stem triplet are highlighted: the node-based Titanosauriformes, and the branch-based sister clades Brachiosauridae and Somphospondyli.

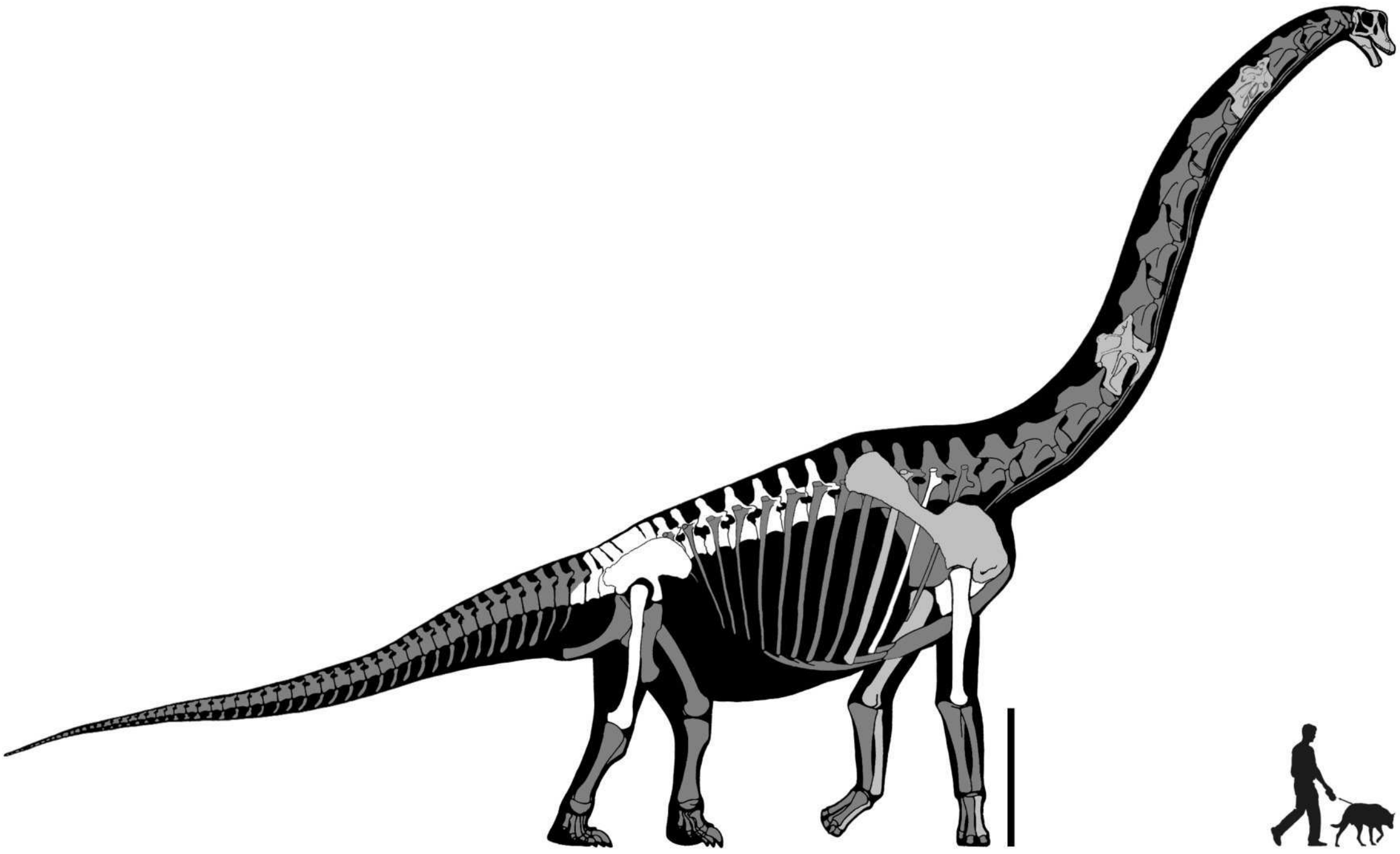

FIGURE 7. Skeletal reconstruction of *Brachiosaurus altithorax*. White bones represent the elements of the holotype FMNH P 25107. Light grey bones represent material referred to *B*. *altithorax*: the Felch Quarry skull USNM 5730, the cervical vertebrae BYU 12866 (C?5) and BYU 12867 (C?10), the "*Ultrasauros*" scapulocoracoid BYU 9462, the Potter Creek left humerus USNM 21903, left radius and right metacarpal III BYU 4744, and the left metacarpal II OMNH 01138. Dark grey bones modified from Paul's (1988) reconstruction of *Giraffatitan brancai*. Scale bar equals 2 m.

Chapter 3 follows. This paper was published in Palaeontology 50(6):1547–1564, published by The Palaeontological Society.

# AN UNUSUAL NEW NEOSAUROPOD DINOSAUR FROM THE LOWER CRETACEOUS HASTINGS BEDS GROUP OF EAST SUSSEX, ENGLAND

*by* MICHAEL P. TAYLOR *and* DARREN NAISH

Palaeobiology Research Group, School of Earth and Environmental Sciences, University of Portsmouth, Burnaby Road, Portsmouth PO1 3QL, UK; e-mails dino@miketaylor.org.uk; darren.naish@port.ac.uk

**Abstract:** *Xenoposeidon proneneukos* gen. et sp. nov. is a neosauropod represented by BMNH R2095, a well-preserved partial mid-to-posterior dorsal vertebra from the Berriasian–Valanginian Hastings Beds Group of Ecclesbourne Glen, East Sussex, England. It was briefly described by Lydekker in 1893, but it has subsequently been overlooked. This specimen's concave cotyle, large lateral pneumatic fossae, complex system of bony laminae and camerate internal structure show that it represents a neosauropod dinosaur. However, it differs from all other sauropods in the form of its neural arch, which is taller than the centrum, covers the entire dorsal surface of the centrum, has its posterior margin continuous with that of the cotyle, and slopes forward at 35 degrees relative to the vertical. Also unique is a broad, flat area of featureless bone on the lateral face of the arch; the accessory infraparapophyseal and postzygapophyseal laminae which meet in a 'V'; and the asymmetric neural canal, small and round posteriorly but large and teardrop-shaped anteriorly, bounded by arched supporting laminae. The specimen cannot be referred to any known sauropod genus, and clearly represents a new genus and possibly a new 'family'. Other sauropod remains from the Hastings Beds Group represent basal Titanosauriformes, Titanosauria and Diplodocidae; *X. proneneukos* may bring to four the number of sauropod 'families' represented in this unit. Sauropods may in general have been much less morphologically conservative than is usually assumed. Since neurocentral fusion is complete in R2095, it is probably from a mature or nearly mature animal. Nevertheless, size comparisons of R2095 with corresponding vertebrae in the *Brachiosaurus brancai* holotype HMN SII and *Diplodocus carnegii* holotype CM 84 suggest a rather small sauropod: perhaps 15 m long and 7600 kg in mass if built like a brachiosaurid, or 20 m and 2800 kg if built like a diplodocid.

THE remains of sauropod dinosaurs have been known from the Lower Cretaceous Wealden strata of the English mainland since the 1840s. Although sauropods were not recognised as a distinct dinosaurian group until somewhat later (Phillips 1871; Marsh 1878*a*), the first named sauropod, *Cetiosaurus brevis* Owen, 1841, was coined for Wealden material (Naish and Martill 2001; Upchurch and Martin 2003).

Most Wealden sauropods are from the Barremian Wessex Formation of the Isle of Wight. Far less well represented are the sauropods of the older Berriasian–Valanginian (Allen and Wimbledon 1991) Hastings Beds Group of the mainland Wealden. Specimens have been collected from Cuckfield, West Sussex (Owen 1841; Mantell 1850); Hastings, East Sussex (Mantell 1852); and most recently from Bexhill, East Sussex (Anonymous 2005). There are indications that a taxonomic diversity similar to that of the Wessex Formation is present among these forms, as discussed below.

Here we describe a Hastings Beds Group specimen first reported, briefly, by Lydekker (1893*a*). This specimen was collected by Philip James Rufford and subsequently acquired by the British Museum (Natural History), now the Natural History Museum, where it is deposited as specimen BMNH R2095.

Though consisting only of a single incomplete vertebra, R2095 preserves many phylogenetically informative characters that allow it to be confidently identified as a neosauropod. Furthermore, it is highly distinctive, possessing several autapomorphies. While it is generally difficult to assess the affinities of isolated bones, sauropod vertebrae, especially dorsal vertebrae, are highly diagnostic (Berman and McIntosh 1978, p. 33; Bonaparte 1986*a*, p. 247; McIntosh 1990, p. 345), and this is particularly true of the specimen described here.

Lydekker (1893*a*, p. 276) reported that this specimen was discovered in 'the Wealden of Hastings' (Text-fig. 1), but beyond that no locality or stratigraphic data was recorded. Watson and Cusack (2005, p. 4) confirmed that Rufford generally collected 'from the Wealden beds of the Hastings area, East Sussex'. Specific plant fossils known to have been collected by Rufford came from East Cliff (Watson and Cusack 2005, p. 75) and from the Fairlight Clays of Ecclesbourne Glen (Watson and Cusack 2005, pp. 64, 80, 87, 107, 112, 125, 128, 138, 152-3), both in the Fairlight area. The units exposed at both East Cliff and Ecclesbourne Glen are part of the Ashdown Beds

Formation, which straddles the Berriasian–Valanginian boundary (Text-fig. 2). The vertebra was probably collected from Ecclesbourne Glen since (1) it is closer to Hastings than is East Cliff, and Lydekker (1893*a*) stated that the specimen was collected near Hastings; and (2) the majority of Rufford's documented specimens came from there. The part of the Ashdown Beds Formation exposed at Ecclesbourne Glen is Berriasian in age (Watson and Cusack 2005), so this is the most likely age of R2095.

*Anatomical nomenclature.* The term 'pleurocoel' has been widely used to refer to the lateral excavations in the centra of sauropods and other saurischian dinosaurs. However, the blanket use of this term obscures the morphological diversity of these cavities, which varies considerably between taxa, encompassing everything from broad, shallow fossae to small, deep foramina; and some taxa have both of these. Furthermore, the term has been used inconsistently in the literature, so that characters such as 'pleurocoels present' in cladistic analyses are difficult to interpret. For example, in the analysis of Wilson (2002), character 78 is defined as 'Presacral centra, pneumatopores (pleurocoels): absent (0); present (1)' (Wilson 2002, p. 261), and *Barapasaurus* Jain, Kutty and Roy-Chowdhury, 1975 is scored as 0 ('pleurocoels absent'). While *Barapasaurus* does indeed lack pneumatic foramina, it has shallow lateral fossae (Jain *et al.* 1979, pl. 101-102), a feature that is not conveyed by the traditional terminology. Accordingly, we recommend that the ambiguous term 'pleurocoel' (and Wilson's equivalent 'pneumatopore') be deprecated in favour of the more explicit alternatives 'lateral fossa' and 'lateral foramen' (Britt 1993, 1997; Wedel *et al.* 2000*b*; Wedel 2003, 2005). The EI (elongation index) of Upchurch (1998) is here used as redefined by Wedel *et al.* (2000*b*), being the length of the centrum divided by the height of the cotyle.

*Anatomical abbreviations.* ACDL, anterior centrodiapophyseal lamina; ACPL, anterior centroparapophyseal lamina; CPOL, centropostzygapophyseal lamina; CPRL, centroprezygapophyseal lamina; PCDL, posterior centrodiapophyseal lamina; PCPL, posterior centroparapophyseal lamina; PODL, postzygodiapophyseal lamina; PPDL, paradiapophyseal lamina; PRPL, prezygoparapophyseal lamina. We follow the vertebral lamina nomenclature of Janensch (1929) as translated by Wilson (1999) except in using capital letters for the abbreviations, a convention that allows plurals to be more clearly formed.

*Institutional abbreviations*. BMNH, The Natural History Museum, London, England; CM, Carnegie Museum of Natural History, Pittsburgh, Pennsylvania, USA; FMNH, Field Museum of Natural History, Chicago, Illinois, USA; HMN, Humboldt Museum für Naturkunde, Berlin, Germany; MIWG, Museum of Isle of Wight Geology (now Dinosaur Isle Visitor Centre), Sandown, Isle of Wight, England; MPEF, Museo Paleontológico Egidio Feruglio, Trelew, Argentina.

# SYSTEMATIC PALAEONTOLOGY

**DINOSAURIA Owen, 1842**

**SAURISCHIA Seeley, 1888**

**SAUROPODOMORPHA Huene, 1932**

**SAUROPODA Marsh, 1878*a***

**NEOSAUROPODA Bonaparte, 1986*b***

**Genus XENOPOSEIDON gen. nov.**

*Derivation of name.* From Xenos (strange or alien, Greek) and Poseidon (the god of earthquakes and the sea in Greek mythology), the latter in reference to the sauropod *Sauroposeidon* Wedel, Cifelli and Sanders, 2000*a*. Intended pronunciation: ZEE-no-puh-SYE-d'n.

*Type species*. *Xenoposeidon proneneukos* sp. nov.

*Diagnosis*. As for the type and only species, *X. proneneukos*.

***Xenoposeidon proneneukos* sp. nov.**

**Plates 1–2; Text-figure 3; Table 1.**

*Derivation of name*. Forward sloping (Latin) describing the characteristic morphology of the neural arch. Intended pronunciation: pro-nen-YOO-koss.

*Holotype.* BMNH R2095, Natural History Museum, London, England. A mid-to-posterior dorsal vertebra consisting of partial centrum and neural arch.

*Type locality*. Near Hastings, East Sussex, United Kingdom; probably Ecclesbourne Glen, about two km east of Hastings. Precise locality information either has been lost or was never recorded.

*Type horizon*. Hastings Beds Group (Berriasian–Valanginian, earliest Cretaceous); probably Berriasian part of the Ashdown Beds Formation. Precise stratigraphic information either has been lost or was never recorded.

*Diagnosis.* Differs from all other sauropods in the following characters: (1) neural arch covers dorsal surface of centrum, with its posterior margin continuous with that of the cotyle; (2) neural arch slopes anteriorly 35 degrees relative to the vertical; (3) broad, flat area of featureless bone on lateral face of neural arch; (4) accessory infraparapophyseal and postzygapophyseal laminae meeting ventrally to form a 'V'; (5) neural canal is asymmetric: small and circular posteriorly but tall and teardrop-shaped anteriorly; (6) supporting laminae form vaulted arch over anterior neural canal.

# DESCRIPTION

BMNH R2095 (Pl. 1, 2) is a partial dorsal vertebra from the middle or posterior portion of the dorsal column. Most of the centrum and neural arch are preserved, but the condyle is broken, and the neural spine and dorsal part of the neural arch are missing, as are the pre- and postzygapophyses and diapophyses. However, sufficient laminae remain to allow the positions of the processes to be inferred with some certainty (Text-fig. 3). Measurements are summarised in Table 1.

The most striking features of this specimen are the extreme height, anteroposterior length and anterodorsal inclination of the neural arch. These are clearly genuine osteological features and not the result of post-mortem distortion. Although the dorsalmost preserved part of the neural arch is ventral to the diapophyses, the height even of the remaining portion (160 mm above the anterodorsal margin of the centrum, measured perpendicular to the anteroposterior axis of the centrum) is equal to that of the cotyle. The centrum is 190 mm long measured along its dorsal margin; its anteroventral portion is missing but a maximum length of 200 mm is indicated, assuming that the curvature of the condyle is approximately equal to that of the cotyle. The base of the neural arch is 170 mm in anteroposterior length, 85 per cent of the estimated total length of the centrum, and its posterior margin is continuous with that of the cotyle, forming a single smooth curve when viewed laterally. The angle of the neural arch's inclination relative to the vertical cannot be precisely ascertained due to the absence of the condyle, but was approximately 35 degrees and cannot have varied from this by more than five degrees or so unless the condyle was shaped very differently from that of other sauropods.

A clean break of the condyle exposes within the centrum the dorsal part of a median septum and a pair of ventromedially directed lateral septa, indicative of an extensively pneumatised centrum with camerate, rather than camellate to somphospondylous, internal structure. The ventral portion of the broken condyle cannot be described as it is obscured by a catalogue note. The cotyle is slightly concave, its central portion indented 10-15 mm relative to its margin. It is 160 mm tall and 170 mm wide. A very subtle keel is present on the ventral surface of the centrum, and the ventral border of the centrum is gently arched in lateral view.

On the better preserved left side of the vertebra, a shallow lateral fossa is positioned dorsally on the centrum, and about midway between the anterior and posterior margins of the neural arch, onto which it intrudes. It is very roughly triangular in shape, taller anteriorly than posteriorly, with a maximum height of 80 mm and a total length of 95 mm. Set within this is a deeper lateral foramen, oval, anteroposteriorly elongate and measuring 80 by 40 mm. The fossa and foramen share their ventral borders. On the right side, the lateral fossa is situated even more dorsally, but is taller posteriorly than anteriorly, with a maximum height of 55 mm and a total length of 90 mm. The lateral foramen is much smaller on this side, measuring only 20 by 15 mm, and is anteroventrally placed within the fossa.

On the left side, the dorsal border of the lateral fossa is formed by a prominent sharp-lipped lateral ridge, which extends anterodorsally for 90 mm; this is absent on the right side, seemingly due not to damage but to intravertebral variation. Instead, an irregularly shaped and sharp-lipped border separates the fossa from a more dorsally placed subcircular 'accessory fossa' 30 mm in diameter. On this side, an accessory lamina connects the anterior part of the border between the main and accessory fossae to a prominent boss positioned on the anterior margin of the neural arch, 50 mm above the anterodorsal margin of the centrum. This is not a parapophysis or diapophysis, but seems to be an aberrant feature of this individual. Neither the accessory fossa nor the anterior boss has been reported in any other sauropod vertebra; however, these features are not considered taxonomically significant as their occurrence on only one side of the vertebra suggests that they are either pathological or a developmental aberration. Pneumatic features vary wildly and may be opportunistic, if Witmer (1997, p. 64) is correct that 'Pneumatic diverticula are [...] opportunistic pneumatizing machines, resorbing as much bone as possible within the constraints imposed by local biomechanical loading regimes.'

The remaining features are described from the left side of the vertebra. The right side is consistent with this morphology, although not all features are preserved.

From a point anterior to the anterodorsal margin of the lateral fossa, a vertically oriented ACPL extends dorsally 70 mm to a cross-shaped junction of laminae near the anterior margin of the arch, and may also have extended a similar distance ventrally although damage makes it impossible to establish this. The cross-shaped junction is interpreted as the location of the parapophysis. In sauropods, the position of the parapophysis migrates dorsally in successive dorsal vertebrae, being located ventrally on the centrum of anterior dorsals, dorsally on the centrum in mid-to-anterior dorsals, and on the neural arch of mid-to-posterior dorsals, level with the prezygapophyses in the most posterior dorsals – see for example Hatcher (1901, pl. VII). The high position of the parapophysis on the neural arch of R2095 indicates a mid-to-posterior placement of the vertebra within the dorsal column, but, because the prezygapophyses must have been dorsal to it, it was probably not among the most posterior vertebrae in the sequence.

In addition to the ACPL, three further laminae radiate from the parapophysis:

part of an anteriorly directed PRPL, the ventral portion of a dorsally directed lamina which is interpreted as a PPDL, and a posteroventrally directed accessory lamina supporting the parapophysis. This is presumably homologous with a PCPL, but cannot be so named as it does not approach the centrum, and indeed extends only 30 mm. Where the latter lamina merges with the neural arch, another accessory lamina arises. Directed posterodorsally, it presumably extended to the postzygapophysis and is here regarded as an accessory postzygapophyseal lamina similar to that found in posterior dorsal vertebrae of *Diplodocus carnegii* Hatcher, 1901 (Hatcher 1901, pl. VII). The PPDL, accessory infraparapophyseal and accessory postzygapophyseal lamina form three sides of a quadrilateral fossa; the fourth side, presumably formed by a PODL, is not preserved, although a very low and unobtrusive accessory lamina does join the dorsalmost preserved part of the PPDL to the accessory postzygapophyseal lamina. The near-vertical orientation of the PPDL indicates that the diapophysis was located some distance directly dorsal to the parapophysis, further extending the inferred height of the neural arch and ruling out an interpretation of the accessory postzygapophyseal lamina as the ACDL or as the "accessory PCDL" of Salgado *et al.* (2005). Finally, a broken ridge of bone extends up the posterior margin of the lateral face of the neural arch. Its identity is problematic: it cannot be a PCDL due to the anterior position inferred for the diapophysis.

Between the ACPL and the posterior lamina, above the dorsal margin of the lateral fossa and below the accessory laminae described above, the lateral face of the neural arch is a flat featureless area measuring 90 mm anteroposteriorly and 50 mm dorsoventrally. This feature is not observed in any other sauropod vertebra.

In posterior view, the pedicels of the neural arch are robust pillars, leaning somewhat medially, measuring 30 mm in width, extending at least 130 mm dorsally and merging into the CPOLs before damage obscures their further extent. They enclose a neural canal that is almost exactly circular, 35 mm in diameter. There is no trace of the postzygapophyses or hyposphene, and no indication that these structures were attached to the preserved portion of the arch. It must be assumed, then, that these features were located on the lost, more dorsal, part of the neural arch. The hyposphene, if present, was located at least 90 mm dorsal to the centrum (measured from the floor of the neural canal), and the postzygapophyses at least 140 mm dorsal to the centrum.

In anterior view, too, the pedicels are robust, being 25 mm in width. They merge gradually into the CPRLs and extend dorsally for at least 80 mm, dorsal to which they are broken. In this aspect, however, the neural canal has no roof, instead forming a large teardrop-shaped vacuity 120 mm tall and 55 mm wide. The dorsal portion of this vacuity is bounded by a pair of gently curved, dorsomedially directed laminae unknown in other sauropods, which meet at a 55 degree angle to form an arch dorsal to the neural canal. The vacuity is filled with matrix, so the extent of its penetration posteriorly into the neural arch cannot be assessed. The prezygapophyses are absent; their articular surfaces were probably about 140 mm above the floor of the neural canal, judging by the trajectory of the PRPL.

The most anterodorsal preserved portion of the vertebra is obscured by a flat, anterodorsally directed 'apron' of matrix, 15 mm thick and 120 mm wide, which hampers interpretation of the prezygapophyseal area.

# COMPARISONS AND INTERPRETATION

The large size of the specimen, combined with its concave cotyle, lateral foramina and complex system of bony laminae, indicate that it is a sauropod vertebra (Salgado *et al.* 1997, p. 6; Wilson and Sereno 1998, pp. 42-43). Within this group, the deep excavation of the anterior face of the neural arch and the height of the neural arch exceeding that of the centrum (Upchurch 1998, char. B7, B6) place the specimen within the clade (*Barapasaurus* + Eusauropoda). The deep lateral foramen indicates that the specimen is within or close to Neosauropoda (Salgado *et al.* 1997, pp. 8-9; Wilson and Sereno 1998, p. 44; Upchurch 1998, char. B5), as does the camerate internal structure of the centrum (Wedel 2003:354). Possession of an ACPL suggests placement with Neosauropoda (Upchurch 1998, char. H3), a group of advanced sauropods consisting of diplodocoids, macronarians (camarasaurids, brachiosaurids and titanosaurs), and in some phylogenies *Haplocanthosaurus* Hatcher, 1903*a*. This identification is corroborated by the fact that no definitive non-neosauropod sauropods are known from the Cretaceous (Upchurch and Barrett 2005, p. 119) – *Jobaria tiguidensis* Sereno, Beck, Dutheil, Larsson, Lyon, Moussa, Sadleir, Sidor, Varricchio, Wilson and Wilson, 1999 from the Lower Cretaceous or Cenomanian of Niger, Africa was recovered as a non-neosauropod by Sereno *et al*. (1999) and Wilson (2002), but as a basal macronarian by Upchurch *et al*. (2004).

ACPLs are also present, apparently by way of convergence, in mamenchisaurids, i.e. the mostly Chinese radiation of basal eusauropods including *Mamenchisaurus* Young, 1954 and *Omeisaurus* Young, 1939 (Upchurch 1998, char. D4), suggesting an alternative identity for R2095. (Upchurch terms these animals ‘euhelopodids’, but since *Euhelopus* Romer, 1956 itself is recovered outside this group in some analyses (Wilson and Sereno 1998; Wilson 2002), this name is misleading. Of the other available names for this group, we prefer the older name ‘Mamenchisauridae’ Young and Zhao, 1972 over Wilson’s (2002) ‘Omeisauridae’, as now does Wilson himself (pers. comm. 2006 to MPT).) The posterior dorsal vertebrae of the mamenchisaurid *Mamenchisaurus hochuanensis* Young and Zhao, 1972 indeed have ACPLs, but they do not at all resemble those of R2095, being much shorter and less defined. The vertebrae resemble R2095 in having tall neural arches; however, they lack lateral foramina entirely and their centra are amphiplatyan (Young and Zhao 1972, fig. 7), thereby ruling out a mamenchisaurid identity for R2095.

We now consider each neosauropod group in turn, investigating the possibility of *X. proneneukos*’s membership of these groups.

*Diplodocoidea*

Tall neural arches are not unusual in the dorsal vertebrae of diplodocoids; and forward-sloping neural arches are known in this group, for example in dorsals 6-8 of CM 84, the holotype of *Diplodocus carnegii* (Hatcher 1901, pl. VII). Taken alone, these gross morphological characters of the neural arch suggest that R2095 may represent a diplodocoid. However, the length of the centrum, especially in so posterior a dorsal vertebra, argues against this possibility: the posterior dorsal centra of diplodocoids typically have EI < 1.0, compared with 1.25 for R2095. Furthermore, the lateral foramina of diplodocoids are more anteriorly located on the centrum and not set within fossae (e.g. Hatcher 1901, pl. VII; Ostrom and McIntosh 1966, pl. 19).

Among diplodocoids, rebbachisaurids differ in dorsal morphology from the better known diplodocids and dicraeosaurids, and in some respects R2095 resembles the dorsal vertebra of the type specimen of *Rebbachisaurus garasbae* Lavocat, 1954. As shown by Bonaparte (1999*a*, fig. 39), that vertebra has a tall neural arch whose posterior margin closely approaches, though it is not continuous with, that of the centrum. However, it differs from R2095 in many respects: for example, possession of a very prominent PCPL (LIP of Bonaparte's usage), large and laterally diverging prezygapophyses, depressions at the base of the neural arch (Bonaparte 1999*a*, p. 173), lateral foramina not set within fossae, and a strongly arched ventral border to the centrum. There is, then, no basis for assigning R2095 to this group.

In some phylogenies (e.g. Wilson 2002, fig. 13A), *Haplocanthosaurus* is recovered as a basal diplodocoid close to Rebbachisauridae, and its dorsal vertebrae are quite similar to those of *Rebbachisaurus* (compare Hatcher 1903*b*, pl. I with Bonaparte 1999*a*, fig. 39). R2095 therefore bears a superficial resemblance to the dorsal vertebrae of *Haplocanthosaurus*, but a close relationship with that genus is precluded for the same reasons that R2095 is excluded from Rebbachisauridae. The dorsal vertebrae of *Haplocanthosaurus,* and some rebbachisaurids (e.g. *Limaysaurus* [= "*Rebbachisaurus*"] *tessonei* Calvo and Salgado, 1995), have asymmetric neural canals, but in the opposite sense from R2095: they are circular anteriorly, and tall and arched posteriorly. Furthermore, the posterior arches of the neural canals in these taxa, composed of dorsomedially inclined CPOLs that meet below the zygapophyses, are very different from the anterior arch of R2095, which is composed of novel laminae that enclose the neural canal, laterally bound by the CPRLs.

*Macronaria*

The concave cotyle of R2095 in so posterior a dorsal suggests a macronarian identity (Salgado *et al*. 1997, p. 9). The concavity is sufficiently deep to rule out the possibility of the vertebra being amphicoelous, i.e. it must have had a convex condyle; this is also interpreted as a macronarian synapomorphy (Upchurch 1998, char. J6). However, the shallowness of the cotyle's curvature makes this only a weak indication, since in brachiosaurids, camarasaurids and titanosaurs, even the posterior dorsals are strongly opisthocoelous (Wilson and Sereno 1998, p. 51). Among macronarians, the dorsally arched ventral margin of the centrum in lateral view suggests either a brachiosaurid or camarasaurid identity rather than a titanosaurian one (Wilson and Sereno 1998, p. 51).

*Camarasauridae*

The name Camarasauridae has been widely used (e.g. Bonaparte 1986*a*; McIntosh 1990), even though its membership now seems to be restricted to *Camarasaurus* Cope, 1877. Other putative camarasaurid genera such as *Morosaurus* Marsh, 1878*a* and *Cathetosaurus* Jensen, 1988 are currently considered synonymous with *Camarasaurus* (Osborn and Mook 1921; McIntosh *et al.* 1996), although morphological differences between specimens suggest that the genus may have been over-lumped. Various other genera have been referred to Camarasauridae but most of these are no longer considered to be closely related to *Camarasaurus*: for example, *Opisthocoelicaudia* Borsuk-Bialynicka, 1977 was considered camarasaurid by its describer and by McIntosh (1990), but is now considered titanosaurian (Salgado and Coria 1993; Upchurch 1998); and *Euhelopus* is now considered either a mamenchisaurid (Upchurch 1995, 1998) or closely related to Titanosauria (Wilson and Sereno 1998; Wilson 2002). However, remaining possible camarasaurids include *Janenschia* Wild, 1991, considered camarasaurid by Bonaparte *et al.* (2000) but titanosaurian by Wilson (2002, p. 248) and Upchurch *et al.* (2004, p. 310); the unnamed proximal fibula described by Moser *et al.* (2006, p. 46) as camarasaurid based on the shape of the tibial articular face; and *Datousaurus bashanensis* Dong and Tang, 1984 (Peng *et al.* 2005) and *Dashanpusaurus dongi* Peng, Ye, Gao, Shu and Jiang, 2005. Since *Camarasaurus* morphology differs so characteristically from that of other sauropods, it is useful to refer to ‘camarasaurid’ morphology, and to that end we provisionally use the name Camarasauridae to refer to the clade (*Camarasaurus supremus* Cope, 1877 not *Saltasaurus loricatus* Bonaparte and Powell, 1980): that is, the clade of all organisms sharing more recent ancestry with *Camarasaurus* than with *Saltasaurus*.

The posterior dorsals of *Camarasaurus* have somewhat dorsoventrally elongated neural arches (Osborn and Mook 1921, pl. LXX), and some *Camarasaurus* posterior dorsal vertebrae have a tall infraprezygapophyseal vacuity similar in size to that of R2095 (e.g. Ostrom and McIntosh 1966, pl. 23-25). However, the oval shape of this vacuity is very different, and there are no internal supporting laminae. The neural arches of camarasaurid dorsal vertebrae are typically very close to vertical, giving the vertebrae an ‘upright’ appearance very different from that of R2095 (Osborn and Mook 1921, fig. 37; McIntosh *et al.* 1996, pl. 5, 9); and the small, subcircular, anteriorly placed lateral foramina of camarasaurids contrast with the medium-sized, anteroposteriorly elongate, centrally-placed lateral foramen of R2095. Furthermore, camarasaurid centra are proportionally short, and their neural arches feature prominent infradiapophyseal laminae (Osborn and Mook 1921, pl. LXX) that are absent in R2095. In summary, R2095 does not closely resemble *Camarasaurus*, and a camarasaurid identity may be confidently ruled out.

Instead, the length of the centrum relative to the cotyle height, with an EI of 1.25, suggests a titanosauriform identity for *X. proneneukos* (Upchurch 1998, char. K3). This is corroborated by the shape of the lateral foramen, which is an anteroposteriorly elongate oval (Salgado *et al.* 1997, pp. 18-19) with its posterior margin slightly more acute than its anterior margin (Upchurch 1998, char. M1).

*Brachiosauridae*

The long centrum particularly suggests a brachiosaurid identity, as *Brachiosaurus* Riggs, 1903 has the proportionally longest posterior dorsal centra of all sauropods. Brachiosaurids are the best represented sauropods in the Lower Cretaceous of England (e.g. the '*Eucamerotus*' cotype specimens BMNH R89/90, the unnamed cervical vertebra MIWG 7306 and the undescribed partial skeleton MIWG BP001), so this identity is also supported on palaeobiogeographical grounds.

The cladistic analysis of Salgado *et al.* (1997) recovered a 'Brachiosauridae' that is paraphyletic with respect to Titanosauria, a finding that has been widely quoted (e.g. Wedel *et al.* 2000*b*; Naish *et al.* 2004). However, since only two putative brachiosaurids were included in the analysis (*Brachiosaurus brancai* Janensch, 1914 and *Chubutisaurus* Corro, 1975), this paraphyly amounts to the recovery of *Chubutisaurus* closer to titanosaurs than to *B. brancai*, which is not a particularly surprising result as its brachiosaurid affinity has only ever been tentatively proposed (McIntosh 1990, p. 384), with an alternative titanosaurian identity also mentioned. Furthermore, Salgado *et al.*'s (*Chubutisaurus* + Titanosauria) clade is supported only by a single synapomorphy, 'Distal end of tibia broader transversely than anteroposteriorly (reversal)'. That is, the distal end of the tibia of *Brachiosaurus brancai* is supposed to be longer than broad (Salgado *et al.* 1997, p. 26); but this seems to be contradicted by Salgado *et al.*'s own figure 11. In order to demonstrate that Brachiosauridae as traditionally conceived is paraphyletic, it would be necessary to perform an analysis that includes many putative brachiosaurids, such as *B. altithorax*, *B. brancai*, *Cedarosaurus weiskopfae* Tidwell, Carpenter and Brooks, 1999, *Atlasaurus imelakei* Monbaron, Russell and Taquet, 1999, *Sauroposeidon proteles*, the French '*Bothriospondylus*' material, the '*Eucamerotus*' cotype specimens BMNH R89/90, *Pleurocoelus* Marsh, 1888, the Texan '*Pleurocoelus*' material, *Lapparentosaurus madagascariensis* Bonaparte, 1986*a* and the unnamed Argentinian brachiosaurid MPEF PV 3098/9 (Rauhut 2006). Such an analysis would most likely indicate that some of these taxa are indeed not in the clade Brachiosauridae *sensu* Wilson and Sereno (1998) = (*Brachiosaurus* not *Saltasaurus*), but that a core remains. So far, the analysis that has included most putative brachiosaurids is that of Upchurch *et al.* (2004), which recovered a *Brachiosaurus*-*Cedarosaurus* clade, *Atlasaurus* as a basal macronarian and *Lapparentosaurus* as an indeterminate titanosauriform. Pending restudy of this group, we assess likely membership of Brachiosauridae primarily by morphological similarity to the two *Brachiosaurus* species.

While the overall proportions of R2095 are a good match for those of brachiosaurid dorsals, its lateral excavations are not characteristic of brachiosaurids. In this specimen, a deep foramen is located within a large, shallow fossa, a character usually associated with titanosaurs (Bonaparte and Coria 1993, p. 272), and not found in the *Brachiosaurus altithorax* holotype FMNH P25107 (Riggs 1904, pl. LXXII; MPT, pers. obs. 2005). Only two dorsal vertebrae belonging to *Brachiosaurus brancai* can be interpreted as having this feature: dorsal 7 of the *B. brancai* holotype HMN SII appears to have its lateral foramina located within slightly broader fossae, but its centrum is so reconstructed that this apparent morphology cannot be trusted; and the isolated dorsal vertebra HMN AR1 has a complex divided excavation that could be interpreted in this way, but this vertebra is different from the other *B. brancai* material in several ways and may have been incorrectly referred (MPT, pers. obs. 2005). R2095 also differs from brachiosaurid dorsal vertebrae in the dorsal placement of its foramina and its lack of infradiapophyseal laminae.

*Titanosauria*

Although the lateral fossae and contained foramina of R2095 are a good match for those of titanosaurs (Bonaparte and Coria 1993, p. 272), the specimen is in most other respects incompatible with a titanosaurian identification. The neural spines of titanosaurs are posteriorly inclined by as much as 45 degrees and although the neural spine of R2095 is not preserved, the 35 degree anterior inclination of the neural arch makes such a posterior slope of the spine very unlikely. What remains of the neural arch does not have the 'inflated' appearance characteristic of titanosaurs: the laminae are gracile and clearly delineated, whereas those of titanosaurs are more robust and tend to merge into the wall of the neural arch. The sharp-edged, vertical ACPL of R2095, for example, does not at all resemble the more robust and posteroventrally oriented centroparapophyseal lamina of titanosaurs (Salgado *et al.* 1997, p. 19, fig. 2). *X. proneneukos* also lacks the thick, ventrally forked infradiapophyseal laminae of titanosaurs (Salgado *et al.* 1997, p. 19). Finally, the camerate internal structure of the centrum does not resemble the 'spongy' somphospondylous structure characteristic of titanosaurs, although Wedel (2003, p. 351) pointed out that there are exceptions such as *Gondwanatitan* Kellner and Azevedo, 1999, a seemingly camerate titanosaur. The overall evidence contradicts a titanosaurian identity for R2095.

The origin of titanosaurs has traditionally been interpreted as a vicariance event precipitated by the Late Jurassic breakup of Pangaea into the northern supercontinent of Laurasia and the southern supercontinent of Gondwana (e.g. Lydekker 1893*b*, p. 3; Bonaparte 1984, 1999*c*; Bonaparte and Kielan-Jaworowska 1987; Le Loeuff 1993). Wilson and Upchurch (2003, p. 156) rejected this model, in part on the basis that titanosaur fossils are known from before the Pangaean breakup. However, the pre-Late Jurassic record of titanosaurs is dominated by trace fossils – 'wide-gauge' trackways (Santos *et al.* 1994; Day *et al.* 2002, 2004; see Wilson and Carrano 1999). Titanosaurian body fossils from this era are in short supply and very fragmentary: the earliest titanosarian body-fossil known from adequate material is *Janenschia* from the Kimmeridgian Tendaguru Formation of Tanzania, Africa. We therefore have very little idea what the Middle Jurassic ur-titanosaur, or its Laurasian descendants, looked like. Good Cretaceous titanosaur body fossils are known from Laurasian continents (e.g. *Alamosaurus* Gilmore, 1922 from North America and *Opisthocoelicaudia* from Mongolia); but only from the Maastrichtian, and these may be interpreted as end-Mesozoic immigrants from Gondwana. The body-fossil record of endemic Laurasian Early Cretaceous titanosaurs remains extremely poor, consisting only of suggestive scraps. In this context, it is possible that *Xenoposeidon proneneukos* may represent a titanosaur belonging to the hypothetical endemic Laurasian radiation, in which case it would be the first such known from presacral vertebral material.

In conclusion, while R2095 can be confidently identified as a member of Neosauropoda, its unusual combination of characters, its wholly unique characters and the paucity of comparable Wealden or other Laurasian material preclude assignment to any more specific group within that clade.

*Phylogenetic analysis*

In light of the uncertain result of group-by-group comparisons, and despite the fragmentary material, a preliminary phylogenetic analysis was performed in the hope of elucidating the phylogenetic position of *Xenoposeidon*. We used the data of Harris (2006) and added the new taxon, yielding a matrix of 31 taxa (29 ingroups and two outgroups) and 331 characters. Due to the paucity of material, *Xenoposeidon* could be scored for only thirteen characters, 4% of the total (Table 2). Following Harris (2006), PAUP* 4.0b10 (Swofford 2002) was used to perform a heuristic search using random stepwise addition with 50 replicates and with maximum trees = 500000. The analysis yielded 1089 equally parsimonious trees with length = 785, consistency index (CI) = 0.5248, retention index (RI) = 0.6871 and rescaled consistency index (RC) = 0.3606.

The strict consensus tree (Text-fig. 4A) is poorly resolved, with Neosauropoda, Diplodocoidea and Macronaria all collapsing, and only Flagellicaudata and its subclades differentiated within Neosauropoda. This represents a dramatic loss of resolution compared to the results without *Xenoposeidon* (Harris 2006, fig. 5A), indicating the instability of the new taxon's position. In the 50% majority rule tree (Text-fig. 4B), all the standard sauropod clades were recovered. This majority rule tree recovers *Xenoposeidon* as a non-brachiosaurid basal titanosauriform, the outgroup to the (*Euhelopus* + Titanosauria) clade. However, various most-parsimonious trees also recover *Xenoposeidon* in many other positions, including as a brachiosaurid, basal titanosaur, basal lithostrotian, saltasaurid and rebbachisaurid. In none of the most parsimonious trees does *Xenoposeidon* occur as a non-neosauropod, a camarasaurid or a flagellicaudatan, although in 24 trees it is the outgroup to Flagellicaudata. Two further steps are required if *Xenoposeidon* is constrained to fall outside of Neosauropoda, and one further step if it is constrained to be a camarasaurid. Comparison to the 50% majority rule tree calculated without *Xenoposeidon* (Harris 2006, fig. 6) shows that the inclusion of the new taxon greatly reduces the support for all neosauropod groups outside Flagellicaudata. The phylogenetic instability of *Xenoposeidon* is due not only to the large amount of missing data but also to the unusual combination of character states which, together with its autapomorphies, prevent it from sitting comfortably within any known group.

*Conclusion*

While *X. proneneukos* is clearly a neosauropod, it cannot be referred to any existing neosauropod genus, nor even to any 'family'-level or 'superfamily'-level group – a conclusion first reached by means of group-by-group comparisons and then verified by the phylogenetic analysis. Its unique characters indicate that it is either a highly derived member of one of the known groups, or, more likely, the first representative of a previously unknown group. While we consider this specimen to represent a new 'family'-level clade, raising a new monogeneric family name would be premature; and the new genus's indeterminate position within Neosauropoda means that no useful phylogenetic definition could be formulated.

Although we are reluctant to inflict another vertebra-based taxon upon fellow sauropod workers, BMNH R2095 is highly distinctive and can be separated from other sauropods, and so formal systematic recognition is appropriate. Although some workers have preferred not to raise new names for specimens represented only by limited material, a better criterion is how autapomorphic the preserved portion of the specimen is; and R2095's suite of unique characters emphatically establishes it as distinct. In the light of its separation from all recognised major sauropod clades, failure to recognise it as a separate taxonomic entry would be actively misleading, as typically it is only named genera that participate in diversity surveys such as those of Holmes and Dodson (1997), Fastovsky *et al*. (2004) and Taylor (2006).

# DISCUSSION

*Historical taxonomy*

While the specimen described here represents a diagnosable taxon, the possibility that it is referable to one of the named Hastings Beds Group sauropod taxa must be considered. Two named sauropods are known from the Hastings Beds Group. '*Pelorosaurus*' *becklesii* Mantell, 1852 is based on a humerus, ulna and radius with associated skin, discovered at Hastings. On the basis of the robustness of its limb bones, this taxon appears to be a titanosaur (Upchurch 1995, p. 380; Upchurch *et al*. 2004, p. 308), and one of the earliest reported members of that clade. BMNH R2095 therefore cannot be referred to it. (Since '*P.*' *becklesii* is not congeneric with the *Pelorosaurus* type species *P. conybeari* (see below) it should be given a new name, if it is sufficiently diagnostic. This decision falls outside the scope of the current work.)

The second Hastings Beds Group taxon has a complex nomenclatural history. Four proximal caudal vertebrae (BMNH R2544-2547) and three chevrons (BMNH R2548-2550) from the Hastings Beds Group of Cuckfield, together with specimens from Sandown Bay on the Isle of Wight, were named *Cetiosaurus brevis* Owen, 1842. This is the first named *Cetiosaurus* species that is not a *nomen dubium* and thus is technically the type species. However, because the name *Cetiosaurus* is historically associated with the Middle Jurassic Oxfordshire species *C. oxoniensis* Phillips, 1871, Upchurch and Martin (2003, p. 215) plan to petition the ICZN to make this the type species. *Cetiosaurus brevis* is clearly not congeneric with *C. oxoniensis*: accordingly, the former is referred to as '*C.*' *brevis* from here on. The Isle of Wight '*C.*' *brevis* material was demonstrated to be iguanodontian by Melville (1849) who went on to provide the new name '*C.*' *conybeari* Melville, 1849 for the Cuckfield sauropod component of '*C.*' *brevis*. As has been widely recognized, Melville's (1849) course of action was inadmissible as '*C.*' *brevis* was still available for this material (Ostrom 1970; Steel 1970; Naish and Martill 2001; Upchurch and Martin 2003) and, accordingly, '*C.*' *conybeari* is a junior objective synonym of '*C.*' *brevis*.

Discovered adjacent to the Cuckfield '*C.*' *brevis* vertebrae and chevrons was a large humerus. Mantell (1850) referred this to Melville's (1849) name '*C.*' *conybeari*, but decided that the taxon was distinct enough for its own genus, *Pelorosaurus* Mantell, 1850. (As shown by Torrens (1999, p. 186), Mantell considered the name *Colossosaurus* for this humerus). Though still discussed apart in most taxonomic reviews (e.g. Naish and Martill 2001; Upchurch and Martin 2003), it is therefore clear that *Pelorosaurus conybeari* and '*C.*' *brevis* are objective synonyms, with the latter having priority. As part of the previously mentioned ICZN petition, it is planned to suppress the latter name, and instead conserve the more widely used *Pelorosaurus conybeari*; for now, though, we continue to use '*C.*' *brevis*. The identity and validity of this material remains problematic. The humerus lacks autapomorphies and, though it is brachiosaurid-like and hence conventionally identified as representing a member of that group (e.g. McIntosh 1990), it differs in having a less prominent deltopectoral crest. Furthermore the '*C.*' *brevis* caudal vertebrae are titanosaur-like in at least one feature, the absence of a hyposphenal ridge. On this basis, Upchurch and Martin (2003) proposed that the material be referred to Titanosauriformes *incertae sedis*. It can be seen to be distinct from '*Pelorosaurus*' *becklesii* as the humeri of both species are preserved.

Since R2095 is similar in age and geography to '*C.*' *brevis*, it is conceivable that it might belong to this species; indeed, Lydekker (1893*a*) assumed this to be the case, based on its being distinct from '*Eucamerotus*' ('*Hoplosaurus*' of his usage) and on the unjustified assumption that there were no more than two Wealden sauropods. However, this assignment cannot be supported due to the lack of overlapping material.

To confuse matters further, during part of the 19th and 20th centuries, '*C.*' *brevis* was referred to by the name *Morosaurus brevis*; and it is under this name that R2095 is catalogued. The description of *Morosaurus impar* Marsh, 1878*a* from the Morrison Formation of Como Bluff in Wyoming initiated the naming of several new *Morosaurus* species, and the referral to this genus of species previously classified elsewhere (Marsh 1878*b*, 1889). Marsh (1889) evidently thought that *Morosaurus* might occur in Europe, as '*Pelorosaurus*' *becklesii* was among the species he referred to *Morosaurus*. Nicholson and Lydekker (1889), regarding '*P.*' *becklesii* as a junior synonym of '*Cetiosaurus*' *brevis* and agreeing with Marsh's referral of '*P.*' *becklesii* to *Morosaurus*, then used the new combination *Morosaurus brevis*. This name was now being used for assorted Lower Cretaceous English sauropods belonging to quite different taxa. Use of *Morosaurus brevis* was perpetuated by Lydekker (1890, 1893*a*) and Swinton (1934, 1936). However, Marsh's (1889) original referral of '*Pelorosaurus*' *becklesii* to *Morosaurus* was unsubstantiated as no unique characters shared by the two were identified. The name *Morosaurus* was later shown to be a junior synonym of *Camarasaurus* (Osborn and Mook 1921), so this name is not available for R2095 because it is tied to a holotype now regarded as a junior subjective synonym.

In addition to the named taxa discussed above, a large sauropod metacarpal from Bexhill beach, derived from the Hastings Beds Group, has been identified as diplodocid (Anonymous 2005), an identification confirmed by Matthew F. Bonnan (pers. comm. 2006 to DN). If correctly identified, this specimen indicates the presence of at least three higher sauropod taxa in the Hastings Beds Group (diplodocids, basal titanosauriforms and titanosaurs), or four if *X. proneneukos* indeed represents a new group. The presence of these several different taxa in coeval or near-coeval sediments is not unexpected given the high genus-level sauropod diversity present in many other sauropod-bearing units (e.g. Morrison Formation, Tendaguru Formation).

*Length and mass*

Table 1 shows comparative measurements of R2095 and the dorsal vertebrae of other neosauropods. We can reach some conclusions about the probable size of *X. proneneukos* by comparing its measurements with those of a typical brachiosaurid and a typical diplodocid, reference taxa which bracket the known range of sauropod shapes.

The estimated total centrum length of R2095 including the missing condyle is 200 mm, compared with 330 mm for the seventh dorsal vertebra of *Brachiosaurus brancai* HMN SII (Janensch 1950, p. 44): about 60 per cent as long. If R2095 were built like a brachiosaurid, then, it would be 60 per cent as long as HMN SII, yielding a length of 15 m based on Paul's (1988) estimate of 25 m for that specimen.

The average cotyle diameter of R2095 is 165 mm, compared with 270 mm for HMN SII: again, about 60 per cent. If the two animals were isometrically similar, R2095's mass would have been about $0.6^3 = 22$ per cent that of HMN SII. SII's mass has been variously estimated as 78258 kg (Colbert 1962), 14900 kg (Russell *et al.* 1980), 46600 kg (Alexander 1985), 29000 kg (Anderson *et al.* 1985), 31500 kg (Paul 1988), 74420 kg (Gunga *et al.* 1995), 37400 kg (Christiansen 1997) and 25789 kg (Henderson 2004). Of these estimates, those of Russell *et al.* (1980) and Anderson *et al.* (1985) can be discarded, as they were extrapolated by limb-bone allometry rather than calculated from the volume of models. The estimates of Colbert (1962) and Gunga *et al.* (1995) can also be discarded, as they are based on obviously overweight models. The average of the remaining four estimates is 35322 kg. Based on this figure, the mass of R2095 might have been in the region of 7600 kg, about the weight of a large African bush elephant (*Loxodonta africana*).

R2095 would have been longer and lighter if it were built like a diplodocid. Its centrum length and average cotyle diameter of 200 mm and 165 mm compare with measurements of 270 mm and 295 mm for corresponding vertebrae in *Diplodocus carnegii* CM 84. Therefore, if *X. proneneukos* were diplodocid-like it would be perhaps 74 per cent as long as a 27 m *Diplodocus*, that is, 20 m. Its volume can be estimated as proportional to its centrum length times the square of its average cotyle diameter, under which assumption it would have been 23 per cent as heavy as *Diplodocus*: 2800 kg, based on Wedel's (2005) mass estimate of 12000 kg for CM 84.

While R2095 represents an animal that is small by sauropod standards, neurocentral fusion is complete and the sutures completely obliterated, indicating that it belonged to an individual that was mostly or fully grown (Brochu 1996).

*Sauropod diversity*

Historically, Sauropoda has been considered a morphologically conservative group, showing less diversity in body shape than the other major dinosaurian groups, Theropoda and Ornithischia (e.g. Wilson and Curry Rogers 2005, pp. 1-2). For many decades, the basic division of sauropods into cetiosaurs, mamenchisaurs, diplodocoids, camarasaurs, brachiosaurs and titanosaurs seemed established, and as recently as thirty years ago, Coombs (1975, p. 1) could write that 'little information in the form of startling new specimens has been forthcoming for sauropods over the last forty years'. Recent discoveries are changing this perception, with the discovery of previously unknown morphology in the square-jawed rebbachisaurid *Nigersaurus* Sereno, Beck, Dutheil, Larsson, Lyon, Moussa, Sadleir, Sidor, Varricchio, Wilson and Wilson, 1999, the long-legged titanosaur *Isisaurus* Wilson and Upchurch, 2003 (originally '*Titanosaurus*' *colberti* Jain and Bandyopadhyay, 1997), the short-necked dicraeosaurid *Brachytrachelopan* Rauhut, Remes, Fechner, Cladera and Puerta, 2005, and the truly massive titanosaurs *Argentinosaurus* Bonaparte and Coria, 1993, *Paralititan* Smith,

Lamanna, Lacovara, Dodson, Smith, Poole, Giegengack and Attia, 2001 and *Puertasaurus* Novas, Salgado, Calvo and Agnolin, 2005. During the same period, Rebbachisauridae has emerged as an important group (Calvo and Salgado 1995; Pereda Suberbiola *et al*. 2003; Salgado *et al*. 2004).

Perhaps most interesting of all is the recent erection of two sauropod genera that arguably do not fit into any established group: *Agustinia* Bonaparte, 1999*b* and *Tendaguria* Bonaparte, Heinrich and Wild, 2000. Both of these genera are represented by specimens so different from other sauropods that they have been placed by their authors into new monogeneric 'families', Agustiniidae and Tendaguriidae. Together with *X. proneneukos*, these taxa emphasise just how much remains to be discovered about the Sauropoda and how little of the full sauropod diversity we presently understand. It is hoped that the discovery of new specimens will allow the anatomy and relationships of these enigmatic new sauropods to be elucidated.

# CONCLUSIONS

BMNH R2095 is a highly distinctive dorsal vertebra with several features unique within Sauropoda, and as such warrants a formal name: *Xenoposeidon proneneukos*. It does not seem to belong to any established sauropod group more specific than Neosauropoda, and may represent a new 'family'. *Xenoposeidon* adds to a growing understanding of the richness of sauropod diversity, both within the Hastings Beds Group of the Wealden, and globally.

# ACKNOWLEDGEMENTS

We thank Sandra D. Chapman (Natural History Museum) for access to the specimen, and Nick Pharris (University of Michigan) for etymological assistance. Matthew F. Bonnan (Western Illinois University) and Jeffrey A. Wilson (University of Michigan) gave permission to cite personal communications. We used English translations of several papers from the very useful Polyglot Paleontologist web-site http://ravenel.si.edu/paleo/paleoglot/index.cfm and gratefully acknowledge the efforts of the site maintainer Matthew T. Carrano. Specific thanks are due to the following translators: Sebastián Apesteguía (Bonaparte 1999a), Matthew T. Carrano (Bonaparte 1986b), the late William R. Downs (Young and Zhao 1972), Matthew C. Lamanna (Bonaparte and Coria 1993, Corro 1975 and Lavocat 1954) and Jeffrey A. Wilson (Salgado and Coria 1993). In addition, portions of Janensch 1914 were translated by Gerhard Maier. David M. Martill (University of Portsmouth), Jerald D. Harris (Dixie State College), Leonardo Salgado (Museo de Geología y Paleontología, Buenos Aires) and two anonymous reviewers provided thorough reviews of the manuscript which greatly improved its quality. Fiona Taylor's careful proofreading enabled us to correct several minor errors. Finally, we thank handling editor Oliver W. M. Rauhut (Bayerische Staatssammlung für Paläontologie und Geologie) and Editor-in-Chief David J. Batten (University of Manchester) for their thoroughness and flexibility. even when there have been differences of opinion between us.

# REFERENCES

ALEXANDER, R. M. 1985. Mechanics of posture and gait of some large dinosaurs. *Zoological Journal of the Linnean Society*, **83**, 1–25.

ALLEN, P. and WIMBLEDON, W. A. 1991. Correlation of NW European Purbeck-Wealden (non-marine Lower Cretaceous) as seen from the English type areas. *Cretaceous Research*, **12**, 511–526.

ANDERSON, J. F., HALL-MARTIN, A. and RUSSELL, D. A. 1985. Long-bone circumference and weight in mammals, birds and dinosaurs. *Journal of Zoology*, **207**, 53–61.

ANONYMOUS. 2005. Bexhill's largest dinosaur. *Wealden News: Newsletter of Wealden Geology*, **6**, 1–2.

BERMAN, D. S. and MCINTOSH, J. S. 1978. Skull and Relationships of the Upper Jurassic sauropod *Apatosaurus* (Reptilia, Saurischia). *Bulletin of the Carnegie Museum*, **8**, 1–35.

BONAPARTE, J. F. 1984. Late Cretaceous faunal interchange of terrestrial vertebrates between the Americas. 19–24. *In* REIF, W.-E. and WESTPHAL, F. (eds). *Third Symposium on Mesozoic Terrestrial Ecosystems, Tübingen, Germany*. Attempto Verlag, Tübingen, 259 pp.

——. 1986*a*. The early radiation and phylogenetic relationships of the Jurassic sauropod dinosaurs, based on vertebral anatomy. 247–258. *In* PADIAN, K. (ed). *The Beginning of the Age of Dinosaurs*. Cambridge University Press, Cambridge,

UK, xii+378 pp.

——. 1986*b*. Les dinosaures (Carnosaures, Allosauridés, Sauropodes, Cétiosauridés) du Jurassique moyen de Cerro Cóndor (Chubut, Argentina). *Annales de Paléontologie*, **72**, 325–386.

——. 1999*a*. Evolucion de las vertebras presacras en Sauropodomorpha. *Ameghiniana*, **36**, 115–187.

——. 1999*b*. An armoured sauropod from the Aptian of northern Patagonia, Argentina. 1–12. *In* TOMIDA, Y., RICH, T. H. and VICKERS-RICH, P. (eds). *Proceedings of the Second Gondwanan Dinosaur Symposium*. National Science Museum, Tokyo, x+296 pp.

——. 1999*c*. Tetrapod faunas from South America and India: a palaeobiogeographic interpretation. *Proceedings of the Indian National Science Academy*, **65A**, 427–437.

—— and POWELL, J. E. 1980. A continental assemblage of tetrapods from the Upper Cretaceous beds of El Brete, northwestern Argentina (Sauropoda-Coelurosauria-Carnosauria-Aves). *Memoires de la Société Geologique de France, Nouvelle Série*, **139**, 19–28.

—— and KIELAN-JAWOROWSKA, Z. 1987. Late Cretaceous dinosaur and mammal faunas of Laurasia and Gondwana. 22–29. *In* CURRIE, P. J. and KOSTER, E. H. (eds). *Fourth symposium on Mesozoic terrestrial ecosystems, Drumheller, Canada.* Tyrell Museum of Paleontology, Drumheller, 239 pp.

—— and CORIA, R. A. 1993. Un nuevo y gigantesco sauropodo titanosaurio de la Formacion Río Limay (Albiano-Cenomaniano) de la Provincia de Neuquén, Argentina. *Ameghiniana*, **30**, 271–282.

——, HEINRICH, W.-D. and WILD, R. 2000. Review of *Janenschia* Wild, with the description of a new sauropod from the Tendaguru beds of Tanzania and a discussion on the systematic value of procoelous caudal vertebrae in the Sauropoda. *Palaeontographica A*, **256**, 25–76.

BORSUK-BIALYNICKA, M. 1977. A new camarasaurid sauropod *Opisthocoelicaudia skarzynskii*, gen. n., sp. n., from the Upper Cretaceous of Mongolia.

*Palaeontologica Polonica*, **37**, 5–64.

BRITT, B. B. 1993. *Pneumatic postcranial bones in dinosaurs and other archosaurs (Ph.D. dissertation).* University of Calgary, Calgary, 383 pp.

——. 1997. Postcranial pneumaticity. 590–593. *In* CURRIE, P. J. and PADIAN, K. (eds). *The Encyclopedia of Dinosaurs*. Academic Press, San Diego, xxx+869 pp.

BROCHU, C. A. 1996. Closure of neurocentral sutures during crocodilian ontogeny: implications for maturity assessment in fossil archosaurs. *Journal of Vertebrate Paleontology*, **16**, 49–62.

CALVO, J. O. and SALGADO, L. 1995. *Rebbachisaurus tessonei* sp. nov. a new Sauropoda from the Albian-Cenomanian of Argentina; new evidence on the origin of the Diplodocidae. *Gaia*, **11**, 13–33.

CHRISTIANSEN, P. 1997. Locomotion in sauropod dinosaurs. *Gaia*, **14**, 45–75.

COLBERT, E. H. 1962. The weights of dinosaurs. *American Museum Novitates*, **2076**, 1–16.

COOMBS, W. P. 1975. Sauropod habits and habitats. *Palaeogeography, Palaeoclimatology, Palaeoecology*, **17**, 1–33.

COPE, E. D. 1877. On a gigantic saurian from the Dakota epoch of Colorado. *Paleontology Bulletin*, **25**, 5–10.

CORRO, G. D. 1975. Un nuevo sauropodo del Cretácico Superior. *Chubutisaurus insignis* gen. et sp. nov. (Saurischia, Chubutisauridae nov.) del Cretácico Superior (Chubutiano), Chubut, Argentina. *Actas I Congreso Argentino de Paleontologia y Bioestratigrafia*, **2**, 229–240.

DAY, J. J., UPCHURCH, P., NORMAN, D. B., GALE, A. S. and POWELL, H. P. 2002. Sauropod trackways, evolution and behavior. *Science*, **296**, 1659.

——, NORMAN, D. B., GALE, A. S., UPCHURCH, P. and POWELL, H. P. 2004. A Middle Jurassic dinosaur trackway site from Oxfordshire, UK. *Palaeontology*, **47**, 319–348.

DONG, Z. and TANG, Z. 1984. (In Chinese with English summary). *Vertebrata PalAsiatica*, **22**, 69–74.

FASTOVSKY, D. E., HUANG, Y., HSU, J., MARTIN-MCNAUGHTON, J., SHEEHAN, P. M. and WEISHAMPEL, D. B. 2004. Shape of Mesozoic dinosaur richness. *Geology*, **32**, 877–880.

GILMORE, C. W. 1922. Discovery of a sauropod dinosaur from the Ojo Alamo formation of New Mexico. *Smithsonian Miscellaneous Collections*, **81**, 1–9.

GUNGA, H.-C., KIRSCH, K. A., BAARTZ, F., ROCKER, L., HEINRICH, W.-D., LISOWSKI, W., WIEDEMANN, A. and ALBERTZ, J. 1995. New data on the dimensions of *Brachiosaurus brancai* and their physiological implications. *Naturwissenschaften*, **82**, 190–192.

HARRIS, J. D. 2006. The significance of *Suuwassea emiliae* (Dinosauria: Sauropoda) for flagellicaudatan intrarelationships and evolution. *Journal of Systematic Palaeontology*, **4**, 185–198.

HATCHER, J. B. 1901. *Diplodocus* (Marsh): its osteology, taxonomy and probable habits, with a restoration of the skeleton. *Memoirs of the Carnegie Museum*, **1**, 1–63.

——. 1903*a*. A new name for the dinosaur *Haplocanthus* Hatcher. *Proceedings of the Biological Society of Washington*, **16**, 100.

——. 1903*b*. Osteology of *Haplocanthosaurus* with description of a new species, and remarks on the probable habits of the Sauropoda and the age and origin of the Atlantosaurus beds. *Memoirs of the Carnegie Museum*, **2**, 1–72 and plates I-V.

HENDERSON, D. M. 2004. Tipsy punters: sauropod dinosaur pneumaticity, bouyancy and aquatic habits. *Proceedings of the Royal Society of London, B (Supplement)*, **271**, S180–S183.

HOLMES, T. and DODSON, P. 1997. Counting more dinosaurs: how many kinds are there. 125–128. *In* WOHLBERG, D. L., STUMP, E. and ROSENBERG, G. D. (eds). *Dinofest International: Proceedings of a Symposium Held at Arizona State*

*University*, 587 pp.

HUENE, F. V. 1932. Die fossile Reptile-Ordnung Saurischia, ihre Entwicklung und Geschichte. *Monographien zur Geologie und Palaeontologie (Serie 1)*, **4**, 1–361.

JAIN, S. L. and BANDYOPADHYAY, S. 1997. New titanosaurid (Dinosauria: Sauropoda) from the Late Cretaceous of central India. *Journal of Vertebrate Paleontology*, **17**, 114–136.

——, KUTTY, T. S. and ROY-CHOWDHURY, T. K. 1975. The sauropod dinosaur from the Lower Jurassic Kota Formation of India. *Proceedings of the Royal Society of London A*, **188**, 221–228.

——, KUTTY, T. S. and ROY-CHOWDHURY, T. K. 1979. Some characteristics of *Barapasaurus tagorei*, a sauropod dinosaur from the Lower Jurassic of Deccan, India. *Proceedings of the IV International Gondwana Symposium, Calcutta*, **1**, 204–216.

JANENSCH, W. 1914. Übersicht über der Wirbeltierfauna der Tendaguru-Schichten nebst einer kurzen Charakterisierung der neu aufgefuhrten Arten von Sauropoden. *Archiv fur Biontologie*, **3**, 81–110.

——. 1929. Die Wirbelsaule der Gattung *Dicraeosaurus*. *Palaeontographica (Suppl. 7)*, **2**, 37–133.

——. 1950. Die Wirbelsaule von *Brachiosaurus brancai*. *Palaeontographica (Suppl. 7)*, **3**, 27–93.

JENSEN, J. A. 1988. A fourth new sauropod dinosaur from the Upper Jurassic of the Colorado Plateau and sauropod bipedalism. *Great Basin Naturalist*, **48**, 121–145.

KELLNER, A. W. A. and AZEVEDO, S. A. K. 1999. A new sauropod dinosaur (Titanosauria) from the Late Cretaceous of Brazil. 111–142. *In* TOMIDA, Y., RICH, T. H. and VICKERS-RICH, P. (eds). *Proceedings of the Second Gondwanan Dinosaur Symposium, Tokyo National Science Museum Monograph 15*. Tokyo National Science Museum, Tokyo, x+296 pp.

LAVOCAT, R. 1954. Sur les Dinosauriens du continental intercalaire des Kem-Kem de

la Daoura. *Comptes Rendus 19th Intenational Geological Congress 1952*, **1**, 65–68.

LE LOEUFF, J. 1993. European titanosaurs. *Revue de Paléobiologie*, **7**, 105–117.

LYDEKKER, R. 1890. *Catalogue of the fossil Reptilia and Amphibia in the British Museum (Natural History). Part IV containing the orders Anomodontia, Ecaudata, Caudata and Labyrinthodontia; and Supplement.* British Museum (Natural History), London, 295 pp.

——. 1893*a*. On a sauropodous dinosaurian vertebra from the Wealden of Hastings. *Quarterly Journal of the Geological Society, London*, **49**, 276–280.

——. 1893*b*. The dinosaurs of Patagonia. *Anales del Museo de La Plata*, **2**, 1–14.

MANTELL, G. A. 1850. On the *Pelorosaurus*: an undescribed gigantic terrestrial reptile, whose remains are associated with those of the *Iguanodon* and other saurians in the strata of Tilgate Forest, in Sussex. *Philosophical Transactions of the Royal Society of London*, **140**, 379–390.

——. 1852. On the structure of the *Iguanodon* and on the fauna and flora of the Wealden Formation. *Notice: Proceedings of the Royal institute of Great Britain*, **1**, 141–146.

MARSH, O. C. 1878*a*. Principal characters of American Jurassic dinosaurs. Part I. *American Journal of Science, Series 3*, **16**, 411–416.

——. 1878*b*. Notice of new dinosaurian reptiles. *American Journal of Science, Series 3*, **15**, 241–244.

——. 1888. Notice of a new genus of Sauropoda and other new dinosaurs from the Potomac Formation. *American Journal of Science, Series 3*, **35**, 89–94.

——. 1889. Notice of new American dinosaurs. *American Journal of Science, Series 3*, **37**, 331–336.

MCINTOSH, J. S. 1990. Sauropoda. 345–401. *In* WEISHAMPEL, D. B., DODSON, P. and OSMóLSKA, H. (eds). *The Dinosauria*. University of California Press, Berkeley and Los Angeles, 733 pp.

——, MILLER, W. E., STADTMAN, K. L. and GILLETTE, D. D. 1996. The osteology of *Camarasaurus lewisi* (Jensen, 1988). *BYU Geology Studies*, **41**, 73–115.

MELVILLE, A. G. 1849. Notes on the vertebral column of Iguanodon. *Philosophical Transactions of the Royal Society of London*, **139**, 285–300.

MONBARON, M., RUSSELL, D. A. and TAQUET, P. 1999. *Atlasaurus imelakei* n.g., n.sp., a brachiosaurid-like sauropod from the Middle Jurassic of Morocco. *Comptes Rendus de l'Académie des Sciences. Science de la terre and des planètes*, **329**, 519–526.

MOSER, M., MATHUR, U. B., FüRSICH, F. T., PANDEY, D. K. and MATHUR, N. 2006. Oldest camarasauromorph sauropod (Dinosauria) discovered in the Middle Jurassic (Bajocian) of the Khadir Island, Kachchh, western India. *Paläontologische Zeitschrift*, **80**, 34–51.

NAISH, D. and MARTILL, D. M. 2001. Saurischian dinosaurs 1: Sauropods. 185–241. *In* MARTILL, D. M. and NAISH, D. (eds). *Dinosaurs of the Isle of Wight*. The Palaeontological Association, London, 433 pp.

——, MARTILL, D. M., COOPER, D. and STEVENS, K. A. 2004. Europe's largest dinosaur? A giant brachiosaurid cervical vertebra from the Wessex Formation (Early Cretaceous) of southern England. *Cretaceous Research*, **25**, 787–795.

NICHOLSON, H. A. and LYDEKKER, R. 1889. *A manual of palaeontology for the use of students, with a general introduction on the principles of palaeontology. Third edition. Vol. ii.* William Blackwood and Sons, Edinburgh and London, xi+889 pp.

NOVAS, F. E., SALGADO, L., CALVO, J. and AGNOLIN, F. 2005. Giant titanosaur (Dinosauria, Sauropoda) from the Late Cretaceous of Patagonia. *Revista del Museo Argentino dei Ciencias Naturales, n.s.*, **7**, 37–41.

OSBORN, H. F. and MOOK, C. C. 1921. *Camarasaurus*, *Amphicoelias* and other sauropods of Cope. *Memoirs of the American Museum of Natural History, n.s.*, **3**, 247–387 and plates LX-LXXXV.

OSTROM, J. H. 1970. Stratigraphy and paleontology of the Cloverly Formation (Lower Cretaceous) of the Bighorn Basin area, Wyoming and Montana. *Bulletin of the Peabody Museum of Natural History*, **35**, 1–234.

—— and MCINTOSH, J. S. 1966. *Marsh's Dinosaurs: the Collections from Como Bluff.* Yale University Press, New Haven, CT, 388 pp.

OWEN, R. 1841. A description of a portion of the skeleton of *Cetiosaurus*, a gigantic extinct saurian occurring in the Oolitic Formation of different parts of England. *Proceedings of the Geological Society of London*, **3**, 457–462.

——. 1842. Report on British fossil reptiles, Part II. *Reports of the British Association for the Advancement of Sciences*, **11**, 60–204.

PAUL, G. S. 1988. The brachiosaur giants of the Morrison and Tendaguru with a description of a new subgenus, *Giraffatitan*, and a comparison of the world's largest dinosaurs. *Hunteria*, **2**, 1–14.

PENG, G., YE, Y., GAO, Y., SHU, C. and JIANG, S. 2005. *Jurassic dinosaur faunas in Zigong.* Sichuan People's Publishing House, Zigong, 236 pp.

PEREDA SUBERBIOLA, X., TORCIDA, F., IZQUIERDO, L. A., HUERTA, P., MONTERO, D. and PEREZ, G. 2003. First rebbachisaurid dinosaur (Sauropoda, Diplodocoidea) from the early Cretaceous of Spain: palaeobiogeographical implications. *Bulletin de la Société Geologique de France*, **174**, 471–479.

PHILLIPS, J. 1871. *Geology of Oxford and the Valley of the Thames.* Clarendon Press, Oxford, 529 pp.

RAUHUT, O. W. M. 2006. A brachiosaurid sauropod from the Late Jurassic Cañadón Calcáreo Formation of Chubut, Argentina. *Fossil Record*, **9**, 226–237.

——, REMES, K., FECHNER, R., CLADERA, G. and PUERTA, P. 2005. Discovery of a short-necked sauropod dinosaur from the Late Jurassic period of Patagonia. *Nature*, **435**, 670–672.

RIGGS, E. S. 1903. *Brachiosaurus altithorax*, the largest known dinosaur. *American Journal of Science*, **15**, 299–306.

——. 1904. Structure and relationships of opisthocoelian dinosaurs. Part II, the Brachiosauridae. *Field Columbian Museum, Geological Series 2*, **6**, 229–247.

ROMER, A. S. 1956. *Osteology of the Reptiles.* University of Chicago Press, Chicago, 772 pp.

RUSSELL, D., BELAND, P. and MCINTOSH, J. S. 1980. Paleoecology of the dinosaurs of Tendaguru (Tanzania). *Memoires de la Société Geologique de France*, **139**, 169–175.

SALGADO, L. and CORIA, R. A. 1993. Consideraciones sobre las relaciones filogeneticas de Opisthocoelicaudia skarzynskii (Sauropoda) del Cretacio superior de Mongolia. *Ameghiniana*, **30**, 339.

——, CORIA, R. A. and CALVO, J. O. 1997. Evolution of titanosaurid sauropods. I: Phylogenetic analysis based on the postcranial evidence. *Ameghiniana*, **34**, 3–32.

——, APESTEGUIA, S. and HEREDIA, S. E. 2005. A new specimen of *Neuquensaurus australis*, a Late Cretaceous saltasaurine titanosaur from North Patagonia. *Journal of Vertebrate Paleontology*, **25**, 623–634.

——, GARRIDO, A., COCCA, S. E. and COCCA, J. R. 2004. Lower Cretaceous rebbachisaurid sauropods from Cerro Aguada Del Leon (Lohan Cura Formation), Neuquen Province, Northwestern Patagonia, Argentina. *Journal of Vertebrate Paleontology*, **24**, 903–912.

SANTOS, V. F., LOCKLEY, M. G., MEYER, C. A., CARVALHO, J., GALOPIM, A. M. and MORATALLA, J. J. 1994. A new sauropod tracksite from the Middle Jurassic of Portugal. *Gaia*, **10**, 5–14.

SEELEY, H. G. 1888. On the classification of the fossil animals commonly named Dinosauria. *Proceedings of the Royal Society of London*, **43**, 165–171.

SERENO, P. C., BECK, A. L., DUTHEIL, D. B., LARSSON, H. C. E., LYON, G. H., MOUSSA, B., SADLEIR, R. W., SIDOR, C. A., VARRICCHIO, D. J., WILSON, G. P. and WILSON, J. A. 1999. Cretaceous sauropods from the Sahara and the uneven rate of skeletal evolution among dinosaurs. *Science*, **282**, 1342–1347.

SMITH, J. B., LAMANNA, M. C., LACOVARA, K. J., DODSON, P., SMITH, J. R., POOLE, J. C., GIEGENGACK, R. and ATTIA, Y. 2001. A giant sauropod dinosaur from an Upper Cretaceous mangrove deposit in Egypt. *Science*, **292**, 1704–1706.

STEEL, R. 1970. *Handbuch der Paläoherpetologie. Part 14. Saurischia.* Gustav Fischer Verlag, Stuttgart, 87 pp.

SWINTON, W. E. 1934. *The Dinosaurs: A Short History of a Great Group of Reptiles.* Murby, London, 233 pp.

——. 1936. The dinosaurs of the Isle of Wight. *Proceedings of the Geologists' Association*, **47**, 204–220.

SWOFFORD, D. L. 2002. *PAUP*: phylogenetic analysis using parsimony (* and other methods).* Sinauer Associates, Sunderland, Mass.

TAYLOR, M. P. 2006. Dinosaur diversity analysed by clade, age, place and year of description. 134–138. *In* BARRETT, P. M. (ed). *Ninth international symposium on Mesozoic terrestrial ecosystems and biota, Manchester, UK*. Cambridge Publications, Cambridge, UK, 187 pp.

TIDWELL, V., CARPENTER, K. and BROOKS, W. 1999. New sauropod from the Lower Cretaceous of Utah, USA. *Oryctos*, **2**, 21–37.

TORRENS, H. 1999. Politics and paleontology: Richard Owen and the invention of dinosaurs. 175–190. *In* FARLOW, J. O. and BRETT-SURMAN, M. K. (eds). *The Complete Dinosaur*. Indiana University Press, Bloomington, Indiana, 754 pp.

UPCHURCH, P. 1995. The evolutionary history of sauropod dinosaurs. *Philosophical Transactions of the Royal Society of London Series B*, **349**, 365–390.

——. 1998. The phylogenetic relationships of sauropod dinosaurs. *Zoological Journal of the Linnean Society*, **124**, 43–103.

—— and MARTIN, J. 2003. The anatomy and taxonomy of *Cetiosaurus* (Saurischia, Sauropoda) from the Middle Jurassic of England. *Journal of Vertebrate Paleontology*, **23**, 208–231.

—— and BARRETT, P. M. 2005. Phylogenetic and Taxic Perspectives on Sauropod

Diversity. 104–124. *In* CURRY ROGERS, K. and WILSON, J. A. (eds). *The Sauropods: Evolution and Paleobiology*. University of California Press, Berkeley, Los Angeles and London, ix+349 pp.

——, BARRETT, P. M. and DODSON, P. 2004. Sauropoda. 259–322. *In* WEISHAMPEL, D. B., DODSON, P. and OSMóLSKA, H. (eds). *The Dinosauria, 2nd edition*. University of California Press, Berkeley and Los Angeles, 861 pp.

WATSON, J. and CUSACK, H. A. 2005. Cycadales of the English Wealden. *Monograph of the Palaeontographical Society*, **158**, 1–189 and plates 1-10.

WEDEL, M. J. 2003. The evolution of vertebral pneumaticity in sauropod dinosaurs. *Journal of Vertebrate Paleontology*, **23**, 344–357.

——. 2005. Postcranial skeletal pneumaticity in sauropods and its implications for mass estimates. 201–228. *In* WILSON, J. A. and CURRY-ROGERS, K. (eds). *The Sauropods: Evolution and Paleobiology*. University of California Press, Berkeley, ix+349 pp.

——, CIFELLI, R. L. and SANDERS, R. K. 2000*a*. *Sauroposeidon proteles*, a new sauropod from the Early Cretaceous of Oklahoma. *Journal of Vertebrate Paleontology*, **20**, 109–114.

——, CIFELLI, R. L. and SANDERS, R. K. 2000*b*. Osteology, paleobiology, and relationships of the sauropod dinosaur *Sauroposeidon*. *Acta Palaeontologica Polonica*, **45**, 343–388.

WILD, R. 1991. *Janenschia* n. g. *robusta* (E. Fraas 1908) pro *Tornieria robusta* (E. Fraas 1908) (Reptilia, Saurischia, Sauropodomorpha). *Stuttgarter Beiträge zur Naturkunde, Serie B (Geologie und Paläontologie)*, **173**, 4–1-4.

WILSON, J. A. 1999. A nomenclature for vertebral laminae in sauropods and other saurischian dinosaurs. *Journal of Vertebrate Paleontology*, **19**, 639–653.

——. 2002. Sauropod dinosaur phylogeny: critique and cladistic analysis. *Zoological Journal of the Linnean Society*, **136**, 217–276.

—— and SERENO, P. C. 1998. Early evolution and higher-level phylogeny of sauropod dinosaurs. *Society of Vertebrate Paleontology Memoir*, **5**, 1–68.

—— and CARRANO, M. T. 1999. Titanosaurs and the origin of "wide-gauge" trackways: a biomechanical and systematic perspective on sauropod locomotion. *Paleobiology*, **25**, 252–267.

—— and UPCHURCH, P. 2003. A revision of *Titanosaurus* Lydekker (Dinosauria - Sauropoda), the first dinosaur genus with a 'Gondwanan' distribution. *Journal of Systematic Palaeontology*, **1**, 125–160.

—— and CURRY ROGERS, K. 2005. Introduction: monoliths of the Mesozoic. 1–14. *In* CURRY ROGERS, K. and WILSON, J. A. (eds). *The sauropods: evolution and paleobiology*. University of California Press, Berkeley and Los Angeles, x+349 pp.

WITMER, L. M. 1997. The evolution of the antorbital cavity of archosaurs: a study in soft-tissue reconstruction in the fossil record with an analysis of the function of pneumaticity. *Society of Vertebrate Paleontology Memoir*, **3**, 1–73.

YOUNG, C.-C. 1939. On the new Sauropoda, with notes on other fragmentary reptiles from Szechuan. *Bulletin of the Geological Society of China*, **19**, 279–315.

——. 1954. On a new sauropod from Yiping, Szechuan, China. *Acta Scientia Sinica*, **3**, 491–504.

—— and ZHAO, X. 1972. [*Mamenchisaurus*. In Chinese: description of the type material of *Mamenchisaurus hochuanensis*]. *Institute of Vertebrate Paleontology and Paleoanthropology Monograph Series I*, **8**, 1–30.

**TABLE 1.** Measurements of *Xenoposeidon proneneukos* gen. et sp. nov. holotype BMNH R2095, and comparison with mid-posterior dorsal vertebrae of other neosauropods. All measurements are in mm. The suffix 'e' indicates an estimation, and '+' indicates a minimum possible value, e.g. the length of the preserved portion of a broken element. Measurements for *Brachiosaurus altithorax* FMNH P25107 are taken from Riggs (1904, p. 234): D?7 and D?11 are the vertebrae described by Riggs as presacrals VI and II respectively, on the assumption than *B. altithorax* had 12 dorsal vertebrae. Measurements for *Brachiosaurus brancai* HMN SII are taken from Janensch (1950, p. 44), except those suffixed 't' which were omitted from Janensch's account and so measured by MPT. Measurements for *Diplodocus carnegii* CM 84 are taken from Hatcher (1901, p. 38). Measurements suffixed 'i' were interpolated by measuring from Riggs (1904, pl. LXXII) for *B. altithorax*, Janensch (1950, fig. 56) for *B. brancai* and Hatcher (1901, pl. VII) for *D. carnegii*.

| | *Xenoposeidon* | *Brachiosaurus altithorax* | | *B. brancai* | *Diplodocus carnegii* | |
|---|---|---|---|---|---|---|
| | BMNH R2095 | FMNH P25107 | | HMN SII | CM 84 | |
| | | D?7 | D?11 | D7 | D7 | D8 |
| Total height of vertebra | 300+ | 900 | 800 | 770+ | 980i | 970i |
| Total centrum length including condyle | 200e | 440 | 350 | 330 | 264 | 275 |
| Total centrum length excluding condyle | 190 | | | 294 | | |
| Cotyle height | 160 | 270 | 280 | 220t | | |
| Cotyle width | 170 | 300 | 310 | 320t | | |
| Average cotyle diameter | 165 | 285 | 295 | 270t | 280 | 309 |
| Centrum length / cotyle height (EI) | 1.25 | 1.63 | 1.25 | 1.50 | 0.94 | 0.89 |
| Depth of cotylar depression | 10 | 80 | 70 | | | |
| Anteroposterior length of lateral fossa | 95 | -- | -- | --? | -- | -- |
| Dorsoventral height of lateral fossa | 80 | -- | -- | --? | -- | -- |
| Anteroposterior length of lateral foramen | 80 | 190 | 160 | 97i | 120i | 130i |
| Dorsoventral height of lateral foramen | 40 | 100 | 70 | 58i | 85i | 95i |

| | *Xenoposeidon* | *Brachiosaurus altithorax* | | *B. brancai* | *Diplodocus carnegii* | |
|---|---|---|---|---|---|---|
| | BMNH R2095 | FMNH P25107 | | HMN SII | CM 84 | |
| | | D?7 | D?11 | D7 | D7 | D8 |
| Anteroposterior length of base of neural arch | 170 | 220i | 155i | 170i | 180i | 165i |
| Neural arch base length / centrum length | 0.85 | 0.50 | 0.44 | 0.52 | 0.68 | 0.60 |
| Height of neural arch above centrum | 160+ | | | | | |
| Height of neural arch pedicels, posterior | 130+ | | | | | |
| Thickness of neural arch pedicels, posterior | 30 | | | | | |
| Height of neural canal, posterior | 35 | | | | | |
| Width of neural canal, posterior | 35 | | | | | |
| Height of neural arch pedicels, anterior | 80+ | | | | | |
| Thickness of neural arch pedicels, anterior | 25 | | | | | |
| Height of neural canal, anterior | 120 | | | | | |
| Width of neural canal, anterior | 55 | | | | | |
| Height of hyposphene above centrum | 90+ | | | | | |
| Height of postzygapophyses above centrum | 140e | | | | | |
| Height of prezygapophyses above centrum | 140e | | | | | |

**TABLE 2.** Character scores for *Xenoposeidon* in the matrix used for the phylogenetic analysis in this paper. Apart from the addition of *Xenoposeidon*, the matrix is identical to that of Harris (2006). *Xenoposeidon* is unscored for all characters except those listed. Conventional anatomical nomenclature is here used in place of the avian nomenclature of Harris.

| Character | | Score | |
|---|---|---|---|
| 123 | Lateral fossae in majority of dorsal centra | 2 | Present as deep excavations that ramify into centrum and into base of neural arch (leaving only thin septum in body midline) |
| 124 | Position of lateral foramina on dorsal centra | 2 | Set within lateral fossa |
| 125 | Anterior face of dorsal neural arches | 1 | Deeply excavated |
| 127 | Single midline lamina extending ventrally from hyposphene in dorsal vertebrae | 0 | Absent |
| 134 | Morphology of ventral surfaces of anterior dorsal centra | 0 | Ventrally convex [inferred from posterior dorsal] |
| 137 | Ratio of dorsoventral height of neural arch:dorsoventral height of dorsal centrum | 1 | >1.0 |
| 139 | Anterior centroparapophyseal lamina on middle and posterior dorsal neural arches | 1 | Present |
| 140 | Prezygaparapophyseal lamina on middle and posterior dorsal neural arches | 1 | Present |

| Character | | Score | |
|---|---|---|---|
| 141 | Posterior centroparapophyseal lamina on middle and posterior dorsal neural arches | 1 | Present [as the homologous accessory infraparapophyseal lamina] |
| 149 | Orientation of middle and posterior dorsal neural spines | 0 | Vertical [rather than posterodorsally inclined] |
| 150 | Morphology of articular face of posterior dorsal centra | 1 | Opisthocoelous |
| 151 | Cross-sectional morphology of posterior dorsal centra | 1 | Dorsoventrally compressed |
| 153 | Position of diapophysis on posterior dorsal vertebrae | 1 | Dorsal to parapophysis |

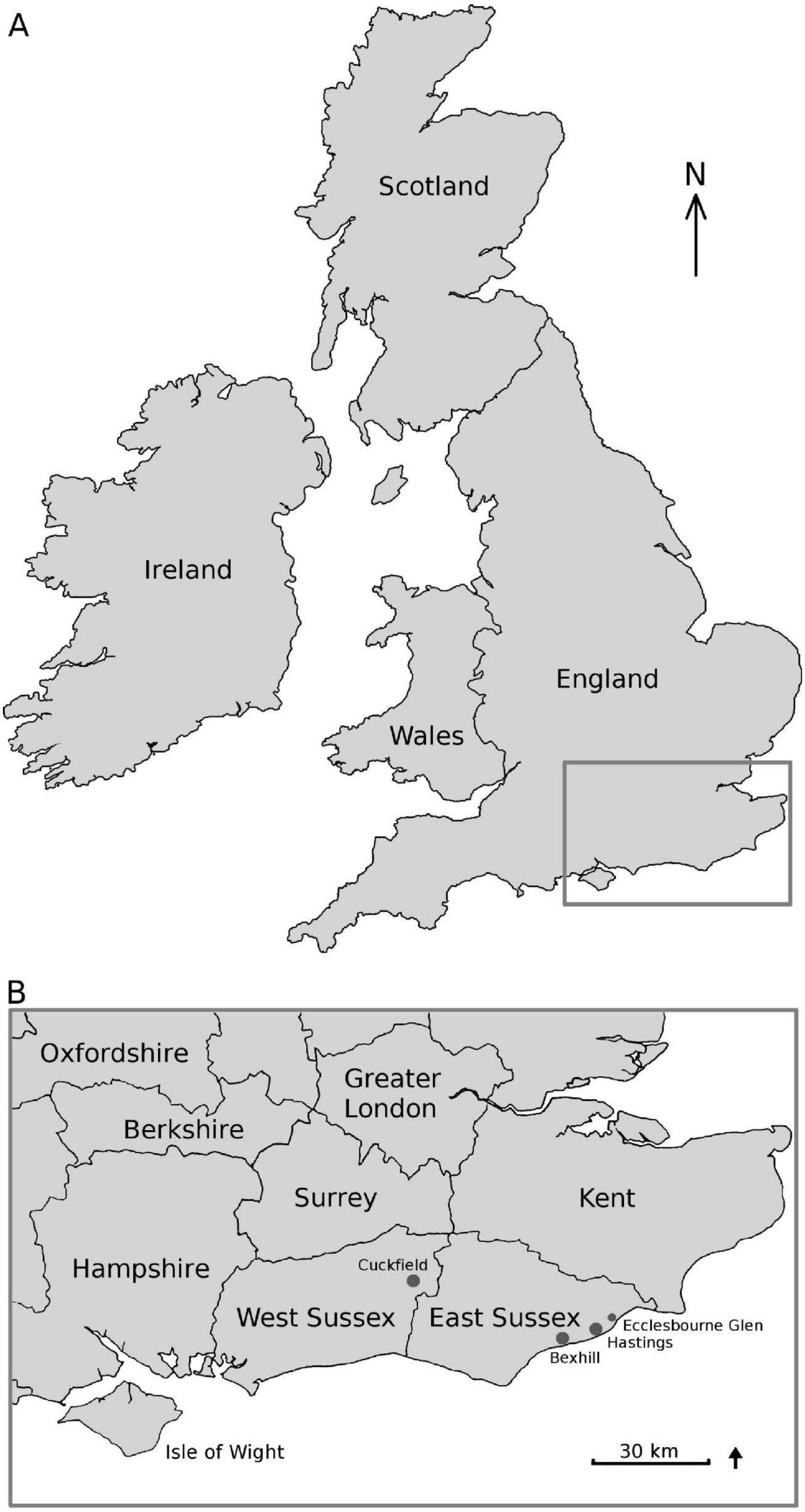

**TEXT-FIG 1**. Map indicating Ecclesbourne Glen and locale, near Hastings, East Sussex, England, the probable discovery location of *Xenoposeidon proneneukos* gen. et sp. nov. holotype BMNH R2095. A. Small-scale map of Great Britain, with box indicating position of larger-scale map. B. Larger-scale map of south-east England.

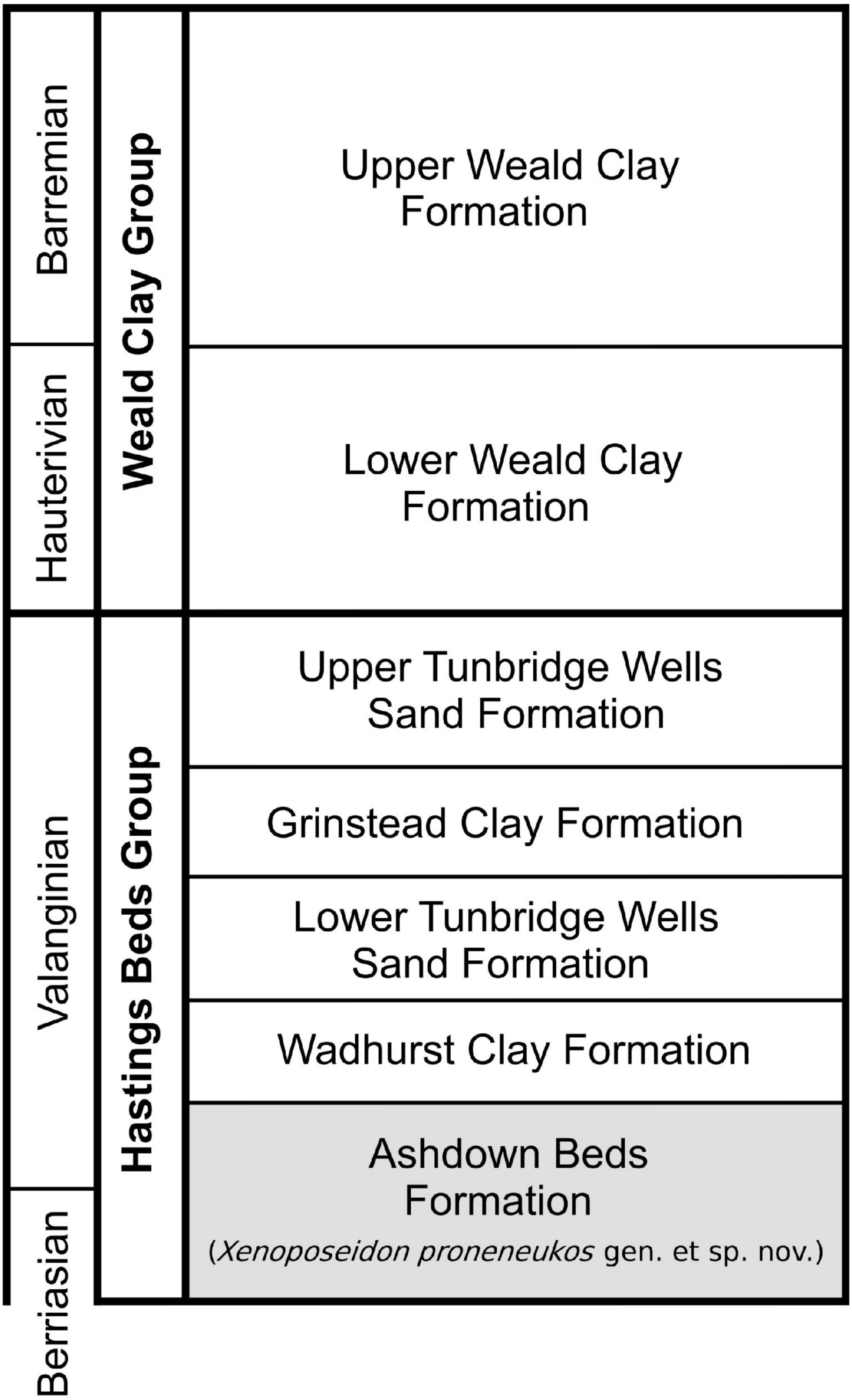

**TEXT-FIG 2**. Schematic lithostratigraphy of the Wealden, indicating the origin of *Xenoposeidon proneneukos* gen. et sp. nov. holotype BMNH R2095 from within the Ashdown Beds Formation of the Hastings Beds Group.

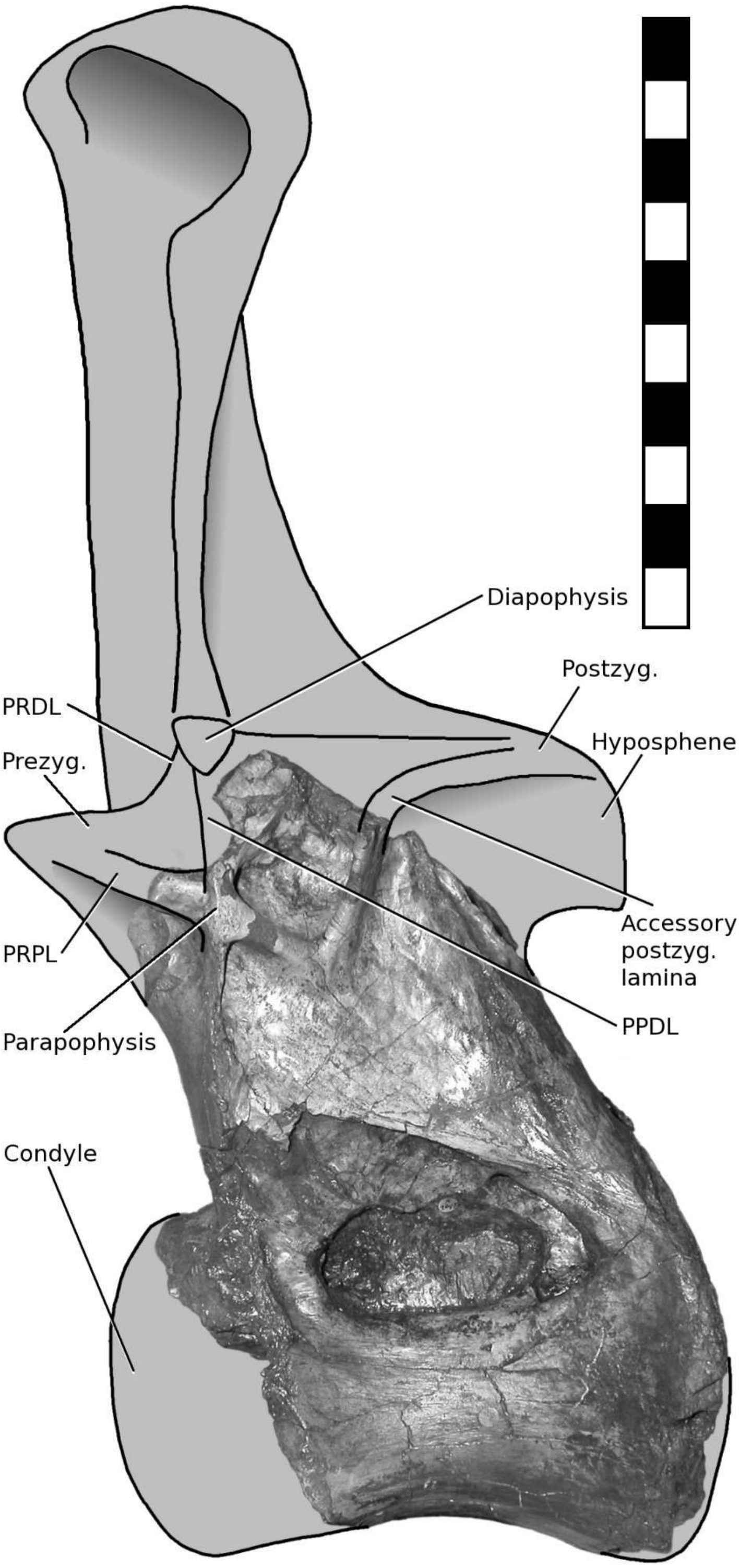

**TEXT-FIG 3**. *Xenoposeidon proneneukos* gen. et sp. nov. holotype mid-to-posterior dorsal vertebra BMNH R2095, speculative reconstruction, in left lateral view. The location of the prezygapophyses, postzygapophyses and diapophyses are inferred with some confidence from the preserved laminae; the neural spine is based on an idealised slender neosauropod neural spine. Scale bar 200 mm.

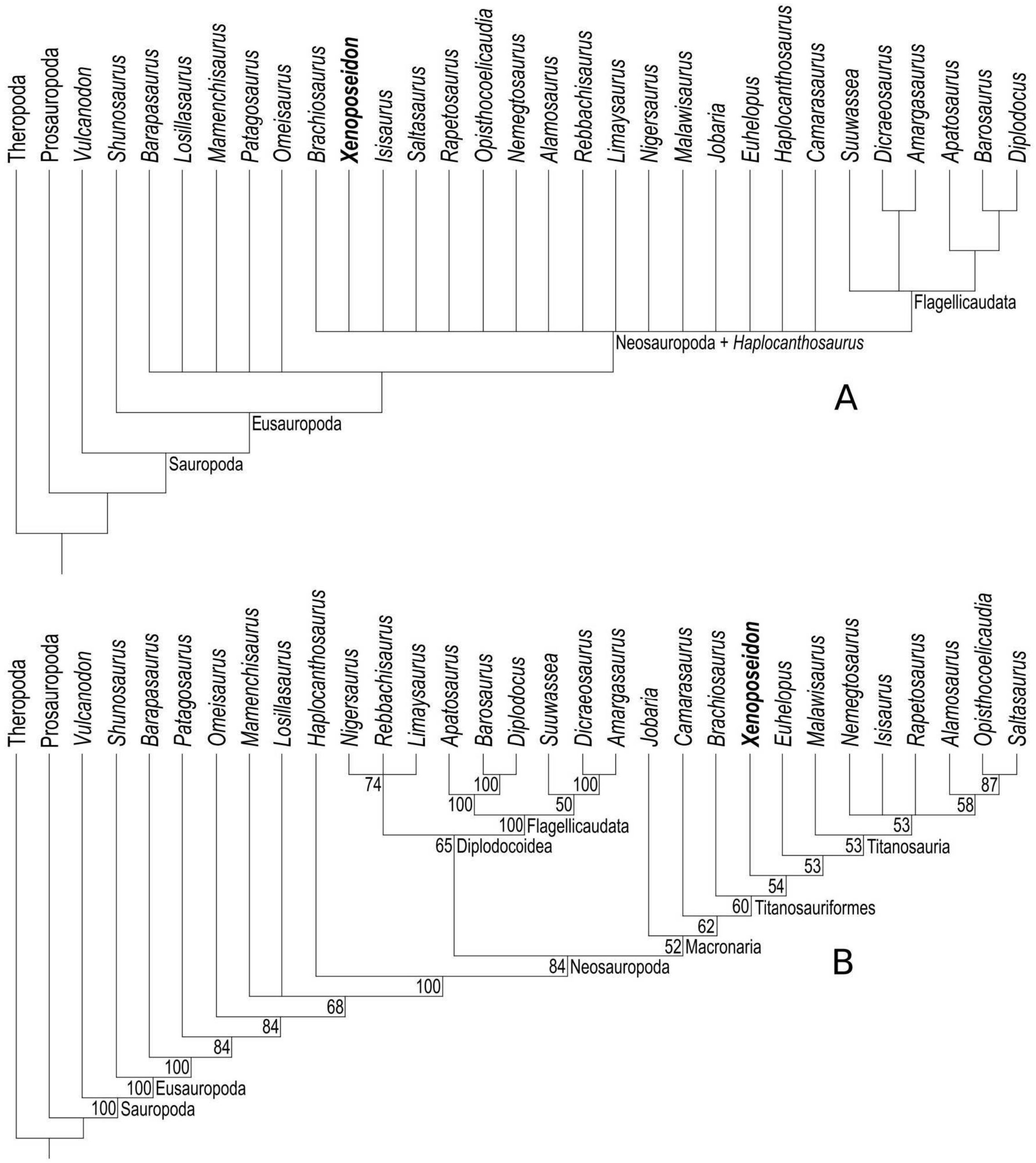

**TEXT-FIG 4**. Phylogenetic relationships of *Xenoposeidon proneneukos* gen. et sp. nov., produced using PAUP* 4.0b10 on the matrix of Harris (2006) augmented by *Xenoposeidon*, having 31 taxa and 331 characters. A. Strict consensus of 1089 most parsimonious trees (length = 785, CI = 0.5248, RU = 0.6871, RC = 0.3606). B. 50% majority rule consensus. Clade names are positioned to the right of the branches that they label; occurrence percentages are positioned to the left of these branches.

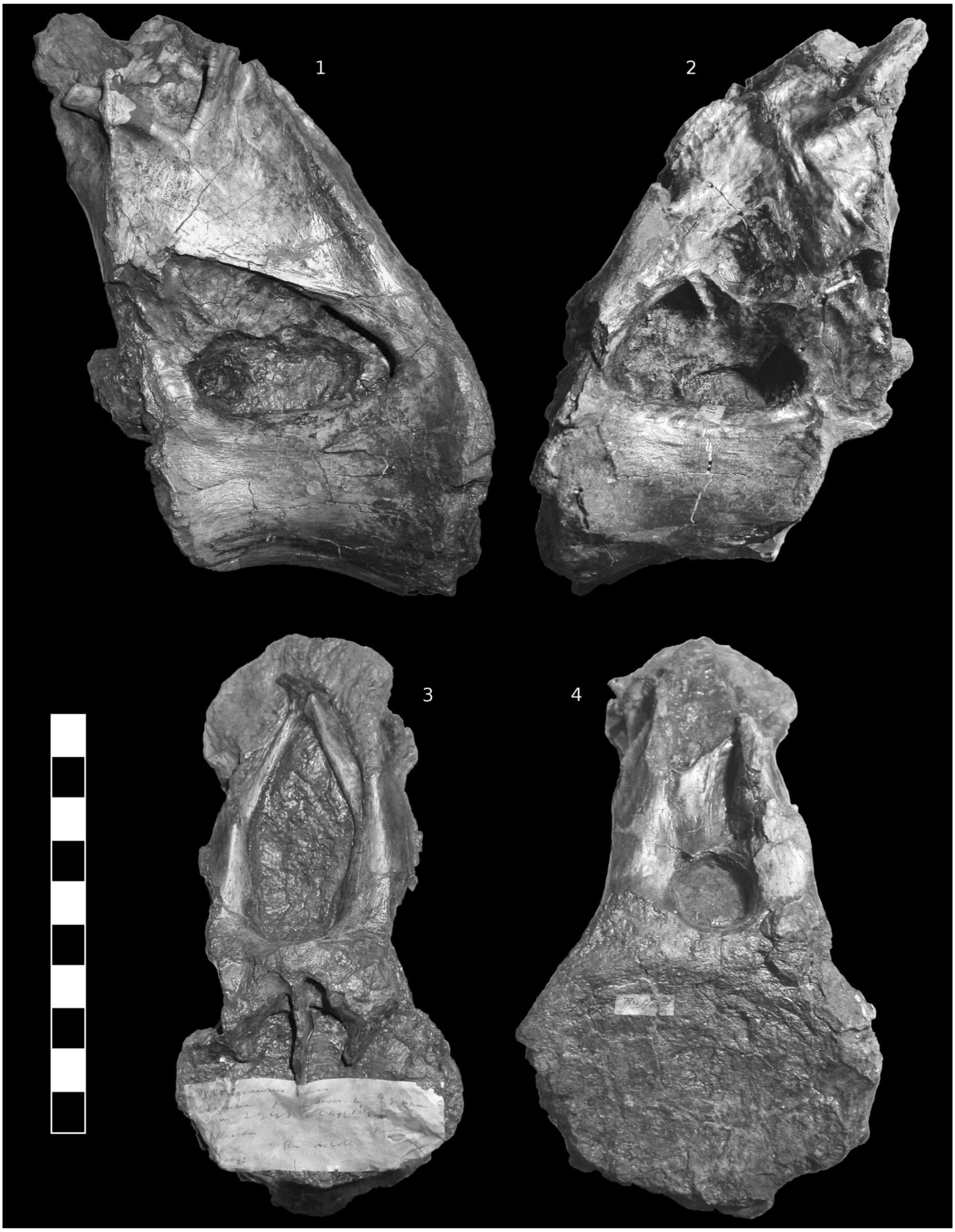

## EXPLANATION OF PLATE 1

Figs 1-4. *Xenoposeidon proneneukos* gen. et sp. nov. holotype mid-to-posterior dorsal vertebra BMNH R2095. 1, left lateral view. 2, right lateral. 3, anterior. 4, posterior. Scale bar 200 mm.

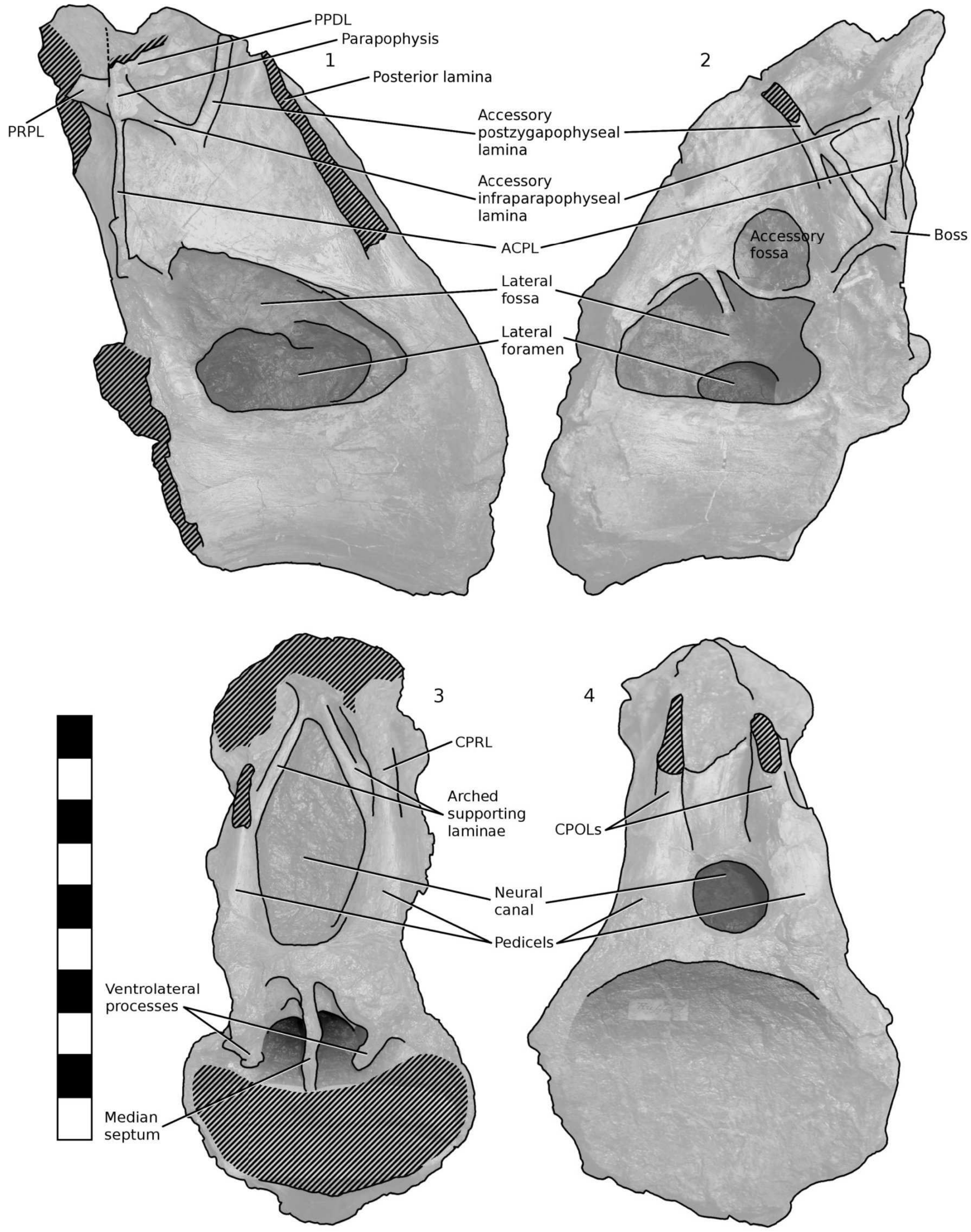

## EXPLANATION OF PLATE 2

Figs 1-4. Interpretive drawing of *Xenoposeidon proneneukos* gen. et sp. nov. holotype mid-to-posterior dorsal vertebra BMNH R2095. 1, left lateral view. 2, right lateral. 3, anterior. 4, posterior. Scale bar 200 mm. Breakage is indicated by diagonal hatching. The PPDL (preserved only on the left side) is a sheet of bone projecting anterolaterally

from the neural arch and with its anterolateral margin running dorsoventrally, but which is broken off just dorsal to the parapophysis.

Chapter 4 follows. This paper has been formatted for submission to the Journal of Vertebrate Paleontology, published by The Society of Vertebrate Paleontology.

# A NEW SAUROPOD DINOSAUR FROM THE LOWER CRETACEOUS CEDAR MOUNTAIN FORMATION, UTAH, U.S.A.

MICHAEL P. TAYLOR

Palaeobiology Research Group, School of Earth and Environmental Sciences,
University of Portsmouth, Burnaby Road, Portsmouth PO1 3QL, United Kingdom,
dino@miketaylor.org.uk;

ABSTRACT—The Hotel Mesa sauropod is a new genus and species of sauropod dinosaur from the Hotel Mesa quarry in Grand County, Utah, U.S.A., in the upper (Albian) part of the Ruby Ranch Member of the Lower Cretaceous Cedar Mountain Formation. It is known from at least two fragmentary specimens of different sizes. The type specimen is OMNH 66430, the left ilium of a juvenile individual; referred specimens include a crushed presacral centrum, a complete and well-preserved mid-to-posterior caudal vertebra, the partial centrum of a distal caudal vertebra, a complete pneumatic anterior dorsal rib from the right side, the nearly complete left scapula of a much larger, presumably adult, individual, and two partial sternal plates. It is diagnosed by five autapomorphies of the type specimen: preacetabular lobe 55% of total ilium length, longer than in any other sauropod; preacetabular lobe directed anterolaterally at 30 degrees to the sagittal, but straight in dorsal view and vertically oriented; postacetabular lobe reduced to near absence; ischiadic peduncle reduced to very low bulge; ilium proportionally tall. The Hotel Mesa sauropod cannot be congeneric with either of the Cedar Mountain Formation sauropods *Cedarosaurus* or *Venenosaurus* – the former due to differences in their scapulae and a 15 Ma age difference, the latter due to significant differences in their caudal vertebrae and scapulae. The Hotel Mesa sauropod was scored for 20 of 331 characters in a phylogenetic analysis. It was recovered as a camarasauromorph in all most parsimonious trees, but with little resolution within that clade and with only weak support.

## INTRODUCTION

The record of Early Cretaceous sauropod dinosaurs in North America was for many years poorly represented, with the only recognized genera being *Astrodon* Leidy, 1865 and *Pleurocoelus* Marsh, 1888, the former represented only by teeth and often considered synonymous with the latter (e.g. Hatcher, 1903a; Gilmore, 1921; Carpenter and Tidwell, 2005). In recent years, this record has been greatly expanded and clarified by the discovery and description of *Sonorasaurus* Ratkevich, 1998, *Cedarosaurus* Tidwell, Carpenter and Brooks, 1999, *Sauroposeidon* Wedel, Cifelli and Sanders, 2000a, *Venenosaurus* Tidwell, Carpenter and Meyer, 2001, and *Paluxysaurus* Rose, 2007. Further material, representing yet more new sauropod taxa, is known and awaits description: for example, two new taxa, a camarasaurid and a titanosaur, in the Dalton Wells quarry (Eberth et al., 2006:220) and a titanosaur from the Yellow Cat Member of the Cedar Mountain Formation represented by an articulated sequence of five presacral vertebrae (Tidwell and Carpenter, 2007).

The Hotel Mesa quarry, Oklahoma Museum of Natural History locality V857, is located in Grand County, eastern Utah, just east of the Colorado river (Fig. 1), and exposes the Ruby Ranch Member of the Lower Cretaceous Cedar Mountain Formation (Kirkland et al., 1997:96–97). (East of the Colorado river, this formation is sometimes referred to as the Burro Canyon Formation, but I follow the more widely used nomenclature.) Because the Hotel Mesa quarry is located only a few meters below the boundary with the Dakota Formation, it can be dated to the Albian (latest Early Cretaceous).

This quarry contains remains of a previously unrecognised sauropod taxon which I describe here. The site was initially uncovered by vandals. The first sauropod element, a scapula, was collected by Hayes in 1994; he returned in 1995 and collected an ilium; in the same year, Kirkland collected many further elements, listed below. During the latter expedition, Cifelli collected a mid-caudal vertebra. The ilium and scapula of this sauropod were figured, but not described, by Kirkland et al. (1997:93), who described them as “comparable to *Pleurocoelus*” (at that time the only known Early Cretaceous sauropod from North America). Although the quarry has not been worked in several years, it is not exhausted and may contain additional relevant material (Kirkland, pers.

comm., October, 2007).

**Anatomical Nomenclature**—I follow Upchurch et al. (2004a) in describing scapulae as though oriented horizontally: the coracoid articular surface is designated anterior and the 'distal' end posterior.

**Institutional Abbreviations**—**AMNH**, American Museum of Natural History, New York City, New York, U.S.A.; **BMNH**, the Natural History Museum, London, England; **CCG**, Chengdu College of Geology, China; **CM**, Carnegie Museum of Natural History, Pittsburgh, Pennsylvania, U.S.A.; **DMNH**, Denver Museum of Natural History, Denver, Colorado, U.S.A.; **FMNH**, Field Museum of Natural History, Chicago, Illinois; **FWMSH**, Fort Worth Museum of Science and History, Fort Worth, Texas; **HMN**, Humboldt Museum für Naturkunde, Berlin, Germany; **OMNH**, Oklahoma Museum of Natural History, Norman, Oklahoma, U.S.A.; **ZDM**, Zigong Dinosaur Museum, Zigong, China.

Names of clades are used as summarized in Table 1.

## SYSTEMATIC PALEONTOLOGY

DINOSAURIA Owen, 1842
SAURISCHIA Seeley, 1888
SAUROPODA Marsh, 1878
NEOSAUROPODA Bonaparte, 1986
CAMARASAUROMORPHA Salgado, Coria and Calvo, 1997
gen. nov. (Name to be announced in published version)

**Type species**—sp. nov.

**Diagnosis**—as for type species (see below).

sp. nov. (Name to be announced in published version)
(Figs. 2–8; Table 2)

**Holotype**—OMNH 66430, a left ilium.

**Referred Specimens**—OMNH 66429, crushed presacral centrum; OMNH 61248, mid-to-posterior caudal vertebra; OMNH 27794, partial distal caudal centrum; OMNH 27766, anterior right dorsal rib; OMNH 27761, nearly complete left scapula missing anterior portion; OMNH 66431 and 66432, two partial sternal plates; other fragments as detailed in Table 2.

**Type Locality and Horizon**—Hotel Mesa quarry (OMNH locality V857), Grand County, eastern Utah. Top of the Ruby Ranch Member of the Cedar Mountain Formation (Lower Cretaceous, Albian).

**Diagnosis**—Preacetabular lobe 55% of total ilium length, longer than in any other sauropod; preacetabular lobe directed anterolaterally at 30 degrees to the sagittal, but straight in dorsal view and vertically oriented; postacetabular lobe reduced to near absence; ischiadic peduncle reduced to very low bulge; ilium proportionally tall; presacral vertebrae camellate; mid-to-posterior caudal vertebrae with elongate pre- and postzygapophyseal rami, having the postzygapophyseal facets hanging below the level of the ramus; first dorsal rib with expanded, dorsally oriented articular facts, laterally curving shaft, and distally directed pneumatic foramen in head; acromion expansion of scapula pronounced and steep, but not forming acromion fossa; dorsal and ventral margins of scapular blade 'stepped'; sternal plates crescentic, and three times as long as broad.

Unambiguous autapomorphies distinguishing the Hotel Mesa sauropod from the root of the polytomy in which it is recovered in the strict consensus of most parsimonious trees in the phylogenetic analysis below: character 184, ratio of centrum length:height in middle caudal vertebrae $\geq 2.0$; 185, sharp ridge on lateral surface of middle caudal centra at arch-body junction absent; 212, posterior end of scapular body racquet-shaped (dorsoventrally expanded); 261, in lateral view, the most anteroventral point on the iliac preacetabular lobe is also the most anterior point (preacetabular lobe is pointed); 264, projected line connecting articular surfaces of ischiadic and pubic peduncles of ilium passes ventral to ventral margin of postacetabular lobe of ilium.

DESCRIPTION

This taxon is based on a collection of elements all from the same quarry, all of them consistent with assignment to a single taxon (Fig. 2). However, the elements were not found articulated, and their differing sizes do not permit interpretation as belonging to a single individual. For example, the partial scapula is 98 cm long. Reconstruction after the scapula of *Giraffatitan brancai* (Janensch, 1914) indicates that the complete element was about 121 cm long. ("*Brachiosaurus*" *brancai* has been shown to be generically distinct from the type species *Brachiosaurus altithorax* Riggs, 1903 by Taylor, in press, so the genus name *Giraffatitan* Paul, 1988 must be used.) In *Rapetosaurus* Curry Rogers and Forster, 2001, the scapula and ilium are about the same length (Curry Rogers and Forster, 2001:fig. 3) but the ilium of the Hotel Mesa sauropod is only one third the reconstructed length of the scapula. The quarry therefore contains at least two individuals of widely differing sizes. Since all the informative elements from Hotel Mesa have characters that indicate a titanosauriform identity, the null hypothesis is that they represent a single taxon, although this hypothesis is subject to revision pending the recovery of more material.

The assignment of specimen numbers to the material described here is complex (Table 2). Specimen number OMNH 27773 comprises three elements; OMNH 27784 consists of 21 small fragments of bone, none of them informative; all other elements have their own specimen numbers, in the range 27761–27800 apart from the mid-caudal vertebra OMNH 61248 and the reassigned numbers 66429-66432, for elements extracted from OMNH 27773.

**Ilium**

The most informative element is OMNH 66430, a left ilium (Fig. 3). The ilium was preserved complete, but lay hidden beneath the scapula, and so was damaged in the field (Kirkland, pers. comm., March, 2008). The ilium is preserved in three parts: one provides most of the bone, including the well preserved preacetabular lobe, pubic and ischiadic peduncles and acetabular margin, and the other two provide most of the dorsal margin, giving a good indication of the degree of curvature. The relative positions and orientation of the two smaller fragments can not be definitely ascertained, but they appear to be parts of a single large fragment broken at the point where I have reconstructed them as touching; and if this interpretation is correct then the curvature of the pair indicates which side must be oriented laterally.

The ilium is remarkable in that the preacetabular lobe is relatively larger than in any other sauropod (Fig. 10) and the postacetabular lobe is reduced almost to the point of absence. The ischiadic peduncle is reduced to a very low ventral projection from almost the most posterior point of the ilium. The near absence of the ischiadic peduncle cannot be attributed to damage as the iliac articular surface is preserved. Immediately posterodorsal to this surface is a subtle notch between the peduncle and the very reduced postacetabular lobe. This notch and the areas either side of it are composed of finished bone, demonstrating that the great reduction of the postacetabular lobe, too, is a genuine osteological feature and not due to damage. In regard to the proportionally large preacetabular lobe, the ilium of the Hotel Mesa sauropod resembles that of *Rapetosaurus* (Table 3, Figure 10E.). However, that taxon has a normal postacetabular lobe. In overall proportions, the ilium of the Hotel Mesa sauropod is most similar to the left ilium HMN J1 assigned to *Giraffatitan brancai* (Janensch, 1961:pl. E, fig. 2) which also has a reduced postacetabular lobe – see also Fig. 10D. However, the ilium of the Hotel Mesa sauropod is proportionally taller than that of *G. brancai*, and its anterior margin comes to a point rather than being smoothly rounded as in that taxon.

As with many sauropods, the preacetabular lobe of the ilium flares laterally. However, in most sauropods this flaring is progressive, so that in dorsal or ventral view the most posterior part of the preacetabular lobe is nearly parallel with a line drawn between the pubic and ischiadic peduncles, and smooth lateral curvature inclines the more anterior parts increasingly laterally, so that the more anterior part is almost at right angles to this line and the ilium appears smoothly curved in dorsal or ventral view – for example, *Apatosaurus* Marsh, 1877 (Upchurch et al., 2004a:pl. XXX, figs. D–E), *Haplocanthosaurus* Hatcher, 1903b (Hatcher, 1903a:pl. V, fig. 1) and *Saltasaurus* Bonaparte and Powell, 1980 (Powell, 1992:fig. 17). In the Hotel Mesa sauropod, by contrast, the blade of the ilium appears to be 'hinged' – deflected laterally directly anterior to the pubic peduncle – so that the preacetabular lobe is straight in dorsal or ventral view, and directed anterolaterally by an angle of about 30 degrees to the sagittal. In this respect, it more closely resembles the ilium of *Camarasaurus* Cope, 1877 (Osborn and Mook, 1921:fig. 49) although it differs in other respects.

The Hotel Mesa sauropod's ilium is laterally compressed, and unlike most sauropod ilia the dorsal margin is not deflected laterally relative to the more ventral part, so that in ventral view it appears very thin (Fig. 3b). It is well established that the long bones of sauropods grow isometrically through ontogeny (Carpenter and McIntosh, 1994; Wilhite, 1999, 2003; Ikejiri et al., 2005; Tidwell and Wilhite, 2005) while their vertebrae undergo significant changes in proportions, lamination, and pneumatic excavations and neurocentral fusion (Wedel, 2003a:248, b:352–354). Ontogenetic change in limb-girdle elements such as the ilium are less well understood due to a paucity of sufficiently well preserved specimens (Wilhite, pers. comm., October, 2007). Therefore the lateral compression of the Hotel Mesa ilium may be a juvenile character, with the ilium thickening through ontogeny to support the growing weight of the animal, or may be phylogenetically significant.

**Presacral centrum**

A single presacral centrum, OMNH 66429, has been found (Fig. 4). Unfortunately, preservation is very poor: the neural arch and all processes have been lost, and the centrum has been greatly crushed dorsoventrally so that the remaining part is essentially flat: the element is 14 cm in both anteroposterior length and transverse width, but no more than 3 cm in dorsoventral depth. The small size indicates that this element belonged to a juvenile. The internal structure of the centrum is visible, however, and consists of fine septa dividing a hollow internal space irregularly into many small camellae. This morphology is characteristic of titanosauriforms (Wedel, 2003b:354–355). Highly camellate internal structure has not previously been observed in juvenile sauropod vertebrae, but this may be due to sampling bias: so far, all juvenile sauropod vertebrae that have been studied for internal structure have been those of *Camarasaurus* and diplodocoids, which follow an ontogenetic trajectory in which large, shallow lateral fossae eventually develop into camerae from which smaller accessory camerae and camellae develop (Wedel et al., 2000b:fig. 11; Wedel, 2003b:349). The Hotel Mesa presacral suggests that camellate vertebrae may have developed differently, possibly by in-situ formation of camellae during pneumatization.

**Caudal vertebrae**

OMNH 61248 is a distinctive caudal vertebra with elongated pre- and postzygapophyseal rami (Fig. 5). Apart from the tip of the right prezygapophysis, the element is complete and well preserved. While the centrum is only 11 cm in length, the distance from the prezygapophysis to postzygapophysis is 14.5 cm. The centrum is slightly broader than tall (6 cm compared with 5.5 cm anteriorly, 6.5 cm compared with 5 cm posteriorly) and gently waisted. The neural arch is set forward on the centrum but does not reach the anterior margin. The neural spine is so reduced and so strongly inclined posteriorly as to be all but indistinguishable, and is apparent only as a very low eminence above the postzygapophyses. The postzygapophyseal facets themselves are set on the posterolateral faces of a low process that hangs below the main postzygapophyseal ramus. Chevron facets are weakly present on the posterior margin of the ventral surface of the centrum, but not on the anterior margin. The elongation index of 2.2 indicates a mid-to-posterior position in the caudal sequence for this element, as similar centrum proportions do not appear until about caudal 30 in *Giraffatitan brancai* (Janensch, 1950:pl. III).

This vertebra most closely resembles the indeterminate sauropod vertebra BMNH 27500 from the Wessex Formation of the Isle of Wight, figured by Naish and Martill (2001:pl. 33). The Barremian age of that specimen places it about fifteen million years earlier than OMNH 61248. Its neural arch is less elevated than that of the Hotel Mesa specimen, its postzygapophyses project yet further posteriorly and its prezygapophyses less far anteriorly, and it is very mildly biconvex rather than procoelous; but in other respects, including absolute size, it is a good match for the Hotel Mesa sauropod caudal.

Also included in the Hotel Mesa material is OMNH 27794, a partial distal caudal centrum figured by Wedel (2005:fig. 7.7). This centrum is approximately round in cross-section, about 4 cm in diameter, and internally consists of apneumatic cancellous bone.

**Ribs**

The Hotel Mesa material contains several dorsal ribs in various states of preservation but no readily identifiable cervical ribs. The dorsal rib elements include the shaft of a large, flat rib (OMNH 27762), portions of several smaller rib shafts (OMNH 27763–27765, 27768 and others), a flattened rib head (OMNH 27767) and most informatively a small complete rib (OMNH 27766, Fig. 6). Despite its excellent preservation and apparent lack of distortion, this element is difficult to interpret. Its shaft is straight for almost its whole length and both articular facets are directed dorsally rather than being inclined medially. The tuberculum is directly in line with the main part of the shaft of the rib, and the capitulum is at an angle of about 30 degrees to it. In these respects the rib resembles the most anterior dorsal rib of the *Diplodocus carnegii* holotype CM 84 (pers. obs., MPT; this element was not figured by Hatcher (1901)). I therefore interpret this rib as having probably belonged to the right side of the first dorsal vertebra.

The rib is unusual in other respects, however, most notably that the distal part of the shaft curves laterally rather than medially. Careful inspection of the bone reveals no indication of distortion or of incorrect reconstruction. It may be possible that in life the thorax was transversely compressed so that the proximal part of the rib shaft was directed ventromedially and the more distal part was vertical. Both articular facets are subcircular in proximal view, and significantly expanded compared with the rami that bear them. On the anterior face of the head, a low ridge arises just below the capitulum and extends down the medial edge of the rib for about 40% of its length.

The head of the rib is also unusual in that a thin sheet of bone connects the rami that support the articular facets, and this sheet extends much further proximally than in most sauropod ribs. Its precise extent cannot be ascertained due to breakage. The sheet of bone is perforated close to the capitular ramus, and from this perforation a pneumatic cavity invades the shaft of the rib, extending distally from within a shallow fossa in the posterior face. Pneumatization of the dorsal ribs is a synapomorphy of Titanosauriformes (Wilson and Sereno, 1998), although pneumatic dorsal ribs are also infrequently present in diplodocids (Gilmore, 1936; Lovelace et al., 2003).

**Pectoral girdle**

OMNH 27761 is a partial scapula, consisting of the blade and part of the anterior expansion, and missing the glenoid region and the remainder of the anterior expansion (Fig. 7). As preserved, the element is nearly flat; but this may be due to post-mortem distortion, and in any case the most strongly curved part of most sauropod scapulae is the anterior part that is is missing from this specimen. The gentle curvature preserved in the posterior part of the blade indicates that the element was from the left side. The bone is surprisingly thin in all preserved parts, never exceeding a few cm, in contrast to for example the scapula of *Camarasaurus supremus*, which is thick even in mid-blade (Osborn and Mook, 1921:fig. 74b). This suggests that the glenoid thickening and the acromial ridge may have been located some distance anterior to the preserved portion, and a reconstruction after the proportions of *Giraffatitan brancai* (Fig. 7) suggests that about 80% of the scapula's full length is preserved. The posterior part of the acromion expansion is preserved, however, and is sufficient to show that this expansion was pronounced, so that the maximum dorsoventral height of the scapula was more than two and half times its minimum height, at the midpoint of the blade. The dorsal margin slopes up towards the anterior expansion rather than forming a ventrally directed 'hook' or a distinct fossa between the blade and the acromion process.

The scapula is distinctive in the nature of its posterior expansion. In some sauropods, the posterior part of the scapula blade is expanded not at all or only slightly: for example in *Omeisaurus* Young, 1939 (He et al., 1988:fig. 41), *Apatosaurus* Marsh, 1877 (Upchurch et al., 2004b:fig. 4) and *Rapetosaurus* (Fig. 11E). In others, the ventral margin of the scapular blade is straight or nearly so while the dorsal margin is deflected dorsally to create an asymmetric expansion: for example in *Camarasaurus* (Fig. 11C) and *Giraffatitan brancai* (Fig. 11D). In a few sauropods, however, the ventral margin of the blade is also deflected ventrally, to form a 'racquet-shaped' distal expansion. This is seen in rebbachisaurids and some titanosaurs, e.g. some specimens of *Alamosaurus* Gilmore, 1922 (Gilmore, 1946:fig. 6). In the Hotel Mesa sauropod, the posterior part of the scapular blade is expanded in a characteristic manner: the ventral margin is straight except for a posteroventral excursion two thirds of the way along the preserved portion, after which the margin continues parallel to its original trajectory, so that the excursion appears as a gentle 'step'. The dorsal margin is also 'stepped' in this manner, though with two distinct steps rather than one, of which the more anterior is most strongly pronounced. The net result of these features is that the dorsal and ventral borders of the scapula are both straight near the posterior extremity, and that they are subparallel, diverging by only about five degrees in the region just anterior to the rounded end of the posterior expansion. The step in the ventral border is not known in any other sauropod; however the scapula of *Neuquensaurus* Powell, 1992 has a stepped dorsal border similar to that of the Hotel Mesa sauropod (Huene, 1929 via Glut, 1997:275).

These characters of the scapula must be treated with some caution, however, since this bone appears subject to more variation than any other in the sauropod skeleton: see for example the range of shapes in scapulae of *Giraffatitan brancai* (Janensch, 1961:pl. XV, figs. 1–3) and in *Camarasaurus supremus* (Osborn and Mook, 1921:figs. 74–80).

Two small, flat elements OMNH 26631 and 26632 are interpreted as partial sternal plates (Fig. 8). The medial edge of each is identifiable due to its rugose texture which formed the attachment site for cartilage joining the plates to each other and to the sternal ribs. The sternals are anteroposteriorly elongate and mediolaterally narrow: when complete, they were probably at least three times as long as broad, as in "*Saltasaurus*" *robustus* Huene, 1929 (Huene, 1929 via McIntosh, 1990:fig. 16.9L) and proportionally longer than in any other sauropod including *Saltasaurus loricatus* (McIntosh, 1990:fig. 16.9; Powell, 1992:fig. 30). The sternals are crescentic in shape, the anterior and posterior extremities curving laterally away from the midline. This state was considered a titanosaurian synapomorphy by Wilson (2002:268) but its distribution is more complex in the current analysis, being synapomorphic for Neosauropoda with losses in Flagellicaudata and *Camarasaurus*.

## COMPARISON WITH OTHER EARLY CRETACEOUS NORTH AMERICAN SAUROPODS

The Hotel Mesa sauropod cannot be directly compared with *Astrodon*/*Pleurocoelus*, *Sonorasaurus* or *Sauroposeidon* due to the absence of overlapping material between these genera. Two other sauropod dinosaurs are already known from the Cedar Mountain Formation: *Cedarosaurus* from the Yellow Cat Member and *Venenosaurus* from the Poison Strip Member. The Hotel Mesa sauropod can be distinguished from both of these taxa as discussed below, and is from the stratigraphically higher Ruby Ranch Member. Since the Yellow Cat Member is Barremian in age, the Poison Strip Member is Aptian, and the upper part of the Ruby Ranch Member (where the Hotel Mesa quarry is located) is Albian (Kirkland et al., 1997:fig. 1), these three sauropods together span the last three ages of the Early Cretaceous. The Hotel Mesa sauropod is also distinct from *Paluxysaurus*, as shown below.

***Cedarosaurus***

*Cedarosaurus* is known from a single partial, semi-disarticulated skeleton, DMNH 39045, described by Tidwell et al. (1999). Although much of the skeleton is preserved, relatively few elements overlap with the material of the Hotel Mesa sauropod described above: dorsal vertebrae, mid-to-posterior caudal vertebrae, dorsal ribs, partial scapulae, and sternal plates. The single crushed presacral centrum of the Hotel Mesa sauropod cannot be usefully compared with the dorsal vertebrae of *Cedarosaurus* beyond the observation that the presacral bone texture described by Tidwell et al. (1999:23) as "numerous matrix filled chambers which are separated by thin walls of bone" is a good match. Tidwell et al. (1999) did not figure either the ribs or sternal plates of *Cedarosaurus*, but photographs supplied by V. Tidwell show that its sternals are generally similar in shape to those of the Hotel Mesa sauropod, though much larger and somewhat less elongate. Tidwell et al. (1999:25) noted the absence of pneumatic foramina in the two preserved rib heads while recognizing the possibility that anterior ribs might be pneumatic while posterior ribs of the same individual lack this feature. Photographs of a rib head were supplied by V. Tidwell and show little resemblance to that of the Hotel Mesa sauropod, but damage to both articular facets hinders comparison. The preserved portions of *Cedarosaurus* scapulae are from the anterior end and therefore do not greatly overlap with the more posterior preserved portion of the Hotel Mesa sauropod's scapula: however, the mid scapular region of *Cedarosaurus* differs in the possession of a more pronounced acromion process, less straight ventral border and relatively narrower scapular shaft. Finally, the mid-to-posterior caudal vertebra of the Hotel Mesa sauropod lacks the distinctive sharp ridge extending along the edge of the neural arch described by Tidwell et al. (1999:25, fig. 5); but other differences such as its greater elongation and greatly reduced neural spine are not inconsistent with the caudals of *Cedarosaurus*, taking into account that the the Hotel Mesa sauropod's caudal is from a more distal position in the caudal sequence than any of those figured by Tidwell et al. (1999). In conclusion, the preponderance of the scant morphological evidence supports the generic separation of the Hotel Mesa sauropod from *Cedarosaurus*. Furthermore, the Yellow Cat Member is Barremian in age, giving *Cedarosaurus* a minimum age of 121 Ma, while the mid-Albian position of the Hotel Mesa quarry at the top of the Ruby Ranch Member suggests an age of about 106 Ma.

While this 15 Ma gap does not in itself prove generic separation, it strongly corroborates this conclusion.

### *Venenosaurus*

*Venenosaurus* was originally described from a single small adult specimen, DMNH 40932, although elements from one or more juveniles were also present in the quarry (Tidwell et al., 2001:140). Some of the juvenile material was subsequently described by Tidwell and Wilhite (2005), but none of this material overlaps with that of the Hotel Mesa sauropod, so that comparisons with *Venenosaurus* must be on the basis of the type material alone. The overlapping material consists of caudal vertebrae, dorsal ribs, and a left scapula. The “distal caudal” of *Venenosaurus* figured by Tidwell et al. (2001:fig. 11.4C) is similar to the mid-to-posterior caudal of the Hotel Mesa sauropod in the proportions of the centrum and in the elongation of the posteriorly directed postzygapophyseal ramus. However, the *Venenosaurus* distal caudal has very much shorter prezygapophyses, and a much less tall neural arch which is set forward almost to the margin of the centrum rather than set back 10% of the centrum's length. It also lacks the characteristic ventral process that hangs from the postzygapophyseal ramus in the Hotel Mesa sauropod and bears the postzygapophyseal facets. The scapula of *Venenosaurus* figured by Tidwell et al. (2001:fig. 11.5A) does not resemble that of the Hotel Mesa sauropod, having a more curved ventral border, a much less steep ascent of the dorsal border towards the anterior expansion, a less expanded posterior blade, and no sign of the 'steps' apparent on both borders of the blade of the Hotel Mesa sauropod's scapula. The illustrated dorsal rib head of *Venenosaurus* (Tidwell et al., 2001:fig. 11.9) differs from that of the Hotel Mesa sauropod, having a very short tuberculum, a capitulum no broader than the ramus that supports it, and a very different pneumatic excavation which invades the bone in a proximal direction, penetrating the capitulum, rather than distally, penetrating the shaft. These differences of the ribs, however, may be less significant than they appear: the tuberculum of the *Venenosaurus* rib is “somewhat eroded” (Tidwell et al., 2001:153) which may explain its shortness; the degree of expansion of the capitular head may vary serially, with the *Venenosaurus* rib being from a more posterior position than the Hotel Mesa sauropod's rib; and pneumatic features tend to vary both serially and between individuals, and even on occasion between the two sides of a single element (e.g. in *Xenoposeidon*: see Taylor and Naish, 2007:1552–1553). Nevertheless, the balance of evidence strongly indicates that the Hotel Mesa sauropod is distinct from *Venenosaurus*.

***Paluxysaurus***

The internal structure of the presacral vertebrae of *Paluxysaurus* seems to be camellate, like that of the Hotel Mesa sauropod, based on the referred isolated dorsal centrum FWMSH 93B-10-48 (Rose, 2007:17, 44-45). Some of the 'distal' caudal vertebrae of *Paluxysaurus* (Rose, 2007:fig. 17) somewhat resemble the Hotel Mesa sauropod's mid-to-posterior caudal, but none has the very elongate prezygapophyses or ventral process of the postzygapophyseal ramus that characterize the latter, and the *Paluxysaurus* caudals have distinct, dorsally projecting neural spines. The figured ribs of *Paluxysaurus* (Rose, 2007:fig. 15) differ from that of the Hotel Mesa sauropod in every respect save the pneumatic invasion of the rib-head in the direction of the shaft, but these ribs are too badly damaged for useful comparison and in any case are probably from a less anterior position. The prepared scapulae of *Paluxysaurus* (Rose, 2007:fig. 20) differ from that of the Hotel Mesa sauropod in their more concave ventral border, narrower blade, less expanded posterior extremity, and lack of 'steps' on the anterior and posterior borders. The sternal plates of *Paluxysaurus* are much less proportionally elongate than those of the Hotel Mesa sauropod. The ilium of *Paluxysaurus* is not figured, and the description does not permit detailed comparison with that of the Hotel Mesa sauropod. In conclusion, significant differences in the mid-caudal vertebrae, scapulae and sternal plates demonstrate that these two genera are separate, supported by likely differences in the ribs.

## PHYLOGENETIC ANALYSIS

I attempted to establish the phylogenetic position of the Hotel Mesa sauropod by means of a phylogenetic analysis. I used the matrix of Harris (2006) as a basis, adding the single new taxon to yield a matrix of 31 taxa (29 ingroups and two outgroups) and 331 characters. The only change made was the rescoring of character 261 for *Rapetosaurus* ("In lateral view, the anteroventralmost point on the iliac preacetabular process") changing it from state 1 ("is posterior to the anteriormost part of the process (process is semicircular with posteroventral excursion of cartilage cap”) to state 0 (“is also the anteriormost point (preacetabular process is pointed”). The Hotel Mesa sauropod could be scored for 20 of the 331 characters, 6% of the total (Table 4).

Following Harris (2006), PAUP* 4.0b10 (Swofford, 2002) was used to perform a heuristic search using random stepwise addition with 50 replicates and with maximum trees = 500,000. The analysis yielded 180 equally parsimonious trees with length = 788, consistency index (CI) = 0.5228, retention index (RI) = 0.6848, and rescaled consistency index (RC) = 0.3581.

The strict consensus tree (Fig. 9A) is poorly resolved, with Titanosauriformes collapsing into a broad polytomy within which only Saltasauridae is differentiated. This represents a dramatic loss of resolution compared to the results without the Hotel Mesa sauropod (Harris, 2006: fig. 5A). A posteriori deletion of the Hotel Mesa sauropod, however, yields a well resolved Macronaria similar to that of Harris's analysis, with *Camarasaurus*, *Brachiosaurus* Riggs, 1903, *Euhelopus* Romer, 1956 and *Malawisaurus* Jacobs, Winkler, Downs and Gomani, 1993 as successive singleton outgroups to a group of more derived titanosaurs. This demonstrates that the addition of the Hotel Mesa sauropod to the matrix does not cause instability in the relationships between these more fully represented taxa, and that it is only the position of the Hotel Mesa sauropod itself that is unstable. Among the equally most parsimonious positions of the Hotel Mesa sauropod are as a non-titanosauriform camarasauromorph, a basal titanosauriform, a basal somphospondyl, the sister taxon to *Euhelopus*, a basal titanosaur, a basal lithostrotian and a derived non-saltasaurid lithostrotian. One further step is sufficient to place the Hotel Mesa sauropod as a brachiosaurid, a basal (non-camarasauromorph) macronarian, a basal (non-diplodocid) diplodocoid or even a non-neosauropod. Three further steps are required for the Hotel Mesa sauropod to be recovered as a saltasaurid, specifically an opisthocoelicaudiine.

In the 50 per cent majority rule tree (Fig. 9B) all the standard sauropod clades are recovered. This tree shows the most likely position of the Hotel Mesa sauropod as a basal somphospondyl, in a trichotomy with *Euhelopus* and Titanosauria.

In order to investigate a possible source of instability, I also re-ran the analysis with the *Rapetosaurus* ilium character restored to the state given by Harris (2006), and found that the value of this character made no difference to the results: all trees are one step longer with the new value, but the topology of all consensus trees (strict, semistrict, and majority rule) is unaffected by the changed scoring.

## DISCUSSION

### Functional Anatomy

The functional significance of the unusual ilium of the Hotel Mesa sauropod is difficult to interpret due to the absence of functionally related skeletal elements such as the pubis and ischium, posterior dorsal and anterior caudal vertebrae, and femur. In life, these elements work together as a functional complex, each affecting the function of the others. Some speculation is warranted, however.

The large preacetabular blade of the ilium provides an anchor for large protraction muscles, which would have allowed the leg to be moved forwards powerfully; this is surprising as femoral retraction is required for forward locomotion, requiring large muscles to pull the femur backwards, and the ilium of the Hotel Mesa sauropod offers almost no attachment area for posteriorly located muscles. The caudofemoralis muscle is the main power generator in locomotion for extant reptiles, and osteological correlates indicate that was also true of dinosaurs. This muscle connects the femur to the tail, so in the absence of proximal caudal vertebrae of the Hotel Mesa sauropod it is not possible to determine whether the femoral retractors were enlarged in proportion with the protractors. If so, then this increase in musculature would indicate that the Hotel Mesa sauropod may have been unusually athletic for a sauropod.

The preacetabular blade of the ilium also anchors abductors (i.e. muscles which draw the leg laterally away from the median plane). These muscles are important for creating abduction torque when standing, and may have facilitated bipedal stance or even bipedal walking.

A third possibility is that the proportionally large leg muscles were required to drive unusually long legs. The large anterior expansion of the scapula provides weak additional support for this hypothesis. If this interpretation were correct, the Hotel Mesa sauropod might resemble a giraffe in gross morphology.

**Sauropods in the Earliest Cretaceous**

Understanding of the history and evolution of sauropods in the mid-Mesozoic is impaired by the unavailability of rocks from the earliest Cretaceous in many parts of the world. For example, since the Barremian–Cenomanian Cedar Mountain Formation directly overlies the Kimmeridgian–Tithonian Morrison Formation in North America, the first three ages of the Cretaceous (Berriasian, Valanginian and Hauterivian) are all missing from the North American fossil record. For this reason, the fossil record of diplodocids is extremely limited, with all known diplodocid genera having arisen in the Kimmeridgian–Tithonian (Taylor, 2006:137). However, it is quite possible that, rather than becoming extinct at the end of the Jurassic, diplodocids continued to thrive during the 17 Ma gap in the fossil record, dying out only towards the end of that interval and being succeeded gradually by the macronarian sauropod fauna that characterizes the Cedar Mountain Formation.

This idea can be investigated by searching for late-surviving diplodocids in earliest Cretaceous strata outside North America. Until the recognition of *Lourinhasaurus* Dantas, Sanz, Silva, Ortega, Santos and Cachão, 1998 from the Late Jurassic of Portugal, no diplodocid genus had been named from outside North America, although the type species of *Lourinhasaurus*, "*Apatosaurus*" *alenquerensis* Lapparent and Zbyszewski, 1957 was considered by its describers to represent a diplodocid and the referred species "*Barosaurus*" *africanus* Fraas, 1908 was known from Tendaguru in Tanzania. The African "*Barosaurus*" material is now recognised as comprising two distinct new diplodocid genera, *Tornieria* Sternfeld, 1911 (Remes, 2006) and *Australodocus* Remes, 2007, both in fact belonging to Diplodocinae, so the existence of Late Jurassic diplodocids is now well established outside North America, with representatives in both Europe and Africa. Both the Portuguese Lourinhã Formation and the African Tendaguru Formation end at the Jurassic/Cretaceous boundary, but other latest-Jurassic formations in Portugal are conformably overlain by Early Cretaceous strata correlative with the Wealden Supergroup of England. It is in these strata that Early Cretaceous diplodocids may most usefully be sought, and there are signs that diplodocids may indeed have been present in the Wealden: Taylor and Naish (2007:1560) reported the presence of a large sauropod metacarpal from the Hastings Beds Group of the Wealden which has been identified as diplodocid, and Naish and Martill (2001:232-234) discussed other putative, though not definitive, Wealden diplodocid material. Thus it seems likely that diplodocids did indeed survive into the Cretaceous, at least in Europe and probably also in North America, and that their apparent end-Jurassic extinction is actually an artefact produced by the lack of representative strata from the earliest Cretaceous.

The most striking differences between Late Jurassic and Early Cretaceous sauropods in North America is that the former are abundant and dominated by diplodocids, whereas the latter are comparatively scarce and dominated by macronarians. It is currently impossible to determine whether this shift happened suddenly, or gradually over many millions of years in the earliest Cretaceous. It is natural to assume that if the shift was sudden, it happened at the end of the Jurassic, but that is not necessarily the case. It is possible that the diplodocid-dominated fauna persisted through the early ages of the Cretaceous and collapsed just before the earliest preserved Cretaceous sediments were deposited. The age and tempo of this faunal shift cannot be determined on the basis of the North American record; future inferences will have to be based on improved understanding of global changes in conditions in the earliest Cretaceous, and careful analysis of faunal changes on neighboring continents, especially Europe.

## CONCLUSIONS

The improving record of Early Cretaceous sauropods in North America is extended by the new Hotel Mesa sauropod, so that generic-level diversity of sauropods in this epoch now approaches that of the Late Jurassic. The new taxon is represented by at least two individuals of different sizes, probably representing a juvenile and an adult. It it clearly separate from all previously known Cedar Mountain Formation sauropods, and is distinguished from all other sauropods by several unique characters of the ilium and the scapula. The new taxon is probably a fairly basal camarasauromorph, although resolution is poor due to the incompleteness of the material. The distinctive characters of the ilium (e.g. huge preacetabular blade, no postacetabular blade, very tall overall, transversely thin) probably have some functional significance, although in the absence of other pelvic elements, femora and proximal caudals, it is not possible to interpret with certainty.

## ACKNOWLEDGMENTS

I thank R. L. Cifelli, N. Czaplewski and J. Person for access to the Hotel Mesa sauropod type material, and K. Davies for assistance with logistical matters. V. Tidwell kindly provided unpublished photographs of *Cedarosaurus* material. I thank J. I. Kirkland and D. R. Wilhite for permission to cite personal communications, J. R. Hutchinson for discussions on the possible functional significance of the new taxon's anatomy, and M. J. Wedel for discussions on Early Cretaceous North American sauropod diversity.

## LITERATURE CITED

Bird, J. 2005. New finds at the Price River II site, Cedar Mountain Formation in eastern Utah. Journal of Vertebrate Paleontology 25, Supplement to Number 3:37A.

Bonaparte, J. F. 1986. Les dinosaures (Carnosaures, Allosauridés, Sauropodes, Cétiosauridés) du Jurassique moyen de Cerro Cóndor (Chubut, Argentina). Annales de Paléontologie 72:325–386.

Bonaparte, J. F., and J. E. Powell. 1980. A continental assemblage of tetrapods from the Upper Cretaceous beds of El Brete, northwestern Argentina (Sauropoda–Coelurosauria–Carnosauria–Aves). Mémoires de la Société Géologique de France, Nouvelle Série 139:19–28.

Britt, B. B., and K. L. Stadtman. 1996. The Early Cretaceous Dalton Wells dinosaur fauna and the earliest North American titanosaurid sauropod. Journal of Vertebrate Paleontology 16, Supplement to Number 3:24A.

Britt, B. B., and K. L. Stadtman. 1997. Dalton Wells Quarry; pp. 165–166 in P. J. Currie and K. Padian (eds.), The Encyclopedia of Dinosaurs. Academic Press, San Diego.

Britt, B., D. Eberth, R. Scheetz, and B. Greenhalgh. 2004. Taphonomy of the Dalton Wells dinosaur quarry (Cedar Mountain Formation, Lower Cretaceous, Utah). Journal of Vertebrate Paleontology 24, Supplement to Number 3:41A.

Britt, B. B., R. D. Scheetz, J. S. McIntosh, and K. L. Stadtman. 1998. Osteological characters of an Early Cretaceous titanosaurid sauropod dinosaur from the Cedar Mountain Formation of Utah. Journal of Vertebrate Paleontology 18, Supplement to Number 3:29A.

Britt, B. B., K. L. Stadtman, R. D. Scheetz, and J. S. McIntosh. 1997. Camarasaurid and titanosaurid sauropods from the Early Cretaceous Dalton Wells Quarry (Cedar Mountain Formation), Utah. Journal of Vertebrate Paleontology 17, Supplement to Number 3:34A.

Britt, B. D. Burton, G. Gehrels, E. Christiansen, and D. Chure. 2007. Laser ablation zircon U-Pb geochronology of the Cedar Mountain and Dakota Formations of Dinosaur National Monument, Utah. Journal of Vertebrate Paleontology 27,

Supplement to Number 3:53A.

Burge, D. L., and J. H. Bird. 2001. Fauna of the Price River II Quarry of eastern Utah. Journal of Vertebrate Paleontology 21, Supplement to Number 3:37A.

Burge, D. L., J. H. Bird, B. B. Britt, D. J. Chure, and R. L. Scheetz. 2000. A brachiosaurid from the Ruby Ranch Mbr. (Cedar Mountain Fm.) near Price, Utah, and sauropod faunal change across the Jurassic-Cretaceous boundary of North America. Journal of Vertebrate Paleontology 20, Supplement to Number 3:32A.

Carpenter, K., and J. S. McIntosh. 1994. Upper Jurassic sauropod babies from the Morrison Formation; pp. 265–278 in K. Carpenter, K. F. Hirsch, and J. R. Horner (eds.), Dinosaur Eggs and Babies. Cambridge University Press, Cambridge.

Carpenter, K., and V. Tidwell. 2005. Reassessment of the Early Cretaceous Sauropod *Astrodon johnsoni* Leidy 1865 (Titanosauriformes); pp. 78–114 in V. Tidwell, and K. Carpenter (eds.), Thunder Lizards: the Sauropodomorph Dinosaurs. Indiana University Press, Bloomington, Indiana.

Chure, D. J. 2001. A new sauropod with a well preserved skull from the Cedar Mountain Fm. (Cretaceous) of Dinosaur National Monument, UT. Journal of Vertebrate Paleontology 21, Supplement to Number 3:40A.

Chure, D., B. Britt, and B. Greenhalgh. 2006. A new titanosauriform sauropod with abundant skull material from the Cedar Mountain Formation, Dinosaur National Monument. Journal of Vertebrate Paleontology 26, Supplement to Number 3:50A.

Cifelli, R. L., J. D. Gardner, R. L. Nydam, and D. L. Brinkman. 1997. Additions to the vertebrate fauna of the Antlers Formation (Lower Cretaceous), southeastern Oklahoma. Oklahoma Geology Notes 57:124–131.

Cope, E. D. 1877. On a gigantic saurian from the Dakota epoch of Colorado. Paleontology Bulletin 25:5–10.

Coulson, A., R. Barrick, W. Straight, S. Decherd, and J. Bird. 2004. Description of the new brachiosaurid (Dinosauria: Sauropoda) from the Ruby Ranch Member (Cretaceous: Albian) of the Cedar Mountain Formation, Utah. Journal of Vertebrate Paleontology 24, Supplement to Number 3:48A.

Curry Rogers, K., and C. A. Forster. 2001. Last of the dinosaur titans: a new sauropod from Madagascar. Nature 412:530–534.

Curtice, B. D. 2000. The axial skeleton of *Sonorasaurus thompsoni* Ratkevich 1998. Mesa Southwest Museum Bulletin 7:83–87.

Dantas, P., J. L. Sanz, C. M. Silva, F. Ortega, V. F. Santos, and M. Cachão. 1998. *Lourinhasaurus* n. gen. novo dinossáurio saurópode do Jurássico superior (Kimmeridgiano superiorTitoniano inferior) de Portugal. Actas do V Congresso de Geologia 84:91–94.

DeCourten, F. L. 1991. New data on Early Cretaceous dinosaurs from the Long Walk Quarry and tracksite, Emery County, Utah; pp. 311–325 in T. C. Chidsey, Jr. (ed.), Geology of East-Central Utah, Utah Geological Association Publication 19.

Eberth, D. A., B. B. Britt, R. Scheetz, K. L. Stadtman, and D. B. Brinkman. 2006. Dalton Wells – geology and significance of debris-flow-hosted dinosaur bonebeds in the Cedar Mountain Formation (Lower Cretaceous) of eastern Utah, U.S.A. Paleogeography, Paleoclimatology, Paleoecology 236:217–245.

Fraas, E. 1908. Ostafrikanische Dinosaurier. Palaeontographica 55:105–144.

Gallup, M. R. 1989. Functional morphology of the hindfoot of the Texas sauropod *Pleurocoelus* sp. indet.; pp. 71–74 in J. O. Farlow (ed.), Paleobiology of the Dinosaurs, Geological Society of America Special Paper 238.

Gilmore, C. W. 1921. The fauna of the Arundel Formation of Maryland. U.S. National Museum Proceedings 59:581–594.

Gilmore, C. W. 1922. Discovery of a sauropod dinosaur from the Ojo Alamo Formation of New Mexico. Smithsonian Miscellaneous Collections 81:1–9.

Gilmore, C. W. 1936. Osteology of *Apatosaurus* with special reference to specimens in the Carnegie Museum. Memoirs of the Carnegie Museum 11:175–300.

Gilmore, C. W. 1946. Reptilian fauna from the North Horn Formation of central Utah. U. S. Geological Survey Professional Paper 210-C:29–52 and plates 3–14.

Glut, D. F. 1997. Dinosaurs: the Encyclopedia. McFarland & Company, Inc., Jefferson, NC, 1076 pp.

Gomani, E. M., L. L. Jacobs, and D. A. Winkler. 1999. Comparison of the African titanosaurian, *Malawisaurus*, with a North America Early Cretaceous sauropod; pp. 223–233 in Y. Tomida, T. H. Rich, and P. Vickers-Rich (eds.), Proceedings of the Second Gondwanan Dinosaur Symposium. Tokyo National Science Museum Monograph 15.

Harris, J. D. 2006. The significance of *Suuwassea emiliae* (Dinosauria: Sauropoda) for flagellicaudatan intrarelationships and evolution. Journal of Systematic Palaeontology 4:185–198.

Hatcher, J. B. 1901. *Diplodocus* (Marsh): its osteology, taxonomy and probable habits, with a restoration of the skeleton. Memoirs of the Carnegie Museum 1:1–63.

Hatcher, J. B. 1903a. Osteology of *Haplocanthosaurus* with description of a new species, and remarks on the probable habits of the Sauropoda and the age and origin of the Atlantosaurus beds. Memoirs of the Carnegie Museum 2:1–72 and plates I–V.

Hatcher, J. B. 1903b. A new name for the dinosaur *Haplocanthus* Hatcher. Proceedings of the Biological Society of Washington 16:100.

He, X., K. Li, and K. Cai. 1988. The Middle Jurassic dinosaur fauna from Dashanpu, Zigong, Sichuan, vol. IV: sauropod dinosaurs (2): *Omeisaurus tianfuensis*. Sichuan Publishing House of Science and Technology, Chengdu, China, 143 + 20 plates pp.

Huene, F. v. 1929. Los Saurisquios y Ornitisquios del Cretaceo Argentina. Annales Museo de La Plata, 3, Serie 2a. Museo de La Plata, Argentina, 196 pp.

Ikejiri, T., V. Tidwell, and D. L. Trexler. 2005. new adult specimens of *Camarasaurus lentus* highlight ontogenetic variation within the species; pp. 154–179 in V. Tidwell, and K. Carpenter (eds.), Thunder Lizards: the Sauropodomorph Dinosaurs. Indiana University Press, Bloomington, Indiana.

Jacobs, L. L., D. A. Winkler, W. R. Downs, and E. M. Gomani. 1993. New material of an Early Cretaceous titanosaurid sauropod dinosaur from Malawi. Palaeontology 36:523–534.

Janensch, W. 1914. Übersicht über der Wirbeltierfauna der Tendaguru-Schichten nebst einer kurzen Charakterisierung der neu aufgefuhrten Arten von Sauropoden.

Archiv fur Biontologie 3:81–110.

Janensch, W. 1950. Die Wirbelsaule von *Brachiosaurus brancai*. Palaeontographica (Suppl. 7) 3:27–93.

Janensch, W. 1961. Die Gliedmaszen und Gliedmaszengürtel der Sauropoden der Tendaguru-Schichten. Palaeontographica (Suppl. 7) 3:177–235 and plates XV–XXIII.

Kirkland, J. I., and S. K. Madsen. 2007. The Lower Cretaceous Cedar Mountain Formation, eastern Utah: the view up an always interesting learning curve. Fieldtrip Guidebook, Geological Society of America, Rocky Mountain Section. 108 pp.

Kirkland, J. I., B. Britt, D. Burge, K. Carpenter, R. Cifelli, F. DeCourten, J. Eaton, S. Hasiotis, and T. Lawton. 1997. Lower to middle Cretaceous dinosaur faunas of the Central Colorado Plateau: a key to understanding 35 million years of tectonics, sedimentology, evolution, and biogeography. Brigham Young University Geology Studies 42:69–103.

Langston, W., Jr. 1974. Non-mammalian Commanchean tetrapods. Geoscience and Man 8:77–102.

Lapparent, A. F. d., and G. Zbyszewski. 1957. Mémoire no. 2 (nouvelle série): les dinosauriens du Portugal. Services Géologiques du Portugal, Lisbon, Portugal, 63 pp.

Larkin, P. 1910. The occurrence of a sauropod dinosaur in the Trinity Cretaceous of Oklahoma. Journal of Geology 18:93–98.

Leidy, J. 1865. Cretaceous reptiles of the United States. Smithsonian Contribution to Knowledge 192:1–135.

Lovelace, D., W. R. Wahl, and S. A. Hartman. 2003. Evidence for costal pneumaticity in a diplodocid dinosaur (*Supersaurus vivianae*). Journal of Vertebrate Paleontology 23, Supplement to Number 3:73A.

Marsh, O. C. 1877. Notice of new dinosaurian reptiles from the Jurassic Formation. American Journal of Science and Arts 14:514–516.

Marsh, O. C. 1878. Principal characters of American Jurassic dinosaurs. Part I.

American Journal of Science, Series 3, 16:411–416.

Marsh, O. C. 1888. Notice of a new genus of Sauropoda and other new dinosaurs from the Potomac Formation. American Journal of Science, Series 3, 35:89–94.

Maxwell, W. D., and R. L. Cifelli, R. L. 2000. Last evidence of sauropod dinosaurs (Saurischia: Sauropodomorpha) in the North American mid-Cretaceous. Brigham Young University Geology Studies 45:19–24.

McIntosh, J. S. 1990. Sauropoda; pp. 345–401 in D. B. Weishampel, P. Dodson, and H. Osmólska (eds.), The Dinosauria. University of California Press, Berkeley and Los Angeles.

Naish, D., and D. M. Martill. 2001. Saurischian dinosaurs 1: Sauropods; pp. 185–241 in D. M. Martill, and D. Naish (eds.), Dinosaurs of the Isle of Wight. The Palaeontological Association, London.

Osborn, H. F., and C. C. Mook. 1921. *Camarasaurus*, *Amphicoelias* and other sauropods of Cope. Memoirs of the American Museum of Natural History, n.s. 3:247–387 and plates LX–LXXXV.

Ostrom, J. H. 1970. Stratigraphy and paleontology of the Cloverly Formation (Lower Cretaceous) of the Bighorn Basin area, Wyoming and Montana. Bulletin of the Peabody Museum of Natural History 35:1–234.

Ouyang, H., and Y. Ye. 2002. The first mamenchisaurian skeleton with complete skull: *Mamenchisaurus youngi*. Sichuan Science and Technology Press, Chengdu, China, 111 pp.

Owen, R. 1842. Report on British fossil reptiles, Part II. Reports of the British Association for the Advancement of Sciences 11:60–204.

Paul, G. S. 1988. The brachiosaur giants of the Morrison and Tendaguru with a description of a new subgenus, *Giraffatitan*, and a comparison of the world's largest dinosaurs. Hunteria 2:1–14.

Powell, J. E. 1992. Osteología de *Saltasaurus loricatus* (Sauropoda–Titanosauridae) del Cretácico Superior del Noroeste Argentino; pp. 165–230 in J. L. Sanz, and A. D. Buscalioni (eds.), Los Dinosaurios y su Entorno Biotico. Actas del Segundo Curso de Paleontologia en Cuenca. Instituto Juan de Valdés, Ayuntamiento de Cuenca.

Ratkevich, R. 1998. The Cretaceous brachiosaurid dinosaur, *Sonorasaurus thompsoni* gen. et sp. nov, from Arizona. Journal of the Arizona-Nevada Acedemy of Science 31:71–82.

Remes, K. 2006. Revision of the Tendaguru sauropod dinosaur *Tornieria africana* (Fraas) and its relevance for sauropod paleobiogeography. Journal of Vertebrate Paleontology 26:651–669.

Remes, K. 2007. A second Gondwanan diplodocid dinosaur from the Upper Jurassic Tendaguru beds of Tanzania, East Africa. Palaeontology 50:653–667.

Riggs, E. S. 1903. *Brachiosaurus altithorax*, the largest known dinosaur. American Journal of Science 15:299–306.

Romer, A. S. 1956. Osteology of the Reptiles. University of Chicago Press, Chicago, 772 pp.

Rose, P. J. 2007. A new titanosauriform sauropod (Dinosauria: Saurischia) from the Early Cretaceous of Central Texas and its phylogenetic relationships. Palaeontologia Electronica 10:8A: 1–65.

Salgado, L., R. A. Coria, and J. O. Calvo. 1997. Evolution of titanosaurid sauropods. I: Phylogenetic analysis based on the postcranial evidence. Ameghiniana 34:3–32.

Salgado, L., A. Garrido, S. E. Cocca, and J. R. Cocca. 2004. Lower Cretaceous rebbachisaurid sauropods from Cerro Aguada Del Leon (Lohan Cura Formation), Neuquen Province, Northwestern Patagonia, Argentina. Journal of Vertebrate Paleontology 24:903–912.

Seeley, H. G. 1888. On the classification of the fossil animals commonly named Dinosauria. Proceedings of the Royal Society of London 43:165–171.

Sereno, P. C. 1998. A rationale for phylogenetic definitions, with application to the higher-level taxonomy of Dinosauria. Neues Jahrbuch für Geologie und Paläontologie, Abhandlungen 210:41–83.

Sternfeld, R. 1911. Zur Nomenklatur der Gattung *Gigantosaurus* Fraas. Sitzungesberichte der Gesellschaft Naturforschender Freunde zu Berlin 1911:398–

398.

Swofford, D. L. 2002. PAUP*: phylogenetic analysis using parsimony (* and other methods). Sinauer Associates, Sunderland, Mass.

Taylor, M. P. 2006. Dinosaur diversity analysed by clade, age, place and year of description; pp. 134–138 in P. M. Barrett (ed.), Ninth international symposium on Mesozoic terrestrial ecosystems and biota, Manchester, UK. Cambridge Publications, Cambridge, UK.

Taylor, M. P. In press. A re-evaluation of *Brachiosaurus altithorax* Riggs 1903 (Dinosauria, Sauropoda) and its generic separation from *Giraffatitan brancai* (Janensch 1914). Journal of Vertebrae Palaeontology.

Taylor, M. P., and D. Naish. 2005. The phylogenetic taxonomy of Diplodocoidea (Dinosauria: Sauropoda). PaleoBios 25:1–7.

Taylor, M. P., and D. Naish. 2007. An unusual new neosauropod dinosaur from the Lower Cretaceous Hastings Beds Group of East Sussex, England. Palaeontology 50:1547–1564.

Tidwell, V., and K. Carpenter. 2007. First description of cervical vertebrae for an Early Cretaceous titanosaur from North America. Journal of Vertebrate Paleontology 27, Supplement to Number 3:158A–158A.

Tidwell, V., and D. R. Wilhite. 2005. Ontogenetic variation and isometric growth in the forelimb of the Early Cretaceous sauropod *Venenosaurus*; pp. 187–198 in V. Tidwell, and K. Carpenter (eds.), Thunder Lizards: the Sauropodomorph Dinosaurs. Indiana University Press, Bloomington, Indiana.

Tidwell, V., K. Carpenter, and W. Brooks. 1999. New sauropod from the Lower Cretaceous of Utah, U.S.A. Oryctos 2:21–37.

Tidwell, V., K. Carpenter, and S. Meyer. 2001. New Titanosauriform (Sauropoda) from the Poison Strip Member of the Cedar Mountain Formation (Lower Cretaceous), Utah; pp. 139–165 in D. H. Tanke, and K. Carpenter (eds.), Mesozoic Vertebrate Life: New Research inspired by the Paleontology of Philip J. Currie. Indiana University Press, Bloomington and Indianapolis, Indiana.

Upchurch, P., P. M. Barrett, and P. Dodson. 2004a. Sauropoda; pp. 259–322 in D. B. Weishampel, P. Dodson, and H. Osmólska (eds.), The Dinosauria, 2nd edition. University of California Press, Berkeley and Los Angeles.

Upchurch, P., Y. Tomida, and P. M. Barrett. 2004b. A new specimen of *Apatosaurus ajax* (Sauropoda: Diplodocidae) from the Morrison Formation (Upper Jurassic) of Wyoming, U.S.A. National Science Museum Monographs 26:1–110 and plates 1–10.

Wedel, M. J. 2003a. Vertebral pneumaticity, air sacs, and the physiology of sauropod dinosaurs. Paleobiology 29:243–255.

Wedel, M. J. 2003b. The evolution of vertebral pneumaticity in sauropod dinosaurs. Journal of Vertebrate Paleontology 23:344–357.

Wedel, M. J. 2005. Postcranial skeletal pneumaticity in sauropods and its implications for mass estimates; pp. 201–228 in J. A. Wilson, and K. Curry-Rogers (eds.), The Sauropods: Evolution and Paleobiology. University of California Press, Berkeley.

Wedel, M. J., and R. L. Cifelli. 2005. *Sauroposeidon*: Oklahoma's native giant. Oklahoma Geology Notes 65:40–57.

Wedel, M. J., R. L. Cifelli, and R. K. Sanders. 2000a. *Sauroposeidon proteles*, a new sauropod from the Early Cretaceous of Oklahoma. Journal of Vertebrate Paleontology 20:109–114.

Wedel, M.J., R. L. Cifelli, and R. K. Sanders. 2000b. Osteology, paleobiology, and relationships of the sauropod dinosaur *Sauroposeidon*. Acta Palaeontologica Polonica 45:343–388.

Wilhite, R. 1999. Ontogenetic variation in the appendicular skeleton of the genus *Camarasaurus*. Master's thesis, Brigham Young University.

Wilhite, R. 2003. Biomechanical reconstruction of the appendicular skeleton in three North American Jurassic sauropods. Ph.D. dissertation, Louisiana State University. http://etd.lsu.edu/docs/available/etd-0408103-003549/

Wilson, J. A. 2002. Sauropod dinosaur phylogeny: critique and cladistic analysis. Zoological Journal of the Linnean Society 136:217–276.

Wilson, J. A., and P. C. Sereno. 1998. Early evolution and higher-level phylogeny of sauropod dinosaurs. Society of Vertebrate Paleontology Memoir 5:1–68.

Wilson, J. A., and P. Upchurch. 2003. A revision of *Titanosaurus* Lydekker (Dinosauria – Sauropoda), the first dinosaur genus with a 'Gondwanan' distribution. Journal of Systematic Palaeontology 1:125–160.

Winkler, D. A., L. L. Jacobs, and P. A. Murry. 1997. Jones Ranch: an Early Cretaceous sauropod bone-bed in Texas. Journal of Vertebrate Paleontology 17, Supplement to Number 3:85A.

Yates, A. M. 2006. Solving a dinosaurian puzzle: the identity of *Aliwalia rex* Galton. Historical Biology 19:93–123.

Young, C.-C. 1939. On the new Sauropoda, with notes on other fragmentary reptiles from Szechuan. Bulletin of the Geological Society of China 19:279–315.

Young, C.-C., and X. Zhao. 1972. [Description of the type material of *Mamenchisaurus hochuanensis*]. Institute of Vertebrate Paleontology and Paleoanthropology Monograph Series I 8:1–30. [Chinese]

TABLE 1. Clade names used in this study and the definitions used. For simplicity, specifiers are indicated by genus rather than species; in each case, the type species of the genus is intended. Node-based clades are indicated by "+", branch-based clades by "not".

| Clade name | As defined by | Definition |
|---|---|---|
| Sauropoda | Yates (2006:12) | *Saltasaurus* not *Melanorosaurus* |
| Neosauropoda | Wilson and Sereno (1998:46) | *Diplodocus* + *Saltasaurus* |
| Diplodocoidea | Wilson and Sereno (1998:17) | *Diplodocus* not *Saltasaurus* |
| Rebbachisauridae | Salgado et al. (2004:910) | *Rebbachisaurus* not *Diplodocus* |
| Diplodocidae | Sereno (1998:63) | *Diplodocus* not *Dicraeosaurus* |
| Diplodocinae | Taylor and Naish (2005:5) | *Diplodocus* not *Apatosaurus* |
| Macronaria | Wilson and Sereno (1998:49) | *Saltasaurus* not *Diplodocus* |
| Camarasauromorpha | Upchurch et al. (2004a:306) | *Camarasaurus* + *Saltasaurus* |
| Camarasauridae | Taylor and Naish (2007:1555) | *Camarasaurus* not *Saltasaurus* |
| Titanosauriformes | Wilson and Sereno (1998:51) | *Brachiosaurus* + *Saltasaurus* |
| Brachiosauridae | Wilson and Sereno (1998:20–21) | *Brachiosaurus* not *Saltasaurus* |
| Somphospondyli | Wilson and Sereno (1998:53) | *Saltasaurus* not *Brachiosaurus* |
| Titanosauria | Wilson and Upchurch (2003:156) | *Andesaurus* + *Saltasaurus* |
| Lithostrotia | Wilson and Upchurch | *Malawisaurus* + *Saltasaurus* |
| | (2003:156) | |
| Saltasauridae | Sereno (1998:63) | *Saltasaurus* + *Opisthocoelicaudia* |
| Opisthocoelicaudiinae | Sereno (1998:63) | *Opisthocoelicaudia* not *Saltasaurus* |

TABLE 2. Material referred to the Hotel Mesa sauropod.

| Specimen number | Description | Length (cm) | Width (cm) | Thickness (cm) |
|---|---|---|---|---|
| 27761 | Partial scapula, missing anterior part | 98 | 55 | ?5 |
| 27762 | Large, flat rib shaft | 101 | 7.5 | 2 |
| 27763 | Rib fragment | 63 | 8 | 5 |
| 27764 | Rib fragment | 60 | 2.5 | 2 |
| 27765 | Rib fragment | 58 | 6 | 3 |
| 27766 | Complete pneumatic rib | 76 | 12 | 4 |
| 27767 | Proximal part of rib, flattened | 26 | 17 | 2 |
| 27768 | Rib fragment | 31 | 6 | 1.5 |
| 27769 | Fragment of ?ischium | 10 | 5 | 2.5 |
| 27770 | Rib fragment | 14 | 3.5 | 1 |
| 27771 | Rib fragment | 17 | 3 | 2 |
| 27772 | Rib fragment (in two parts) | 13 | 7 | 2 |
| 27773 | Flat scrap | 7 | 5 | 0.5 |
| 27773 | Flat scrap | 11.5 | 9 | 1 |
| 27773 | Rib fragment | 13.5 | 4 | 2 |
| 27774–27783 | Fragments | | | |
| 27784 | Collection of 21 fragments | | | |
| 27785–27793 | Fragments | | | |
| 27794 | Partial distal caudal centrum | | | |
| 27795–27800 | Fragments | | | |
| 61248 | Nearly complete mid-caudal vertebra | 11 | 6 | 5.5 |
| 66429 | Crushed presacral centrum | 14 | 14 | 3 |
| 66430 | Ilium | 40.5 | 31 | 8 |
| 66430 | Dorsal fragment of ilium | 14 | 3 | 1 |
| 66430 | Dorsal fragment of ilium | 15.5 | 9 | 1 |
| 66431 | Incomplete sternal plate | 15 | 7.5 | 1.5 |
| 66432 | Incomplete sternal plate | 11.5 | 7 | 1 |

TABLE 3. Relative measurements of ilia in sauropods. Total length is measured along the longest axis of the ilium; lengths of preacetabular and postacetabular lobes are measured parallel to this axis, and extend from the extremity of the lobe to the anterior margin of the pubic peduncle and posterior margin of the ischiadic peduncle respectively.

| Taxon | Specimen | Reference | Total length (cm) | Length of preacetabular lobe (cm) | | Length of postacetabular lobe (cm) | |
|---|---|---|---|---|---|---|---|
| *Mamenchisaurus hochuanensis* | CCG V 20401 | Young and Zhao (1972:pl. 6, fig. 1a) | 102 | 39 | 38% | 18 | 18% |
| *Diplodocus carnegii* | CM 94 | Hatcher (1901:pl. X, fig. 1) | 109[1] | 41 | 38% | 20 | 18% |
| *Camarasaurus supremus* | AMNH 5761 Il. 1 | Osborn and Mook (1921:fig. 87) | 115 | 36 | 31% | 19 | 17% |
| *Giraffatitan brancai*[2] | HMN Aa 13 | Janensch (1961:pl. E, fig. 1a) | 119 | 47 | 39% | 21 | 18% |
| | HMN J1 | Janensch (1961:pl. E, fig. 2) | 105.5 | 55 | 52% | 16 | 15% |
| *Rapetosaurus krausi* | FMNH PR 2209 | Curry Rogers and Forster (2001:fig. 3h) | 42 | 20 | 48% | 7 | 17% |
| The Hotel Mesa sauropod | OMNH 66430 | (This study) | 40.5 | 22.3 | 55% | 0 | 0% |

**Notes.**

[1]Hatcher did not state the length of the ilium of CM 94 and did not figure that of CM 84.

I have assumed that the ilium of the figured specimen CM 94 is the same size as that of CM 84 (Hatcher, 1901:46) and calculated the proportions from the figured specimen.

[2]Janensch's (1961: pl. E) figures of the two ilia of *Giraffatitan brancai* are not executed from an orthogonal perspective: the ilium of HMN Aa 13 is illustrated from a slightly anterolateral position, foreshortening the preacetabular lobe, and that of HMN J 1 from a slight posterolateral position, foreshortening the postacetabular lobe. The true lengths of the two lobes are probably somewhere between the two percentages calculated from the figures: about 45% for the former, and 16% for the latter.

TABLE 4. Character scores for the Hotel Mesa sauropod in the matrix used for the phylogenetic analysis of this paper. Apart from the addition of the Hotel Mesa sauropod and the rescoring of character 261 for *Rapetosaurus*, the matrix is identical to that of Harris (2006). The Hotel Mesa sauropod is unscored for all characters except those listed. Conventional anatomical nomenclature is here used in place of the avian nomenclature of Harris (2006).

| Character | | Score | |
|---|---|---|---|
| 184 | Ratio of centrum length:height in middle caudal vertebrae | 1 | ≥ 2.0 |
| 185 | Sharp ridge on lateral surface of middle caudal centra at arch-body junction | 0 | absent |
| 186 | Morphology of articular surfaces in middle caudal centra | 0 | subcircular |
| 187 | Ventral longitudinal excavation on anterior and middle caudal centra | 0 | absent |
| 188 | Morphology of anterior articular face of middle and posterior caudal centra | 0 | amphicoelous/amphiplatyan |
| 189 | Position of neural arches over centra on middle caudal vertebrae | 1 | located mostly or entirely over anterior half of centrum |
| 191 | Morphology of posterior caudal centra | 0 | cylindrical |
| 197 | Proximal pneumatic foramina on dorsal ribs | 1 | present |
| 198 | Morphology of proximal ends of anterior dorsal ribs | 1 | strongly convex anteriorly and deeply concave posteriorly |
| 199 | Cross-sectional shape of anterior dorsal ribs | 0 | subcircular |
| 208 | Size of scapular acromion | 1 | broad (dorsoventral width more than 150% minimum width of scapular body) |
| 210 | Morphology of portion of acromion posterior to deltoid crest | 0 | flat or convex and decreases in mediolateral thickness toward posterior margin |
| 212 | Morphology of scapular body | 2 | posterior end racquet-shaped (dorsoventrally expanded) |
| 221 | Morphology of sternal plate | 2 | elliptical with concave lateral margin |
| 259 | Morphology of dorsal margin of ilium body (in lateral view) | 1 | semicircular (markedly convex) |
| 260 | Position of dorsalmost point on ilium | 1 | anterior to base of pubic process |
| 261 | In lateral view, the most anteroventral point on the iliac preacetabular process | 0 | is also the most anterior point (preacetabular lobe is pointed) |
| 262 | Orientation of preacetabular lobe of ilium with respect to axis of body | 1 | anterolateral in vertical plane |
| 263 | Size of ischiadic peduncle of ilium | 1 | low and rounded (long axis of ilium oriented anterodorsally-posteroventrally) |
| 264 | Projected line connecting articular surfaces of ischiadic and pubic peduncles of ilium | 0 | passes ventral to ventral margin of postacetabular lobe of ilium |

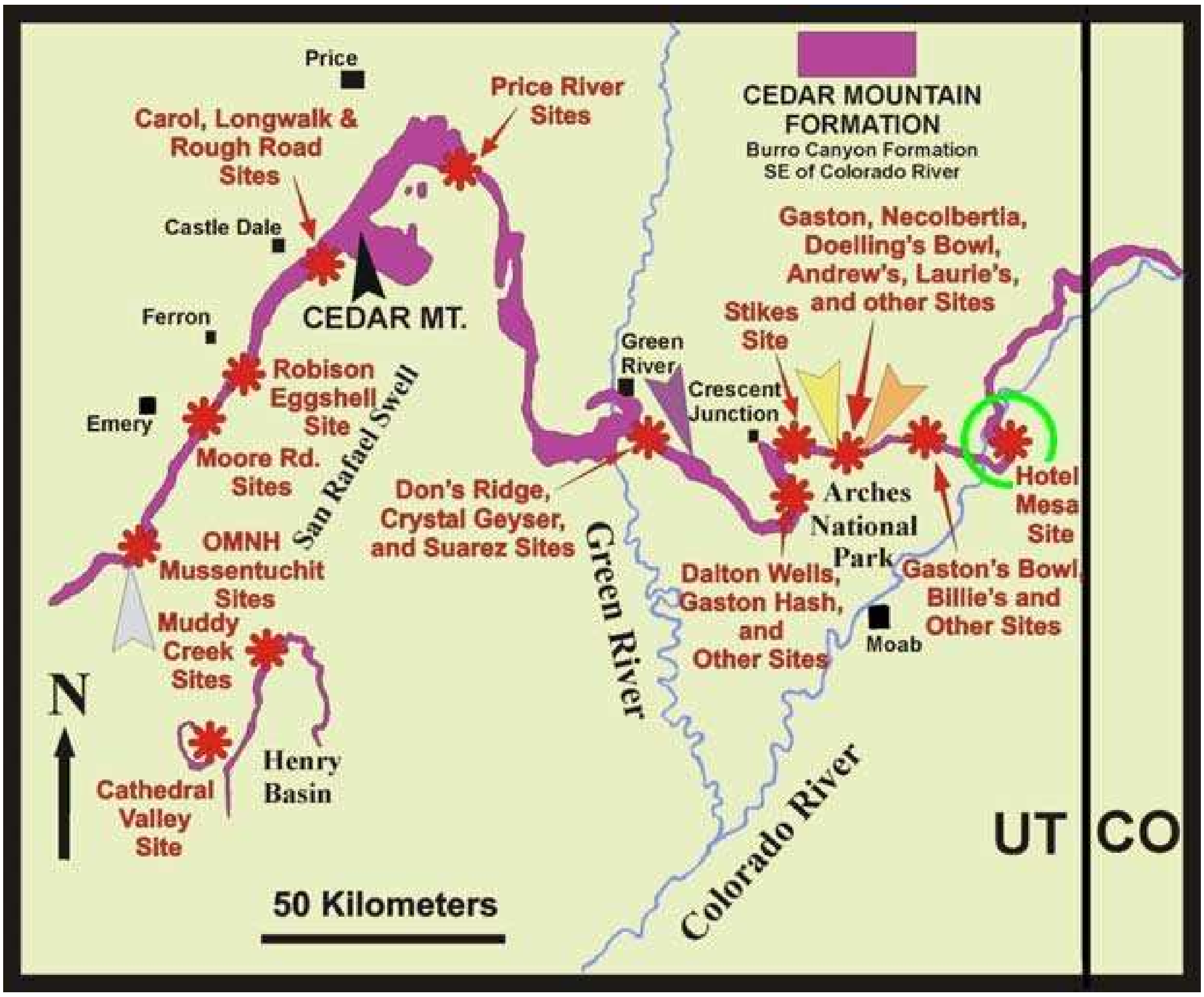

FIGURE 1. Map showing the type locality of the Hotel Mesa sauropod, OMNH locality V857, Grand County, eastern Utah. Reproduced from Kirkland and Madsen (2007:fig. 2).

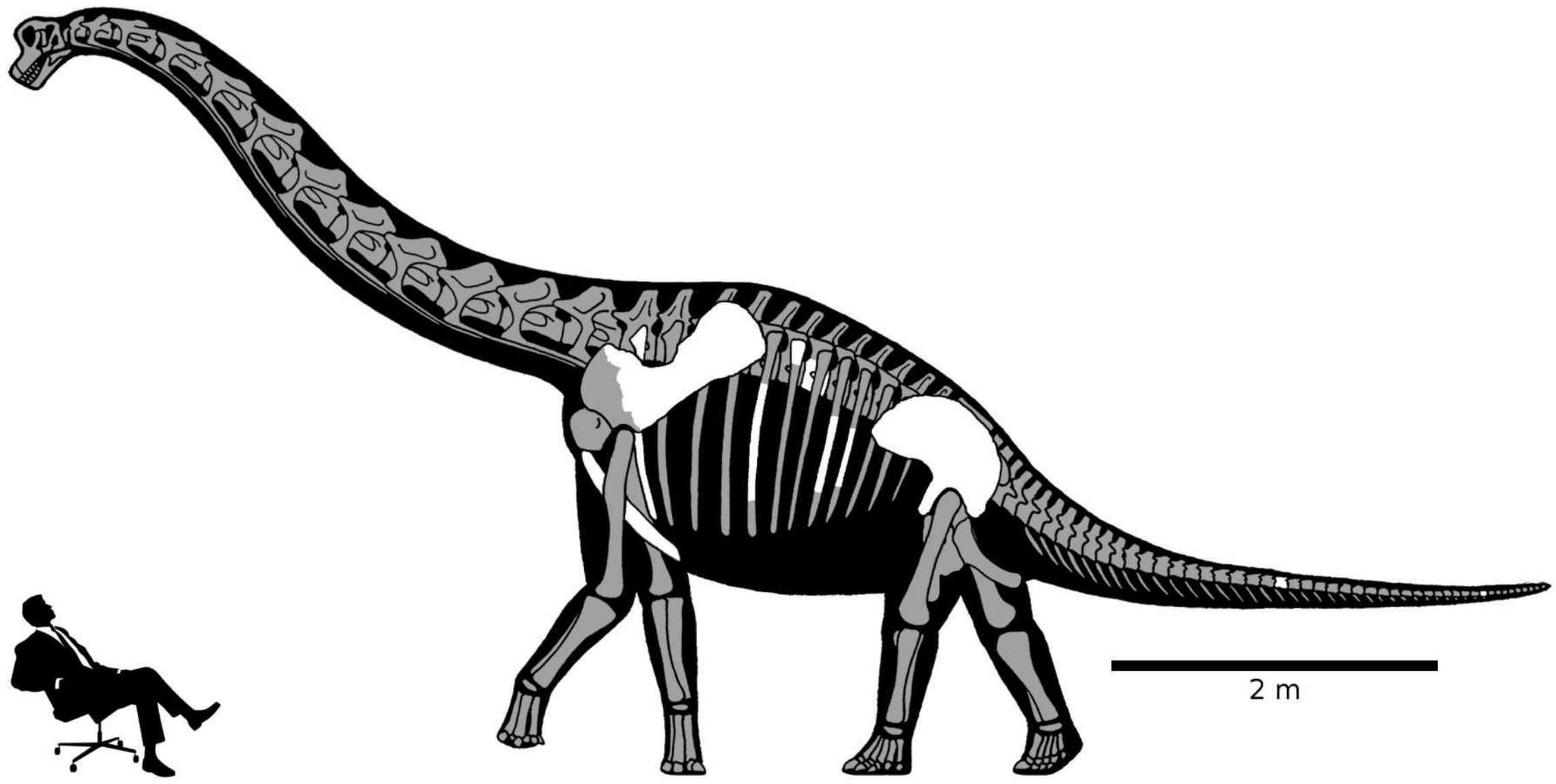

FIGURE 2. Skeletal atlas of the Hotel Mesa sauropod in left lateral view. Preserved elements are white, missing elements are reconstructed in gray after the *Giraffatitan* reconstruction of Wedel and Cifelli (2005:fig. 15B). Scale bar equals 2 m.

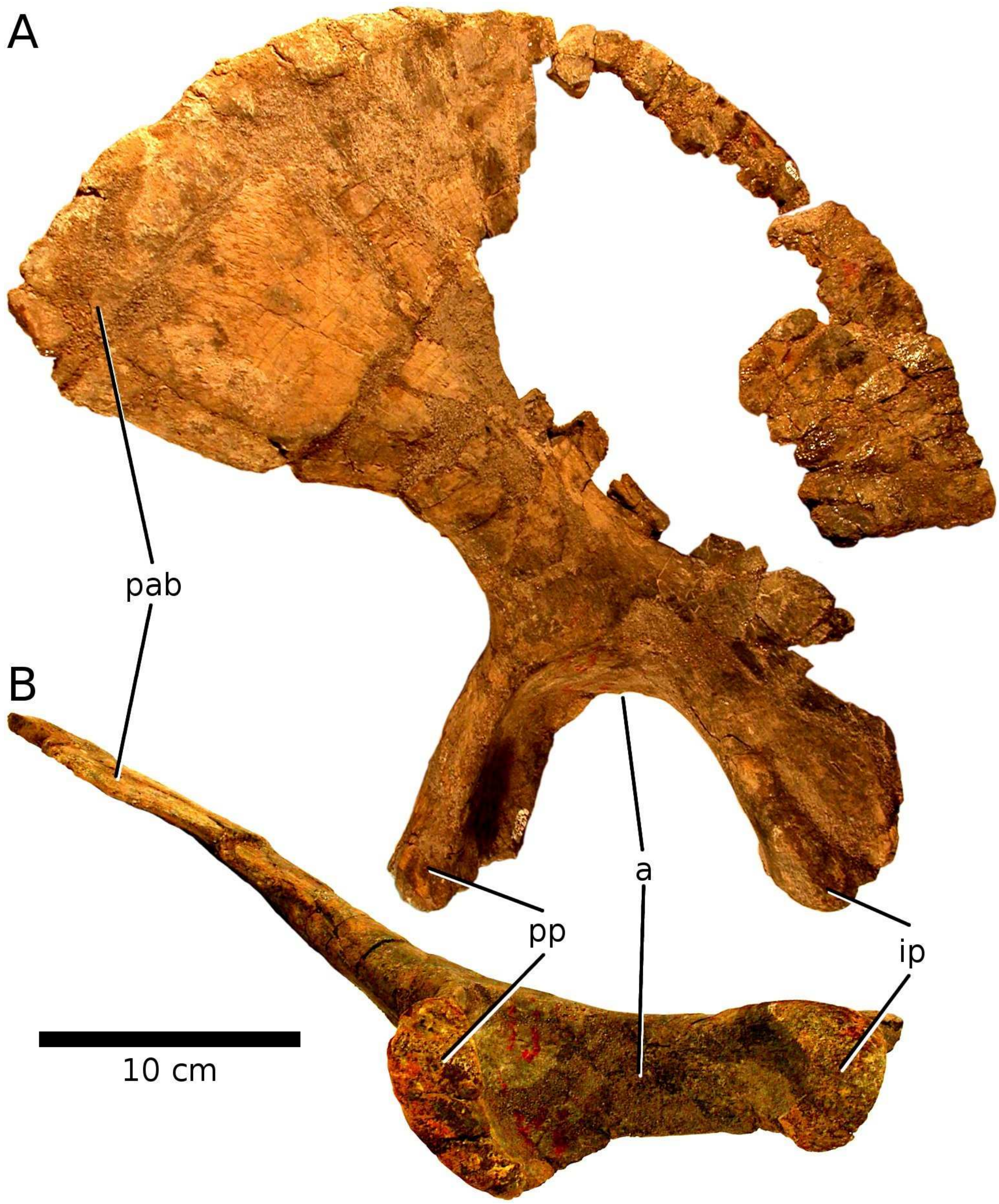

FIGURE 3. Left ilium of the Hotel Mesa sauropod (type specimen) OMNH 66430. **A**, lateral view reconstructed from the three fragments; **B**, ventral view. a, acetabulum; ip, ischiadic peduncle; pab, pre-acetabular blade; pp, pubic peduncle. Scale bar equals 10 cm.

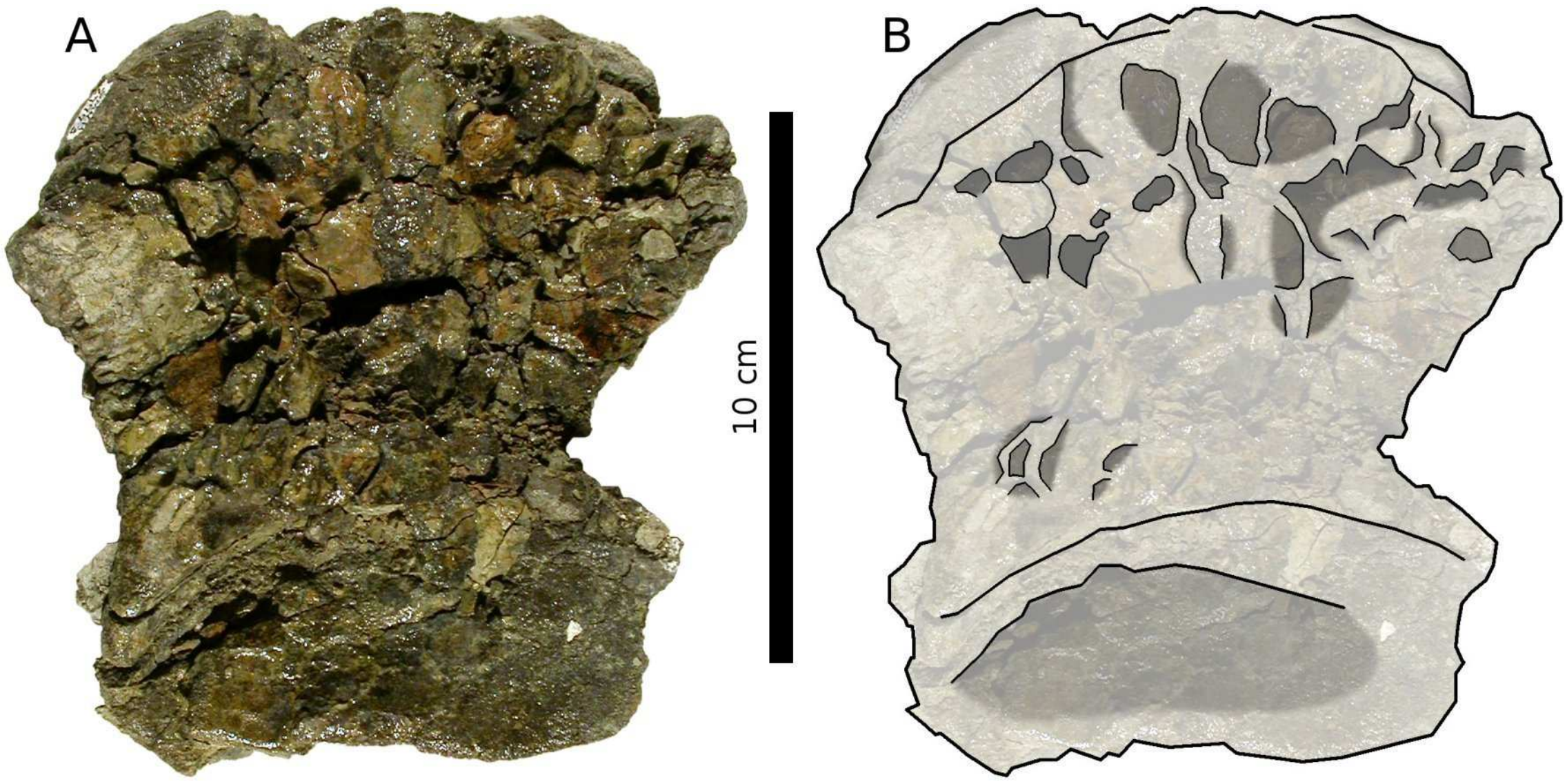

FIGURE 4. Damaged presacral vertebra of the Hotel Mesa sauropod, OMNH 66429, in dorsal view. **A**, photograph; **B**, interpretive drawing. Shading indicates air spaces. Scale bar equals 10 cm.

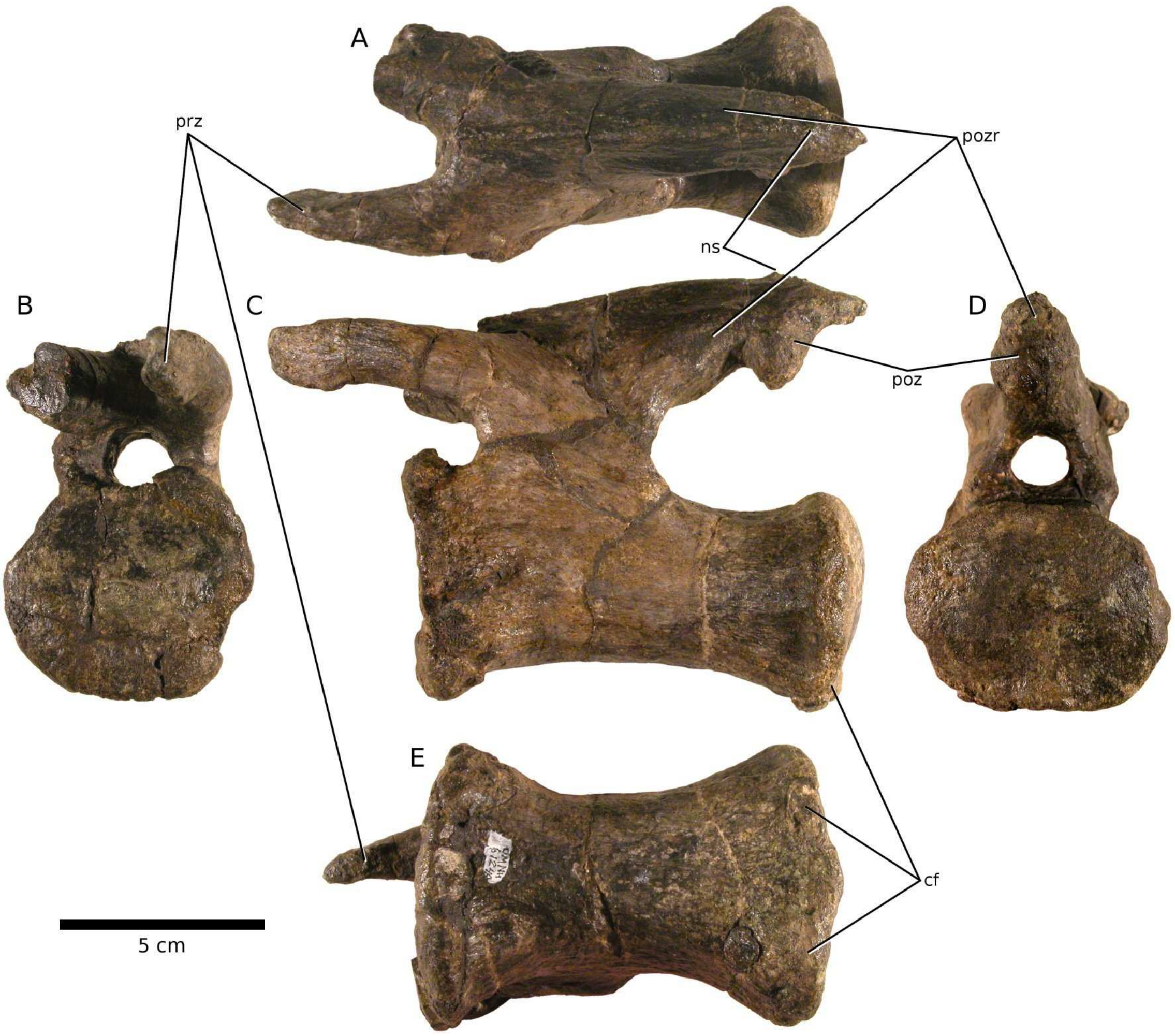

FIGURE 5. Mid-caudal vertebra of the Hotel Mesa sauropod, OMNH 61248. **A**, dorsal view; **B**, anterior; **C**, left lateral; **D**, posterior; **E**, ventral view. cf, chevron facets; ns, neural spine; poz, postzygapophyses; pozr, postzygapophyseal ramus; prz, prezygapophyses. Scale bar equals 5 cm.

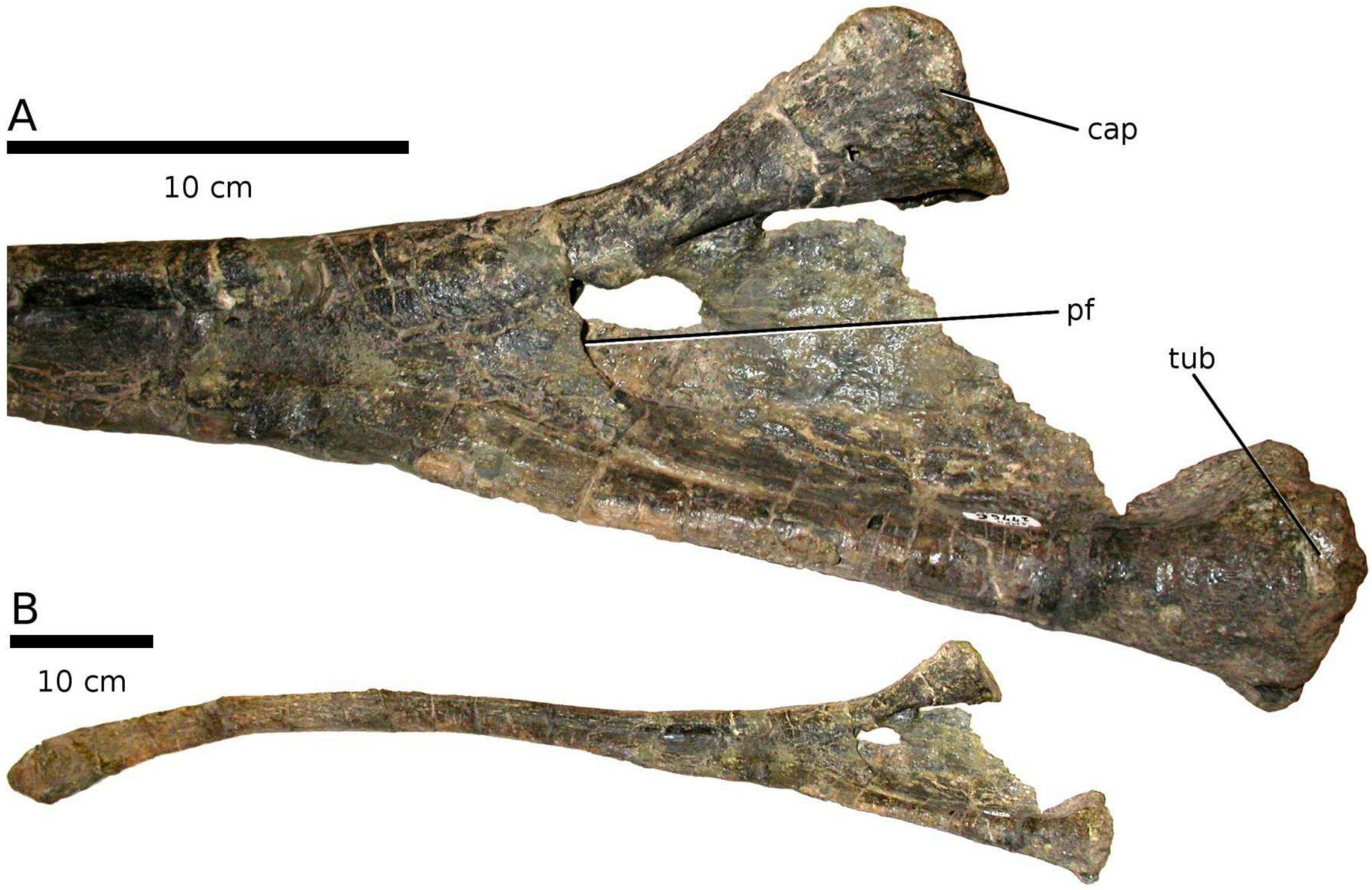

FIGURE 6. First right dorsal rib of the Hotel Mesa sauropod, OMNH 27766 in posterior view. **A**, head of rib, showing pneumatic invasion of shaft; **B**, complete rib, showing laterally directed curvature of shaft. cap, capitulum; pf, pneumatic fossa; tub, tuberculum. Scale bar equals 10 cm.

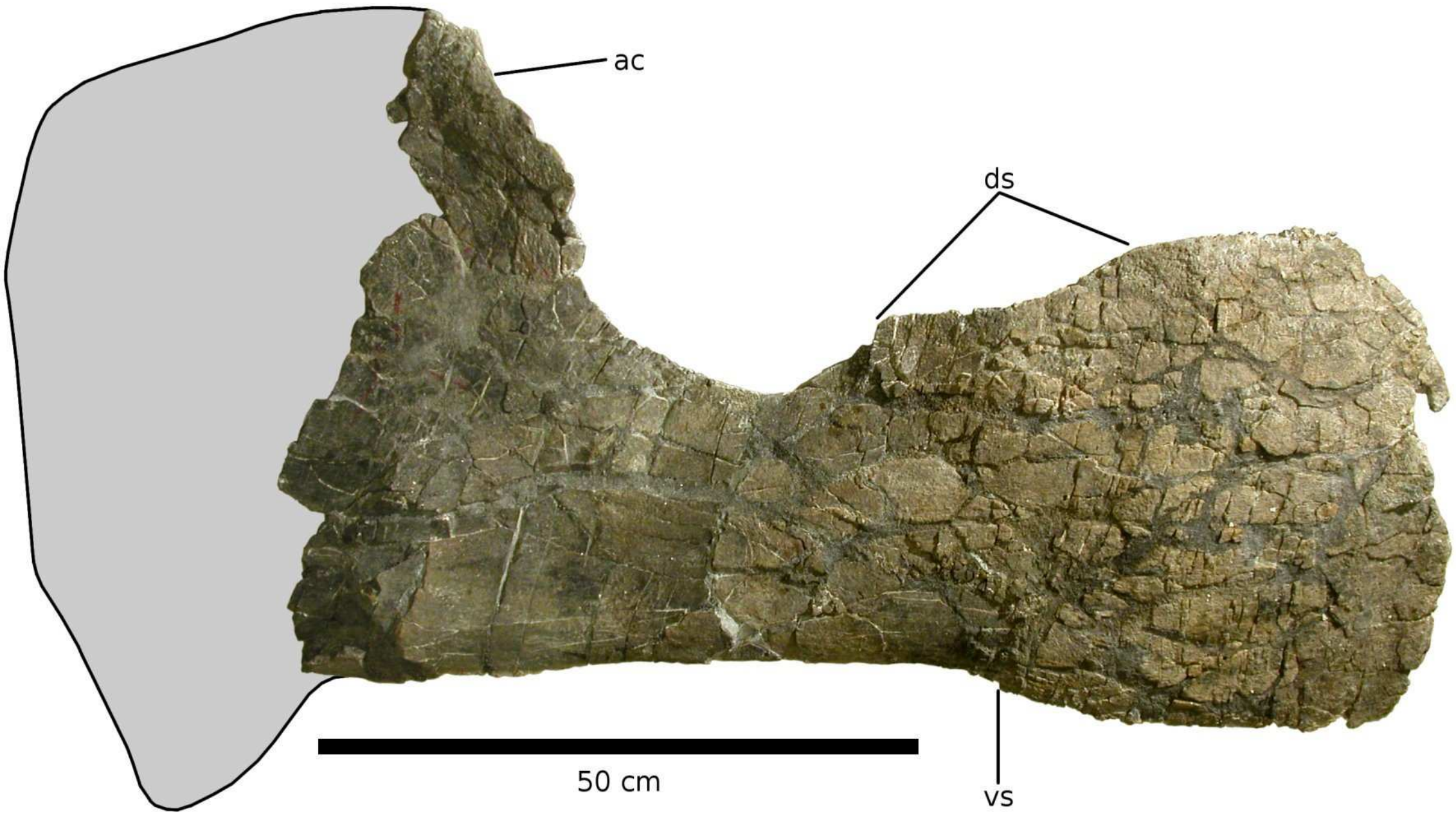

FIGURE 7. Partial left scapula of the Hotel Mesa sauropod, OMNH 27761, in lateral view, reconstructed after *Giraffatitan brancai* HMN Sa 9 (Janensch 1961:pl. 15, fig. 1). ac, acromion expansion; ds, dorsal 'steps'; vs, ventral 'step'. Scale bar equals 50 cm.

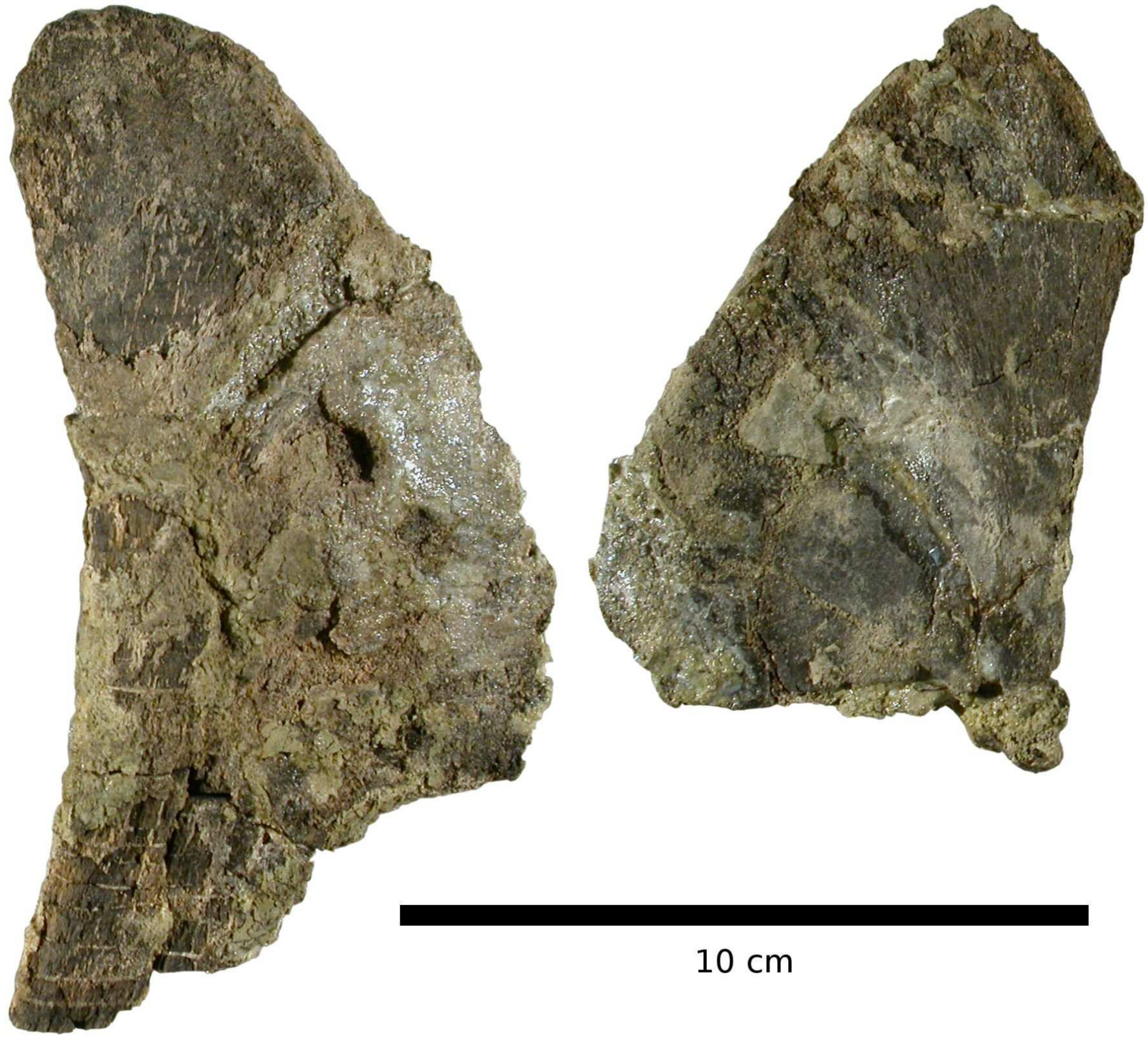

FIGURE 8. Partial paired sternal plates of the Hotel Mesa sauropod, OMNH 66431 and 66432, in ?ventral view. Scale bar equals 10 cm.

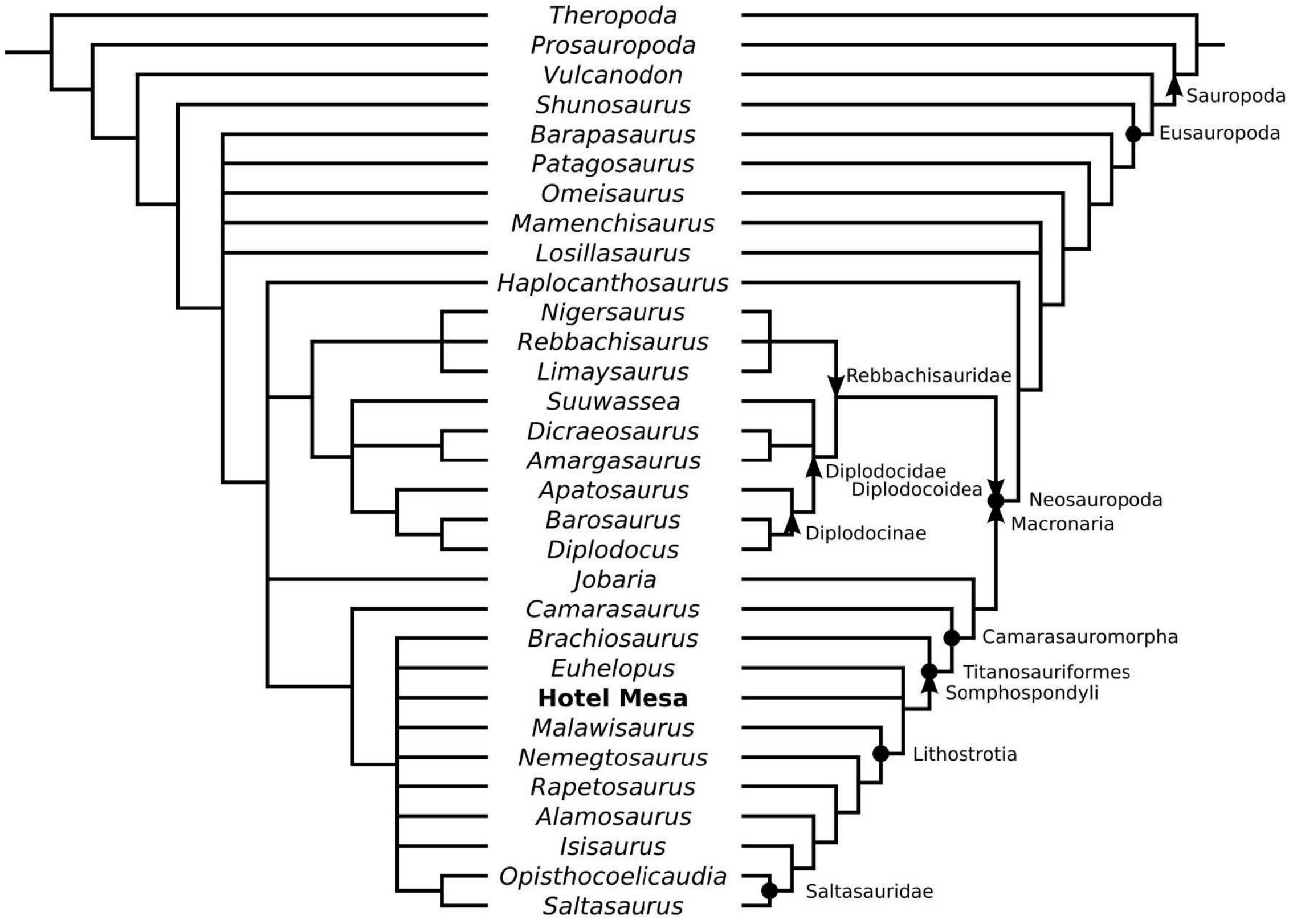

FIGURE 9. Phylogenetic relationships of the Hotel Mesa sauropod produced using PAUP* 4.0b10 on the matrix of Harris (2006) augmented by the Hotel Mesa sauropod, having 31 taxa and 331 characters. Left side, strict consensus of 180 most parsimonious trees (length = 788; CI = 0.5228; RI = 0.6848; RC = 0.3581); Right side, 50 per cent majority rule consensus.

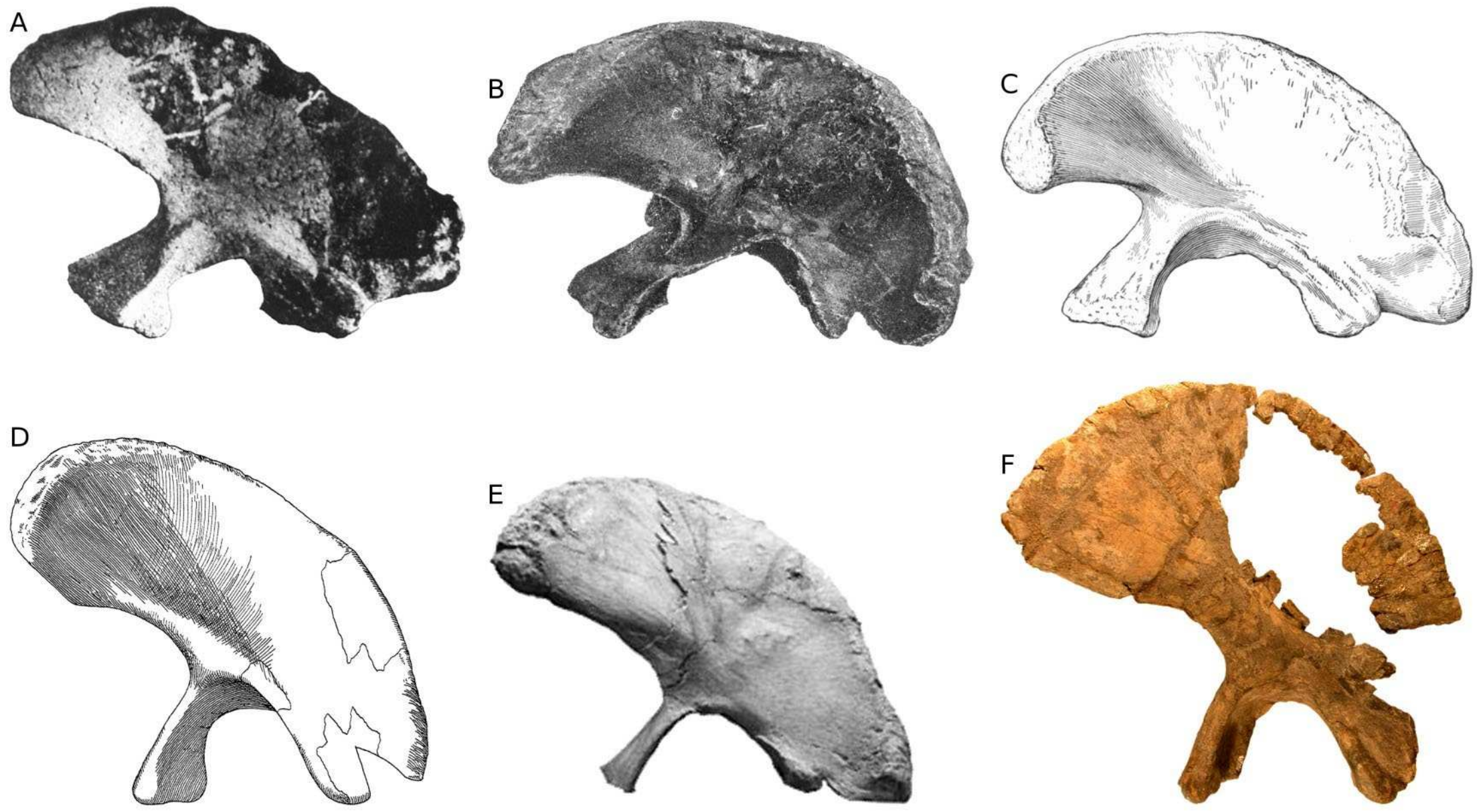

FIGURE 10. Ilia of sauropod dinosaurs, scaled to same total length. **A**, *Mamenchisaurus hochuanensis* holotype CCG V 20401, right ilium reversed, modified from Young and Zhao (1972:pl. 6, fig. 1a); **B**, *Diplodocus carnegii* CM 94, right ilium reversed, modified from Hatcher (1901:pl. X, fig. 1); **C**, *Camarasaurus supremus* AMNH 5761 Il. 1, left ilium, modified from Osborn and Mook (1921:fig. 87); **D**, *Giraffatitan brancai* HMN J1, left ilium, modified from Janensch (1961:pl. E, fig. 2); **E**, *Rapetosaurus krausi* holotype FMNH PR 2209, left ilium, modified from Curry Rogers and Forster (2001:fig. 3h); **F**, The Hotel Mesa sauropod holotype OMNH 66430, left ilium.

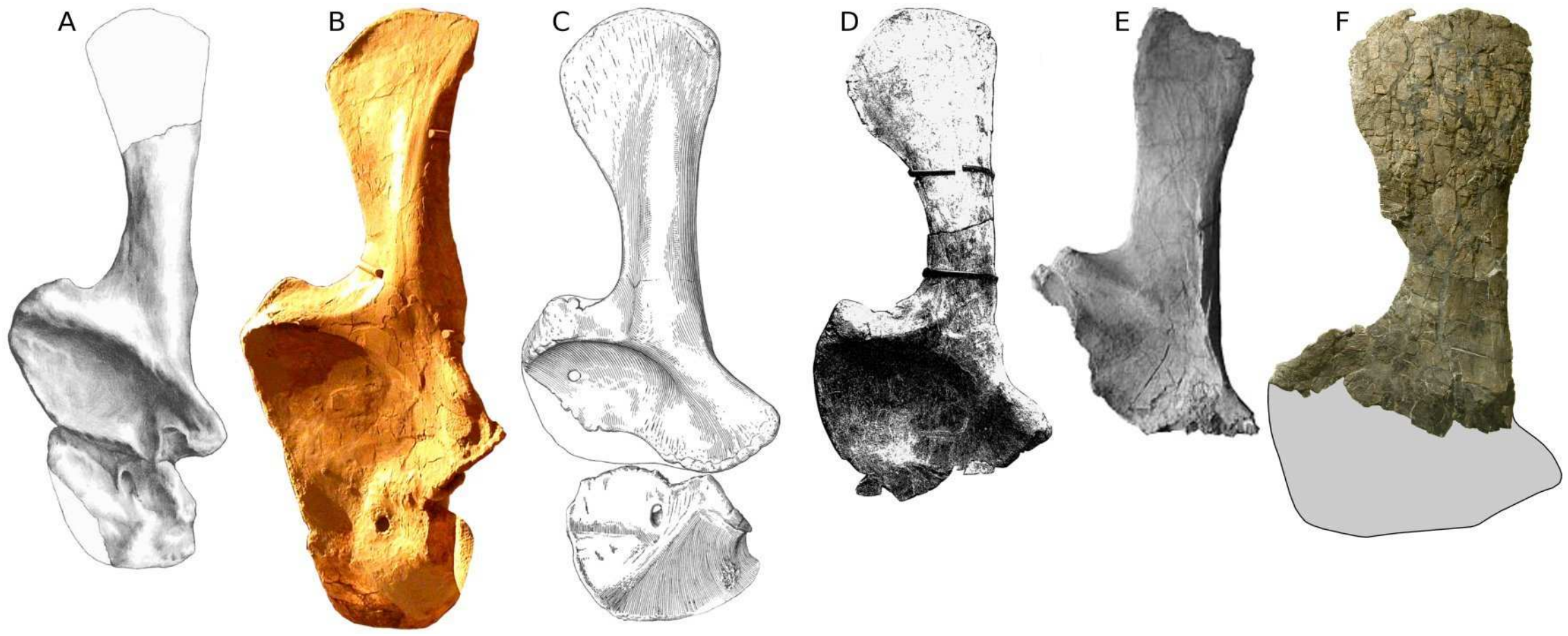

FIGURE 11. Scapulocoracoids and scapulae of sauropod dinosaurs, scaled to same length of scapular blade from posterior point of glenoid to posterior margin of blade. **A**, *Mamenchisaurus youngi* holotype ZDM0083, left scapulocoracoid, modified from Ouyang and Ye (2002:fig. 22); **B**, *Diplodocus longus* USNM 10865, right scapulocoracoid reversed, photograph by MPT; **C**, *Camarasaurus supremus* AMNH 5761 Sc. 1, left scapula, and AMNH 5761 Cor. 1, left coracoid, probably associated, modified from Osborn and Mook (1921:figs. 75, 81a); **D**, *Giraffatitan brancai* HMN Sa 9, left scapula, modified from Janensch (1961:pl. 15, fig. 1); **E**, *Rapetosaurus krausi* holotype FMNH PR 2209, right scapula reversed, modified from Curry Rogers and Forster (2001:fig. 3d); **F**, the Hotel Mesa sauropod OMNH 27761, left scapula, reconstructed after *Giraffatitan brancai*.

Chapter 5 follows. This paper is in review at Paleobiology, published by The Paleontological Society.

# Vertebral morphology and the evolution of long necks in sauropod dinosaurs

Michael P. Taylor and Mathew J. Wedel

RHR: LONG NECKS OF SAUROPOD DINOSAURS

LRH: MICHAEL. P. TAYLOR AND MATHEW. J. WEDEL

*Abstract.* — The necks of sauropod dinosaurs were by far the longest of any terrestrial animals, attaining lengths in excess of 15 m. By contrast, the neck of the giraffe, the longest of any extant animal, does not exceed 2.4 m. The long necks of sauropods were made possible by distinctive anatomical innovations. The presence of pneumaticity, cervical ribs and dorsal tubercles in sauropods shows that their cervical anatomy most resembled that of birds, although the retention of prominent neural spines indicates similarities to crocodilians. Bifid neural spines evolved several times among sauropods and were never secondarily lost. They may have served to shift epaxial muscles laterally. Both pneumatic diverticula and ligament scars are found in the clefts of bifid spines. The elongate ossified cervical ribs of most sauropods allowed hypaxial muscles to be shifted posteriorly and may also have helped to stabilize the neck, preventing inadvertent lateral and dorsal flexion. They could not have functioned as compression members in ventral bracing, for a variety of anatomical and mechanical reasons. Some aspects of cervical anatomy are mechanically puzzling: posterior elongation of the dorsal tubercles, as seen in the caudal vertebrae of some theropods, would have enabled the epaxial muscles to be shifted posteriorly, yet these do not exist in any sauropod. Tall neural spines allow the epaxial tension members to act with a long lever arm, yet the spines of the longest necked sauropods are apomorphically short. The cervical ribs of diplodocoids are apomorphically short; those of *Apatosaurus* are shorter still, absurdly robust, and positioned very low beneath the centrum. Four lineages with very different cervical morphologies evolved ten-meter necks (mamenchisaurids, diplodocids, brachiosaurids, and titanosaurs). Six other groups of terrestrial animals (giraffes, indricotheres, therizinosaurs, ornithomimids, oviraptorosaurs and azhdarchid pterosaurs) all attained necks in the 2–3 m range, but none exceeded this. Three factors contributed to sauropod neck length: sheer size, skeletal pneumaticity, and small heads that merely gathered, rather than processing, food.

*Michael P. Taylor. Palaeobiology Research Group, School of Earth and Environmental Sciences, University of Portsmouth, Portsmouth PO1 3QL, United Kingdom. E-mail: dino@miketaylor.org.uk*

*Mathew J. Wedel. College of Osteopathic Medicine of the Pacific and College of Podiatric Medicine, Western University of Health Sciences, 309 E. Second Street,*

*Pomona, California 91766-1854. E-mail: mathew.wedel@gmail.com*

## Introduction

The necks of sauropod dinosaurs such as *Mamenchisaurus* Young 1954, *Supersaurus* Jensen 1985, *Sauroposeidon* Wedel, Cifelli and Sanders 2000a, and *Puertasaurus* Novas, Salgado, Calvo and Agnolin 2005 were by far the longest of any terrestrial animals, attaining lengths in excess of 15 m (Wedel 2006a). By contrast, the neck of the giraffe, the longest of any extant animal, does not exceed 2.4 m (Toon and Toon 2003: p. 399). The long necks of sauropods were made possible by distinctive anatomical innovations, most notably extensive pneumaticity of the cervical vertebrae which typically consist of about 60% air and 40% bone (Wedel 2005: p. 213) and in some cases attain 75% air or more (Wedel 2005: table 7.2). However, other aspects of the cervical osteology of sauropods appear ill-suited to support long necks. It is difficult to imagine how these apparently maladaptive features can have evolved and survived when selection pressure must have been extreme. We consider the biomechanics of sauropod necks in terms of what is known, what is considered likely and what is possible under the headings Facts, Interpretation and Speculation. Finally, we look at the evolution of long necks in sauropods and other animals, and consider the factors that allowed sauropod necks to grow five times as long as those of any other terrestrial animal.

## Museum Abbreviations

**CM** Carnegie Museum of Natural History, Pittsburgh, Pennsylvania, USA

**HMN** Humboldt Museum für Naturkunde, Berlin, Germany

**IGM** Geological Institute of the Mongolian Academy of Sciences, Ulaan Baatar, Mongolia

**ISI** Geology Museum, Indian Statistical Institute, Calcutta, India

**MCZ** Museum of Comparative Zoology, Harvard University, Cambridge, Massachusetts

**MIWG** Dinosaur Isle, Sandown, Isle of Wight, UK

**OMNH** Oklahoma Museum of Natural History, Norman, Oklahoma, USA

**PMU** Palaeontological Museum, Uppsala, Sweden

**UA** Université d'Antananarivo, Antananarivo, Madagascar

**UJF** University of Jordan Department of Geology Collections, Amman, Jordan

**ZMNH** Zhejiang Museum of Natural History, Hangzhou, China.

## Facts

In extant animals, the mechanically significant soft tissues of the neck (muscles, tendons and ligaments) can be examined and their osteological correlates identified. In extinct animals, except in a very few cases of exceptional preservation, only the fossilized bones are available: using extant animals as guides, osteological features can be interpreted as correlates of the absent soft tissue, so that the ligaments and musculature of the extinct animal can be tentatively reconstructed (Bryant and Russell 1992; Witmer 1995). In order to do this for sauropods, it is necessary first to examine their extant outgroups, the birds and crocodilians.

In all vertebrates, axial musculature is divided both into left and right sides and into epaxial and hypaxial (i.e. dorsal and ventral to the vertebral column) domains, yielding four quadrants. In birds, the largest and mechanically most important epaxial muscles (M. longus colli dorsalis and M. cervicalis ascendens) insert on the dorsal tubercles of the cervical vertebrae, and the large hypaxial muscles (M. flexor colli lateralis, M. flexor colli medialis, and M. longus colli ventralis) insert on the cervical ribs (Baumel et al. 1993; Tsuihiji 2004). The osteology of the cervical vertebrae makes mechanical sense; the major muscle insertions are prominent osteological features located at the four "corners" of the vertebrae (Fig. 1A). Non-avian theropods resembled birds in this respect, having prominent dorsal tubercles and sizable cervical ribs, which point in the four expected directions (Fig. 1B).

The cervical architecture is rather different in crocodilians, and in non-archosaurian diapsids such as lizards, snakes, ichthyosaurs and plesiosaurs: there are no dorsal tubercles, and the long neck muscles attach to the neural spines rather than the dorsal tubercles (Fig. 1C). In most sauropods, the cervical vertebrae do have dorsal tubercles, but the neural spines are as prominent or more so (Fig. 1D). In this respect, sauropod osteology is intermediate between the conditions of crocodilians and birds, perhaps more closely resembling that of crocodilians and other non-theropod diapsids than that of birds – so the widely recognized similarity of sauropod cervicals to those of birds (e.g. Wedel and Sanders 2002; Tsuihiji 2004), while significant, should not be accepted unreservedly. Since the prominent neural spine serves as the primary attachment site for epaxial muscles in most theropod outgroups, the condition in birds and other theropods is derived; that of sauropods retains aspects of the primitive condition.

Although sauropods shared a common bauplan, their morphological disparity was much greater than has usually been assumed (Taylor and Naish 2007: pp. 1560–1561). This disparity is particularly evident in the cervical vertebrae (Fig. 2). Those of *Apatosaurus* Marsh 1877, for example, are anteroposteriorly short and dorsoventrally tall, and have short, robust cervical ribs mounted far ventral to the centra; the cervical centra of *Isisaurus colberti* Jain and Bandyopadhyay 1997 are even shorter anteroposteriorly, but have more dorsally located cervical ribs; by contrast, the cervical vertebrae of *Erketu ellisoni* Ksepka and Norell 2006 are relatively much longer and lower, and have long, thin cervical ribs mounted somewhat ventral to the centra. Towards the middle ground of these extremes fall the cervical vertebrae of *Brachiosaurus brancai* Janensch 1914, which are anteroposteriorly longer and dorsoventrally shorter than those of *Apatosaurus,* but not as anteroposteriorly long or as dorsoventrally short as those of *Erketu* Ksepka and Norell 2006. In light of the demanding mechanical constraints that were imposed on sauropods, it is surprising that their necks vary so much morphologically, with different lineages having evolved dramatically different solutions to the problem of neck elongation and elevation.

Because sauropods were so much bigger than their relatives, and their necks so much longer, mechanical considerations in the construction of their necks were significantly more important than in their outgroups. Furthermore, the great size and shape disparity between sauropods and their outgroups means that interpretations of cervical soft-tissue anatomy in sauropods cannot be based purely on the extant phylogenetic bracket method: this alone would be no more informative than trying to determine the anatomy of elephants from that of manatees and hyraxes. For example, in most extant vertebrates including birds and crocodilians, the diameter of the neck is three or four times that of the cervical vertebrae that form its core (e.g. Wedel 2003: Fig. 2). This cannot have been the case in sauropods, as such over-muscled necks would have been too heavy to lift; and the various published reconstructions of sauropod neck cross sections (e.g. Paul 1997: Fig. 4; Schwarz et al. 2007: Fig. 7, pp. 8A, 9E) all agree in making the total diameter including soft-tissue only 105–125% that of the vertebrae alone.

## Interpretation

Interpretation of sauropods as living animals is made especially difficult by the lack of good extant analogues. Among animals with long necks, giraffes, camels, and other artiodactyls have very different cervical osteology and (we assume) myology, and even the longest of their necks, at about 2.4 m, is less than one sixth the length attained by some sauropods. Birds are phylogenetically closer to sauropods, and some birds (e.g. swans and ostriches) have proportionally very long necks. Furthermore, the presence in most sauropods of dorsal tubercles similar to those of birds suggests that sauropods were myologically similar to birds. However, the small absolute size of birds means that the sets of forces acting on their necks are so different that we can't assume that sauropod necks functioned in the same ways – just as the problems involved in flight through air for high-Reynolds number fliers such as birds are very different than than they are for low-Reynolds number fliers such as fruit-flies, whose aerodynamics are dominated by friction drag rather than form drag.

With all these caveats in mind, the best extant analogues for sauropod necks nevertheless remain those of birds: they are the only extant animals that share with sauropods dorsal tubercles above their postzygapophyses, pronounced cervical ribs, and pneumatic foramina (Fig. 1A, D). The first two of these features were inherited from a common saurischian ancestor. The foramina seem to have been independently derived in birds, but this was possible because air sacs and soft-tissue pneumatic diverticula were likely present in the common saurischian ancestor (Wedel 2006b, 2007). These observations enable us to draw conclusions about sauropod neck soft tissue beyond what the extant phylogenetic bracket would allow. Specifically, the dorsal tubercles are osteological correlates of the M. longus colli dorsalis and M. cervicalis ascendens epaxial muscles, which must therefore have been present in sauropods, although we can not conclude from this that they were necessarily the dominant epaxial muscles as they are in birds.

Neural Spines

The neural spines and dorsal tubercles of sauropods both anchored epaxial muscles, but as they were differently developed in different taxa, they were probably of varying mechanical importance in different taxa. For example, based on their relative heights, dorsal tubercles may have dominated neural spines in *Apatosaurus* (Fig. 1A) but neural spines may have dominated in *Isisaurus* Wilson and Upchurch 2003 and *Brachiosaurus brancai* Riggs 1903a (Fig. 1C, D). In some sauropods, including *Erketu* and *Mamenchisaurus*, which were long-necked even by sauropod standards, the neural spines are strikingly low, and the dorsal tubercles no higher – a surprising arrangement, as low spines would have reduced the lever arm with which the epaxial tension members worked. Among these sauropods with low neural spines, some have rugose neurapophyses with spurs directed anteriorly and posteriorly from the tip of the spine (Fig. 3). These appear either to have anchored discontinuous interspinous ligaments, as found in all birds (see Wedel et al. 2000b: Fig. 20), or to have been embedded in a continuous supraspinous ligament, as found in the ostrich (Dzemski and Christian 2007: pp. 701–702).

In some sauropods, the cervical neural spines are bifid (i.e. having separate left and right metapophyses and a trough between them). This morphology appears to have evolved at least five times (in *Mamenchisaurus*, flagellicaudatans, *Camarasaurus* Cope 1877, *Erketu* and *Opisthocoelicaudia* Borsuk-Bialynicka 1977) with no apparent reversals. This morphology, then, seems to have been easy for sauropods to gain, but difficult or perhaps impossible to lose. Bifid cervical vertebrae are extremely uncommon in other taxa, and among extant animals they are found only in ratite birds, e.g. *Rhea americana* Linnaeus 1758 (Tsuihiji 2004: Fig. 2B), *Casuarius casuarius* Brisson 1760 (Schwarz et al. 2007: Fig. 5B) and *Dromaius novaehollandiae* Latham 1790 (Osborn 1898: Fig. 1). It has often been assumed that in sauropods with bifid cervical spines, the intermetapophyseal trough housed a large ligament analogous to the nuchal ligament of artiodactyl mammals (e.g. Janensch 1929: Plate 4; Alexander 1985: pp. 13–14; Wilson and Sereno 1998: p. 60). Such an arrangement seems unlikely, as lowering the ligament into the trough would reduce its mechanical advantage; however, this is similar to the arrangement seen in *Rhea americana*, in which branches of the "nuchal ligament" attach to the base of the trough (Tsuihiji 2004: Fig. 3). More direct evidence is found in ligament scars in the troughs of some diplodocids: these can be prominent, as in the doorknob-sized attachment site in the *Apatosaurus* sp. cervical OMNH 01341 (Fig. 4A).

Ligament cannot have filled the trough, as envisaged by Alexander (1985: Fig. 4C), however, because pneumatic foramina are often found in the base of the troughs of presacral vertebrae, for example in the dorsal vertebrae of *Camarasaurus* sp. CM 584 (Fig. 4B). In some specimens, a ligament scar and pneumatic foramen occur together in the intermetapophyseal trough (Fig. 4A, Schwarz et al. 2007: Fig. 6E). Pneumatic diverticula are sometimes found between the postzygapophyses, invading the medial aspect of the centropostzygapophyseal laminae, even in sauropods with non-bifid spines, as shown by the isolated brachiosaurid cervical MIWG 7306 from the Isle of Wight (D. Naish personal communication 2008), so the presence of soft-tissue diverticula in this location is probably primitive for Neosauropoda at least.

One possible advantage of bifid spines would be to increase the lateral leverage of the ligaments and muscles that attach to the metapophyses, enabling them to contribute to lateral motion as well as vertical. A cantilevered beam, such as a sauropod neck, requires only a single dorsal tension member to stabilize it vertically, but two (one on each side) to stabilize it horizontally. A sauropod neck that was supported from above only by a single midline tension member would need additional horizontal stabilization from muscles and ligaments not directly involved in support.

Whatever the advantages of bifid spines, they were clearly not indispensable, as some sauropod lineages evolved very long necks with unsplit spines (e.g. brachiosaurids, culminating in *Sauroposeidon*, and most titanosaurs, including the very long-necked *Puertasaurus*). Even in taxa that do have bifid spines, they are never split through the whole series: for example, the first eight cervicals of *Barosaurus* Marsh 1890 do not have bifid spines (McIntosh 2005; MJW, pers. obs). If bifid spines conferred a great advantage, they would presumably be found throughout the neck – although the importance of stability, and the difficulty of attaining it, is greater in the posterior part of the neck, which bears greater forces than the anterior part. Since bifid spines always occur together with unsplit spines, it seems likely that however they were used mechanically, it was probably not radically different from neural spine function in vertebrae with unsplit spines.

Dorsal Tubercles

As noted above, the dorsal tubercles are the insertion points of the largest and longest epaxial muscles in birds, whereas in crocodilians the dorsal tubercles are non-existent, and no major muscles insert above the postzygapophyses (Tsuihiji 2004). Dorsal tubercles are found in most, though not all, sauropods and theropods. For example, they seem to be absent in the titanosaurs *Malawisaurus* Jacobs, Winkler, Downs and Gomani 1993 (Gomani 2005: Fig. 8) and *Isisaurus*, (Fig. 2C); but their presence in other titanosaurs such as *Rapetosaurus* Curry Rogers and Forster 2001 (Curry Rogers and Forster 2001: Fig. 3A) and *Saltasaurus* Bonaparte and Powell 1980 (Powell 1992: Fig. 5) and in outgroups such as *Brachiosaurus* (Fig 2D) and *Camarasaurus* (Osborn and Mook 1921: Plate LXVII, Fig. 9; McIntosh et al. 1996: Fig. 29) indicates that their absences in *Malawisaurus* and *Isisaurus* represent secondary losses.

The existence of dorsal tubercles on the cervical vertebrae of most sauropods, together with those in theropods and birds, suggests that epaxial muscles were inserting above the postzygapophyses at least by the origin of Saurischia. Dorsal tubercles are also known in basal ornithischians, e.g. *Lesothosaurus* Galton 1978 (Sereno 1991: Fig. 8A) and *Heterodontosaurus* Crompton and Charig 1962 (Santa Luca 1980: Fig. 5A), and also in the basal pterosaur *Rhamphorhynchus* Meyer 1846 (Bonde and Christiansen 2003: Fig. 6–9), suggesting that these insertion points were in use at the base of Ornithodira.

In sauropods, the size and location of the dorsal tubercles is variable: in C8 of *Brachiosaurus brancai*, the dorsal tubercles are approximately half as high above the centrum as the neurapophysis (Fig. 2D); in anterior cervicals of *Erketu*, the dorsal tubercles are equally as high as the tips of the neural spines (Fig. 2E), although the spines are higher in posterior cervicals. It is possible that in the posterior cervicals of some *Apatosaurus ajax* Marsh 1877 specimens, the dorsal tubercles are higher than the metapophyses (Fig. 2A), but it is difficult to be sure as the vertebrae that seem to most closely approach this condition are at least partly reconstructed in plaster (Barbour 1890: Fig. 1). In any event, it is clear from preserved sequences of *Apatosaurus* cervicals (Gilmore 1936: Plate XXIV; Upchurch et al. 2004: Plate 1) that in this genus the neural spines are proportionally higher in the anterior cervicals than in the posterior. The trend is opposite in *Erketu*, in which the dorsal tubercles increasingly dominate neural spines anteriorly. This further demonstrates the variety of different mechanical strategies used by different sauropods to support their long necks. In those sauropods without ostensible dorsal tubercles, it can not necessarily be concluded that muscles did not insert above the postzygapophyses: phylogenetic bracketing suggests that they did, but the insertions are not marked by obvious scars or processes and these muscles were probably less important than those attached to the spine.

Cervical Ribs

In extant birds, cervical ribs are the insertion points for the M longus colli ventralis hypaxial muscles. No bird has cervical ribs long enough to overlap, but the tendons that insert on the cervical ribs do overlap and are free to slide past each other longitudinally. In less derived saurischians, including sauropods, the long ventral tendons are ossified into long, overlapping cervical ribs which are secondarily shortened in Diplodocoidea and in Maniraptoriformes, including birds. The null hypothesis is that the long cervical ribs of theropods and sauropods functioned similarly to the short cervical ribs and long tendons of birds, as the insertions of long hypaxial muscles.

Ossification of the hypaxial tendons into long cervical ribs may have provided several benefits for sauropods:

- Long tendons move the bulk of the hypaxial neck muscles closer to the base of the neck, which reduces the lever arm of the neck mass. However, tendon has a much lower Young's modulus than bone, so that contraction of the hypaxial muscles in sauropods would waste energy in stretching the tendon rather than shifting the vertebra to which is is attached. Ossification of this tendon would have resulted in a stiffer material and so more efficient deployment of hypaxial musculature.
- It has been suggested (Wedel et al. 2000b: p. 380) that elongate cervical ribs may have played a role in ventrally stabilizing the neck, i.e. preventing involuntary dorsal extension.

- Stiff cervical ribs would have helped provide lateral stabilization for the neck, which would have been especially important in taxa with epaxial tension members concentrated on the midline as discussed above.

If either of the first two hypotheses is accurate, it is difficult to understand why diplodocids evolved apomorphically short cervical ribs, especially long-necked forms such as *Barosaurus* and *Supersaurus.* If the primary role of long cervical ribs was in providing lateral stabilization for taxa with midline epaxial tension members, then the need for this stabilization would be reduced in forms with bifid spines, such as diplodocids, which shifted their epaxial tension members laterally as they were attached to the metapophyses. This, however, would raise the question of why other taxa with bifid spines (e.g. *Camarasaurus*) also retained elongate cervical ribs, and in the case of *Mamenchisaurus* apparently evolved apomorphically long cervical ribs (Russell and Zheng 1993: pp. 2089–2090). It may be that these taxa retained their epaxial tension members primarily on the midline, in the intermetapophyseal trough, while diplodocids shifted theirs laterally; but we know from osteological evidence (see above) that at least some diplodocids did have ligaments or muscles anchored within the trough.

Ventral compression bracing

Another function that has been suggested for the cervical ribs of sauropods is in ventral bracing (Frey and Martin 1997; Martin et al. 1998). In order to maintain the neck as a cantilevered beam anchored at only one end, either dorsally located tension members or ventrally located compression members are necessary. In the case of a horizontal sauropod neck, it has been generally assumed that the incompressible bony centra of the vertebrae acted in compression, and epaxial ligaments and muscles acted as tension members (e.g. Alexander 1985). However, Martin et al. (1998) proposed an alternative mechanism whereby, instead, ventral compression bracing was provided by incompressible members, namely the cervical ribs. This was proposed as the primary or only neck bracing in *Mamenchisaurus*, *Euhelopus* Romer 1956, *Omeisaurus* Young 1939, *Brachiosaurus* and *Camarasaurus*; and as a significant factor in *Diplodocus* Marsh 1878, *Cetiosaurus* Owen 1841 and *Haplocanthosaurus* Hatcher 1903.

In many sauropods, including diplodocids, the cervical ribs were simply too short to span their centra, and so could not form a continuous incompressible brace, but even in other sauropods, there are several problems with this proposal:

1. In those sauropods whose cervical ribs are long enough to overlap multiple centra, the overlapping ribs appear much too slender to support the mass of the neck. Consider for example the neck of the *Brachiosaurus brancai* lectotype specimen HMN SII. Using graphic double integration (Jerison 1973; Hurlburt 1999; Murray and Vickers-Rich 2004), one of us estimated the volume of the head as 0.14 $m^3$, and that of the anterior part of the neck (as far as the posterior end of C8) as 1.38 $m^3$ (Taylor, in press). These are conservative figures, based on a reconstruction in which the entire neck volume is 4.12 $m^3$, compared with 11.2 $m^3$ in Gunga et al. (1995) and 7.3 $m^3$ in Gunga et al. (2008). Using a density of 0.6 $kg/dm^3$ (Paul 1988: p. 10), this yields a total mass of 912 kg for the head and anterior part of the neck, equivalent to a weight of 8943 newtons. The head-and-anterior-neck segment is about 5.5 m long. Assuming that the center of mass of this segment is about one third of the way out from the C8–C9 joint, the weight acts 1.83 m out, yielding a moment of 16366 Nm. As the cervical ribs of C8 are 30 cm below the centroid of its cotyle, the compression force acting through the ribs at this point would be 54553 N. Careful measurement and cross-scaling of the cervical-rib cross-sections in Janensch (1950: Fig. 85) indicates that the total cross-section of the cervical ribs at the posterior end of C8, including tapering ends of the ribs of C6 and C7 would have been about 5 $cm^2$ on each side, for a total of 10 $cm^2$ or 0.001 $m^2$. The 54553 N force corresponding to the weight of the head and anterior part of the neck would have exerted a stress of 54 MPa on the cervical ribs.

This is well below the ultimate failure stress of compact bone, which is variously given as 193 MPa for longitudinal compression in human bone (Reilly and Burstein 1975: p. 404), 195–217 MPa for the radius of a horse (McGowan 1999: p. 88), 270 MPa for unspecified mammal bone (Alexander 1989: p. 46), and 130–205 MPa for unspecified bone loaded parallel to the grain (Hildebrand 1988: p. 423). However, continuous compression loading of 54 MPa is unrealistic for five reasons. First, this figure is based on static forces only (i.e. a stationary standing animal), whereas locomotory stress is typically about twice that of standing (Jayes and Alexander 1978); second, bending stresses would inevitably also have arisen, and are typically much greater than the accompanying compression stresses (Alexander 1989: p. 52); third, compression loading of long thin elements such as cervical ribs would likely result in catastrophic buckling; fourth, in live animals the persistent application of stresses well below ultimate failure stress causes eventual failure through bone fatigue, in which microdamage accumulates more quickly than it can be repaired by bone remodelling – for example, Taylor (1998) calculated that human long bones will fail within $10^5$ cycles of loading to 23–30 MPa; fifth, live animals typically operate within a safety factor of 2–4 (Biewener 1990), that is, under stresses much lower than would cause failure, and we can assume sauropods also would have done so. In conclusion, the cervical ribs of *Brachiosaurus brancai* would not have been strong enough to ventrally brace its neck in life.

2. In extant birds, the M longus colli ventralis tendons that are homologous to the cervical ribs of sauropods are free to slide past each other. In order to form an incompressible mass, Martin et al. (1998: Fig. 3) were forced to postulate ligaments binding consecutive cervical ribs together. Such ligaments do not exist in birds, and there is no osteological evidence for their existence in sauropods. It is true that ligaments connect the first three cervical ribs in crocodiles; however, these ribs are dorsoventrally tall and blade-like, and overlap over broad contacts, unlike the condition in the thin bony rods that are sauropod cervical ribs. Ligaments binding cervical ribs together would represent an evolutionary novelty in sauropods, a hypothesis not supported by available evidence.

3. If, despite this, the cervical ribs were able somehow to act as an incompressible bundle, then their inability to slide past each other would make it impossible for them to fulfill the function they have in birds, as insertion points for the ventral muscles.

4. Even if the anterior cervicals were ventrally braced against the posterior cervicals, there would be no mechanism by which the compression could be transferred to the torso, as in all sauropods the cervical ribs in the most posterior cervicals are very short and do not extend as far back as the head of the next rib. In some sauropods, including *Mamenchisaurus*, *Omeisaurus*, and *Euhelopus*, even the most posterior cervical vertebrae have low neural spines, as well as short cervical ribs, yet since ventral compression bracing is impossible with short cervical ribs, the dorsal tension members must have been sufficient to suspend the neck even with these low spines; so the same mechanisms could surely also have worked in the more anterior cervicals.

5. If the cervical ribs were the primary compression members in the necks of sauropods, then the construction of the cervical centra is puzzling, as they are much more able to sustain compressive stress than the ribs, and feature an I-beam-like cross section that ideally suits them to act as beams. Similarly, in the scheme of Martin et al. (1998), it is not clear what would provide the tension dorsal to the cervical ribs if not the epaxial muscles and ligaments conventionally envisaged. The centra themselves would make poor tension members, as this would put undue stress on the intervertebral joints.

6. Compression members can only be passive, not active, since muscles can only pull and not push. If ventral bracing alone were used, then, the neck would sag to the lowest possible posture and not be able to rise. Therefore, even if ventral compression were significant in supporting the necks of some sauropods, dorsal tension support would also be necessary in order to raise the neck.

In conclusion, the ventral bracing hypotheses requires the cervical ribs to tolerate unrealistic stresses, depends on the existence of novel ligaments for which there is no evidence, makes the ribs useless for the purpose they serve in birds, is powerless to explain how the most proximal part of the neck was supported, renders the construction of the centra mystifying, and results in a neck that can not be raised above its lowest pose. We therefore reject the ventral bracing hypothesis.

As noted by Schwarz et al. (2007: p. 184), even if ventral bracing were used in some sauropods, it would not have worked for diplodocids or dicraeosaurids, which had short cervical ribs. This suggests that ventral bracing cannot have been indispensable, as the longest sauropod neck known from from actual bones is that of the diplodocid *Supersaurus*, whose cervical ribs are sub-equal in length to their centra (Lovelace et al. 2008: p. 530).

*Apatosaurus* presents a final riddle regarding cervical ribs. Even among diplodocids, it had extraordinary cervical ribs: very short, very robust, and positioned very low, far below the centra on extremely long parapophyses (Fig. 2A, 2B), so that the neck of *Apatosaurus* must have been triangular in cross-section. What function can the ribs have evolved to perform? They were much too short to have functioned in horizontal or vertical stabilization or ventral bracing, and in any case seem over-engineered for most of these functions. It is tempting to infer that the autapomorphies of the neck in *Apatosaurus* are adaptations for some unique aspect of its lifestyle, perhaps violent intraspecific combat similar to the "necking" of giraffes. Even if this were so, however, it is difficult to see the benefit in *Apatosaurus excelsus* (Marsh 1879a) of cervical ribs held so far below the centrum – an arrangement that seems to make little sense from any mechanical perspective, and may have to be written off as an inexplicable consequence of sexual selection or species recognition.

## Speculation

Several important questions about the construction of sauropod necks remain difficult to answer with the evidence currently available:

1. The central paradox of sauropod cervical morphology is that the vertebrae appear better suited for anchoring hypaxial than epaxial musculature, even though holding the neck up was important and, due to gravity, much more difficult than drawing it down. First, the cervical ribs present a greater area for muscle attachment than the dorsal tubercles do; and second, the much greater length of the cervical ribs in most sauropods enabled the hypaxial musculature to be shifted backwards much further than the epaxial musculature, as the dorsal tubercles are not elongate in any known sauropod cervical. We know that posterior elongation of the dorsal tubercles is developmentally possible in saurischians, because those in the tail of *Deinonychus* Ostrom 1969a are extended to the length of a centrum (Ostrom 1969b: Fig. 37). Fig. 5 shows the cervical skeleton of *Euhelopus* as it actually is, and reconstructed with speculative muscle attachments that would have been more mechanically efficient: why did sauropod necks not evolve this way?

2. Whatever the purpose of elongate ossified cervical ribs in sauropods and other saurischians, the question arises of why they are apomorphically short in diplodocids. And why, within Diplodocidae, did *Apatosaurus* shorten the ribs yet further, greatly increase their robustness, and displace them so far ventrally? Reduction in cervical rib length may have increased neck flexibility, but the DinoMorph project of Stevens and Parrish (1999) seems to show that the necks of diplodocids were nevertheless relatively inflexible, due the limits on motion imposed by the need to avoid disarticulating the relatively small zygapophyseal facets.

3. The sauropods with the proportionally longest necks tend to be those whose necks make the least apparent mechanical sense. It is particularly notable that mamenchisaurids (*Mamenchisaurus* and *Omeisaurus*) have very low neural spines, as does *Erketu* in the preserved, anterior, cervicals. These low spines would have reduced the lever arm with which epaxial tension members acted. A speculative explanation, at least, can be offered: although tension members had to exert greater force to allow for the shorter lever arm, they would have needed to contract a shorter distance in order to raise the neck. Might the neural spines, then, have been connected by strongly pennate muscles, able to contract very powerfully but only over a short distance? If so, this would suggest that the necks of mamenchisaurids were rather inflexible in life, not because of osteological limits to movement as shown by Stevens and Parrish (1999) for diplodocids, but because the muscles would simply not be mechanically capable of contracting enough to lift the neck very far above its lowest position.

## Evolution of Long Necks

Table 1 lists a selection of long-necked sauropods, mostly known from complete or nearly complete necks, showing how they vary in length, cervical count, centrum length, cervical rib length, and elongation index (sensu Wedel et al. [2000b: p. 346], the anteroposterior length of the centrum divided by the midline height of the cotyle). At least four different sauropod lineages (Mamenchisauridae, Diplodocidae, Brachiosauridae and Titanosauria) evolved necks 10 meters long or longer, four times as long as those of the next longest-necked terrestrial animals. We now consider the longest necks that evolved in vertebrates other than sauropods.

### Extant Animals

Among extant animals, adult bull giraffes can attain 2.4m (Toon and Toon 2003: p. 399), and no other extant terrestrial animal exceeds 1 m.

### Extinct Mammals

The largest terrestrial mammal of all time was the long-necked rhinoceratoid *Paraceratherium* Forster-Cooper 1911 (= *Baluchitherium* Forster-Cooper 1913, *Indricotherium* Borissiak 1915). Measurements of this animal are hard to come by, and the definitive monograph remains to be written, but the length of its neck can be estimated from the skeletal reconstructions of Paul (1997: p. 151). According to the scale-bar, the neck of the fairly complete specimen AMNH 26387 is 1.5 m long. The larger but less complete specimen AMNH 26168/75 is estimated to have weighed 2.1 times as much, which is consistent with isometric scaling by a factor of 1.28, so the larger specimen may have had a neck about 2.0 m long. This is in accord with a measurement of 2.04 m taken from the composite reconstruction of Osborn (1923: Fig. 9), and with the measurements of cervicals 1, 3 and 6 (Osborn 1923: p. 7).

### Theropods

At least three lineages of theropod dinosaurs evolved long necks.

*Therizinosaurus cheloniformis* Maleev 1954 is a bizarre, long-necked giant theropod,

known from incomplete remains. Measuring from Barsbold (1976: Fig. 1), its humerus was about 75 cm long. In a skeletal reconstruction of the therizinosauroid *Nanshiungosaurus* Dong 1979 by Paul (1997: p. 145), the neck is 2.9 times the length of the humerus. If *Therizinosaurus* were similarly proportioned, its neck would have been about 2.2 m long.

Another giant theropod, *Deinocheirus mirificus* Osmólska and Roniewicz 1969, is known only from a pair of forelimbs, of which the left humerus is 938 mm long (Osmólska and Roniewicz 1969: p. 9). *Deinocheirus* probably belongs to the long-necked ornithomimid group of theropods (Kobayashi and Barsbold 2006) and thus may have had roughly the same proportions as *Struthiomimus* Osborn 1916. Osborn (1916: pp. 474–475) gives a humerus length of 310 mm for *Struthiomimus*, and a total neck length 2.5 times as long, at 770 mm. If it was similarly proportioned, *Deinocheirus* would have had a neck about 2.35 m long.

A third giant theropod, *Gigantoraptor erlianensis* Xu, Tan, Wang, Zhao and Tan 2007 belongs to another long-necked group, Oviraptorosauria. Measured from the skeletal reconstruction of Xu et al. (2007: Fig. 1A), it appears to have had a neck 2.15 m in length – although this is conjectural as almost no cervical material is known.

## Pterosaurs

Although it is often noted in general terms that azhdarchid pterosaurs had long necks (e.g. Howse 1986; Witton and Naish 2008), there are no published numeric estimates of neck length in this group. *Zhejiangopterus linhaiensis* Cai and Wei 1994 is the only azhdarchid for which a substantially complete neck has been described, so we will base our estimates on this species. Cai and Wei (1994: table 7) give the lengths of cervicals 3–7 for three specimens, ZMNH M1323, M1324 and M1328. In all three, C5 is the longest cervical, as is generally true of pterodacyloid pterosaurs such as azhdarchids (Howse 1986: p. 323). Cai and Wei (1994) do not give lengths for C1 and C2, stating only that “the atlas-axis is completely fused and extremely short but morphological details are indistinct due to being obscured by the cranium” (p. 183, translation by Will Downs). Their Figure 6, a reconstruction of *Zhejiangopterus linhaiensis*, bears this out, showing the atlas-axis as about one quarter the length of C3. Using this ratio to estimate the C1–2 lengths for each specimen, we find by adding the lengths of the individual cervicals that the three specimens had necks measuring approximately 511, 339 and 398 mm. These lengths are 3.60, 4.04 and 4.06 times the lengths of their respective C5s. On average, then, total neck length in *Zhejiangopterus* was about 3.85 times that of C5.

The azhdarchid *Arambourgiana philadelphiae* (Arambourg 1959) is the largest pterosaur for which cervical material is known. Its type specimen, UJF VF1, a single near-complete cervical vertebra, has been damaged and is missing its central portion, but plaster replicas made before the damage indicate the extent of the missing portion. The preserved part of the vertebra was 620 mm long before the damage, and when complete it would have been about 780 mm long (Martill et al. 1998: p. 72). Assuming that the preserved element is C5, as considered likely by Howse (1986: p. 318) and Frey and Martill (1996: p. 240), the length of the whole neck can be estimated as 3.85 times that length, which is 3.0 m.

Another azhdarchid, *Hatzegopteryx thambema* Buffetaut, Grigorescu and Csiki 2002, may have been even larger than *Arambourgiana*, but no cervical material is known. Since its skull was much more robust that those of other azhdarchids (Buffetaut et al. 2002a: p. 183), it was probably carried on a proportionally shorter and stronger neck.

Plesiosaurs

As marine reptiles, plesiosaurs benefited from the support of water. The long necks of elasmosaurid plesiosaurs were constructed very differently from those of sauropods, consisting of many very short cervicals – as many as 71 in the neck of *Elasmosaurus platyurus* Cope 1868 (Sachs 2005: p. 92). Despite their very numerous cervicals, even elasmosaurids did not attain neck lengths even half those of the longest-necked sauropods. The cervical vertebrae of *Elasmosaurus platyurus* holotype ANSP 10081 sum to 610.5 cm, based on individual cervical lengths listed by Sachs (2005: p. 95). For other plesiosaurs, Evans (1993) estimated that the thickness of intercervical cartilage amounted to 14% of centrum length in *Muraenosaurus* Seeley 1874 and 20% in *Cryptoclidus* Seeley 1892. Using the average of 17%, we can estimate the total neck length of *Elasmosaurus* as 7.1 m.

Discounting the aquatic plesiosaurs, neck-length limits in the range of two to three meters seem to apply to every group except sauropods, which exceeded this limit by a factor of five. Whatever mechanical barriers prevented the evolution of truly long necks in other terrestrial vertebrates, sauropods did not just break that barrier – they smashed it. Since four separate sauropod lineages evolved necks three or four times longer than those of any of their rivals, it seems likely that sauropods shared a suite of features that facilitated the evolution of such long necks. What were these features?

### Large Body Size

It is obviously impossible for an animal with a torso the size of a giraffe's to carry a 10 m neck. Sheer size is probably a necessary, but not sufficient, condition for evolving an absolutely long neck. Mere isometric scaling would of course suffice for larger animals to have longer necks, but Parrish (2006: p. 213) found a stronger result: that neck length is positively allometric with respect to body size in sauropods, varying with torso length to the power 1.35. This suggests that the necks of super-giant sauropods may have been even longer than imagined: Carpenter (2006: p. 133) estimated the neck length of the apocryphal giant *Amphicoelias fragillimus* Cope 1878 as 16.75 m, 2.21 times the length of 7.5 m used for *Diplodocus*, but if Parrish's allometric curve pertained then the true value would have been $2.21^{1.35} = 2.92$ times as long as the neck of *Diplodocus*, or 21.9 m; and the longest single vertebra would have been 187 cm long.

The allometric equation of Parrish (2006) is descriptive, but does not in itself suggest a causal link between size and neck length. As noted by Wedel et al. (2000b: p. 377), one possible explanation is that, because of their size, sauropods were under strong selection for larger feeding envelopes, which drove them to evolve longer necks.

### Pneumaticity

The pneumatic spaces in both the bones and soft tissues of sauropod necks greatly decreased their weight: in the extreme case of the prezygapophyseal rami of *Sauroposeidon*, 89% of the bone volume was air (Wedel 2005: table 7.2), and while the impact of soft-tissue diverticula is more difficult to assess, it is easy to imagine that the density of the entire neck may have been less than 0.5 kg/dm$^3$. While pneumaticity was undoubtedly an important adaptation for increasing the length of the neck without greatly increasing its mass, a longer neck remains more mechanically demanding than a shorter neck of the same mass, because that mass acts further from the fulcrum of the cervicodorsal joint, increasing the moment that must be counteracted by the epaxial tension members. Also, longer trachea and blood vessels cause physiological difficulties: weight support is only one of the problems imposed by a long neck.

While pneumaticity may be necessary for the development of a long neck, it is clearly not sufficient: while three groups of theropods, all pneumatic, evolved necks in the 2–2.5 m range, and pneumatic pterosaurs attained 3 m, these remain well short of even the less impressive sauropod necks.

Small Heads

The heads of sauropods were small relative to body mass, and in many clades further lightened by reduced dentition, because, unlike other large-bodied animals such as hadrosaurs, ceratopsians and elephants, they did not orally process their food. Sauropod heads were simple cropping devices with a brain and sense organs, and did not require special equipment for obtaining food, such as the long beaks of azhdarchids. The reduction in head weight would have reduced the required lifting power of the necks that carried them, and therefore the muscle and ligament mass could be reduced, allowing the necks to be longer than would have been possible with heavier heads. Other groups of large animals have not evolved long necks, instead either developing large heads on short necks (ceratopsians, proboscideans) or a compromise of a medium-sized head on a medium-length neck (hadrosaurs, indricotheres)

However, the three theropod clades mentioned above (ornithomimosaurs, therizinosaurs and to a lesser extent oviraptorosaurs) also appear to have had small heads, proportionally similar in size to those of sauropods. Why did they not evolve necks as long as those of sauropods? Possible reasons include the following:

- All theropods were bipedal, and the demands of bipedal locomotion may have prevented them from evolving the giant body sizes that are correlated with very long necks.
- The long-necked theropods may not have been under the same selection pressure to evolve long necks as were sauropods. If they were omnivorous, for example, then their use of more nutritious food may have mitigated the need for increased feeding envelopes. Among extant theropods, the ostrich is very long-necked but feeds mostly from the ground (Dzemski and Christian 2007), and so has no selective pressure to evolve a longer neck than it has.

- All of the largest long-necked theropods lived in the Late Cretaceous, two of them in the Campanian–Maastrichtian. Had they not died out at the end of the Cretaceous, they might have gone on to attain larger size. On the other hand, sauropods attained large size very quickly in evolutionary terms, with a 104 cm humerus from the late Norian or Rhaetian indicating a *Camarasaurus*-sized sauropod only about ten million years after the first known dinosaurs (Buffetaut et al. 2002b): if theropods had the potential to evolve large body size, they had ample time in which to do so.
- Finally, it should be noted that all three of the long-necked theropods discussed above are known from incomplete remains that do not include any informative cervical material. It is possible that neck length was positively allometric in these clades, as in sauropods, and they may have had necks somewhat longer than isometric scaling suggests.

In conclusion, no other clade has all three of the suggested adaptations for long necks that are found in sauropods: birds have pneumaticity and small heads, but are small; proboscideans are large, but lack postcranial pneumaticity and have large heads; azhdarchid pterosaurs were quite large and very pneumatic, but had large heads. Were it not for the end-Cretaceous extinction, non-avian theropods would have been the best candidates for evolving sauropod-like long necks, due to the combination of pneumaticity, small heads in some clades. and potential for large body size.

## Conclusions

The presence of dorsal tubercles together with prominent neural spines in the cervical vertebrae of most sauropods shows that their cervical musculature was intermediate between that of birds and that of crocodilians, although in other respects (pneumaticity, cervical ribs) they more closely resembled birds. Detailed interpretation of sauropod cervical osteology and soft-tissue reconstruction is difficult because of the lack of extant analogues, but both the dorsal tubercles and neural spines would have anchored epaxial tension members (muscles and ligaments),

Bifid neural spines evolved several times among sauropods (and do not ever seem to have been secondarily lost). They may have served to improve the lateral leverage of epaxial tension members, but whatever they were doing was probably not dramatically different from how unsplit spines functioned in their relatives. There is evidence of both pneumatic diverticula and ligament attachment in the spinal cleft – in some cases, both occurring in the same element.

The elongate ossified cervical ribs of most sauropods would have allowed hypaxial muscles to be shifted posteriorly, reducing the lever arm with which their weight drew the neck down. They may also have helped to stabilize the neck, preventing inadvertent lateral and dorsal flexion. They could not, however, have functioned as compression members in ventral bracing, for a variety of anatomical and mechanical reasons.

Several aspects of sauropod cervical anatomy do not seem to make mechanical sense. Posterior elongation of the dorsal tubercles, as seen in the caudal vertebrae of *Deinonychus*, would have enabled the big epaxial muscles to be shifted posteriorly just as cervical ribs do for the smaller hypaxial muscles, yet these do not exist in any sauropod. Tall neural spines allow the epaxial tension members to act with a long lever arm, yet the spines of the longest necked sauropods, including *Mamenchisaurus*, *Erketu* and *Sauroposeidon*, are apomorphically short. The cervical ribs of diplodocoids (which include the longest necked of all sauropods) are apomorphically short; those of *Apatosaurus* are shorter still, absurdly robust, and positioned very low beneath the centrum.

There is tremendous morphological disparity between the cervical vertebrae of different sauropods, as shown in Fig. 2. The four lineages that evolved ten-meter necks (mamenchisaurids, diplodocids, brachiosaurids, and titanosaurs) all have distinctive cervicals: although they converge in some obvious ways (e.g., high elongation index in the longest-necked taxa), their cervicals remain characteristic and easy to tell apart.

Despite this disparity, all these lineages of sauropods attained necks many times longer than those of all other terrestrial animals, and longer even than those of aquatic animals. Six groups of terrestrial animals (giraffes, indricotheres, therizinosaurs, ornithomimids, oviraptorosaurs and azhdarchid pterosaurs) all attained necks in the 2–3 m range, but none exceeded this.

Saropods probably evolved such long necks due to a combination of three factors:

sheer size, skeletal pneumaticity, and small heads that merely gathered, rather than processing, food.

**Acknowledgments**

We thank D. T. Ksepka (American Museum of Natural History) for providing high-resolution versions of the figures from his description of *Erketu* and L. P. A. M. Claessens (College of the Holy Cross) for providing unpublished images of alligator vertebrae. Discussions with J. R. Hutchinson (Royal Veterinary College) and R. M. Alexander (University of Leeds) improved our understanding of bone stress. D. Naish (University of Portsmouth) allowed us to quote a personal communication and furnished photographs supporting it. D. W. E. Hone investigated the status of the *Omeisaurus junghsiensis* material and allowed us to note his conclusion. We used translations of several papers from the Polyglot Paleontologist web-site (http://www.paleoglot.org/index.cfm).

**Literature Cited**

Alexander, R. M. 1985. Mechanics of posture and gait of some large dinosaurs. Zoological Journal of the Linnean Society 83:1–25.

Alexander, R. M. 1989. Dynamics of Dinosaurs and Other Extinct Giants. Columbia University Press, New York.

Arambourg, C. 1959. *Titanopteryx philadelphiae* nov. gen., nov. sp., ptérosaurien géant. Notes et Mémoires sur le Moyen-Orient 7:229–234.

Barbour, E. H. 1890. Scientific News: 5. Notes on the Paleontological Laboratory of the United States Geological Survey under Professor Marsh. The American Naturalist 24:388–400.

Barsbold, R. 1976. New information on *Therizinosaurus* (Therizinosauridae, Theropoda) [in Russian]. Pp. 76–92 *in* N. N. Kramarenko, B. Luvsandansan, Y. I. Voronin, R. Barsbold, A. K. Rozhdestvensky, B. A. Trofimov, and V. Y. Reshetov, eds. Paleontology and Biostratigraphy of Mongolia. Joint Soviet-Mongolian Paleontological Expedition, transactions 3. Nauka Press, Moscow.

Baumel, J. J., A. S. King, J. E. Breazile, H. E. Evans, and J. C. V. Berge. 1993. Handbook of Avian Anatomy: Nomina Anatomica Avium, Second Edition. Nuttall Ornithological Club, Cambridge, Massachusetts.

Biewener, A. A. 1990. Biomechanics of Mammalian Terrestrian Locomotion. Science 250:1097–1103.

Bonaparte, J. F., and J. E. Powell. 1980. A continental assemblage of tetrapods from the Upper Cretaceous beds of El Brete, northwestern Argentina (Sauropoda–Coelurosauria–Carnosauria–Aves). Mémoires de la Société Géologique de France, Nouvelle Série 139:19–28.

Bonde, N., and P. Christiansen. 2003. The detailed anatomy of *Rhamphorynchus*: axial pneumaticity and its implications. Pp. 217–232 *in* E. Buffetaut, and J.-M. Mazin, eds. Evolution and Palaeobiology of Pterosaurs. Geological Society, London.

Borissiak, A. A. 1915. Ob indrikoterii (*Indricotherium* n.g.). Geologiskie Vestnik 1:131–134.

Borsuk-Bialynicka, M. 1977. A new camarasaurid sauropod *Opisthocoelicaudia skarzynskii*, gen. n., sp. n., from the Upper Cretaceous of Mongolia. Palaeontologica Polonica 37:5–64.

Brisson, M. J. 1760. Ornithologie ou méthode contenant la division des oiseaux en ordres, sections, genres, especes & leurs variétés. A laquelle on a joint une description exacte de chaque espece, avec les citations des auteurs qui en ont traité, les noms qu'ils leur ont donnés, ceux que leur ont donnés les différentes nations, & les noms vulgaires. Ouvrage enrichi de figures en taille-douce. Tome V. Bauche, Paris.

Bryant, H. N., and A. P. Russell. 1992. The role of phylogenetic analysis in the inference of unpreserved attributes of extinct taxa. Philosophical Transactions: Biological Sciences 337:405–418.

Buffetaut, E., D., Grigorescu, and Z. Csiki. 2002a. A new giant pterosaur with a robust skull from the latest Cretaceous of Romania. Naturwissenschaften 89:180–184.

Buffetaut, E., V. Suteethorn, J. Le Loeuff, G. Cuny, H. Tong, and S. Khansubha. 2002b. The first giant dinosaurs: a large sauropod from the Late Triassic of Thailand. Comptes Rendus Paleovol 1:103–109.

Cai, Z., and F. Wei. 1994. *Zhejiangopterus linhaiensis* (Pterosauria) from the Upper

Cretaceous of Linhai, Zhejiang, China [in Chinese]. Vertebrata PalAsiatica 32:181–194.

Carpenter, K. 2006. Biggest of the big: a critical re-evaluation of the mega-sauropod *Amphicoelias fragillimus* Cope, 1878. New Mexico Museum of Natural History and Science Bulletin 36:131–137.

Cope, E. D. 1868. Remarks on a new enaliosaurian, *Elasmosaurus platyurus*. Proceedings of the Academy of Natural Sciences of Philadelphia 1868:92–93.

Cope, E. D. 1877. On a gigantic saurian from the Dakota epoch of Colorado. Paleontology Bulletin 25:5–10.

Cope, E. D. 1878. Geology and paleontology: a new species of *Amphicoelias*. The American Naturalist 12:563–566.

Crompton, A. W., and A. J. Charig. 1962. A new ornithischian from the Upper Triassic of South Africa. Nature 196:1074–1077.

Curry Rogers, K., and C. A. Forster. 2001. Last of the dinosaur titans: a new sauropod from Madagascar. Nature 412:530–534.

Daudin,, F. M. 1801. Histoire naturelle, generale et particuliere des reptiles, volume 1. F. Dufart, Paris.

Depéret, C. 1896. Note sur les dinosauriens sauropodes & théropodes du Cretace Superieur de Madagascar. Bulletin de la Société Geologique de France 24:176–194.

Dong, Z. 1979. Cretaceous dinosaurs of Hunan, China. Pp. 342–250 *in* Institute of Vertebrate Paleontology and Paleoanthropology and Nanjing Institute of Paleontology, ed. Mesozoic and Cenozoic Red Beds of South China: Selected Papers from the "Cretaceous–Tertiary Workshop" [in Chinese]. Science Press, Nanxiong.

Dzemski, G., and A. Christian. 2007. Flexibility along the neck of the ostrich (*Struthio camelus*) and consequences for the reconstruction of dinosaurs with extreme neck length. Journal of Morphology 268:701–714.

Evans, M. 1993. An investigation into the neck flexibility of two plesiosauroid plesiosaurs: *Cryptoclidus eurymerus* and *Muraenosaurus leedsii*. unpublished MSc thesis, University College.

Forster-Cooper, C. 1911. *Paraceratherium bugtiense*, a new genus of Rhinocerotidae from the Bugti Hills of Baluchistan, Preliminary notice. Annals and Magazine of Natural History 8:711–716.

Forster-Cooper, C. 1913. Correction of generic name [*Thaumastotherium* to *Baluchitherium*]. Annals and Magazine of Natural History 12:504.

Frey, E., and D. M. Martill. 1996. A reappraisal of *Arambourgiania* (Pterosauria, Pterodactyloidea): one of the world's largest flying animals. Neues Jahrbuch für Geologie und Paläontologie, Abhandlungen 199:221–247.

Frey, E., and J. Martin. 1997. Long necks of sauropods. Pp. 406–409 *in* P. J. Currie, and K. Padian, eds. The Encyclopedia of Dinosaurs. Academic Press, San Diego.

Galton, P. M. 1978. Fabrosauridae, the basal family of ornithischian dinosaurs (Reptilia: Ornithischia). Paläontologische Zeitschrift 52:138–159.

Gilmore, C. W. 1936. Osteology of *Apatosaurus*, with special reference to specimens in the Carnegie Museum. Memoirs of the Carnegie Museum 11:175–298.

Gomani, E. M. 2005. Sauropod dinosaurs from the Early Cretaceous of Malawi, Africa. Palaeontologia Electronica 8:27A: 1–37.

Gunga, H.-C., K. A. Kirsch, F. Baartz, L. Rocker, W.-D. Heinrich, W. Lisowski, A. Wiedemann, and J. Albertz. 1995. New data on the dimensions of *Brachiosaurus brancai* and their physiological implications. Naturwissenschaften 82:190–192.

Gunga, H.-C., T. Suthau, A. Bellmann, S. Stoinski, A. Friedrich, T. Trippel, K. Kirsch, and O. Hellwich. 2008. A new body mass estimation of *Brachiosaurus brancai* Janensch, 1914 mounted and exhibited at the Museum of Natural History (Berlin, Germany). Fossil Record 11:28–33.

Hatcher, J. B. 1903. A new name for the dinosaur *Haplocanthus* Hatcher. Proceedings of the Biological Society of Washington 16:100.

Hildebrand, M. 1988. Analysis of vertebrate structure, 3rd ed. Wiley, New York.

Howse, S. C. B. 1986. On the cervical vertebrae of the Pterodactyloidea (Reptilia: Archosauria). Zoological Journal of the Linnean Society 88:307–328.

Hurlburt, G. R. 1999. Comparison of body mass estimation techniques, using Recent reptiles and the pelycosaur *Edaphosaurus boanerges*. Journal of Vertebrate Paleontology 19:338–350.

Jacobs, L. L., D. A. Winkler, W. R. Downs, and E. M. Gomani. 1993. New material of an Early Cretaceous titanosaurid sauropod dinosaur from Malawi. Palaeontology 36:523–534.

Jain, S. L., and S. Bandyopadhyay. 1997. New titanosaurid (Dinosauria: Sauropoda)

from the Late Cretaceous of central India. Journal of Vertebrate Paleontology 17:114–136.

Janensch, W. 1914. Übersicht über der Wirbeltierfauna der Tendaguru-Schichten nebst einer kurzen Charakterisierung der neu aufgefuhrten Arten von Sauropoden. Archiv fur Biontologie 3:81–110.

Janensch, W. 1929. Die Wirbelsaule der Gattung *Dicraeosaurus*. Palaeontographica (Suppl. 7) 2:35–133.

Janensch, W. 1950. Die Wirbelsaule von *Brachiosaurus brancai*. Palaeontographica (Suppl. 7) 3:27–93.

Jayes, A. S., and R. M. Alexander. 1978. Mechanics of locomotion of dogs (*Canis familiaris*) and sheep (*Ovis aries*). Journal of Zoology, London 185:289–308.

Jensen, J. A. 1985. Three new sauropod dinosaurs from the Upper Jurassic of Colorado. Great Basin Naturalist 45:697–709.

Jerison, H. J. 1973. Evolution of the brain and intelligence. Academic Press, New York.

Kobayashi, Y., and R. Barsbold. 2006. Ornithomimids from the Nemegt Formation of Mongolia. Journal of the Paleontological Society of Korea 22:195–207.

Ksepka, D. T., and M. A. Norell. 2006. *Erketu ellisoni*, a long-necked sauropod from Bor Guve (Dornogov Aimag, Mongolia). American Museum Novitates 3508:1–16.

Latham, J. 1790. Index ornithologicus, sive systema ornithologiae; complectens avium divisionem in classes, ordines, genera, species, ipsarumque varietates, adjectis synonymis, locis, descriptionibus, &c. Leigh and Sotheby, London.

Linnaeus, C. 1758. Systema naturae per regnum tria naturae secundum classes, ordines, genera, species, cum characteribus, differentiis, synonimis, loci, Edition 10, Volume 1. Salvius, Stockholm.

Lovelace, D. M., S. A. Hartman, and W. R. Wahl. 2008. Morphology of a specimen of *Supersaurus* (Dinosauria, Sauropoda) from the Morrison Formation of Wyoming, and a re-evaluation of diplodocid phylogeny. Arquivos do Museu Nacional, Rio de Janeiro 65:527–544.

Maleev, E. A. 1954. New turtle-like reptile in Mongolia [in Russian]. Priroda 3:106–108.

Marsh, O. C. 1877. Notice of new dinosaurian reptiles from the Jurassic formation. American Journal of Science, Series 3, 14:514–516.

Marsh, O. C. 1878. Principal characters of American Jurassic dinosaurs, Part I. American Journal of Science, Series 3, 16:411–416.

Marsh, O. C. 1879a. Notice of new Jurassic reptiles. American Journal of Science, Series 3, 18:501–505.

Marsh, O. C. 1879b. Principal characters of American Jurassic dinosaurs, Part II. American Journal of Science, Series 3, 17:86–92.

Marsh, O. C. 1890. Description of new dinosaurian reptiles. American Journal of Science, Series 3, 39:81–86.

Martill, D. M., E. Frey, R. M. Sadaqah, and H. N. Khoury. 1998. Discovery of the holotype of the giant pterosaur *Titanopteryx philadephiae* Arambourg, 1959 and the status of *Arambourgiania* and *Quetzalcoatlus*. Neues Jahrbuch für Geologie und Paläontologie, Abhandlungen 207:57–76.

Martin, J., V. Martin-Rolland, and E. Frey. 1998. Not cranes or masts, but beams: the biomechanics of sauropod necks. Oryctos 1:113–120.

McGowan, C. 1999. A Practical Guide to Vertebrate Mechanics. Cambridge University Press, Cambridge, UK.

McIntosh, J. S. 1995. Remarks on the North American sauropod *Apatosaurus* Marsh. Pp. 119–123 *in* A. Sun, and Y. Wang, eds. Sixth symposium on Mesozoic terrestrial ecosystems and biota, Beijing, China. China Ocean Press, Beijing, China.

McIntosh, J. S. 2005. The genus *Barosaurus* Marsh (Sauropoda, Diplodocidae). Pp. 38–77 *in* V. Tidwell, and K. Carpenter, eds. Thunder Lizards: the Sauropodomorph Dinosaurs. Indiana University Press, Bloomington, Indiana.

McIntosh, J. S., C. A. Miles, K. C. Cloward, and J. R. Parker. 1996. A new nearly complete skeleton of *Camarasaurus*. Bulletin of Gunma Museum of Natural History 1:1–87.

Meyer, H. v. 1846. *Pterodactylus* (*Rhamphorhynchus*) *gemmingi* aus dem Kalkschiefer von Solenhofen. Palaeontographica 1:1–20.

Murray, P. F., and P. Vickers-Rich. 2004. Magnificent mihirungs. Indiana University Press, Bloomington, Indiana.

Novas, F. E., L. Salgado, J. Calvo, and F. Agnolin. 2005. Giant titanosaur (Dinosauria, Sauropoda) from the Late Cretaceous of Patagonia. Revista del Museo Argentino dei Ciencias Naturales, Nuevo Serie 7:37–41.

O'Connor, P. M. 2007. The postcranial axial skeleton of *Majungasaurus crenatissimus* (Theropoda: Abelisauridae) from the Late Cretaceous of Madagascar. Pp. 127–162 *in* S. D. Sampson, and D. W. Krause, eds. *Majungasaurus crenatissimus* (Theropoda: Abelisauridae) from the Late Cretaceous of Madagascar (Society of Vertebrate Paleontology Memoir 8). Society of Vertebrate Paleontology, Northbrook, Illinois.

Osborn, H. F. 1898. Additional characters of the great herbivorous dinosaur *Camarasaurus*. Bulletin of the American Museum of Natural History 10:219–233.

Osborn, H. F. 1916. Skeletal adaptations of *Ornitholestes*, *Struthiomimus*, *Tyrannosaurus*. Bulletin of the American Museum of Natural History 35:733–771.

Osborn, H. F. 1923. *Baluchitherium grangeri*, a giant hornless rhinoceros from Mongolia. American Museum Novitates 78:1–15.

Osborn, H. F., and C. C. Mook. 1921. *Camarasaurus*, *Amphicoelias* and other sauropods of Cope. Memoirs of the American Museum of Natural History, new series 3:247–387.

Osmólska, H., and E. Roniewicz. 1969. Deinocheiridae, a new family of theropod dinosaurs. Palaeontologia Polonica 21:5–19.

Ostrom, J. H. 1969a. A new theropod dinosaur from the Lower Cretaceous of Montana. Postilla 128:1–17.

Ostrom, J. H. 1969b. Osteology of *Deinonychus antirrhopus*, an unusual theropod from the Lower Cretaceous of Montana. Bulletin of the Peabody Museum of Natural History 30:1–165.

Ostrom, J. H., and J. S. McIntosh. 1966. Marsh's Dinosaurs: the collections from Como Bluff. Yale University Press, New Haven, Connecticut.

Owen, R. 1841. A description of a portion of the skeleton of the *Cetiosaurus*, a gigantic extinct Saurian Reptile occurring in the Oolitic formations of different portions of England. Proceedings of the Geological Society of London 3:457–462.

Parrish, J. M. 2006. The origins of high browsing and the effects of phylogeny and scaling on neck length in sauropodomorphs. Pp. 201–224 *in* M. T. Carrano, T. J. Gaudin, R. W. Blob, and J. R. Wible, eds. Amniote Paleobiology. University of Chicago Press, Chicago.

Paul, G. S. 1988. The brachiosaur giants of the Morrison and Tendaguru with a description of a new subgenus, *Giraffatitan*, and a comparison of the world's largest dinosaurs. Hunteria 2:1–14.

Paul, G. S. 1997. Dinosaur models: the good, the bad, and using them to estimate the mass of dinosaurs. Pp. 129–154 *in* D. L. Wohlberg, E. Stump, and G. D. Rosenberg, eds. Dinofest International: proceedings of a symposium held at Arizona State University. Academy of Natural Sciences, Philadelphia.

Powell, J. E. 1992. Osteología de *Saltasaurus loricatus* (Sauropoda–Titanosauridae) del Cretácico Superior del Noroeste Argentino. Pp. 165–230 *in* J. L. Sanz, and A. D. Buscalioni, eds. Los Dinosaurios y su Entorno Biotico. Actas del Segundo Curso de Paleontologia en Cuenca. Instituto Juan de Valdés, Ayuntamiento de Cuenca.

Reilly, D. T., and A. H. Burstein. 1975. The elastic and ultimate properties of compact bone tissue. Journal of Biomechanics 8:393–405.

Riggs, E. S. 1903a. *Brachiosaurus altithorax*, the largest known dinosaur. American Journal of Science 15:299–306.

Riggs, E. S. 1903b. Structure and relationships of opisthocoelian dinosaurs. Part I, *Apatosaurus* Marsh. Field Columbian Museum, Geological Series 2:165–196.

Romer, A. S. 1956. Osteology of the Reptiles. University of Chicago Press, Chicago.

Russell, D. A., and Z. Zheng. 1993. A large mamenchisaurid from the Junggar Basin, Xinjiang, People's Republic of China. Canadian Journal of Earth Sciences 30:2082–2095.

Sachs, S. 2005. Redescription of *Elasmosaurus platyurus* Cope 1868 (Plesiosauria: Elasmosauridae) from the Upper Cretaceous (Lower Campanian) of Kansas, U.S.A. Paludicola 5:92–106.

Santa Luca, A. P. 1980. The postcranial skeleton of *Heterodontosaurus tucki* (Reptilia, Ornithischia) from the Stormberg of South Africa. Annals of the South African Museum 79:159–211.

Schwarz, D., E. Frey, and C. A. Meyer. 2007. Pneumaticity and soft-tissue reconstructions in the neck of diplodocid and dicraeosaurid sauropods. Acta Palaeontologica Polonica 52:167–188.

Seeley, H. G. 1874. On *Muraenosaurus leedsii*, a plesosaurian from the Oxford Clay. Quarterly Journal of the Geological Society, London 30:197–208.

Seeley, H. G. 1892. The nature of the shoulder girdle and clavicular arch in the Sauropterygia. Proceedings of the Royal Society, London 51:119–151.

Sereno, P. C. 1991. *Lesothosaurus*, 'Fabrosaurids,' and the early evolution of Ornithischia. Society of Vertebrate Paleontology 11:168–197.

Stevens, K. A., and J. M. Parrish. 1999. Neck posture and feeding habits of two Jurassic sauropod dinosaurs. Science 284:798–800.

Taylor, D. 1998. Fatigue of bone and bones: an analysis based on stressed volume. Journal of Orthopaedic Research 16:163–169.

Taylor, M. P., and D. Naish. 2007. An unusual new neosauropod dinosaur from the Lower Cretaceous Hastings Beds Group of East Sussex, England. Palaeontology 50:1547–1564.

Toon, A., and S. B. Toon. 2003. Okapis and giraffes. Pp. 399–409 *in* M. Hutchins, D. Kleiman, V. Geist, and M. McDade, eds. Grzimek's Animal Life Encyclopedia, 2nd ed., vol 15: Mammals IV. Gale Group, Farmington Hills, Michigan.

Tsuihiji, T. 2004. The ligament system in the neck of *Rhea americana* and its implication for the bifurcated neural spines of sauropod dinosaurs. Journal of Vertebrate Paleontology 24:165–172.

Upchurch, P., Y. Tomida, and P. M. Barrett. 2004. A new specimen of *Apatosaurus ajax* (Sauropoda: Diplodocidae) from the Morrison Formation (Upper Jurassic) of Wyoming, USA. National Science Museum Monographs 26:1–110.

Wedel, M. J. 2003. Vertebral pneumaticity, air sacs, and the physiology of sauropod dinosaurs. Paleobiology 29:243–255.

Wedel, M. J. 2005. Postcranial skeletal pneumaticity in sauropods and its implications for mass estimates. Pp. 201–228 *in* J. A. Wilson, and K. Curry-Rogers, eds. The Sauropods: Evolution and Paleobiology. University of California Press, Berkeley.

Wedel, M. J. 2006a. Pneumaticity, neck length, and body size in sauropods. Journal of Vertebrate Paleontology 26:3–137A.

Wedel, M. J. 2006b. Origin of postcranial skeletal pneumaticity in dinosaurs. Integrative Zoology 2:80–85.

Wedel, M. J. 2007. What pneumaticity tells us about 'prosauropods', and vice versa. Pp. 207–222 *in* P. M. Barrett, and D. J. Batten, eds. Special Papers in Palaeontology 77: Evolution and Palaeobiology of Early Sauropodomorph Dinosaurs. The Palaeontological Association, U.K.

Wedel, M. J., and R. K. Sanders. 2002. Osteological correlates of cervical musculature in Aves and Sauropoda (Dinosauria: Saurischia), with comments on the cervical ribs of *Apatosaurus*. PaleoBios 22(3):1–6.

Wedel, M. J., R. L. Cifelli, and R. K. Sanders. 2000a. *Sauroposeidon proteles*, a new sauropod from the Early Cretaceous of Oklahoma. Journal of Vertebrate Paleontology 20:109–114.

Wedel, M. J., R. L. Cifelli, and R. K. Sanders. 2000b. Osteology, paleobiology, and relationships of the sauropod dinosaur *Sauroposeidon*. Acta Palaeontologica Polonica 45:343–388.

Wilson, J. A., and P. C. Sereno. 1998. Early evolution and higher-level phylogeny of sauropod dinosaurs. Society of Vertebrate Paleontology Memoir 5:1–68.

Wilson, J. A., and P. Upchurch. 2003. A revision of *Titanosaurus* Lydekker (Dinosauria – Sauropoda), the first dinosaur genus with a 'Gondwanan' distribution. Journal of Systematic Palaeontology 1:125–160.

Wiman, C. 1929. Die Kreide-Dinosaurier aus Shantung. Palaeontologia Sinica (Series C) 6:1–67.

Witmer, L. M. 1995. The extant phylogenetic bracket and the importance of reconstructing soft tissues in fossils. Pp. 19–33 *in* J. J. Thomason, ed. Functional morphology in vertebrate paleontology. Cambridge University Press, Cambridge, UK.

Witton, M. P., and D. Naish. 2008. A reappraisal of azhdarchid pterosaur functional morphology and paleoecology. PLoS ONE 3:e2271 (16 pages).

Xu, X., Q. Tan, J. Wang, X. Zhao, and L. Tan. 2007. A gigantic bird-like dinosaur from the Late Cretaceous of China. Nature 447:844–847.

Young, C.-C. 1939. On a new Sauropoda, with notes on other fragmentary reptiles from Szechuan. Bulletin of the Geological Society of China 19:279–315.

Young, C.-C. 1954. On a new sauropod from Yiping, Szechuan, China. Acta Scientia Sinica 3:491–504.

Young, C.-C., and X. Zhao. 1972. *Mamenchisaurus hochuanensis* sp. nov. [in Chinese]. Institute of Vertebrate Paleontology and Paleoanthropology Monograph Series I 8:1–30.

TABLE 1. Neck statistics of some sauropods, chosen because of unusually long, short or complete necks. *Mamenchisaurus sinocanadorum* Russell and Zheng 1993 is known only from skull elefments and anterior cervicals, but its neck is estimated to have been about 12 m long by comparison with *M. hochuanensis* Young and Zhao 1972.

| **Taxon** | **Neck length (m)** | **Cervical count** | **Longest centrum (cm)** | | **Longest cervical rib (cm)** | | **Maximum elongation index** | |
|---|---|---|---|---|---|---|---|---|
| *Mamenchisaurus hochuanensis* | 9.5 | 19 | 73 | C11 | 210 | C14 | 2.9 | C6 |
| *Mamenchisaurus sinocanadorum* | 12 est. | 19? | | | ≥ 410 | | | |
| *Brachytrachelopan mesai* | 1.1 est. | 12? | 10 | | ≤ centrum | | ≤ 1 | |
| *Apatosaurus louisae* | 5.9 | 15 | 55 | C11 | 39 | C11 | 3.7 | C4 |
| *Diplodocus carnegii* | 6.5 | 15 | 64 | C14 | 48 | C11 | 4.9 | C7 |
| *Barosaurus lentus* | 8.5 est. | 16? | 87 | C14 | < centrum | | 5.4 | C8 |
| *Supersaurus vivianae* | 15.0 est. | 15? | ≥ 138 | | ≤ centrum | | | |
| *Brachiosaurus brancai* | 8.5 | 13 | 100 | C10 | 290 | C7 | 5.4 | C5 |
| *Sauroposeidon proteles* | 11.5 est. | 13? | 125 | C8 | 342 | C6 | 6.1 | C6 |
| *Euhelopus zdanskyi* | 4.0 | 17 | 28 | C11 | 72 | C14 | 4.0 | C4 |

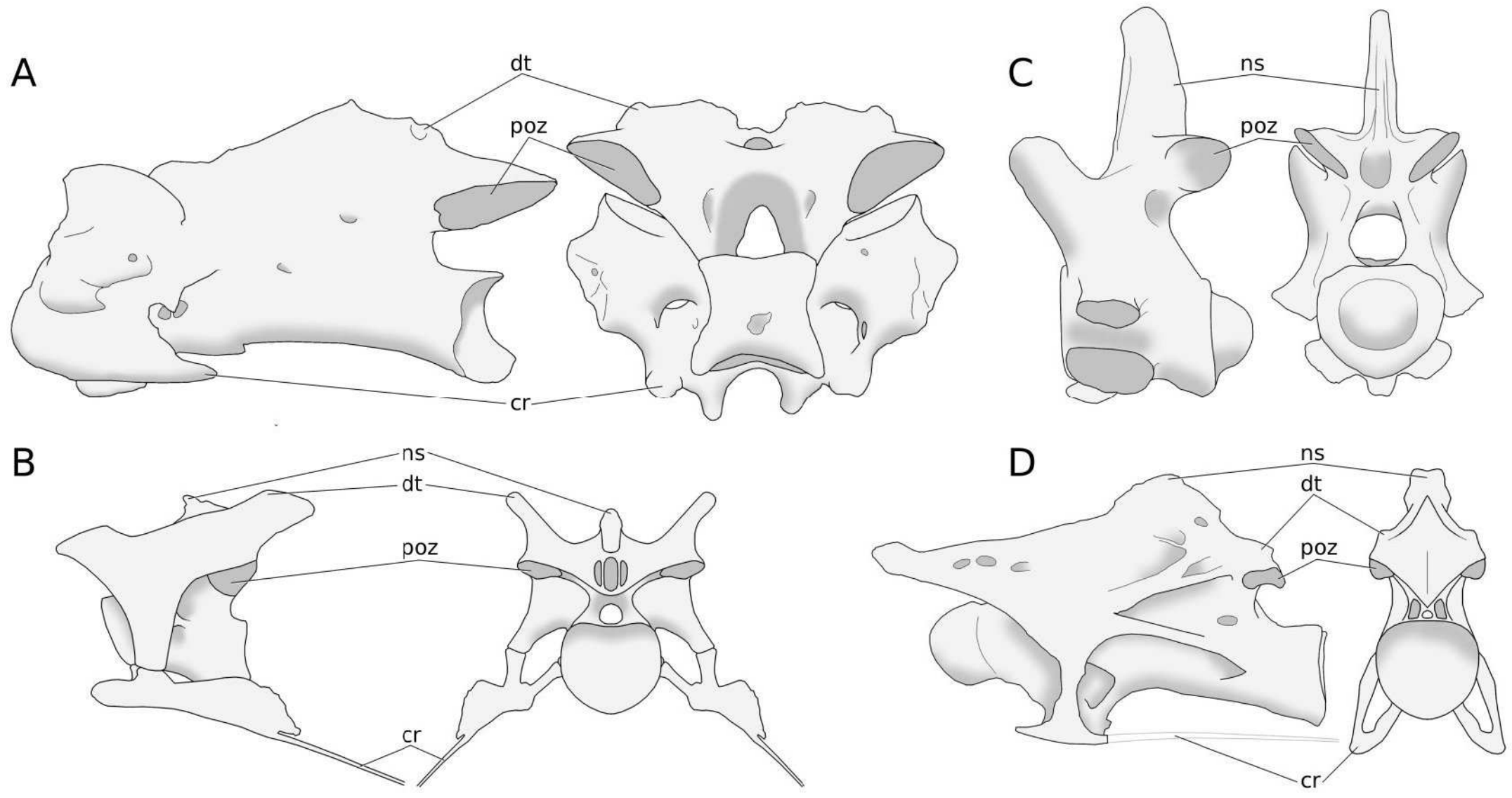

FIGURE 1. Basic cervical vertebral architecture in archosaurs, in posterior and lateral views. A, Seventh cervical vertebra of a turkey, *Meleagris gallopavo* Linnaeus 1758, traced from photographs by MPT. B, Fifth cervical vertebra of the abelisaurid theropod *Majungasaurus crenatissimus* Depéret 1896, UA 8678, traced from O'Connor (2007: Figs. 8 and 20). The dorsal tubercles and cervical ribs are aligned with the expected vectors of muscular forces. The dorsal tubercles are both larger and taller than the neural spine, as expected based on their mechanical importance. The posterior surface of the neurapophysis is covered by a large rugosity, which is interpreted as an interspinous ligament scar like that of birds (O'Connor 2007). Because this scar covers the entire posterior surface of the neurapophysis, it leaves little room for muscle attachments to the spine. C, Fifth cervical vertebra of *Alligator mississippiensis* Daudin, 1801, MCZ 81457, traced from 3D scans by Leon Claessens, courtesy of MCZ. Dorsal tubercles are absent. D, Eighth cervical vertebra of *Brachiosaurus brancai* lectotype HMN SII, traced from Janensch (1950: Fig. 43 and 46). Abbreviations: cr, cervical rib; dt, dorsal tubercle; ns, neural spine; poz, postzygapophysis.

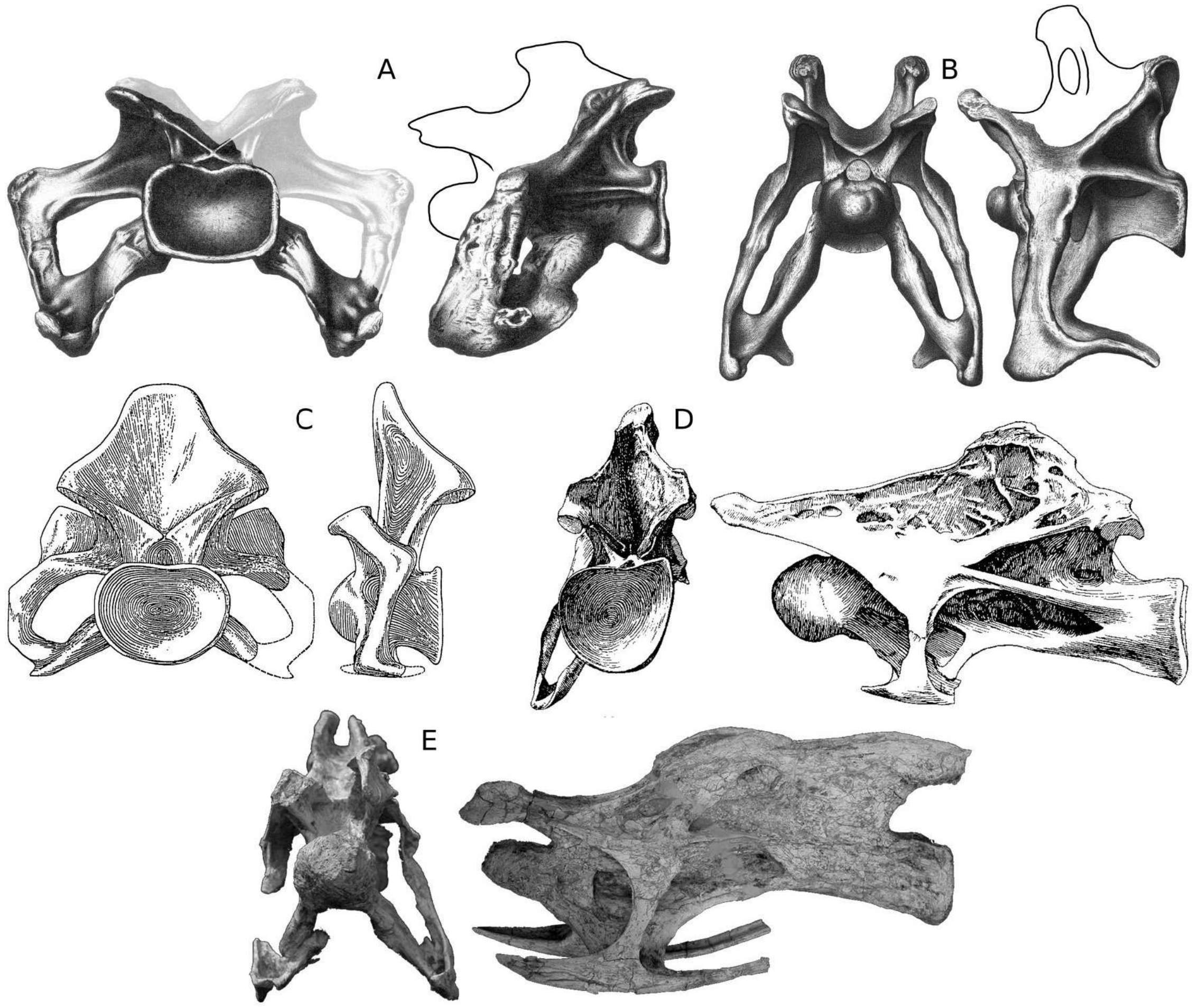

FIGURE 2. Disparity of sauropod cervical vertebrae. A, *Apatosaurus "laticollis"* Marsh 1879b holotype YPM 1861, cervical ?13, now referred to *Apatosaurus ajax* (see McIntosh 1995), in posterior and left lateral views, after Ostrom and McIntosh (1966: Plate 15); the portion reconstructed in plaster (Barbour 1890: Fig. 1) is grayed out in posterior view; lateral view reconstructed after *Apatosaurus louisae* Gilmore 1936 (Gilmore 1936: Plate XXIV). B, "*Brontosaurus*" *excelsus* Marsh 1879a holotype YPM 1980, cervical 8, now referred to *Apatosaurus excelsus* (see Riggs 1903b), in anterior and left lateral views, after Ostrom and McIntosh (1966: Plate 12); lateral view reconstructed after *Apatosaurus louisae* (Gilmore 1936: Plate XXIV). C, "*Titanosaurus" colberti* Jain and Bandyopadhyay 1997 holotype ISIR 335/2, mid-cervical vertebra, now referred to *Isisaurus* (See Wilson and Upchurch 2003), in posterior and left lateral views, after Jain and Bandyopadhyay (1997: Fig. 4). D, *Brachiosaurus brancai* lectotype HMN SII, cervical 8 in posterior and left lateral views, modified from Janensch (1950: Fig. 43–46). E, *Erketu ellisoni* holotype IGM 100/1803,

cervical 4 in anterior and left lateral views, modified from Ksepka and Norell (2006: Fig. 5a–d).

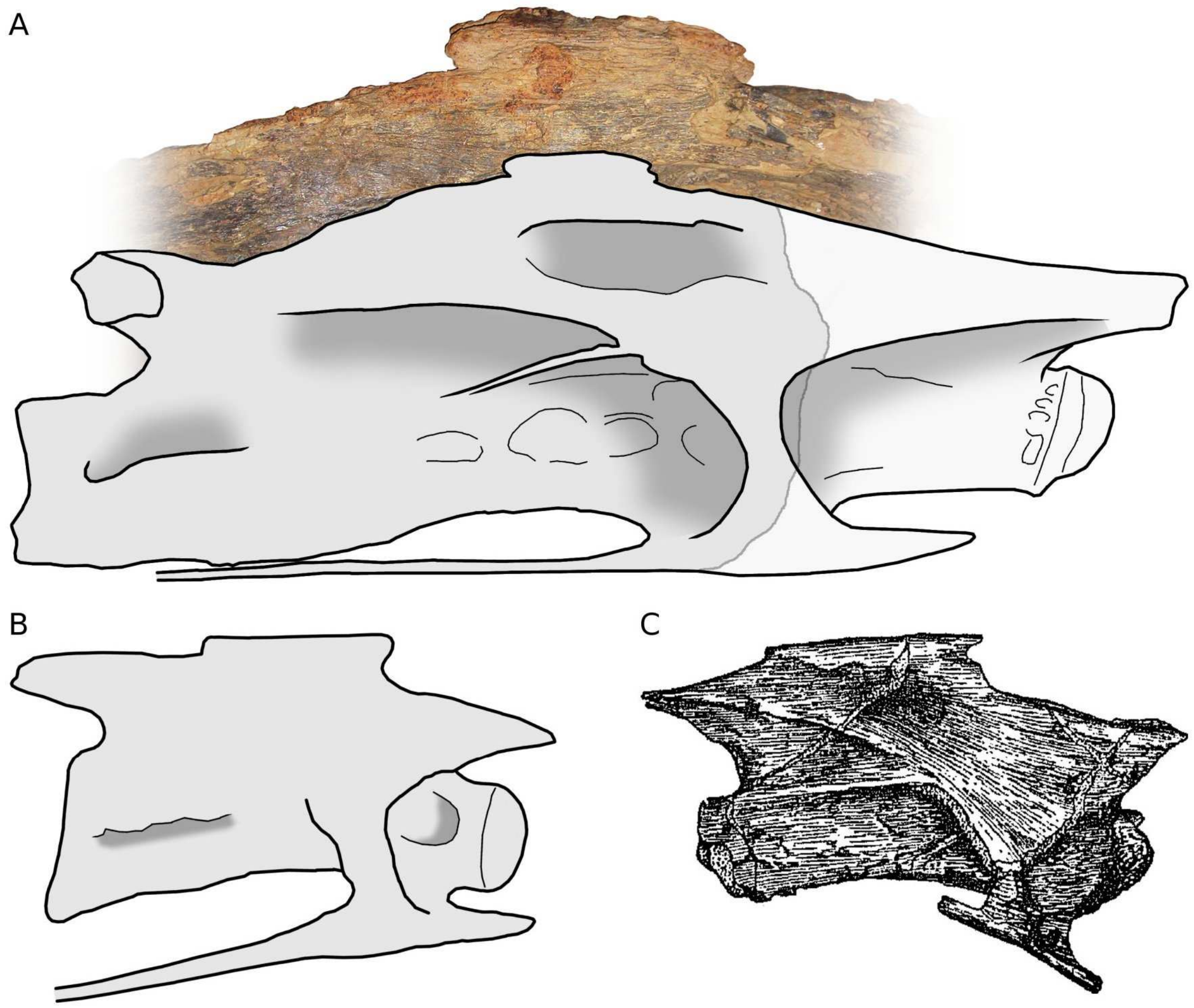

FIGURE 3. Sauropod cervical vertebrae showing anteriorly and posteriorly directed spurs projecting from neurapophyses. A, cervical 5 of *Sauroposeidon* holotype OMNH 53062 in right lateral view, photograph by MJW. B, cervical 9 of *Mamenchisaurus hochuanensis* holotype CCG V 20401 in left lateral view, reversed, from photograph by MPT. C, cervical 7 or 8 of *Omeisaurus junghsiensis* Young 1939 holotype in right lateral view, after Young 1939: Fig. 2. (No specimen number was assigned to this material, which has since been lost. D. W. E. Hone personal communication 2008.)

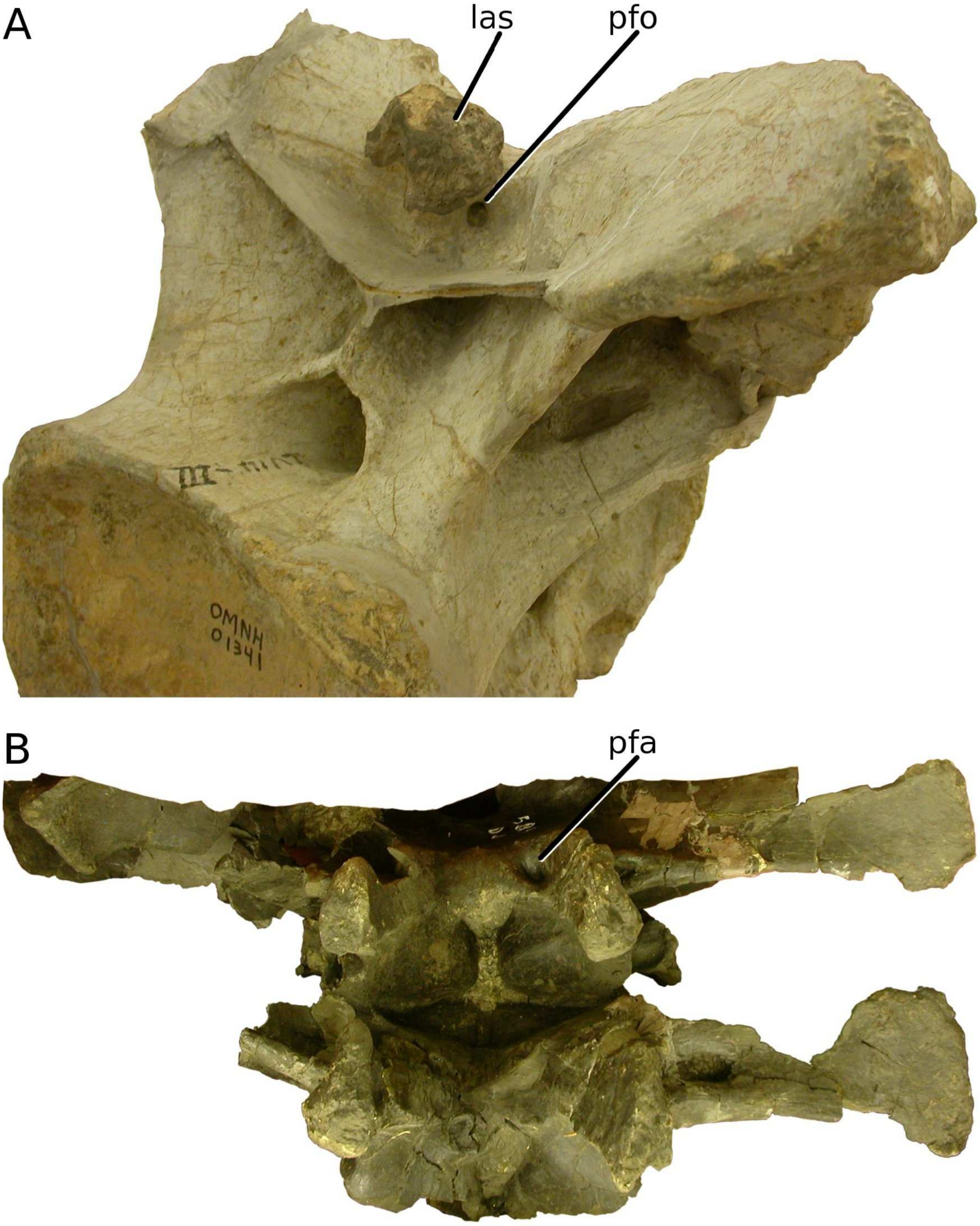

FIGURE 4. Bifid presacral vertebrae of sauropods showing ligament scars and pneumatic foramina in the intermetapophyseal trough. A, *Apatosaurus* sp. cervical vertebra OMNH 01341 in right posterodorsolateral view, photograph by MJW. B, *Camarasaurus* sp. dorsal vertebrae CM 584 in dorsal view, photograph by MJW. Abbreviations: las, ligament attachment site; pfa, pneumatic fossa; pfo, pneumatic foramen.

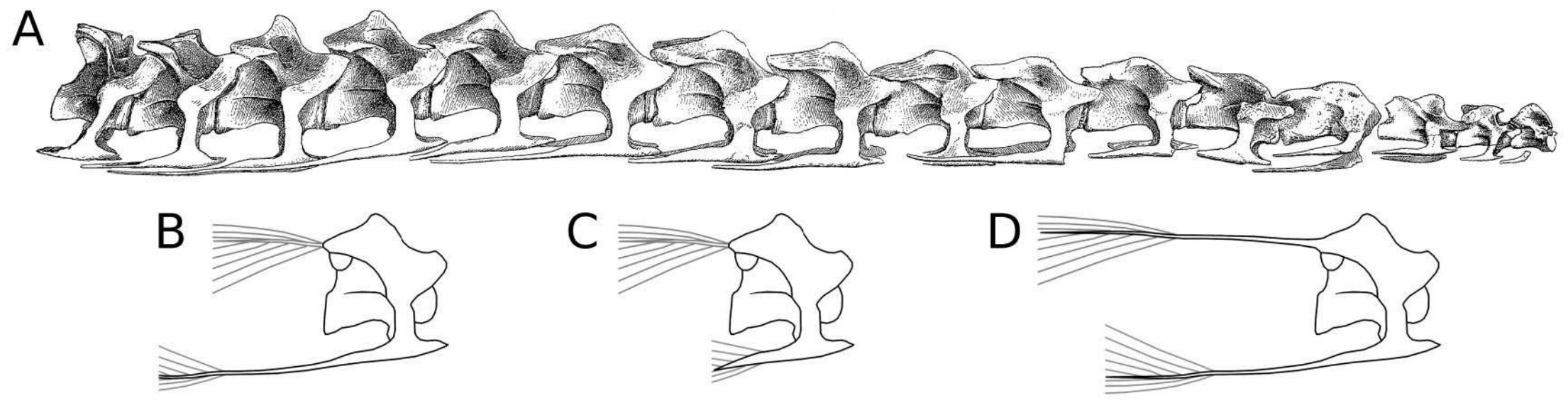

FIGURE 5. Real and speculative muscle attachments in sauropod cervical vertebrae. A, The second through seventeenth cervical vertebrae of *Euhelopus zdanskyi* Wiman 1929 cotype specimen PMU R233a-δ ("Exemplar a"). B, Cervical 14 as it actually exists, with prominent but very short dorsal tubercles and long cervical ribs. C, Cervical 14 as it would appear with short cervical ribs. The long ventral neck muscles would have to attach close to the centrum. D, Speculative version of cervical 14 with the dorsal tubercles extended posteriorly as long bony processes. Such processes would allow the bulk of both the dorsal and ventral neck muscles to be located more posteriorly in the neck, but they are not present in any known sauropod or other non-avian dinosaur. Modified from Wiman (1929: Plate 3).

This page intentionally left blank.

## Future work

Following on from the research presented here, I intend to develop the following projects:

- An updated analysis of dinosaur diversity, tracking the changing sizes of clades through time, using a new form of information-rich diagram generated by my own diversity analysis program.
- The role of articular cartilage as a limiting factor on sauropod gigantism is a previously overlooked area worthy of investigation.
- The results of the analysis of *Brachiosaurus* and *Giraffatitan*, and photographs and notes taken from the Humboldt Museum in November 2008, will enable a more complete description of the Natural History Museum's Tendaguru brachiosaur, which may represent a distinct taxon.
- Together with an excellent collaborator, I intend to work on sauropods of the Wealden. At present the intention is to write an initial paper on the limb elements known from this supergroup, separating the titanosaur "*Pelorosaurus*" *becklesii* from its misassigned genus; and to follow this with a comprehensive analysis of the Wealden's sauropod vertebrae, introducing about thirty new cladistic characters of the dorsal vertebrae to help establish the affinities of the many isolated dorsals.

In addition to these projects, which have already seen the light of day in the form of conference presentations, three further projects are under way: one on caudal pneumaticity in sauropods; one considering the problems of circulation to the heads of erect-necked sauropods; and one evaluating the various factors (limb strength, metabolic cost, feeding rates, reproductive limits, etc.) that can constrain gigantism, and the different ways in which sauropods and elephants deal with them.

# Appendix: specimens inspected

The following specimens were personally inspected in the course of work on this dissertation.

| Museum | Taxon | Specimens |
|---|---|---|
| BMNH | *Ornithopsis hulkei* | R28632 |
| | *Eucamerotus foxi* | R2522 |
| | "*Ornithopsis*"[1] | R88/89 |
| | *Xenoposeidon proneneukos* | R2095 |
| | "*Cetiosaurus*" *brevis*[2] | R2544–R2550 |
| | Sauropoda indet. | R90, R2523, R2239 |
| | *Pelorosaurus conybeari* | R28626 |
| | "*Pelorosaurus*" *becklesii* | R1868 |
| | "*Cetiosaurus*" *humerocristatus* | R44635 |
| | *Dinodocus mackesoni* | R14695 |
| | *Ischyrosaurus manseli* | R41626 |
| | *Oplosaurus armatus* | R964 |
| | *Diplodocus carnegii* (cast) | CM 84/CM 94 |

1 These specimens, a pair of well-preserved dorsal vertebrae, are labelled as "*Ornithopsis*" but were referred to *Eucamerotus* by Blows (1995:190) and indeed listed as paratypes. However, no synapomorphies link these vertebrae to the *Ornithopsis* type specimen, an eroded and crushed centrum, or to the *Eucamerotus* type specimen R2522, a very robust partial neural arch. Since the mislabelled elements are diagnosable, they may subsequently be given their own name.

2 As discussed by Upchurch and Martin (2003:215), although *C. brevis* is technically the type species of *Cetiosaurus*, that genus name has overwhelmingly been used in the literature to refer to the Middle Jurassic basal eusauropod *C. oxoniensis*. Since the Lower Cretaceous titanosauriform *C. brevis* is certainly not congeneric with this species, Upchurch, Martin and I have submitted a formal petition to the ICZN (currently in review) to establish *C. oxoniensis* as the type species of *Cetiosaurus*, and to uphold the widely used junior synonym *Pelorosaurus* for the *C. brevis* material.

| | | |
|---|---|---|
| | *Camarasaurus lentus* (cast) | R12154 |
| FMNH | *Brachiosaurus altithorax* | P25107 |
| | *Giraffa camelopardis angolensis* | 34426 |
| HMN | *Giraffatitan* (= "*Brachiosaurus*") *brancai* | SII, SI, AR1, No8, Aa13, Sa9, T1 |
| | *Dicraeosaurus hansemanni* | m |
| | *Tendaguria tanzaniensis* | NB4, NB5 |
| | *Diplodocus carnegii* (cast) | CM 84/CM 94 |
| LCM | *Cetiosaurus oxoniensis* | G468.1968 |
| MIWG | Brachiosauridae indet. ("Angloposeidon") | 7306 |
| | *?Eucamerotus* | BP001 |
| OMNH | The Hotel Mesa sauropod | 66429, 61248, 27794, 27766, 27761, 66431, 66432 |
| | *Sauroposeidon proteles* | 53062 |
| | *Apatosaurus* sp. | 01670 |
| | The Wolf Creek sauropod | 58304, 58307, 60140, 60233, 60698, 60713, 60722, 61627, 61630 |
| USNM | *Brachiosaurus ?altithorax* | 21903 |
| | *Diplodocus longus* | 10865 |
| Taylor | *Struthio camelus* (ostrich) | N/A |
| | *Meleagris gallopavo* (turkey) | N/A |
| | *Gallus gallus* (chicken) | N/A |

## Museum abbreviations

BMNH – Natural History Museum, London, UK

FMNH – Field Museum of Natural History, Chicago, Illinois, USA

HMN – Humboldt Museum für Naturkunde, Berlin, Germany

MIWG – Dinosaur Isle, Sandown, Isle of Wight, UK

LCM – Leicester City Museum, Leicester, UK

OMNH – Oklahoma Museum of Natural History, Norman, Oklahoma, USA

OUMNH – Oxford University Museum of Natural History, Oxford, UK

USNM – National Museum of Natural History, Smithsonian Institution, Washington D.C., USA

Taylor – my personal collection